\documentstyle[12pt,thesis]{report}
 \catcode`\@=11 
 \@addtoreset{equation}{chapter}
 
\newcommand{\beq}{\begin{equation}}
\newcommand{\be}{\begin{equation}}
\newcommand{\eeq}{\end{equation}}
\newcommand{\ee}{\end{equation}}
\newcommand{\bea}{\begin{eqnarray}}
\newcommand{\eea}{\end{eqnarray}}
\newcommand{\ba}{\beq\begin{array}{lll} }
\newcommand{\ea}{\end{array}\eeq}
\newcommand{\bi}{\bibitem}
\newcommand{\p}{\partial}
\newcommand{\dis}{\displaystyle}
\newtheorem{prop}{Proposition}

\def\itaz{\int_{t_i}^{t_f}}
\def\itif{\int_{\tau_i}^{\tau_f}}
\def\bx{{\bf x}}
\def\bt{{\bf t}}
\def\bp{{\bf p}}
\def\bH{{\bf H}}
\def\bOmega{{\bf \Omega}}

\def\bpi{\mbox{\boldmath $\pi$}}
\def\bchi{\mbox{\boldmath $\chi$}}
\def\bPhi{{\bf \Phi}}
\def\p{\prod_{i=1}^N}
\def\balpha{\mbox{\boldmath $\alpha$}}

\def\bP{{\bf P}}
\def\ov{\over}
\def\r{\rangle}
\def\l{\langle}
\def\para{\parallel}
\def\dag{\dagger}

\def\IR{{\hbox{{\rm I}\kern-.2em\hbox{\rm R}}}}  

\begin{document}
\begin{sloppypar}

%
%
%
%
\def\citen#1{%
\edef\@tempa{\@ignspaftercomma,#1, \@end, }
\edef\@tempa{\expandafter\@ignendcommas\@tempa\@end}%
\if@filesw \immediate \write \@auxout {\string \citation {\@tempa}}\fi
\@tempcntb\m@ne \let\@h@ld\relax \let\@citea\@empty
\@for \@citeb:=\@tempa\do {\@cmpresscites}%
\@h@ld}
%
\def\@ignspaftercomma#1, {\ifx\@end#1\@empty\else
   #1,\expandafter\@ignspaftercomma\fi}
\def\@ignendcommas,#1,\@end{#1}
%
%
\def\@cmpresscites{%
 \expandafter\let \expandafter\@B@citeB \csname b@\@citeb \endcsname
 \ifx\@B@citeB\relax 
    \@h@ld\@citea\@tempcntb\m@ne{\bf ?}%
    \@warning {Citation `\@citeb ' on page \thepage \space undefined}%
 \else
    \@tempcnta\@tempcntb \advance\@tempcnta\@ne
    \setbox\z@\hbox\bgroup 
    \ifnum\z@<0\@B@citeB \relax
       \egroup \@tempcntb\@B@citeB \relax
       \else \egroup \@tempcntb\m@ne \fi
    \ifnum\@tempcnta=\@tempcntb 
       \ifx\@h@ld\relax 
          \edef \@h@ld{\@citea\@B@citeB}%
       \else 
          \edef\@h@ld{\hbox{--}\penalty\@highpenalty \@B@citeB}%
       \fi
    \else   
       \@h@ld \@citea \@B@citeB \let\@h@ld\relax
 \fi\fi%
 \let\@citea\@citepunct
}
%
\def\@citepunct{,\penalty\@highpenalty\hskip.13em plus.1em minus.1em}%
%
%
\def\@citex[#1]#2{\@cite{\citen{#2}}{#1}}%
%
%
\def\@cite#1#2{\leavevmode\unskip
  \ifnum\lastpenalty=\z@ \penalty\@highpenalty \fi 
  \ [{\multiply\@highpenalty 3 #1
      \if@tempswa,\penalty\@highpenalty\ #2\fi 
    }]\spacefactor\@m}
\let\nocitecount\relax  
%



\title{Quantization of Simple Parametrized Systems}

\author{GIULIO RUFFINI}

\beforepreface
\prefacesection{Abstract}I study the  canonical formulation and quantization of some simple parametrized systems  using Dirac's formalism for constrained systems and  the  Becchi-Rouet-Stora-Tyutin (BRST) extended phase space method. These systems include the non-relativistic parametrized particle, the 
relativistic parametrized particle  and minisuperspace.

 Using  Dirac's formalism for constrained systems---including the Dirac bracket---I analyze for each case the  
construction of the classical reduced phase space and  study the dependence on the gauge fixing used.  I show 
that there are two separate features of these systems that may make this construction difficult: 
\begin{quote}
\begin{itemize}
\item[a)] Because of  the  boundary conditions used,
the actions are not gauge invariant at the boundaries. 
\item[b)] The constraints may have a disconnected solution space.
\end{itemize}
\end{quote}
The relativistic particle and minisuperspace have such complicated constraints, 
while the non-relativistic particle displays only the first feature.

After studying the  role of canonical transformations in the reduced phase space,  I  
show that a change of gauge fixing is equivalent to a canonical
transformation in the reduced phase space. This result clarifies the  
problems associated with  the first feature  above, which until now have clouded 
the  understanding of reparametrization invariant theories. 

I then consider the quantization of these systems using several approaches: Dirac's 
method, Dirac-Fock quantization, and  the BRST formalism.  
I pay special attention to  the development of the inner product in the physical 
space. In the cases of the relativistic particle and 
minisuperspace I consider  first the quantization of one branch of the constraint at 
the time and then discuss
the gravitational and electromagenetic backgrounds in  which 
it is  possible to quantize simultaneously both 
branches and
still obtain a unitary quantum theory which respects space-time covariance. I show that
the two branches  represent the particle (universe) going back and forth in time, and
that to preserve unitarity and space-time covariance,  second quantization is in general 
needed. An exception is provided by the flat case with zero electric field.  

 I motivate and define the inner product in all these cases, 
and obtain, for example, the Klein-Gordon inner product for the relativistic 
case.
Then I show how to 
construct  phase space path integral representations for  amplitudes in these 
approaches---the Batalin-Fradkin-Vilkovisky
(BFV) and the Faddeev path integrals---from which one can then derive   the 
path integrals in coordinate space---the Faddeev-Popov path integral
and the geometric path integral. In particular I establish the connection
between the  Hilbert space representation and the range of the lapse in the path 
integrals,
which leads to the Feynman propagator in the BRST-Fock case, for example.
The role of the Faddeev determinant
in the path integrals in providing the interaction between
the branches is established.

I also examine the  class of paths that contribute in the path integrals and how 
they
affect space-time covariance in the presence of an electromagnetic field. I show 
that it is 
consistent
to take paths that move forward in time only when there is no electric field, just 
as
one would expect from studying the conditions for the covariant  factorization of 
the
 Klein-Gordon equation. The key elements in this analysis are the space-like 
paths and the
behavior of the action under the 
non-trivial element of $Z_2$, the disconnected part of the reparametrization group.


\newpage\tableofcontents
\newpage


\prefacesection{Introduction}
The quantization of gravity---the most important parametrized 
system---remains
one of the biggest challenges in physics. In the case of gravity, the 
quantization program  is full of
problems, not the least of which is that the theory is not renormalizable. 
However,
another important feature of gravity is that it is  a constrained
system, and it is   difficult in general to quantize classical systems
with constraints. When the constraints have solution spaces with 
complex
topologies it becomes extremely difficult to produce a unitary 
quantum theory, specially if one is also trying to implement some 
symmetries.
General relativity is such a constrained system, as I will now explain.

The action for gravity is given by
$$
S_H = \int d^4x \, \sqrt{{}^{\scriptscriptstyle(4)}\!g} \, \left(
{}^{\scriptscriptstyle(4)}\!  R-2\Lambda \right)$$ In the ADM (Arnowitt-Deser-Misner)
formalism \cite{ADM,kuchbook,WBook} it is assumed that the topology of space-time is of the form
$\IR\!\times\!\Sigma$, and with the metric expressed by $$
ds^2 = N^2 dt^2 - g_{ij}(dx^i + N^i dt)(dx^j + N^j dt)
$$ 
it becomes, in the canonical formalism ($g$ and $R$ stand for the 3-geometry  metric and curvature in the time
slices, unless specified otherwise by a $(4)$ supersript)
$$
S_H = \int dt\int_\Sigma  d^3x \, ( \pi^{ij} \dot{g}_{ij} - N^i {\cal H}_i - N {\cal
H}) $$
where the constraints are
$$
{\cal H}_i = -2 \nabla_j \pi^j_i , \:\:\:\:\: 
{\cal H} = G_{ijkl}\, \pi^{ij} \pi^{kl} - \sqrt{g}(R-2\Lambda)
$$
with
$$
 G_{ijkl}= {1\over 2\sqrt{g}} (g_{ik} g_{jl}+  g_{il} g_{jk}  -g_{ij} g_{kl}  )
$$
and where $\pi^{ij}(x)$ is the momentum conjugate to $g_{ij}(x)$,
$$\{g_{ij}(x)  ,\pi^{kl}(x')\} = \delta_i^k \,\delta_j^l\, \delta^3(x-x')$$
The linear constraints generate the space diffeomorphisms, while the 
constraint {\em quadratic in the momenta} generates the dynamics---and is the  one that sets gravity apart from
Yang-Mills.  This is a first class system (in the language of Dirac) with  a
complicated algebra \cite{kuchbook}
\begin{eqnarray*}
\{{\cal H}_i(x) ,{\cal H}_j(x') \} &=&  {\cal H}_j \delta^3(x-x')_{,i} - {\cal H}_i \delta^3(x'-x)_{,j}  \\ 
\{{\cal H}_i(x) ,{\cal H} (x') \} &=&  {\cal H} (x ) \delta^3(x-x')_{,i}\\
\{{\cal H}_i(x) ,{\cal H} (x') \} &=&  g^{ij}(x) {\cal H}_j(x)  \delta^3(x-x')_{,i} -
                                       g^{ij}(x') {\cal H}_j(x')  \delta^3(x'-x)_{,i}
 \end{eqnarray*}

Let me compare this system to Yang-Mills
 (see, for example \cite{TeitelBook}, ex. 19.4), another first class
system.  The Yang-Mills action is
given by $$
S_{YM} = -{1\over 4} \int d^4 x \,  F^a_{\mu\nu} F_a^{\mu\nu}$$ with  
$$ F^a_{\mu\nu} = \partial_\mu
A_\nu^a- \partial_\nu A_\mu^a - C_{bc}^{\:\:\: a}   A_\mu^b A_\nu^c
$$ and where $ C_{bc}^{\:\: \:a}$ are the structure constants of a Lie group.
Going to the hamiltonian formalism we find the
 primary constraints $\pi^0_a (x)\approx 0$, and then the 
secondary ones 
$$
{ \Phi}_a= \partial_i \pi^i_a -  C_{\:\: a c }^{b }   A_i^c \pi^i_b \approx 0
$$ again, one per index---including the space index.
They satisfy the first class algebra 
$$
\{ { \Phi}_a (x) , {\Phi}_b(x')\} = C_{\:\:a b}^c { \Phi}_c 
\, \delta^3 (x-x') 
$$
 The hamiltonian is not zero here 
$$
H = \int d^3 x \, {1\over 2} (\stackrel{\rightarrow}{\pi}^2 + \stackrel{\rightarrow}{B}^2 ),\:\:\:\: \:
B^{ai} = {1\over 2} \epsilon^{ijk} \partial_j A_k^a
$$
unlike  the gravitational case.
The extended hamiltonian reads
$$
{\cal H}_E =\int d^3x\, \left( H + \left( A^{0a} +
 \lambda_2^a\right) {\Phi}_a +
\lambda_1 \pi^0_a \right)
$$
 Most importantly, {\em  the constraints are linear in the momenta,
  which,  as we will
see, is related to the fact that 
Yang-Mills is a true gauge theory---fully invariant.}

  Tied to the problems that arise because the constraint
 $\cal H$ is quadratic in the momenta 
are the issues of the definition of the observables and
 inner product in quantum gravity, as well as
the issue of whether it is   necessary to develop the  
  the so-called ``third quantization'' 
programme.
Is there really universe creation and  annihilation?

I  will not investigate the quantization of gravity here, but will look 
instead at some simpler systems that share some of its  problems 
(constraints non-linear in the momenta) while being  manageable---systems with a 
finite
number of degrees of freedom. These include the parametrized 
non-relativistic particle, the relativistic 
particle---in a variety of background fields---and minisuperspace.  
The 
knowledge   gained from the study of these lesser problems should  
help us with the more difficult task.
At any rate, there is no hope to quantize gravity   if the quantization 
of simple 
parametrized systems like the
relativistic particle cannot be accomplished!  And there are some 
problems, 
as we will 
see. 

I   will first  discuss  the quantization of  the parametrized
 non-relativistic 
particle \cite{NR},  and        investigate  the problems associated with the use 
 of a gauge 
dependent 
action.  It is crucial to
 understand this  
simple system before investigating the relativistic case.  It will be very
helpful
 to see clearly what problems are
associated with the  reparametrization invariance alone, as opposed to those 
associated 
with a constraint whose solution space is split in two---as in the 
relativistic case.

The parametrized relativistic
particle will prove to be a surprisingly interesting system from the 
point of view of  constrained systems, and understanding it will be 
helpful 
in a number of  ways. The main reason  for
studying the relativistic particle is that it is a simple parametrized
theory, and it  provides us with what should be  an almost
perfunctory testing ground for some old and some more recent
machinery in constrained system quantization: Dirac's
method  \cite{Dirac,TeitelBook}, BRST theory  \cite{TeitelBook,BFV}  
and 
several path integral methods. On the other hand, the type of 
constraint one 
finds in these systems is neither linear nor topologically 
simple,
so it provides a serious point of departure from   simple constrained 
systems described  essentially by a vanishing momentum, $P\approx 
0$.  Moreover, mathematically 
at least, this
system is very close to simple minisuperspace models, and streching 
things a 
bit, to  full gravity theory itself.

The similarities of this system to minisuperspace and---on a
grander scale---to gravity itself  thus provide the motivation for this 
research, as these toy models may help us in that more difficult 
programme.  
These similarities are the following two: 

First, this system contains an
inherent complication of typical parametrized theories:
all such  systems are 
reparametrization invariant theories, and  they suffer from 
the 
problem that {\em their  actions are  not invariant at the                  
 boundaries} \cite{Teitelpap}---which, as 
we will see, can be understood to be a consequence of  the use of 
boundary 
conditions on the gauge degrees of freedom. In the hamiltonian 
formalism 
this situation is described by  constraints that are not linear in the 
momenta. 

 It is  hard to define an associated reduced
 phase space to a system that is not fully invariant!
This much is just as true of the non-relativistic case, which also 
comes with a non-linear constraint.

   In addition, {\em the constraint surface
 in this case is   disconnected } (unlike  the corresponding  situation in
the non-relativistic case),  and 
this complicates considerably the quantization procedure. The 
disconnectedness
of the constraint surface is related to  particle   creation---as we will 
see---and unitarity will become a relevant issue.
I will  discuss   when 
and why 
one  needs to develop  a field  theory (chapter \ref{sec:second}).

There are two main approaches to quantization of constrained 
systems. In one,  the gauge degrees of freedom are eliminated 
classically, before quantization is considered. The formalism for doing 
this was developed by Dirac \cite{Dirac}. When these degrees of 
freedom have been eliminated the resulting theory can be quantized 
in the usual way. Here the important question is: What is the reduced 
phase space (i.e., what are the physical degrees of freedom)?  The 
main advantadge of this method is that the most difficult part---the 
reduction---is performed at the classical level. The disadvantadge is 
that this classical reduction may force us to brake invariances---or 
perhaps just to  lose sight of them. At any rate, this is the  point of 
view that we will begin with:  for constrained systems  quantization 
means the
quantization of the reduced phase space, i.e., quantization of
some well defined degrees of freedom which are found classically
and then quantized.

In the other approach (also pioneered by Dirac), one first
 quantizes---assigning operators to 
every degree of freedom, both physical and gauge---and then 
reduces to a physical ``subspace'' by demanding that the {\em 
physical states} satisfy some condition. The advantadge  of this 
method is the preservation of invariances, and the disadvantadge is 
that the reduction becomes, mathematically, more difficult. One of 
the biggest problems will be that it is difficult to define an inner 
product in the physical ``subspace''.   The reason is that in general 
the physical ``subspace'' is not a subspace at all.

One may ask if the two approaches lead to the equivalent theories, 
and for basic systems they do---modulo operator 
orderings\footnote{I will ask the reader to always keep in mind that 
this ambiguity is part of the transition from classical to quantum, and 
is not a new feature associated with quantization of constrained 
systems.}, which are always present. This may not be true of the 
systems we will study here. So here is an important question: Do 
reduction and quantization commute? If not, which one should one 
use?

 Let me point out here   the main ingredients in 
standard
quantum mechanics: \begin{quote}
\begin{enumerate}
\item the states (resolution of identity)
\item the inner product
\item a Hamiltonian for unitary {\em time } evolution
\item a probabilistic interpretation 
\end{enumerate}\end{quote}
These are what we are really after, all the technical details
notwithstanding---and there will be many, so it will be good to keep 
this scheme in mind. As a result, some of the  basic issues associated with the 
quantization of {\em constrained} systems that   will need to be addressed   
are the following:
\begin{quote}
\begin{itemize}
\item[A) ] What is the physical Hilbert space?
\item[B) ] What is the physical inner product?  
\item[C) ] What describes ``time'' and ``time evolution''?
\item[D) ] How can we build a path integral from the above?
\item[E) ] Is the  probabilistic interpretation possible?
\end{itemize}
\end{quote}

In the present work I will
consider several different approaches to quantization: reduced phase 
space 
quantization
\cite{TeitelBook,Faddeev}, Dirac's
original method  \cite{TeitelBook},  the  Fock space approach 
\cite{TeitelBook}, and the Becchi-Stora-Rouet-Tyutin (BRST) \cite{TeitelBook,BFV}. 

I will also consider   
several path integral methods:  in phase space we will study the
Faddeev \cite{TeitelBook,Faddeev}  and the Batalin-Fradkin-Vilkovisky (BFV) 
\cite{TeitelBook,BFV}  path 
integrals, and in
configuration space we will look at the Faddeev-Popov and at the
geometric path integrals  \cite{Geo}.  

One basic  goal of this work is
to compare all  these methods. Do they lead to the same quantum 
theories? 

A more ambitious goal is to provide the connection between the 
Hilbert space and operator  formalisms and the path integrals, which 
has been a serious fault in the conceptual basis for these path 
integral approaches. One of the original questions for                                    
my research, which was formulated by Emil Mottola, was indeed the 
following: ``We have a  path integral for quantum gravity: what does 
it mean? How do you compute it?'' Both of these questions---which I 
still haven't answered---will find their resolution within            a 
well-defined canonical formalism for quantum 
gravity from which the path integrals will 
be constructed.

To refresh the reader's memory---and also to point out some 
ambiguities---I will review in the first chapter 
 how
the quantization  picture is constructed in the case of the 
unconstrained non-relativistic particle, including  how the path 
integral in phase space is
built from the Hilbert space and the hamiltonian,  and how the
path integral in configuration space is then obtained from the
phase space path integral by the integration of the momenta.

In general,  the picture we want to see emerge is\footnote{ {\em PS} 
stands for   Phase Space, and {\em CS} for  Configuration Space.} 

{\em Q. Mechanics $\longrightarrow$ Path Integral in PS
$\longrightarrow$ Path Integral in CS} 

 The above issues are related to a series of
technical questions that have appeared repeatedly in the literature, 
among which are the following: 
\begin{quote}
\begin{itemize}
\item[o) ] What form of the constraint  (or gauge fixing) should one 
use? 
Does it matter?
\item[i) ] What are the inner product and resolution of the identity 
in the 
reduced phase space quantization?
\item[ii) ] What are the inner product and resolution of the identity 
in the 
Dirac quatization?
\item[iii) ] What are the inner product and the resolution of the 
identity in 
the BRST state cohomology space?
\item[iv) ] How does on go from these quantum spaces to  their 
respective 
path integrals?
\item[v) ]  What is the range of the ``lapse'' and the interpretation 
of the path integrals?
\item[vi) ] What paths contribute in the path integrals?
\item[vii) ] Do we have a consistent first quantization scheme for 
these 
systems?
\item[viii) ] What is the role of the disconnected part of the
 reparametrization group?
\end{itemize}
\end{quote}

Chapter~\ref{sec:classical} will discuss the classical aspects of   
parametrized
systems: the actions and their invariances---for the free 
and 
interacting cases---the
constraints that arise,  the notions
of gauge-fixing and its relation to the reduced phase space, the 
Dirac bracket, 
as well as the concept of canonical transformations in the reduced 
phase 
space and how they are related to the gauge-fixing. The BRST 
formalism will 
also be introduced in this classical context.
Chapter~\ref{sec:quantum} will take a formal look at the different 
quantization 
schemes I will consider here:
reduced phase space, Dirac, BRST and Fock space quantization. The 
discussion will cease to be formal in chapter~\ref{sec:product}, where
I will discuss in detail the Hilbert spaces involved in the 
quantization
as well as the inner products  defined on them. The path integrals 
will   be derived here, and the Fradkin-Vilkovisky theorem proved.

The path integral formalisms---in phase space and in
configuration space---will be further developed in 
chapter~\ref{sec:PIs}.  
By now it should be clear to the reader that this is the proper place 
for this chapter.  Path integrals are very mysterious objects when 
one doesn't have a state space, an inner product  and operators to 
describe and build them with, and as I already explained  one of the 
goals of my research has been  to find as solid a basis as possible for 
the development and interpretation of the path integrals for 
constrained systems\footnote{As far as I know, one cannot really do 
it  the other way around.}. 

I will also study the connection between the class of paths that appear
in the path integrals and   space-time covariance and unitarity,
and how this, in turn, relates to the behavior of the action under  
the disconnected part of the diffeomorphism group.

Discussion
of equivalence of the different methods will be folded in as we
go along, and also  reviewed in the general discussion at the
conclusion. Second quantization---i.e., the idea of finding and 
quantizing a
 field lagrangian which as the  constraint    as an 
equation 
of motion---will be discussed  in  chapter~\ref{sec:second}.

 \medskip

 [Nov 2005 note: see \cite{four} for a follow up paper on this thesis.]

\afterpreface

\chapter{The unconstrained particle}
\label{sec:unconstrained}
As a reference point I include first the derivation of the
standard phase-space path integral for the unconstrained 
non-relativistic  
particle, starting from the quantum
mechanical expression for the propagator. This example is very
important, as it illustrates what one would like to have in the
other cases, i.e., a well defined quantum mechanical framework
from which to construct the path integral expression for the
propagator, first in phase space and then in configuration space, 
according 
to 
the picture mentioned in the introduction:

{\em Q. Mechanics $\longrightarrow$ Path Integral in PS 
$\longrightarrow$
 Path Integral in CS}
  
This is the model we will try to emulate in the discussion of the
physical spaces in the parametrized systems. 

I will also comment on the issues of unitarity and on the freedom 
in the 
definition of the inner product.

\newpage\section{Quantum mechanics of the unconstrained particle}

Classically, we have that the standard action for the non-relativistic 
particle 
is
\beq
\label{eq:unconstrained}
I =  \itaz  L \, dt =  \itaz  dt\, \left( { m\over 2 }\left({dx \over 
dt}\right)^2 - V(x) 
\right) 
\eeq
and the equation of motion that results from extremizing it is
\beq
m {d^2x \over dt^2} =- {\partial V \over  \partial x}
\eeq
One can also dicuss the dynamics in  the canonical formalism by defining the
momentum
 \beq
p =  {\partial L \over \partial (dx /dt) }= m {dx \over dt}
\eeq
and the hamiltonian $ H = p{ dx \over dt }- L = {p^2 \over 2m }+ V $ 
,
with the equations of motion
\bea
{dx \over dt } & =  \{ x,H \} =  & {p \over m } \\[.5cm]
{ dp \over dt }& =  \{p,H \} = & -{\partial V \over  \partial x }
\eea
which are equivalent to the one in the lagrangian formalism.

The quantum mechanical framework is defined here by a Hilbert
space of states  ${  |\Psi\r}$. The basic operators are $\bx$ and
$\bp$, with the commutator Heisenberg algebra $ [\bx ,\bp]=i$. A 
basis
for the Hilbert space is provided by the eigenstates of either
one of these hermitean operators, $\bx   | x \r     = x   |   x \r    
$,
or $ \bp   |  p\r      = p   |  p \r $.
Completeness of these states is expressed by the following
resolutions of the identity:
\beq
I  =  \int  dx \,   |  x \r \l    x  |   =  \int  dp   \,  |  p\r \l p |    
\eeq
where the integrations are taken over their full ranges.
The following projection can be inferred from the above algebra
\beq   
\l x    | p \r = \sqrt{1 \over 2 \pi  } e^{ipx} 
\eeq
Indeed, consider the action of $ \bp$ on the state  $  |   p\r  $    in
the $x $ representation:
\beq  
\l   x    | \bp     | p\r      = p  \l   x    | p \r  = -i{ \partial
\over  \partial x}   \l   x    | p \r 
\eeq
The solution to this equation is the above projection formula.

Notice that here we are using a representation of the algebra
in which
 $$  \l x  |  \bp  | f\r = -i{ \partial  \over  \partial x}
 \l x |    f\r  $$ 
The positive definite inner product\footnote{By this I mean
that this inner product yields positive norms for the states.} is defined by 
\beq \l \Psi | \Sigma\r =
( \Psi(x) , \Sigma(x)  ) =   \int_{-\infty}^{\infty}  dx \; \Psi ^*(x)   \Sigma(x) 
\eeq
where, for example,   $   \Psi (x) =   \l    x    | \Psi\r   $,
 is the state in the coordinate representation. The
probabilistic interpretation is tied to the existence of this
inner product, which has the key properties of being
positive definite---yielding positive norms---and, as
we will see,  time-independent.

Time evolution is described by Schr\"{o}dinger's equation\footnote{I will refer to 
any equation of the form $i \partial_t \psi = \hat{H} \psi$ for some hamiltonian, as 
a ``Schr\"{o}dinger equation''.}
\beq
i  {\partial  \over  \partial t  } | \Psi\r   = \bH   | \Psi \r     
\eeq

In this case we take the hamiltonian to be $H = {p^2 \over 2m} + 
V(x) $.
The solution to this equation is
\beq   
 | \Psi \r  = e^{ -i\bH (t-t_o)} | \Psi_0  \r  
\eeq
for some initial state $  |\Psi_0 \r  $.
The time independence of the inner product is now easy to see.

The propagator in the coordinate representation is defined to be
\beq
U(x_f,t_f;x_i,t_i) \equiv  \l  x_f  |e^{-i\bH (t_f - t_i)  } |  x_i \r   
\eeq 
(=U for short).  The path integral in phase space is now easily built 
by inserting
the above resolutions of the identity in this definition. First
break the time interval into N steps and use the multiplication
property of the exponential N times,
\beq
\l  x_f  | e^{ -i\bH  (t_f - t_i) } |  x_i \r     = \l  x_f  | e ^{-i\bH  
\epsilon }    
 e ^{-i\bH \epsilon}   ... e^{ -i\bH  \epsilon}     |  x_i \r   
\eeq
where  $\epsilon    ={t_f - t_i \over N}$. Then insert the above
resolutions of the identity in this expression,  
\beq U =
\int  \l  x_f  | e ^{-i\bH \epsilon} |  p_0\r  \l  p_0  |  
x_1\r \l  
x_1 
|
 e ^{-i\bH \epsilon} ... | p_N\r \l p_N  |  e ^{-i\bH \epsilon}    | 
x_i\r 
dp_0   \prod_{i=1}^N {dx_idp_i }
\eeq
Taking the limit  $ \epsilon  \longrightarrow 0  $ we have
\bea \dis 
  \l    x  |   e^{ -i\bH \epsilon    }    | p\r   \approx  \l   x  |  (1 
-i\bH
\epsilon  )  |    p  \r   \\[.5cm]  \dis 
 =   \l    x  |  (1 -iH \epsilon      )|
      p\r    \approx \l    x   |  e^{-iH \epsilon }    | p\r \\[.5cm]  \dis
=  e^{ -iH \epsilon} \l   x   | p\r  = \sqrt{1 \over 2 \pi }
 e^{i(px-H \epsilon)} 
\eea
where we used the projection $ \l x   | p\r  = \sqrt{1 \over 2 \pi } 
e^{ipx}$ 
(which follows from the Heisenberg algebra). 
Here $H$ is {\em defined} to be \beq
H \equiv \l x | {\bf H } | p \r = H(x,p) \eeq

 Now we can  write 
\beq
U =  \itaz  e ^{i  \Sigma_j ( p_j  \Delta     x_j-iH_j
\epsilon     )}   {dp_0 \over 2 \pi} \p{ dx_idp_i \over 2 \pi}
\eeq
The resulting expression for the propagator is symbolized 
by\footnote{In 
this 
chapter only we define $\dot{a} \equiv da/dt$, instead of 
$da/d\tau$ 
---which will be the norm thereafter. }
\beq
U =  \int  DxDp \; e ^{i \itaz  pdx - Hdt} =  \int  DxDp e^{ i \itaz
dt(p{\dot{x} }- H)}
\eeq
where the measure here means
\beq
DxDp \equiv { dp_0  \over 2 \pi}  \p {dx_i dp_i \over 2 \pi}
\eeq
Notice that we could have simply written 
\beq
e^{i {\bf H} \Delta t} = \lim_{N\rightarrow \infty} \left( 1+ i{\bf 
H}{\Delta 
t\over N} \right)^N
\eeq
to insert the resolutions of the identity.
  The right hand side is  in fact the propagator for a general, 
time-dependent hamiltonian \cite{Shankar,Baym} (see equation \ref{eq:prop}
below).
{\em Thus, the expression above for the propagator in path integral
form is in fact general.}  The path integral provides us with a solution to the
Schr\"{o}dinger equation that satisfies the boundary condition (initial
condition) of becoming a delta function at the initial time---provided we
normalize the position eigenstates in such a way.

This is then the path integral in phase space. The momentum
integrations are easily done to yield Feynman's original path
integral---which he obtained from a different point of view,
\beq
U =  \int   {\cal D}x \; e^{i \itaz  dt(  {1 \over 2}m\dot{x}^2 -V(x))}
\eeq
The measure here means  
\beq
{\cal D}x = \left( {m\over 2\pi  i dt  }\right)^{1 \over 2}\p 
dx_i \left( {m\over 2\pi  i dt  }\right)^{1 \over 2}
\eeq
 where $dt$ stands for  $\epsilon$      (or  $\epsilon_i$).

For the free particle $(V=0)$ we obtain
\beq
U( \Delta x, \Delta  t) = \left( {m\over 2\pi  i \Delta t  }\right)^
{1 \over 2}e^{ i { (\Delta x )^2 m \over 2    \Delta t} }
\eeq
At this point we should make a note on the class of paths that
contribute in the coordinate space path integrals. Notice that
the paths are described in the form $ x = x(t)$, so by definition of
the path integral they go forward in time---there is no chance to
describe paths going back in time unless we do something strange like changing
the sign of the hamiltonian.

Also notice that one could compute the expectation value of the
``operator'' $ \Theta     (    \Delta     t)$, where $   \Theta (z)$ is the 
Heaviside theta function, and then obtain the causal
Green's function for the Schr\"{o}dinger equation---see 
section~\ref{sec:BFVpi}
for more on this---
\beq
\Gamma'_G =   {\Theta( \Delta t) \over 2 \pi  }
\int dp_{0} \; \exp\{  i(p_{0}  \Delta  x-  {\Delta  t p_{0}^2
\over 2m})\} =  \Theta (  \Delta  t) U
\eeq 
which satisfies
 \beq
(i {\partial  \over \partial t} - {{\bf p}^2  \over 2})  \Gamma'_G =
\delta (  \Delta   t)   \delta   (  \Delta   x)
\eeq

Let us now look at the case of electromagnetic interaction and at the 
issue of unitarity.
Recall that {\em unitarity of the propagator results  when the
hamiltonian is hermitean, whether it is time independent or not}
(see for example Shankar's book  on quantum mechanics 
\cite{Shankar}).
In general the propagator
is given by
\beq
\label{eq:prop}
U (t_f,t_i ) = {\cal  T}[e^{ -i \itaz dt' H(t') }]\equiv  \p e^{ -iH(t_j) 
\Delta  
t_j }
\eeq
and in the hermitean case it is a product of unitary operators
and therefore unitary. It has the properties  \cite{Shankar}
\ba \dis
U (t_3,t_2 ) U (t_2,t_1 ) = U (t_3,t_1 )\\[.5cm]  \dis
U^\dagger (t_2,t_1 ) = U^{-1} (t_2,t_1 ) = U (t_1,t_2 )
\ea
Any action that we can write in the form 
$ \int dt \, L[\dot{x}( t), x(t),t ]$ is 
therefore going to yield a unitary quantum theory
with the usual procedure for constructing a hermitean
hamiltonian, {\em if ordering problems are not encountered}. 

Consider now the case of  the non-relativistic case with an 
electromagnetic 
interaction.
The lagrangian for such a case is given by
\beq
\label{eq:uelctro}
{\cal L}_{EM }= {1\over 2}m v_i^2 - e\phi+ {e \over c} v_i A_i
\eeq
(see for example \cite{Shankar}), and the hamiltonian turns out
to be
\beq
{\cal H}_{EM } = { {(p_i - e A_i /c)}^2 \over 2 m} + e \phi
\eeq
The point to observe is that this hamiltonian will become, with the usual 
quantization recipe,
 a  hermitean 
operator, as can  
readily
be seen by expanding it:
\beq
{\cal H} = {1 \over 2 m}( p_i^2 + {e^2 \over c^2}A_i^2 -
{e \over c} (p_i A_i + A_i p_i)) + e \phi
\eeq
{\em Unitarity is therefore guaranteed with this action}.

This completes the review of the non-relativistic unconstrained
case. Keep in mind that this is the model we will try to
imitate when we study the constrained case, as it is {\em the} 
example
of well-defined quantum mechanics.

Let us now study a model for the relativistic unconstrained
particle. One point of view is to use the hamiltonian
\beq
h = +\sqrt{p^2 + m^2}
\eeq
as we will see. This follows from the action
\beq
s = - m  \itaz  dt\;  \sqrt{1 - \dot{x}^2} 
\label{eq:uroot}
\eeq
(one could also consider $h' = - h$  which follows from
 $s' = - s $). Notice that faster-than-light paths are allowed,
although they will come in with a real exponential weight\footnote{ This is 
somewhat
deceiving, as there is no such path integral in lagrangian form and with
a simple measure \cite{rel.unc.ple}.}.

 The form
\beq
s'' =  \itaz  dt  \left( {1 - \dot{x}^2 \over  \lambda (t) }+\lambda (t) 
m^2 \right)
\eeq
---where $\lambda  (t)$ is also to be varied---is very similar.
The equation of motion for  $\lambda (t)$ is just  
\beq 
\lambda (t) =  \pm {1 \over 2 m}\sqrt{1 - \dot{x}^2}.
 \eeq
Substituting this
in the action $ s''$ yields either $s$ or $-s$. Notice, though,
that in a path integral $\lambda$ will not be imaginary
unless forced by a rotation of the integration
contours.

What kind of quantum mechanics and path integral do we get from
this hamiltonian? The path integral is easily computed,
\bea
U =  \int DxDp\:  e^{ i \itaz  dt(p\dot{x } - h)} & = &
  \int  DxDp\: e^{ i \itaz  dt  (p \dot{x } - \sqrt{p^2 + m^2})} = \\
  {1 \over 2 \pi    }\int  dp\: e^
{ i (p    \Delta  x - \Delta t \sqrt{p^2 + m^2})} & = &
 \l x_f |  e^{-i{\bf h}    \Delta     t }  | x_i\r 
\eea
Faster than light ``propagation'' is indeed possible, as hinted by
the above lagrangian; see reference \cite{rel.unc.ple}  for more on 
this path
integral. 

Other relevant actions are $ (i =1,2,3 ;\;  \mu  = 0,1,2,3; \; x^\mu = t, 
x)$
\beq
E =  \itaz  dt \; \left( - m \sqrt{1- \dot{x}_i^2} + A_0 - A_i 
\dot{x}_i^2 \right)
\eeq
with $h_E = -A_0  + \sqrt{(p_i + A_i )^2 + m^2}$  for the 
electromagnetic 
interacting case ($ A_\mu = A_\mu (x^\alpha)$), and
\beq
G = - m  \itaz  dt \sqrt{c(x^\mu) - b(x^\mu){(\dot{x }+ a(x^\mu))}^2}
\eeq
with $ h_G = \sqrt{c({p^2 / b} + m^2 )}$ , for the gravitational 
background
case.  These are special cases of the general 
action for the relativistic particle in a curved space-time and in a 
background electromagnetic field---with $ g_{i0} = g_{0i} = 0 $,
\beq
A_{EG} =  \itif  d\tau  \left(-m\sqrt{g(x^\alpha ) _{\mu  \nu}   
{dx^\mu 
\over d
\tau}  {dx^\nu \over d\tau}} - e {dx^\mu \over d\tau}A_\mu 
(x^\alpha )  \right)
\label{eq:action1}
\eeq
or of the other form
\beq
A_{EG}' =  \itif  d\tau  \left({1\over \lambda 
(\tau )}   g(x^\alpha ) _{\mu  \nu}{dx^\mu \over d\tau } 
  {dx^\nu \over d\tau }  + m\lambda (\tau ) - e 
{dx^\mu \over d\tau }A_\mu (x^\alpha )  \right)
\label{eq:action2}
\eeq
in the gauge $t= \tau $, which altogether eliminates traveling
back and forth in time (something as we will see that is related to 
particle 
creation and
unitarity). The hamiltonian in the first case is just 
\beq
h_{EG} = -eA_0 + \sqrt{g_{00} \{m^2- (p_i + A_i) g^{ij} (p_j + A_j)\}}
\eeq


\newpage\section{Other inner products for the unconstrained particle} 

 In this section I want to  point  out an ambiguity in the
standard construction of a quantum formalism from a simple 
classical system.

 Consider again the simple {\em unconstrained} classical system
 described by the coordinate $x$ and the momentum $p_x$, and a 
hamiltonian $H$. How do we quantize it? Introduce the operators
$\bf  x, \;p_x$, with $[\bf  x, p_x] = i$, and a Hilbert space of states
 $ |\psi\rangle $, together with a dual $\langle \psi|$. To be more 
specific
 we have the representation of the above operators in the form
 $\bf\sim x,\; \bp\sim -i\partial_x$ with
$(\psi_a,\psi_b) = \int dx\: \psi_a^*(x)\: \psi_b(x)$, 
where one has, as usual,  defined $\psi(x) \equiv \langle 
x|\psi\rangle $.

Now, we could
define
\beq \tilde{\psi}(x)\equiv\langle x|{\bf A}|\psi\rangle \eeq
as long as the operator $\bf A$ has an inverse, and 
keep, for example, the dual as before, 
\beq \psi^D (x)\equiv \langle \psi| x\rangle \eeq 
In terms of the old inner product, using the identity
 \beq 
{\bf I} = \int dx\: |x\rangle \langle x| 
\eeq  we obtain the expression 
\beq
\langle \psi_a |\psi_b\rangle  = \int dx\; \langle \psi_a  |x\rangle 
\langle 
x|{\bf A^{-1} A}
|\psi_b\rangle = \int dxdx'\: \psi^D_a(x)\langle x|{\bf A^{-
1}}|x'\rangle 
\langle x' |{\bf A}| \psi_b\rangle  \eeq
Now, by definition of the representation we have (modulo ordering) \beq \langle 
x|{\bf A}|x' 
\rangle 
\equiv
[\hat{A}(x,-i\partial_x)]^{-1} \langle x|x'\rangle \eeq  and $\langle 
x|x'\rangle  
= 
\delta(x-x' )$, hence
\beq \langle \psi_a|\psi_b\rangle  = \int dx\; \psi^D_a (x)\;\left[{ 
\hat{A}}\left(x,-
i\partial_x
\right)
\right]^{-1}
\tilde{ \psi}_b(x)\eeq

 As for the matrix elements of operators, they can  be written as
\beq
\langle \psi_a|{\bf O}|\psi_b\rangle  \equiv \int dx \:\psi_a^D (x)\: 
{\hat{O} \; 
\hat{A}^{-1} }\:\tilde{\psi}_b(x)
\eeq
 Physical quantites are of course left unchanged. 

{\em The point is that we can modify the 
representation
 of the states  if we agree
to compensate for the changes when we evaluate physical quantities, 
which 
must be left unchanged.}  

Alternatively, we can also trace the ambiguity to the definition of the observables. Assume we
fixed the  inner product. Is  $\bf O$ the operator we want, or is it some ``multiple'' of it, $\bf O
\, A^{-1} $? If we have have a classical reference point---or experimental data---we will know what
to do. 

As an example, consider the free, unconstrained,
 {\em relativistic} particle. This system is,as we saw, equivalent
 to the unconstrained non-relativistic case,
 except that the hamiltonian is given by the square-root form 
$h = \sqrt{p_x^2 +m^2}$.
  Now, there are two standard quantization schemes\footnote{This
is discussed in the paper by Hartle and Kuchar, \cite{Hartle86}.}.
 One involves using the Klein-Gordon inner product,
 the other is the so-called Newton-Wigner quantization,
 which parallels the non-relativistic case very closely---indeed, the 
only
difference is in  the hamiltonian.

Let us look at these two  cases. We can start in both situations with
our standard quantum   states and base kets, $ |x\rangle ,
\langle x|x'\rangle  = \delta(x-x')$, (with the same for the $p_x$ 
kets). In 
terms 
of 
wave-functions defined with these kets, $\psi(x)\equiv
\langle x|\psi\rangle $,   the inner
product is the usual one. However,
we could choose to work instead with the kets
\beq
 |\tilde{x}\rangle = {\bf (p_x^2 +m^2)}^{-{1\over 4}} |x\rangle \eeq
Following the reasoning as above we will see that the wave-functions
defined by $\tilde{\psi}(x) \equiv \langle \tilde{x}|\psi\rangle $ are 
designed to work 
with the Klein-Gordon inner product:\beq \langle 
\psi_a|\psi_b\rangle 
=
\int dx\: \tilde{\psi}_a^*(x) \:\sqrt{{\bf p_x^2+m^2}} \:\tilde{\psi}_b 
(x)\eeq
(The reason for calling this the Klein-Gordon inner product is that 
the 
action 
of this operator on the states that satisfy the Klein-Gordon equation 
is the 
same as the time derivative operator.) 

[Note: The way this is described  in the paper I mentioned above, \cite{Hartle86},    
is as follows.
First the new momentum states are defined, \beq |\tilde{p_x}\rangle  
\equiv
 \left(\sqrt{{\bf p_x^2+m^2}}\right)^{1\over 2}\; |p_x\rangle \eeq 
Now, 
\beq 
 {\bf I}=\int dp_x\; |p_x\rangle \langle p_x| =
 \int{dp_x\over \sqrt{p_x^2+m^2}}\: |\tilde{p_x}\rangle \langle 
\tilde{p_x}|\eeq
Then the kets \beq |x_{NW}\rangle \equiv (2\pi)^{-{1/2}}
 \int {dp_x\over\left( 2\sqrt{p_x^2+m^2}\right)^{1\over 2}}\;
e^{ip_\alpha x^\alpha} \;|\tilde{p_x}\rangle  = e^{itp_t}\;|x\rangle 
\eeq 
are 
defined. These
are unitarily equivalent to the old ones. The other states,
\beq  |x^\alpha\rangle  \equiv (2\pi)^{-1/2} \int {dp_x\over 2 
\sqrt{p_x^2+m^2}}\;
e^{itp_t}\; |\tilde{p}\rangle \eeq  are essentially my 
$|\tilde{x}\rangle $'s.]


\newpage\section{Comments on  unitarity and causality}

As it has already been remarked, the question of unitarity in this 
case boils down to a question of ordering of the operators that make 
up the hamiltonian. If we can find an ordering that makes the 
hamiltonian hermitean, then  we are set, whether the hamiltonian is 
time dependent or not.  However, in general there are other 
constraints on the possible orderings one can use.  The need for 
space-time covariance, for example, may conflict with hermicity.  The particular 
form of the hamiltonian we get also follows from the ``gauge choice'', 
from the definition of time and the corresponding foliation of         
space-time.  Is there a nice, covariant ordering such that 
\beq \sqrt{
g_{00} \{m^2- (p_i + A_i) g^{ij} (p_j + A_j)\}   }\eeq is hermitean 
when 
made into an operator in the usual way? 
Well, the specific ordering I already used (by just interpreting the 
above 
equation as an operator equation)  makes the hamiltonian 
hermitean, as long as 
\beq [  \hat{g}_{00},\; (\hat{p}_i + \hat{A}_i) 
\hat{g}^{ij} (\hat{p}_j + \hat{A}_j)]=0\eeq
How about space-time covariance? For consistency {\em we need that  the 
hamiltonian operator transform as the zero component of a vector}. If 
this demand is met, then we will have a consistent theory. Of course, 
we will have broken the strongest version of the equivalence 
principle, because our foliation choice picks a special 
direction  in space-time.  

As for {\em causality}, the square root operator leads to trouble, as
may be expected  from the appeareance of arbitrary orders of 
derivatives in the Taylor expansion of 
such an operator. This type of operator is non-local,
as can be
seen from writing the corresponding ``square-root'' Schr\"{o}dinger 
equation\footnote{Again, I will refer to any equation of the form $i \partial_t \psi 
= \hat{H} \psi$ for some hamiltonian, as a ``Schr\"{o}dinger equation''.} for the free 
case in semi-integral 
from \cite{Baym}:
\beq
i\hbar \psi(x,t) = \int d^3x'\; K(x-x')\,\psi(x',t) 
\eeq with \beq
K(x-x') = \int {d^3 p \over (2\pi\hbar)^3}\, e^{i p (x-x') /\hbar} \sqrt{p^2+m^2}
\eeq
The kernel is sizable within a Compton wavelength of the 
particle---and non-local.
This leads to a violation of causality, because $\partial_t \psi$ depends on the 
values of $\psi$ outside of the light cone. 

Now, consider the D'Alembertian \beq   g^{\mu\nu} 
\nabla_\mu \nabla_\nu= {1\over \sqrt{g} } \, \partial_\mu [
 g^{\mu\nu} \sqrt{g} \; \partial_\nu \, \circ ]\eeq
This is a scalar operator. If this operator can be factorized in a hermitean fashion 
we have
an   answer to the above question about ordering. 

Suppose, for example, with $g_{0i} = 0$ as before, that we have \beq
[  \partial_0 , \hat{g}^{\mu\nu} ]=0=[  \partial_i , \hat{g}^{00} ]=0\eeq Then
\beq \sqrt{
\hat{g}_{00} \{m^2- {1\over \sqrt{\hat{g}} }(\hat{p}_i + \hat{A}_i) 
\hat{g}^{ij} \sqrt{\hat{g}}(\hat{p}_j + \hat{A}_j)\}   
} \eeq will be  hermitean with 
respect 
to the space integration. The issue of space-time covariance is far from simple, 
though, as we will see in the next section. 

Also notice that the mass term should really be substituted by the 
more general possibility \beq \tilde{m}^2   = m^2 + \xi R\eeq  For 
small enough 
$\xi$ there is no classical effect from this term. 

All these issues are extremely relevant, because they will help settle 
the question of whether one can pick a branch and save unitarity 
and/or space-time covariance. 

How are these issues related to the factorization of the Klein-Gordon 
equation? Can we bring these together under the particle creation 
point of view? We will discuss these questions in the
next section, and come back to them  later as well.


\newpage\section{The Schr\"{o}dinger equation and space-time  covariance}

In this section I would like to discuss the effect of
Lorentz transformations on the Schr\"{o}dinger equation.

Let us first state how a covariant Schr\"{o}dinger equation should behave under Lorentz
transformations. The wave-function must transform as     a relative scalar 3-density of
weight 1/2, if we
are to construct  a probability  interpretation. What is imply meant by this is that 
\beq
\int_{-\infty}^{\infty}  d^3 x \; \psi (x)^* \; \psi(x) =
\int_{-\infty}^{\infty}  d^3 x \; \gamma \psi (\Lambda^{-1} x)^* \; \psi(\Lambda^{-1} x)
\eeq 
Similarly, if $|\psi(x^\mu)|^2$ is the probability density of finding the particle at $x^\mu$,
then  it must transform as above so that  $\int_{Vol} d^3 x \; \psi (x)^* \; \psi(x)$ remains 
a constant. 
Thus, under a
change of coordinates  the 
wave-function changes by $\Psi(x^\mu)\longrightarrow \gamma^{1/2}  \Psi(\Lambda^{-
1\mu}_\nu
 x^\nu)$.
This would be the ideal invariant behavior, but {\em we must check that under this 
change the wavefunction still satisfies the Schr\"{o}dinger equation.} In terms of 
the
differential operators involved, by a change of coordinates we can
then see that we can get the same equation in the other coordinate
system by requiring that the hamiltonian operator transform as the zero
component of a 4-vector---again up to a factor of $\gamma$. To be
specific consider the ground state of a hydrogen atom
at rest, $\Psi_{00}(x^\mu)$. For another observer this state
will appear as $\gamma^{1/2}  \Psi_{00}(\Lambda^{-1\mu}_\nu x^\nu)$.  This
observer may ask if this state satisfies the Schr\"{o}dinger 
equation---which
he will write just as the other observer did. The only thing to be
careful about is with   {\em background} fields. If there is a background field,
 the correct
covariance statement for an equation $D({\partial\over \partial x^\mu}, A_\mu 
(x^\mu))
f(x^\mu)=0$ is that this equation imply that  $$ D({\partial\over \partial x^\mu}, 
\Lambda A_\mu
(\Lambda^{-1}  x^\mu)) \: f(\Lambda^{-1}  x^\mu)=0 $$ 
in other words, changing variables to $y=\Lambda^{-1} x$
\beq
D(\Lambda_\nu^\mu{\partial\over \partial y^\nu}, 
\Lambda_\nu^\mu A_\nu
(y^\mu)) \: f(y^\mu)=0
\eeq
Let us see how this is true for the Klein-Gordon equation.
Suppose that 
\beq
\left[ \left(\partial_{x^\mu} - A_\mu(x^\alpha)\right) \, \eta^{\mu \nu} \, 
\left(\partial_{x^\nu} - A_\nu(x^\alpha)\right)  -m^2 \right]\; \phi(x^\alpha)=
0  \eeq
It is easy to see that it then holds that 
\beq
[(\partial_{x^\mu} - \Lambda A_\mu(\Lambda^{-1} x^\alpha))\, \eta^{\mu \nu} \, 
(\partial_{x^\nu} - \Lambda A_\nu(\Lambda^{-1} x^\alpha))  -m^2 ]\;
\phi(\Lambda^{-1} x^\alpha)= 0 
\eeq
To check, simply change the varibles to $y= \Lambda^{-1} x^\alpha$. 
The equation becomes the earlier one,  written in terms of $y$, a dummy index.

It would be unreasonable to ask that $f(x^\mu)$ satisfy the exact same equation:
the background field brakes absolute covariance. Thus he will  take the  point
of view   that otherwise the laws of physics should be the same in all inertial
frames, and this includes the Schr\"{o}dinger equation. Thus, he will perform
 a change in variables in his equation and check if the new equation he gets
is true,
\beq
i \Lambda_0^\mu \, {\partial \over \partial x^{'\mu}}
 \Psi({x'}^{\mu})  = \hat{H}(\Lambda^\mu_\nu A^\nu ({x'}^\nu),
\Lambda^\mu_\nu \partial '_\mu) \Psi({x'}^\mu) 
\eeq
For the usual non-relativistic  hydrogen atom hamiltonian,  the appropiate
transformations are those of the Galilean group. Notice that we are not asking
that the energy of the  state be the same, but that the Schr\"{o}dinger equation be
satisfied. What must happen, in the free case, is
that the effect of a boost on a state of definite energy and momentum 
be transformed
into a a state of definite momentum and energy again---the boosted ones. In 
general, solutions of the equation must be boosted into new solutions (modulo 
readjustment of the functional dependence of the potentials).

Consider for simplicity the equation ($i=1,2,3$)
\beq
|p_0| = \sqrt{-p^i p_i  + m^2}
\eeq
This equation is equivalent to $p_0^2 - p_i^2 = m^2$ of course, so if
it is true in one frame of reference it will hold in all of them. Let
us check this explicitely by boosting both sides,
$$
|\Lambda^\mu_0 p_\mu | = \sqrt{ -\Lambda_i^\mu p_\mu \Lambda^i_\nu p^\nu 
+m^2} =
$$ \beq \sqrt{-\Lambda_\alpha^\mu p_\mu \Lambda^\alpha_\nu p^\nu 
+\Lambda_0^\mu p_\mu \Lambda^0_\nu p^\nu +m^2} 
= \sqrt{\Lambda_0^\mu p_\mu \Lambda^0 _\nu p^\nu}
\eeq
so it holds after a boost, as it should.

Consider next the square-root Schr\"{o}dinger equation,
\beq
i \partial_0 \Psi(x^\mu) = \sqrt{\partial ^i \partial_i + m^2}\, \Psi(x^\mu)\eeq
and suppose that $\Psi(x^\mu)$ indeed satisfies this equation. I will now show that 
after
a boost 
this equation is still satisfied. Indeed, after the boost---as discussed before---we 
have
to check that 
$$
i \Lambda_0^\mu \, {\partial \over \partial x^{'\mu}}
 \Psi({x'}^{\mu}) 
= \sqrt{ \Lambda_i^\mu \partial'_\mu \Lambda^i_\nu \partial^{'\nu} +m^2}\, 
\Psi({x'}^\mu)=$$
\beq
\sqrt{ \Lambda_\alpha^\mu \partial'_\mu \Lambda^\alpha_\nu \partial^{'\nu} 
-\Lambda_0^\mu \partial'_\mu \Lambda^0_\nu \partial^{'\nu} +m^2}\, 
\Psi({x'}^\mu)=
\sqrt{-\Lambda_0^\mu \partial'_\mu \Lambda^0 _\nu \partial^{'\nu}}\, 
\Psi({x'}^\mu)
\eeq
which is consistent. Notice that the crucial part of the proof was that \beq
i \partial_0 \Psi(x^\mu) = \sqrt{\partial^i \partial_i  + m^2}\, \Psi(x^\mu) 
\longrightarrow
[\partial_\mu \partial^\mu + m^2]\, \Psi(x^\mu) =0 \eeq
which is true because\beq
\partial_\mu \partial^\mu + m^2 = \left(\partial_0 - \sqrt{\partial^i 
\partial_i +m^2}\right)\left(\partial_0 + \sqrt{\partial^i \partial_i+m^2}\right)\eeq
since 
\beq
[\partial_0, \partial_i] = 0\eeq Notice that we also needed that 
\beq
[\partial_\mu, \Lambda_0^\nu \partial_\nu ]=0
\eeq

How about the case of an electromagnetic interaction?  One needs to define
what the square root means, because a momentum states expansion definition
will no longer work. As long as the operator in the square-root is hermitean
and positive one can build a basis with its eigenstates, though in general these
will be time-dependent. Let us assume that all this is so.

All that changes is that we need to use gauge-covariant derivatives---i.e., minimal
coupling---defined as follows\footnote{Here
 $A_\mu =(\phi, -\vec{A}),\:\: \:\: \partial_\mu =(\partial_t,\vec{\partial}_x)$, with
our
choice of metric convention, time-space $\sim(+,-,-,-)$.}
\beq D_\mu \equiv \partial_\mu +ie
A_\mu
\eeq
 One can show \cite{Samarov} that this equation---with the minimal coupling---is 
gauge covariant, meaning that if $\psi(x^\mu)$ is a solution and one changes \beq
\psi(x^\mu)  \longrightarrow   e^{-ie\Lambda(x^\mu) }\; \psi(x^\mu), \:\:\:\: 
A_\mu\longrightarrow A_\mu +\partial_\mu \Lambda \eeq
the equation is still valid. How about space-time  covariance? This is 
a much trickier issue (see \cite{Sucher}).
 Just as before we will need
$$
i \Lambda_0^\mu \, {D_0 '}
 \Psi({x'}^{\mu}) 
= \sqrt{ \Lambda_i^\mu D'_\mu \Lambda^i_\nu D^{'\nu} +m^2}\; \Psi({x'}^\mu)=$$
\beq
\sqrt{ \Lambda_\alpha^\mu D'_\mu \Lambda^\alpha_\nu D^{'\nu} 
-\Lambda_0^\mu D'_\mu \Lambda^0_\nu D^{'\nu} +m^2}\; \Psi({x'}^\mu)=
\sqrt{D^{'2} +m^2-\Lambda_0^\mu D'_\mu \Lambda^0 _\nu D^{'\nu}}\; 
\Psi({x'}^\mu)
\eeq
The sufficient condition for this equation to be true
is that   the electric field is zero in some frame (no particle creation condition),   
\beq
[D_0, D_i]= F_{0i} = E_i \eeq as I will now show.

Indeed, if this condition is met we know that the Klein-Gordon equation decouples,
because\beq
D_\mu D^\mu + m^2 = \left(D_0 \pm \sqrt{D^i 
D _i +m^2}\right)\left(D_0 \mp \sqrt{D ^i D_i+m^2}\right)\eeq
since 
the electric field is zero. Hence we know that for a solution
of the Klein-Gordon equation
\beq
  [\left(D_0 \pm \sqrt{D^i 
D _i +m^2}\right)\left(D_0 \mp \sqrt{D ^i D_i+m^2}\right)]_x \; \phi(x) =0
\eeq
 Moreover, because this equation is relativistic we also know that\beq
 [\left(D'_0 \pm \sqrt{D^{'i} 
D' _i +m^2}\right)\left(D'_0 \mp \sqrt{D^{'i} D'_i+m^2}\right)]_x \; \phi(x) =0\eeq
where $D' = \Lambda D$, as discussed earlier. This means that
$$
0= [ D^\Lambda_0 , \sqrt{D^{\Lambda i} D^\Lambda_i+m^2}]_x \; \phi(x)  =
$$
\beq
[ D^\Lambda_0 , m^2 ( 1+  {1\over 2} {D^{\Lambda i} D^\Lambda_i  \over m^2} + ... ]_x \; \phi(x)
=0 \eeq
The fact that this equation is true for {\em any boost}  $\Lambda$ now means that 
$$
0= [ D^\Lambda_0 ,   D^{\Lambda i} D^\Lambda_i   ]_x \; \phi(x)
=
$$ 
\beq
   [ D^\Lambda_0 ,    D^{\Lambda\mu} D^\Lambda_\mu   ]_x \; \phi(x) =
  [ D^\Lambda_0 ,    D^{\mu} D_\mu   ]_x \; \phi(x) =0 \eeq
which is all we need to show that 
\beq
\sqrt{D^{'2} +m^2-\Lambda_0^\mu D'_\mu \Lambda^0 _\nu D^{'\nu}}\; 
\Psi({x'}^\mu) =i \Lambda_0^\mu D'_\mu
\eeq
Thus, this equation, the square-root Schrodinger equation, is covariant {\em 
 if there is a frame
in which the electric field is zero. }
  \newpage\section{Conclusions, summary}

In this chapter I have first reviewed the standard quantization of the 
unconstrained particle, emphasizing the fact that
the  construction and interpretation of the  path integral are straightforward when one
knows what the states, the inner product and the hamiltonian are in the theory.

I have  discussed the usee of different hamiltonians: the non-relativistic case as well
as the square-root relativistic one, with or without  interactions. We have 
seen that unitarity of the resulting theory hinges on  whether the hamiltonian
is hermitean---time dependence of the hamiltonian is not a problem {\em per se}.

Then I have discussed the fact that  it is possible---after choosing a set 
of observables and an inner product---to change the
inner product  in the theory, as long as this change is compensated 
 by a change in the
normalization of the states and a  modification of the observables, which must be
hermitean in the  new inner product. Although this is not a new idea, it
has been overlooked in the literature as a source of ambiguities in 
the quantization
of constrained systems, a point which I will come back to in the
next chapters.

Then I have looked at the issue of whether the Schr\"{o}dinger equation is 
space-time covariant, as well as gauge-covariant.  For the square-root
case the answer is that it is always gauge-covariant---a result 
of  Samarov
\cite{Samarov}---and  I have showed that relativistic covariance
demands that there exist a frame in which the electric field is zero. We will
see that this 
is a recurring theme.

\chapter{Classical aspects of parametrized systems}
\label{sec:classical}
Parametrized systems are constrained systems of a special kind. In this 
chapter I will review the formalism developed by Dirac and apply it to
our systems, and see how to obtain the physical---as opposed to gauge---degrees
of freedom and their dynamics.  
\newpage\section{The non-relativistic particle: general considerations}
\label{sec:nrgeneral}
The action for the parametrized non-relativistic particle 
is\footnote{Unless 
otherwise stated we define $ a = da / d\tau$.}
\beq
\label{eq:unc}
S = \itif d\tau \; L = \itif  d\tau \; { m \over 2}{\dot{x}^2\over \dot{t} }   
\eeq
This form follows from the trick suggested by Dirac \cite{Dirac} in 
which 
the 
time coordinate is added to the degrees of freedom in the 
Lagrangian: let 
$x,t=x(\tau),t(\tau) $ so that $ \tau $ is the new ``time'' parameter, 
then
\beq
S = \itif dt\; L[x,{dx \over dt},t]  = \itif d\tau \; L[x,{dx\over d\tau}
{ 1\over \dot{t}},t]  {dt\over d\tau}  
\eeq
This action is
invariant under reparametrizations that do not affect the
boundaries. Indeed, let  $\tau  \longrightarrow  f( \tau ) $ with $f( 
\tau_i)=
\tau_i$ and $f( \tau_f)=  \tau_f$,  and with $ df / d \tau  >0$;
then the action becomes
\beq
S \longrightarrow   \itif  d\tau \; { m \over 2}{ \dot{x}(f(\tau)) ^2\over 
\dot{t}(f(\tau)) }   
= 
\itif  d\tau {df\over d\tau} \; { m \over 2}{ {x}'(f)^2\over  {t}'(f ) }  
\eeq
 so it is invariant. Notice that if $ df / d \tau  <0 $ and  $f(
\tau_i)=  \tau_f$, $ f( \tau_f)=  \tau_i $, then the action is not left 
unchanged  but changes sign,
$S\longrightarrow  -S$,  so for this action the invariance requires $df 
/ d
\tau  >0$. 

We can think of the full reparametrization group as being the direct product 
of $Z_2$ and the reparametrizations connected with the identity. 
We can say that this action carries a faithful representation of
the $Z_2$ part of the reparametrization group. The action of the  $Z_2$
part of the reparametrization group can be described by two types of 
reparametrization functions: $f_+$, which maps $\tau_i$ and $\tau_f$
into
themselves, and $f_-$, which maps $\tau_i$ into $\tau_f$ and viceversa.
The group multiplication is then described by
\beq
Z_2
= \left\{  \begin{array}{ll}
f_+ \cdot f_+ &= f_+\\
f_+\cdot f_- &= f_-\\
f_- \cdot f_- & = f_+ 
\end{array}
\right.
\eeq
The full diffeomorphism group is given by ${\cal G} = Z_2 \otimes {\cal F}_+$
where $ {\cal F}_+$ denotes the part connected to the identity.

The Euler-Lagrange equations of motion that follow from this
action are
\beq
{d \over d \tau}\left( {\dot{x}\over \dot{t} }\right) = 0 =
 {d \over d \tau}\left( {\dot{x}^2\over \dot{t}^2 }\right) 
\eeq
Let us now go to the hamiltonian formulation in the usual way,
by defining the momenta
\bea 
p_x = m { \dot{x}\over \dot{t}} & \: \: 
p_t  = -{m\over 2} {\dot{x}^2\over  \dot{t}^2 } 
\eea
We find the constraint\footnote{ Indeed $ det { \partial L \over 
\partial 
\dot{x}^\mu \partial \dot{x}^\nu } = 0 $.}
\beq
\Phi\equiv p_t +{p_x^2 \over 2m}\approx 0
\eeq
together with the zero hamiltonian $ H = p_t\dot{t} + p_x\dot{x}
- L \equiv 0$. The equations of motion are generated from 
the extended hamiltonian $H_E = v \Phi$ :
\ba \dis 
\dot{x} = \{ x, H_E \} = v {p_x \over m}& \:\: \dot{p}_x= 0\\
\dot{t} = \{t, H_E\} = v &\:\: \dot{p}_t = 0
\ea
This matches the lagrangian formulation above---in the sense that
the dynamics are reproduced---with the identification $v=1$. The
same equations of motion can be obtained from the so-called first 
order
action in the phase space coordinates
\beq
A=\itif d\tau(p_t \dot{t} + p_x \dot{x} - v\Phi)
\eeq
Notice that this action is invariant under the gauge transformations 
\cite{Teitelpap}
\ba
\delta x = \epsilon (\tau) \{x, \Phi\} & \delta p_x = 
\epsilon(\tau)\{p_x,\Phi\} \\
\delta t = \epsilon (\tau) \{t, \Phi\} & \delta p_t = 
\epsilon(\tau)\{p_t,\Phi\} \\
\delta v = \dot{\epsilon}(\tau)
\ea
as long as the gauge parameter vanishes at the boundaries, i.e.,  
$\epsilon 
(\tau_i) = \epsilon (\tau_f) =0$. It is not hard to see also
that this symmetry is the same as the  one in the 
lagrangian form, with the identification $f(\tau) = \tau +\epsilon 
(\tau)$. 
It is very important to be aware that this situation in which  there is 
a 
restriction in the gauge freedom at the boundaries is very different 
from 
the 
standard concept one has of a gauge theory, where there is no such 
restriction. Indeed, one usually understands the quantization of a 
system 
with symmetries as the quantization of the ``true'' and ``underlying'' 
degrees 
of 
freedom one assumes exist. The present situation is from this point 
of 
view
really troublesome. 

As explained in reference \cite{Teitelpap}, this new twist in the 
concept of 
invariance is a 
consequence of the form of the constraint, which is non-linear in the 
coordinates conjugate to what is fixed at the boundaries, i.e., the 
momenta. 
Under the above gauge transformation, the first order action changes, 
as a 
boundary term appears:
\beq
A \longrightarrow A + \epsilon (\tau)\left.  (p_i {\partial \Phi \over 
\partial 
p_i} - 
\Phi) 
\right| _{\tau_i}^{\tau_f}
\eeq
which vanishes when the constraint $\Phi$ is linear in the 
momenta---or 
with some other boundary conditions on the phase space variables. 
Reference \cite{Teitelpap} elaborates more on this point, as well as 
on the 
idea of 
modifying the action at the boundaries so that it becomes fully 
invariant. 
We 
will study these ideas in more detail shortly.

As one would expect the above dynamics match those of the 
unconstrained 
action---equation (\ref{eq:unconstrained})---when the gauge $ t = 
\tau $ 
is 
used. {\em That one has to use a specific gauge to recover the 
``physical''
coordinates is already an indication of trouble to come}, as we will 
see 
when 
we study the reduced phase space of this constrained system. 

The electromagnetic interaction case---which comes from applying 
Dirac's  trick 
to the unconstrained lagrangian in equation (\ref{eq:uelctro})---is 
given 
by 
the action
\beq
\label{eq:celctro}
{\cal S}_{EM} = \itif d\tau  \; {L}_E = \itif d\tau \; \left({1\over 2}m 
{\dot{x}_i^2 
\over 
\dot{t} } - e\phi \dot{t}  + {e \over c} \dot{x}_i A_i \right)
\eeq
and it has the same reparametrization invariance 
properties as the free case:  
$\tau  
\longrightarrow  f( \tau ) $ with $f( \tau_i)=
\tau_i$ and $f( \tau_f)=  \tau_f$,  and with $ df / d \tau  >0$, leaves 
the 
action unchanged. The covariant part, however,  is gone. In the hamiltonian 
formulation we have,
\bea 
p_i = m {\dot{x}_i \over \dot{t} } + {e \over c} A_i & \dis \:\:\:
p_t   = -m { \dot{x}_i^2 \over 2\dot{t}^2 } - e\phi
\eea
and we find the constraint\footnote{ Again, $ det { \partial L \over 
\partial 
\dot{x}^\mu \partial \dot{x}^\nu } = 0 $. }
\beq
\Phi_{EM}\equiv p_t  +e\phi +{(p_i -{e \over c} A_i)^2 \over 2m} 
=p_t + {\cal H}_{EM }\approx 0 
\eeq
together with the zero hamiltonian $ H = p_t\dot{t} + p_i\dot{x}_i
- L \equiv 0$. As usual, the equations of motion are generated by the 
extended hamiltonian $H_E = v \Phi_{EM}$ :
\ba
\dis \dot{x}_i = \{ x_i, H_E \} =\{ x_i, {\cal H}_{EM }\}=
 v ({p_i\over m} - {e \over m c} A_i )              
 & \:\:\dis \dot{p}_i= v[{e \over m c} A_{j,i} \; (p_j -{e\over c } A_j) -e
\phi_{,i}]
 \\\dis \dot{t} = \{t, H_E\} = v  &\:\:\dis
 \dot{p}_t =v[{e \over m c} A_{j,0} \; (p_j -{e\over c } A_j) -e 
\phi_{,0}]
\ea
This matches the lagrangian formulation  with the identification 
$v=1$. 
The 
first order action is $
A=\itif d\tau \; (p_t \dot{t} + p_i\dot{x_i} - v\Phi_{EM})$.
Notice that this action is invariant---as  before---under the gauge 
transformations generated by the constraint  
\cite{Teitelpap} (
$\delta z = \epsilon (\tau) \{z, \Phi_{EM})\}$ and $\delta v = 
\dot{\epsilon}(\tau)$, 
 where $z$ stands for all q's and p's),  with the same condition on the 
gauge 
parameter (i.e., that it vanishes  at the boundaries: $\epsilon 
(\tau_i) = \epsilon (\tau_f) =0$). Again as before, it is not hard to see  
that 
this symmetry is the same as the  one in the 
lagrangian form, with the identification $f(\tau) = \tau +\epsilon 
(\tau)$.

\newpage\section{The relativistic particle: general considerations}
\label{sec:rgeneral}
The action for the free parametrized relativistic particle is $(c = 1)$
\beq
S =  \itif  d \tau\;  L = 
-m  \itif  d \tau\;  \sqrt{\dot{t}( \tau )^2 - \dot{x}( \tau )^2}
\label{eq:freeaction1}
\eeq
(where from now on $\dot{ a} \equiv { da \over d\tau} $). This is 
just a 
possible form---basically the proper time---and it
is not well defined when the mass is zero. Another  form is
\beq
S' =  \itif  d \tau\;   L' =   \itif d \tau \;  \left( \,  {\dot{t}( \tau )^2 - \dot{x}( 
\tau )^2 
\over  \lambda      ( \tau )} + m  \lambda ( \tau )  \, \right)
\label{eq:freeaction2}
\eeq
and it is well defined also in the case $ m = 0.$ This action is
invariant under reparametrizations that do not affect the
boundaries. Indeed, let  $\tau  \longrightarrow  f( \tau ) $ with $f( 
\tau_i)=
\tau_i$ and $f( \tau_f)=  \tau_f$,  and with $ df / d \tau  >0$;
then the action becomes
\beq
S \longrightarrow   -m  \itif  d \tau \;  \sqrt{\dot{t}( f(\tau)) ^2 - 
\dot{x}( 
f(\tau))^2} = -m  \int_{f({\tau}_i)}^{f({\tau}_f)}  d\tau \;
| \dot{f}(\tau) |    \sqrt{t'( f)^2 - x'( f)^2}
\eeq
 so it is invariant. Notice that if $ df / d \tau  <0 $ and  $f(
\tau_i)=  \tau_f$, $ f( \tau_f)=  \tau_i $, then the action is also left 
unchanged,
$S\longrightarrow  S$,  {\em so for this action the invariance allows 
both  $df / d
\tau  >0$ or $df / d \tau  <0$.} We can
say that this action carries the trivial representation
of the $Z_2$ part of the 
reparametrization group. This is a very important point, as
we will see later.

The Euler-Lagrange equations of motion that follow from this
action are
\beq
{d \over d \tau}\left( {\dot{t}\over L}\right) = 0 =
 {d \over d \tau}\left( {{\dot{x}}\over L}\right) 
\eeq
Let us now  go to the hamiltonian formulation in the usual way; 
defining  the momenta 
\bea 
p_x =- m^2 {\dot{x}\over L} &\:\:\:
p_t  = m^2 {\dot{t}\over L} 
\eea
we find the constraint\footnote{ indeed $ det { \partial L \over 
\partial 
\dot{x}^\mu \partial \dot{x}^\nu } = 0 $ }
\beq
\Phi\equiv p_t^2 - p_x^2 \approx 0
\eeq
together with the zero hamiltonian $ H = p_t\dot{t} + p_x\dot{x}
- L \equiv 0$. The equations of motion are generated from 
the extended hamiltonian $H_E = v \Phi$ :
\ba
\dot{x} &= \{ x, H_E \} = -2 v p_x &\:\:\: \dot{p}_x= 0\\
\dot{t} &= \{t, H_E\} = 2 v p_t & \:\:\: \dot{p}_t = 0
\ea
This matches the lagrangian formulation above---in the sense that
the dynamics are reproduced---with the identification $v=-
L/2m^2$. The
same equations of motion can be obtained from the so-called first 
order
action in the phase space coordinates
\beq
A=\itif d\tau(p_t \dot{t} + p_x \dot{x} - v\Phi)
\eeq
Notice that this action is invariant under the gauge transformations 
\cite{Teitelpap}
\ba
\delta x &= \epsilon (\tau) \{x, \Phi\} & \delta p_x = 
\epsilon(\tau)\{p_x,\Phi\} \\[.5cm]
\delta t &= \epsilon (\tau) \{t, \Phi\} & \delta p_t = 
\epsilon(\tau)\{p_t,\Phi\} 
\\[.5cm]
\delta v &= \dot{\epsilon}(\tau)
\ea
as long as the gauge parameter vanishes at the boundaries, i.e. 
$\epsilon 
(\tau_i) = \epsilon (\tau_f) =0$. It is not hard to see also
that this symmetry is the same as the orientation-preserving one in 
the 
lagrangian form, with the identification $f(\tau) = \tau +\epsilon 
(\tau)$. 
Again, it is very important to be aware that this situation in which  there 
is a 
restriction in the gauge freedom at the boundaries is very different 
from the 
standard concept one has of a gauge theory, where there is no such 
restriction. Indeed, one usually understands the quantization of a 
system 
with symmetries as the quantization of the ``true'' and ``underlying'' 
degrees 
of 
freedom one assumes exist, and,  just   as in the
non-relativistic case, the present situation is from this point 
of view
really troublesome. 

As explained in reference \cite{Teitelpap}, this new twist in the 
concept of 
invariance is a 
consequence of the form of the constraint, which is non-linear in 
the 
coordinates conjugate to what is fixed at the boundaries, i.e., the 
momenta. 
Under the above gauge transformation, the first order action 
changes, as a 
boundary term appears:
\beq
A \longrightarrow A + \left. \epsilon (\tau) (p_i {\partial \Phi \over 
\partial p_i} - 
\Phi) 
\right|_{\tau_i}^{\tau_f}
\eeq
which vanishes when the constraint $\Phi$ is linear in the 
momenta---or 
with some other boundary conditions on the phase space variables. 
Reference \cite{Teitelpap} elaborates more on this point, as well as 
on the 
idea of 
modifying the action at the boundaries so that it becomes fully 
invariant. We 
will study these ideas in more detail shortly---see also reference 
\cite{NR}.

As one would expect the above dynamics match those of the 
unconstrained 
action---equation (\ref{eq:uroot})---when the gauge $ t = \tau $ 
is 
used. That one has to use a specific gauge to recover the ``physical'' 
coordinates is already an indication of trouble to come.

Consider next the more general actions for the interacting particle, 
equations (\ref{eq:action1}) and (\ref{eq:action2}).  Notice that  
both actions are  invariant under the {\em connected } part of the 
reparametrization (or diffeomorphism) group: 

 $\tau  \longrightarrow  f( \tau ) $ with $f( \tau_i)=
\tau_i$ and $f( \tau_f)=  \tau_f$,  and with $ df / d \tau  >0$.  

{\em The invariance---or covariance---under the disconnected part of the group is 
lost when going to the interacting case.} 

As mentioned, we can think of the full reparametrization group as being the direct 
product 
of $Z_2$ and reparametrizations connected with the identity. It appears that in the
interacting case we have lost ``half'' of the representation.

Let us first look at the action $A_{EG}$ in equation 
(\ref{eq:action1}).  Because of the reparametrization invariance we 
find---using Dirac's formalism as usual---the constraint
\beq
\Phi_{EG} = (p_\mu -A_\mu) g^{\mu \nu} (p_\nu -A_\nu) -m^2 
\approx 0 , 
\eeq
together with a zero hamiltonian, $ H_{EG} \equiv 0$. 

Using the action $A_{EG} ' $ in equation (\ref{eq:action1}), the 
situation is just a bit
more complicated. The lagrangian equations of motion are
\beq 
{ g_{\mu\nu,\beta} \ov \lambda} \dot{x}^\mu \dot{x}^\nu -e 
\dot{x}^\mu A_{\mu,\beta} - {d \ov d\tau} \left( 2 {g_{\mu\beta} 
\ov \lambda} \dot{x}^\mu - e A_\beta \right)  =0 \eeq
 and from varying $\lambda$ 
\beq
\lambda = \pm \sqrt{ {1 \ov m} g_{\mu\nu} \dot{x}^\mu 
\dot{x}^\nu}
\eeq
In the hamiltonian formalism the situation is the following: first 
we find the constraint $p_\lambda \approx 0 $, since the action is 
independent of $\dot{\lambda}$. The hamiltonian is not zero, $ 
H_{EG}' \equiv \lambda \Phi_{EG}$. However, one now needs---as 
Dirac explains \cite{Dirac}--- $\dot{\lambda} =0$, and this implies 
$\Phi_{EG} \approx 0$, i.e., the constraint above appears here as a 
secondary 
constraint. 

Again, the constraint contains two branches. It can be rewritten as
\beq
\label{eq:branches}
\Pi_0 = -\Pi_i \tilde{g}^{0i} \pm \sqrt{ (\Pi_i \tilde{g}^{0i})^2 - \Pi_i 
\tilde{g}^{ij} \Pi_j +m^2/g^{00}} \eeq
where $ \Pi \equiv p - A  $, and $ \tilde{a} \equiv a/ g^{00} $.
%
\newpage\section{Minisuperspace: general considerations}
\label{sec:minic}
The action we will use for our minisuperspace model is 
\beq
S_{M} = {1\over 2} \itif d\tau \,    \left(    {g_{AB} \dot{Q}^A \dot{Q}^B 
\over 
N}   + N U(Q)   \right)
\eeq
which is again equivalent to
\beq
S_{M} = - {1\ov 2} \itif d\tau\, \sqrt{ U(Q) g_{AB} \dot{Q}^A \dot{Q}^B }
\eeq
where the signature of $g_{AB}$ is Lorentzian. In this homogeneous 
cosmological model we have a lapse $N$, which is essentially the 
time-time 
component of the metric, and a parameter $Q^0$ characterizing
the scale of the universe. The other parameters describe spatial 
anisotropies.
We see that, mathematically, we have been discussing these models all along when 
studying 
the relativistic particle. 

In order to understand the origin of these models it helps to think of 
the 
dynamical field as an infinite number of of variables that depend on 
$\tau$, 
$g\sim g_{x^i} (\tau)$. It is not hard to imagine that  by demanding 
enough 
symmetries on the solutions  most of these infinite variables will be 
fixed. In 
practice one writes the most general form of the metric that satisfies 
the 
symmetries imposed and plugs that in into the action---or Einstein's 
equations.

Since mathematically this system is essentially equivalent to that of 
the 
relativistic particle in a curved background and no electromagnetic 
field, the 
earlier discussions on symmetries etc. will all apply 
here. Notice, in particular, that both the orientation preserving and 
orientation reversing diffeomorphism symmetries are present in this 
case, 
since there is no analog of the electromagnetic field to brake 
them---for this 
model at least.

In fact, the only difference of importance is in the interpretation
of the theory, and it will be essential to remember this when we 
discuss the 
quantum aspects of minisuperspace, in particular when we discuss 
unitarity, 
space-time covariance and causality.

The equations of motion are given by \cite{guven} 
\beq
{d \ov d\tau} \left( {2 g_{AB} \dot{Q}^B \over N}\right) + N{ \partial 
U\ov 
\partial Q^A} = 0
 \eeq
as well as
\beq
N=\pm \sqrt{  {g_{AB} \dot{Q}^A \dot{Q}^B \over U(Q) }}
\eeq
The constraint that follows from this form of the action is 
\beq
\Phi_M \equiv P_A P_B g^{AB} - U(Q) \approx 0
 \eeq
which looks like that of a relativistic particle in a curved background  
with a 
coordinate dependent mass.

  
\newpage\section{Gauge fixing, the Dirac bracket, and the Reduced Phase 
Space}
\label{sec:rpsc}
Next we consider the idea that at the classical level one should be 
able to 
get rid of the gauge, or unphysical degrees of freedom. As Dirac 
explained  
30 years ago \cite{Dirac}, the appearance of a first class constraint in  
the 
process 
of defining the momenta is an indication of gauge freedom, or 
invariance. 
This can be seen when one considers the dynamics of the system: the 
phase space trajectories are defined only up to gauge 
transformations, 
and the constraints themselves are the generators of the gauge 
transformation  (in the canonical sense). This is because of the 
inherent 
ambiguity in the hamiltonian. A possible approach to quantization is to quantize 
and 
then 
 constrain, but one would think that it should be  possible to eliminate the 
unphysical gauge 
degrees of freedom at the classical level. This is the idea behind the 
reduced phase space approach: there is gauge freedom at the 
classical 
level, so why not fix the gauge (or identify gauge-equivalent points) 
immediately? Gauge-fixing can be done by adding an additional 
constraint to the system, as we shall see.

Before studying the reduced phase space  (RPS) associated with this 
system, let us consider  the simple example in which the phase space 
is described by $(q, p, Q, P )$ and where we have the linear 
constraint 
$\Phi = P \approx  0$. This constraint can be thought to generate a 
gauge 
transformation through  $Q\longrightarrow Q + \epsilon  \{Q , P\}$.
This is because the hamiltonian is defined only up to an arbitrary 
term of 
the form $v(\tau) P$. At any rate the point is that the coordinate $Q$ 
describes a 
gauge  fiber, and the RPS, the physical space, is described by $q$ and 
$p$: 
the physical phase space is recovered by the reduction process 
defined by:
\begin{quote}
a) $\Phi \approx 0$,  and \\
b) $ Q \sim Q + \epsilon \{Q, P\} $. 
\end{quote}
This similarity relation means that it 
is sufficient to consider the gauge invariant functions, which  are 
indeed 
described by
\beq 
0 = \{C_\Phi , \Phi\} = \{C_\Phi, P\} = { \partial C_\Phi \over \partial Q}
\eeq
i.e., they are functions independent of $Q$.

Consider next the following similar idea: take any function 
$F(q,p,Q,P)$ on 
the 
constrained surface. Then fix  $Q = Q(q,p)$. The function 
$\tilde{F} \equiv F(Q(q,p),q,p)$ lives on a subspace of the full phase 
space
 but it can be defined anywhere 
by a gauge invariant extension. Indeed, if we think of $\tilde{F} $ as 
living 
on the 
whole  space, we automatically have
\beq
\{\tilde{F} ,\, P\} = \{F(Q(q,p),q,p),\,  P\} =  { \partial \tilde{F}  \over 
\partial 
Q}=0
\eeq
so that $\tilde{F} $ is independent of $Q$, of course.

Now suppose that the original system has dynamics generated by
\beq
H_E = h(q,p) + v(\tau) \Phi =  h(q,p) + v(\tau) P
\eeq
where $v(\tau) $ is an arbitrary function of $\tau$-time. The 
dynamics, in 
the 
language of gauge 
invariant functions, are described by
\beq
\dot{C}_\Phi =  {\partial C_\Phi \over \partial \tau}+ \{C_\Phi ,H_E\} = 
 {\partial C_\Phi\over \partial \tau}+ \{C_\Phi, h\}
\eeq
so that we have dynamics for the physical degrees of freedom only. 
Suppose we decide to go the ``gauge-fixing way'' by adding a 
gauge-fixing 
constraint to the system
\beq
\chi = Q - Q(q,p) = 0
\eeq
The first thing to do is to make sure that both this new constraint 
and the 
original one are preserved in $\tau$-time, and for that purpose the 
hamiltonian is 
``extended'' further: $H_{E'} \equiv  h(q,p) + v \Phi + w \chi$. Then 
one 
demands
$\dot{\Phi} = \dot{\chi} = 0$, and this yields
\bea & &
 0 = \dot{\chi}  = \{\chi,H_{E'}\} = \{\chi,h\} + v  \\ & & 
0 = \dot{P} = w   \label{eq:chidot} \nonumber
\eea
which fixes $  v$ and $w$. Now let 
$F(q,q,Q,P )$  be an arbitrary function in phase space. The 
dynamics are then given by
\beq
\dot{F} =  {\partial F \over \partial\tau}+ \{F,H_{E'}\} \approx 
 {\partial F \over\partial \tau}+ \{F, h\} + v \{F,\Phi\}
\label{eq:dyn}
\eeq 
\subsection{The Dirac bracket}
\label{sec:Diracbracket}
One can also just define $\{ \; , \; \}_*$ , the Dirac bracket 
\cite{Dirac,TeitelBook}:
\beq
 \{Q,F\}_* = \{P,F\}_* = 0  
\eeq
for all $F$. In general, the Dirac bracket is defined by
\beq
\{A,B \}_* = \{A,B\} - \{A,\chi_i\} (c_{ij})^{-1} \{\chi_j,B\}
\eeq
where the matrix  $c_{ij}$ is given by
\beq
c_{ij} = \{\chi_i,\chi_j\}
\eeq
and where $\chi_i$ is short for the constraints, $\chi_i  =
\chi ,\Phi $. This is just
\beq
\{A,B\}_* = \{A,B\} - ( \{ A,\chi \},\{A, \Phi \}) 
{1\over \{\chi, \Phi \} }
\left( \begin{array}{cc} 0 & -1 \\ 1 & 0 \end{array}   \right)
\left( \begin{array}{c} \{\chi ,B\} \\ \{\Phi ,B\} \end{array} 
\right)                        
\eeq
The equation of motion above can then be
written as \ba
\dis \dot{F}&\dis =  {\partial F \over\partial \tau}+ \{F,H_{E'}\}  
\\[.5cm]
&\dis =  {\partial F \over\partial \tau} + \{F,H_{E'}\}_* + \{F,\chi_i\}
 (c_{ij})^{-1} \{\chi_j,H_{E'}\} \label{eq:dynDirac1}\\[.5cm]
&\dis =  {\partial F \over \partial\tau}+ \{F,H_{E'}\}_* = {\partial F 
\over 
\partial\tau}+ \{F,h\}_*  \label{eq:dynDirac2}
\ea
since   $\{\chi_j,H_{E'}\} = 0$ for this case of $\tau$-time 
independent 
gauge-fixing (see equation (\ref{eq:chidot})).  Moreover 
\beq
\{F,H_{E'}\} = \{F,H_{E'}\}_* = \{F,H_{E}\}_* = \{F,h\}_*
\eeq
 which  follows from $H_{E'} - H_{E} = a\Phi + b\chi$, 
$\; H_{E} - h = v\Phi $,
and $\{\chi_j,F \}_* = 0 $  for any\footnote{Indeed, the Dirac
bracket  is designed so that one can set the constraints to zero {\em 
before}
its  computation.} function $F$. 

Let us look now at the case of a more general constraint and also at 
more 
general gauge-fixings, including the possibility of $\tau$-dependent 
ones. 
The goal is to obtain the reduced phase space and describe its 
dynamics. 
We 
can start by looking at the extended hamiltonian $ H_{E'} = h(q,p) + v 
\Phi 
(Q,P) + w\,  \chi (Q,P,\tau)$. Again, we require that $ \dot{\Phi} = 
\dot{\chi} 
= 0 
$. These equations imply $w=0$ and 
\beq
v={1 \over \{ \Phi , \chi\} }{\partial \chi \over \partial \tau} 
\eeq
and the dynamics are again described by   equation (\ref{eq:dyn}). 
What 
happens in the Dirac bracket formalism? Equation 
(\ref{eq:dynDirac1}) is 
still 
correct, but equation (\ref{eq:dynDirac2}) is not since 
$\{\chi_j,H_{E'}\} = 
0$ 
doesn't hold anymore. In fact $ \dot{\chi} = 0 $ now means that 
$\{\chi,H_{E'}\} = -{\partial \chi / \partial \tau}$. One can check after 
some simple algebra that the 
equation of motion is now
\beq
\dot{F}={\partial F \over\partial \tau} + \{F,h\}_* + 
{\partial \chi \over \partial \tau}{ \{F,\Phi\} \over
 \{ \Phi , \chi \} }
\eeq
which looks different than (\ref{eq:dynDirac2}).
In fact, {\em for  entirely general constraints and gauge-fixing 
functions},
 it is easy 
to see that the equation of motion for an arbitrary function in phase 
space 
is
\beq
\label{eq:gdyn}
\dot{F}={\partial F \over\partial \tau} + \{F,h\}_* + 
{\partial \chi \over \partial \tau}{ \{F,\Phi\} \over
 \{ \Phi , \chi \} } + {\partial \Phi \over \partial \tau}{ \{F,\chi\} 
\over
 \{ \chi , \Phi \} }
\eeq
which is equivalent to the extended hamiltonian description with
$$ H_{E'} = h(q,p) + v(\tau)\Phi (q,p,Q,P, \tau) + w(\tau)\chi 
(q,p,Q,P,\tau),$$ 
where 
$v,\; w$ are fixed as usual by $\dot{\Phi}=\dot{\chi}=0$:
\beq
{\partial \Phi \over \partial \tau} +\{ \Phi , H_{E'} \} =0=
{\partial \chi \over \partial \tau}  +\{ \chi , H_{E'} \} 
\eeq
Notice that in general these equations don't have solutions---in 
which case 
equation (\ref{eq:gdyn}) is undefined, of course.  
Since this possibility is in fact going to appear when we consider the 
constant 
of 
the motion coordinate system in section \ref{sec:constant}, let us 
study it 
in 
more detail with a simple toy case which 
illustrates nicely how and when  the dynamics of a
 constrained system
lead to the concept of reduced phase space (see Dirac \cite{Dirac}).
\subsection{$RPS$ vs. $RPS^*$}
\label{sec:RPSandRPS*}
 Recall that the idea of reduced phase space
 comes from the fact that some coordinates in the phase space
of a constrained system have arbitrary dynamics. Consider the 
simple
case in which we have a phase space described by the coordinates
 $Q,P$, and  $q^i, p_i$, with a first class hamitonian $h(q^i,p_i)$,
and a constraint $\Phi = {P^2 / 2}\approx 0$. Now, this form of the
constraint may appear to be  unusual, and perhaps one would 
expect  the system  to be equivalent to one in which the constraint
is $P \approx 0$: the constraint surfaces are definitely the same!
But before jumping to conclusions let us look at the dynamics.

In the original
case they would be described by the extended hamiltonian $H_{E} = 
h + 
v(\tau)
{P^2 / 2}$,  and
the equation of motion for Q would be $\dot{Q} = v(\tau) P =0$ in 
the constraint surface! The constraint and the constrained
dynamics then reduce the phase space to
the space described by $ Sp \sim   [Q,P\!=\!0,q^i,p_i]$ with dynamics
\beq \dot{A}(q^i,p_i,Q,\tau )= {\partial A \over \partial \tau} +
\{A,h(q^i,p_i)\}  \eeq
Let us see how the above methodology fares in this case. 
Introduce an additional constraint, $\chi=Q=0$, and define as usual
$$H_E'= h + v(\tau) \Phi+ w(\tau) \chi.$$
Demanding
$\dot{\chi}=\dot{\Phi}=0$ then yields $vP=0=-wP$. 
As for the Dirac bracket, {\em it cannot be defined, since the matrix}
$\{\chi_i, \chi_j\}$ {\em has no inverse.} This is not due to a bad
choice of gauge-fixing, as any such function will run into this 
problem as long as it is not singular itself (e.g. $ \chi= Q/P$).
Here we see the direct connection between the existence of a 
reduced phase space and the existence of a well-defined Dirac 
bracket.

In the second case we would have
$\dot{Q} = v(\tau)$. This is the usual situation, where we see
clearly that the coordinates $Q,P$ are pure gauge, and that the 
system
{\em reduces} to the system $q^i, p_i, h(q^i, p_i)$. Indeed, the Dirac
bracket allows one to set $Q=P=0$ immediately, $$\{Q,F(Q,P,q^i,p_i) 
\}_* = 
0
=\{P,F( Q,P,q^i,p_i)\}_*$$ 

This system may seem artificial, and one may wonder if it would
ever arise from a lagrangian formulation, say. Both the relativistic 
particle 
and minisuperspace,
for instance, have constraints of the form $P^2- a^2 = 0 $.
Even though this system is not as degenerate as the 
one above, with this constraint
{\em it is not possible to  reduce by the identification procedure 
alone
 the above phase space to the ``physical'' coordinates}
$q^i,p^i$.  One also needs to stay in the constrained surface---and
this is different from 
the usual situation of linear constraints.  Indeed, the 
dynamics for the ``gauge'' coordinate $Q$ are still given by
$ \dot{Q} = v(\tau) P $ ---which is totally arbitrary in the
 constrained surface (unless $a=0$), but is not if
$P=0$. What does this mean?

One aspect of this situation can be understood in the following
terms. Consider the simpler problem in which we want to minimize
a function $f(x)$ in the space spanned by the coordinates $(x,y)$ 
subject to the constraint $x-x_0=0$. The solution is clearly given
by the set $\{ x_0, y, \mbox{ for all } y\}$. This is an example 
of a very simple ``gauge'' system. Using the lagrange multiplier
method (see for example Lanczos' book on classical mechanics 
\cite{Lanczos}), this problem is solved by minimizing the 
function $ F(x) = f(x) + \lambda g(x)  $, where $g(x)$ is some
``version'' of the constraint, and also by demanding $g(x)=0$. This 
means 
that we have to solve the equations 
\ba \dis 
{ \partial f(x) \over \partial x} + \lambda { \partial g(x) \over 
\partial x} = 0 \nonumber \\[.5cm] \dis
{\partial F(x) \over \partial y } = 0 \mbox{ (contains no information)} 
\nonumber \\[.5cm] \dis
g(x) = 0
\ea
Using $ g=x-x_0$ the solution is immediate. However, we can also
check that using $g=(x-x_0)^2$ leads to trouble, as then we find
no solution! The bottom line is that---as any calculus student
 knows--the
function $g(x)$ has to be chosen so that
\beq
\left. \left( {\partial g \over \partial x}, \; {\partial g \over \partial y 
}\right)
\right|_{g=0}
 \neq \vec{0}
\eeq
Similarly, the choice $\Phi = P^2$ is not good, as {\em with it the full
solution space is not found}, just as we saw before.
 The condition on the constraints for
finding the full solution space to the extremization problem is
\beq
\left.{\delta \Phi \over \delta z(\tau)}\right| _{\Phi=0} \neq 0 
\mbox{ for 
all 
}\tau
\eeq
where $z(\tau)$ stands for all $q$'s and $p$'s.
The constraint $P^2=0$ fails this test, and indeed we find less 
solutions when we use it---$\dot{Q}=0$ instead of
 $\dot{Q}=Pv(\tau)=$arbitrary.

Let us now discuss in
more detail the concept of reduced phase space---or $RPS$, for short. 
One
point of view is that, ideally, we can divide the full phase space into 
its
pure gauge and its gauge invariant---or physical---degrees
of freedom. Conventionally we call the gauge invariant subspace the 
reduced
phase space, $RPS$. However, we could also {\em fix the gauge part } 
and 
work with the resulting space. One easy way to think about it is to
go to the ``gauge-invariant'' coordinate system
 (see section~\ref{sec:constant}), fix the gauge coordinates there with
some gauge-fixing function, and then transform
back to the original coordinates. The resulting space---the full
 phase space after gauge fixing---we call $RPS^*$.
Mathematically, this space is the direct product of the two  spaces
\beq
RPS^* = 
\{\mbox{gauge invariant}\}\otimes \{\mbox{pure gauge---fixed}\}
\eeq
and it depends clearly on the gauge-fixing used. Usually this 
distinction is not made, the reason being that one usually
works with actions that are fully gauge invariant and therefore
indifferent  to gauge-fixing choices. However, for the case
of actions that are not fully gauge invariant it is important
to distinguish these two concepts. The reduced phase space proper
is described by the gauge invariant coordinates only, and it is 
different
from $RPS^*$, although isomorphic  to it. 

To denote the constraint and the gauge fixing used we will write  
$RPS_{\Phi}$ and $RPS_{\Phi,\chi}^*$.
Let $Q,P,q,p \equiv   q^a , p_a $ be the coordinates for the full space,  
in
 which $Q$ and $P$ stand as usual for the pure gauge
part---with $P\approx 0$---and $q,p$ for the invariant part. Also, 
let the 
first class extended hamiltonian be given by $H_E= h(q,p) 
+v(\tau)\Phi$.
The space $RPS_\Phi$ is described
by $q,p$ only, and $RPS_{\Phi ,\chi}^*$ by the full set---with $Q,P$ 
fixed by 
$\chi$ and $ \Phi$ respectively. We will consider two simple gauge-
fixings,
\begin{quote}
a) $\chi_1 = Q- f_1 (\tau)$ and \\
b) $\chi_2 = Q- f_2 (\tau)$ \end{quote}
for some arbitrary functions $f_1$ and $f_2$ of the parameter 
$\tau$.
Their respective spaces will be described by\begin{quote}
a) $q_1,p_1,Q_1,P_1, $  and $H_1=h(q,p) + \dot{f_1} P_1$\\
b) $q_2,p_2,Q_2,P_2,  $  and $H_2=h(q,p) + \dot{f_2} 
P_2$\end{quote}
(but we expect  the true physical coordinates $q,p$'s to be  
unchanged.)

Let us now consider the following proposition:
\begin{prop}
\label{prop1}
The spaces $RPS_{\Phi, \chi}^*$ are all isomorphic to each other and 
to
$RPS_\Phi $. In fact , $RPS_{\Phi, \chi_1}^*$ and $RPS_{\Phi, 
\chi_2}^*$ are 
related by
a canonical transformation. The generator
of the canonical transformation depends on ${\chi_1}$ and ${\chi_2}$ 
: it 
is given by \beq
{\cal G}_{\chi_1\rightarrow \chi_2}^\Phi = q_1 p_2 + Q_1 P_2 + (f_2-
f_1) P_2 
={\cal G}(q_1,p_2, Q_1,P_2)
\eeq
The corresponding canonical transformation is
\ba \dis
p_1 ={\partial {\cal G} \over \partial q_1} =p_2 & \dis q_2={\partial 
{\cal G} 
\over 
\partial p_2} = q_1 \\[.5cm] \dis
P_1 ={\partial {\cal G} \over \partial Q_1} =P_2 &\dis  Q_2={\partial 
{\cal G} 
\over 
\partial P_2} = Q_1 +f_2-f_1 \\[.5cm] \dis
H_2 = H_1 + {\partial {\cal G} \over \partial \tau} = H_1 + (\dot{f}_2 - 
\dot{f}_1)P_2
\ea
Symbolically we can write this  result  as $ \dis
   \mbox{Can}_{\cal G} [RPS_{\chi_1}^*] 
=RPS_{\chi_2}^* $.
\end{prop} 
This proposition is trivial in the situation described by the decoupled 
coordinates, but it holds in {\em all } coordinate systems---see the 
next
section---as long as the constraint can be made into a momentum.
Let us review a little bit of the general canonical transformation 
theory (see 
for 
example Goldstein). 
Suppose that two canonical coordinate systems $$
(q,p, h(q,p) ) \mbox{ and } (\tilde{q},\tilde{p}, 
\tilde{h}(\tilde{q},\tilde{p}) )
$$
yield the same dynamics. This will occur if, in fact
$$
   \itif (pq - h(q,p) ) d\tau = \itif (\tilde{p}\tilde{q}- 
\tilde{h}(\tilde{q},\tilde{p})) d\tau  + \itif dW
$$
( careful with the boundary conditions for extremization!) i.e.,
$$
 pq - h(p,q) = \tilde{q}\tilde{p} - \tilde{h}(\tilde{q},\tilde{p}) + 
{dW\over 
d\tau}
$$
The above transformation equations come from these identities.


\subsection{$RPS$ analysis for the non-relativistic particle}
\label{sec:RPSNRPle}
For the  non-relativistic particle, we have a phase space 
described by the coordinates $t, x, p_x $ and $p_t $, and a constraint 
$
\Phi = p_t+p_x^2/2m \approx   0$.
A possible approach is to immediately do a canonical    
transformation such that $\Phi$    becomes a momentum, and to 
proceed 
as 
above. Indeed, for simple constraints this is always 
possible, and is equivalent to what follows (see section 
\ref{sec:constant}).

The gauge invariant functions are those that satisfy\footnote{Greek 
indices 
denote the full space-time range.} 
$\{C_\Phi (q^\alpha, p_\alpha) ,\Phi\} = 0$, i.e.,  
\beq
0 = ({\partial \over \partial t }+ {p_x \over m}{\partial \over \partial 
x }) 
C_\Phi (q^\alpha, p_\alpha) = {\partial C_\Phi  (q^\alpha, p_\alpha) \over \partial 
t } + 
\{ 
C_\Phi (q^\alpha, p_\alpha) ,{p_x^2 \over 2m}\}
\eeq
which is solved by 
\beq
 C_\Phi = C_\Phi (x-(t-t_0){ p_x \over m},\, p_x,p_t)
\eeq
These are the functions, then, of the constants of the motion.

Let us now consider fixing the gauge with a  $\tau$-dependent 
gauge, 
$ \chi_\alpha = t - f(\tau) $, where $\alpha \equiv \dot{f}(\tau)$. As 
before, 
we need 
$
\dot{\chi}_\alpha = 0 = {\partial \chi_\alpha  / \partial \tau} +
 \{\chi_\alpha, H_{E'} \}
$ which implies $ v=\alpha $, so we have
\beq
H_{E'} = \alpha \Phi  =  \alpha ( p_t + {p_x^2 \over 2m})
\eeq
and the equations of motion in this gauge are
\ba
 \dis \dot{t} = \alpha &\dis \dot{x}  = \{x,\alpha {p_x^2 \over 
2m}\}\\
\dis\dot{p}_t   = 0 & \dis \dot{p}_x = 0 
\ea
so the system is reduced to the coordinates $x, p_x$, with a 
hamiltonian 
\beq
H_\alpha =  \alpha {p_x^2 \over 2m}
\eeq
The Dirac bracket computation in this gauge is easily done, since $\{ 
\chi_\alpha, \Phi\} = 1$: let $ \chi_j = \chi_\alpha, \Phi $  as before. 
Then
\beq
\{A,B\}_* = \{A,B\} - ( \{ A,\chi_\alpha \},\{A, \Phi \}) 
\left( \begin{array}{cc} 0 & -1 \\ 1 & 0 \end{array}   \right)
\left( \begin{array}{c} \{\chi_\alpha,B\} \\ \{\Phi ,B\} \end{array} 
\right)                        
\eeq
and this yields
$$ \{x,p_x\}_* = 1,\: \{t,p_t\}_* = 0,\:
\{x,p_t\}_* = -{ p_x\over m}$$
with all others zero.

Recall that we run into a subtlety with the 
Dirac bracket description of the dynamics. In this gauge the 
correct dynamics are described by the equation 
\beq
\label{eq:easydyn}
{dF\over d\tau } = {\partial F \over \partial\tau } + \{F,  \alpha \Phi
\} 
\eeq                                                         
 which follows from equation (\ref{eq:gdyn}) and, of course,  matches the above 
ones 
(\cite{Evans}, and see  ex. 4.8 in reference \cite{TeitelBook}).  

In the gauge $\alpha =1$ the connection to the unconstrained case of 
section~\ref{sec:unconstrained}  is clear, but let us look at the 
interpretation 
of the more general gauge-fixings. First notice that  equation 
(\ref{eq:easydyn}) has a very simple interpretation: for functions of $x, \, p_x$, 
say, one can simply 
factor 
out the term
$ \dot{f} $ to read 
$$
{d F \over d t} = {\partial F \over \partial t} + \{F, {p_x^2 \over 2m} 
\} 
$$
Let us show, anyhow,  that this reduced phase space is the same as 
the 
phase space for the unconstrained non-relativistic particle in 
section~\ref{sec:unconstrained}, in 
some unusual $\tau$-dependent coordinates. Recall that the idea 
here is that 
when one picks a gauge, a coordinate system for the reduced phase 
space is 
implicitly chosen---or an active transformation occurs, depending on 
your 
point of view---and that with $\tau$-dependent choices the 
dynamics will 
look different. Let us go back to the unconstrained system of 
chapter~\ref{sec:unconstrained}: there, $t$ was the time, not $\tau$. 
However, 
as we know from our later point of view $t$ and $\tau$ are one and 
the 
same in the gauge that leads to this
system. So keep in mind in what follows that $t=\tau$. What 
canonical 
transformation corresponds to
$$
H =  {p_x^2\over 2m} \longrightarrow H_\alpha = \alpha {p_x^2\over 
2m} 
\mbox{ ? } 
$$
Consider
\beq
q_\alpha = x + (\alpha-1) {p_x \over m} (t-t_0 ) , \: \: \: 
p_\alpha = p_x 
\eeq
where clearly $ \{ q_\alpha,p_\alpha \}  = 1$. Moreover
$$
{dq_\alpha\over dt} ={dx\over dt} + (\alpha -1) {p_x \over 2m }= \{ 
q_\alpha , H_\alpha \}  = \alpha {p_x \over m}  
$$
since $dp_x /dt = 0$ and $dx/dt = p_x /m $. 

Let us study this picture some more. Let $\epsilon =\alpha -1 $, then
$$
q_\alpha  = x + \epsilon {p_x \over m} (t-t_0 ) = x + \epsilon \{ x, p_x 
^2 {(t-
t_0 )\over 2m} \} 
$$
The generator of this infinitesimal canonical transformation is 
\beq
 G = p_x^2 {(t-t_0 )\over 2m} =p_\alpha^ 2 {(t-t_0 )\over 2m}
\eeq
The full canonical transformation is indeed given by 
\beq
W = x p_\alpha  + \epsilon G = x p_\alpha  + \epsilon p_\alpha^ 2 {(t-
t_0 
)\over 2m}
\eeq
then it follows that
\ba\dis
p_x  = {\partial W \over \partial x} =p_\alpha  &\dis q_\alpha  = 
{\partial 
W\over 
\partial p_\alpha}  = x + \epsilon {\partial G \over \partial 
p_\alpha}\\ \dis
H_\alpha = H + {\partial W\over \partial t} =  \alpha H & \mbox{}
\ea
which is what we had before.

In fact, the choice  $\epsilon = -1$ , i.e., $\alpha = 0$ solves
\beq
H(q,p\!=\!{\partial W/ \partial q})+{ \partial W\over \partial t}=0
\eeq
This is the so-called Principal function equation, the equation that 
defines the 
generating function for the canonical transformation that makes the 
hamiltonian zero (here we have actually  described the method for 
any 
hamiltonian independent of the $q$'s, $H=H(p)$).


Let us now look at the interacting case. With the gauge-fixing choice
$ \chi_\alpha = t - f(\tau) $, where $\alpha \equiv \dot{f}(\tau)$---
as
before. It is easy to see that the equations now read
\ba\dis
\dot{x}_i = \dot{f}\; \{ x_i, {\cal H}_{EM}\}             
 & \:\:\: \dis\dot{p}_i= \dot{f} \; \{ p_i , {\cal H}_{EM}\} =-\dot{f} {\partial 
{\cal
H}_{EM} \over \partial x_i}  \\ \dis\dot{t} = \dot{f}  &\:\:\:\dis
 \dot{p}_t = \dot{f} \; \{ p_t , {\cal H}_{EM}\} = 
-\dot{f} {\partial {\cal H}_{EM} \over \partial t}
\ea
which is not terribly surprising. The Dirac bracket computation is 
easily
done, since again $\{ \chi_\alpha , \Phi \}=1$, and it yields
$$ \{x_i,p_j\}_* = \delta_{ij},\: \: \{t,p_t\}_* = 0,\:\: 
\{x_i,p_t\}_* = -\{ x_i ,{\cal H}_{EM}\}_*$$
with all others zero.
The dynamics for functions of $x_i, \, p_j$ are once more given by $$
 \dot{F} = {\partial F \over \partial\tau} +\dot{f}\; \{ F,{\cal 
H}_{EM}\} 
$$ 
Thus, the behavior with respect to changes of gauge-fixing---and 
their
effects on the $RPS^*$---is just as in the free case. The interacting
case does not bring in any new conceptual problems, although the 
dynamics
are more complicated.

\begin{center} ****************************************** \end{center}

Let us review what the use of a parameter has done and comment on 
the 
peculiarities of this system. First notice that the original 
unconstrained 
case 
is 
recovered through the use of a very special gauge: $t=\tau$. This can 
easily 
be seen 
 in the parametrized second order action. In effect, {\em the original 
system is reached through a gauge choice, not  through the gauge 
invariant 
content 
of the theory}. In fact, {\em the gauge invariant content of the 
parametrized case is described by the constants of the motion}---as 
we 
will 
see---{\em which are true  reparametrization invariants!} The 
``time 
evolution'' effect is provided by the use of $\tau$-dependent gauges, 
that 
is, 
by the use of a different gauge at each $\tau$-time---and of gauge 
dependent 
coordinates, of course. Different $\tau$-time-varying gauge choices 
produce 
different 
systems, which as we have seen, are related by canonical 
transformations.

How is this discussion particular to parametrized systems, i.e., what 
is the 
effect in general of  $\tau$-time dependent 
gauge-fixings? The answer is that the same effect in the dynamics 
will 
occur 
in 
any situation where $\tau$-dependent gauge fixings are used.
Consider again the simple system $Q,P,q,p$ with the constraint $P=0$ 
and
first class hamiltonian $h(q,p)$. Let us use the $\tau$-dependent 
gauge
fixing $\chi = Q-f(\tau)$. It is easy to see that the extended 
hamiltonian
is $H_{E'} = h(q,p) + \dot{f} P$, as follows from the usual requirement
that $\dot{\chi} = 0$. Notice that none of this affects the dynamics in
the $RPS$, the space described by the coordinates $q,p$ and  
hamiltonian
$h(q,p)$. However, if we now did a canonical transformation into 
some 
new canonical pairs  ${\cal Q}^a,{\cal P}_a$ that describe
the full phase space after gauge fixing---or $RPS^*$---we will
have some complicated dynamics.

Parametrized systems are 
special, then, in that  we are forced to use such gauge fixings, 
because the boundary conditions demand it. Indeed, the lack of 
gauge 
freedom 
at the boundaries means that one needs to use gauges that match the 
boundary 
requirements. This, as we saw, is a consequence of the non-linearity 
of the 
constraint. We will see this point in a different and clearer
 light at the end of 
the next section.

\subsection{Constant of the motion coordinate system: an example}
\label{sec:constant}
We will study the non-relativistic free case.
In this coordinate system we take the constraint $\Phi$ to be a new 
coordinate, 
say a momentum, $P\equiv \Phi $. Now we need its conjugate 
coordinate 
$Q$,  
which we can define by
$$
 \{Q,P\}= 1 ={ \partial Q \over \partial t }+ {\partial Q\over \partial 
x} 
{p_x\over m}
$$
 The other coordinates $q,p$ must commute with $Q,P$---and we 
saw 
that 
the 
solution to $\{C_\Phi , P\} = 0 $ is given by any function of $p_x, p_t $  or 
$x-p_x  
t/m 
$---and have the right commutation relation among themselves.
 
A possible solution is given by
\ba \dis
Q= {1 \over 2 }(t+m{x\over p_x})  & \dis\; P= p_t + {p_x^2\over 2m} 
& 
\:\: \mbox{\em gauge 
degrees of freedom }\\ \dis
q= {1 \over 2 }(t-{x\over p_x})  &\dis \; p= p_t - {p_x^2\over 2m}  &  \:\:
\mbox{\em physical 
degrees of freedom }
\ea
There is, however a problem with this solution.  In fact,  in the 
surface $ 
\Phi 
= 
0$,  we have $p_t = - {p_x^2/ 2m} \leq 0$ , so in that surface $p = - 
{p_x^2/ 
m} 
\leq 0 $. The physical momentum coordinate doesn't take values 
over the 
full 
range! It is easy to see that the map 
$$(t,x, p_t , p_x)  \longrightarrow (Q, P, q, p)$$
is not  one-to-one. Indeed the two points $(t,x, p_t , p_x)$, and $(t,-x, 
p_t , 
-p_x)$ are mapped to the same $(Q, P, q, p)$. That there are 
problems 
with 
this 
transformation is not surprising, as it is already undefined at $p_x 
=0$. If 
we 
fix, say, $p_x>0$,  then the map will be  one-to-one, of course. 

This bad choice of 
coordinates is an example of the problem we discussed in section 
\ref{sec:rpsc}. 
Our coordinate choice made the constraint into a momentum, but for 
that 
it 
picked up the ``$x$  part'' of the constraint, which suffers from the 
problem 
illustrated above.  Either we change the constraint to 
$p_x = +\sqrt{-2mp_t}$ ---say---or have to deal with the bad square 
form. 
It is simply not possible to make a constraint like $P^2/2 \approx 0 $ 
into 
a 
momentum. It doesn't cover the full range to begin with. 

This analysis will be of relevance again when we deal with the 
relativistic 
case 
and its quadratic constraint \cite{Rel}. 

A proper solution to the problem of finding a set of canonical 
coordinates 
in 
which the constraint is a momentum is given by 
\ba\dis
Q= t-t_0  & \dis \; P= p_t + {p_x^2\over 2m} &\:\:  \mbox{\em gauge degrees 
of 
freedom}\\ \dis
q= p_x (t-t_0) - mx &\dis \; p= - {p_x \over m}  & \:\: \mbox{\em physical 
degrees 
of 
freedom }
\ea
The system seems to be telling us that to make life simple it is $t$ 
that 
has 
to 
be considered pure gauge. The first order action in this coordinates is 
\beq 
A_{Q,P,q,p}= \itif d\tau\; ( P \dot{Q}  + p \dot{q} - w(\tau)P)  
\eeq 
Now, that is simple! Where did all the invariance-at-the-boundaries 
problems 
go? Well, we need to know what the original boundary conditions 
look like 
in 
this coordinate system. Recall that $t$ and $x$ where fixed at the 
boundaries. 
This corresponds to fixing $Q$ and $-q/m-pQ$ at the boundaries. The 
above 
action doesn't have an extremum with such boundary conditions; a 
surface 
term needs to be added, as explained in reference \cite{Teitelpap}. 
Indeed, 
consider the variation of the action:
\ba \dis
\delta A_{Q,P,q,p}&\dis = \delta  \itif ( P\dot{Q} + p\dot{q} - 
w(\tau)P)
d\tau    \\[.5cm]
&\dis = \itif( \delta P(\dot{Q}- w) + 
{d(P \delta Q)\over d\tau}  - \dot{P} \delta Q - \delta w P+
\delta p\dot{q} + {d(p \delta q)\over d\tau}  -  \dot{p}\delta q ) 
d\tau  
 \\[.5cm]
&\dis =\itif( \delta P(\dot{Q}- w) - \dot{P} \delta Q - \delta w P+
\delta p\dot{q}  -  \dot{p}\delta q ) d\tau     
+\left. (P\delta Q+p\delta q)  \right|_{\tau_i}^{\tau_f}
\ea
With the boundary conditions on $x,t$, the surviving surface term is 
\beq 
\left. (p\delta q)\right|_{\tau_i}^{\tau_f}
 = - \left. p_x (t-t_0) {\delta p_x\over m}\right|_{\tau_i}^{\tau_f}\eeq
To get rid of it we need to add the surface term  $ \left.  B
\right|_{\tau_i}^{\tau_f} 
$ to 
the 
action, with 
\beq \left. (p\delta q + \delta B) \right|_{\tau_i}^{\tau_f} =0 \eeq
A solution is 
\beq
B = (t-t_0) {p_x^2\over 2m} = {m \over 2}Q p^2  
\eeq
 Unsurprisingly, the generator of canonical transformations of the 
previous 
section reappears. This is where the lack of gauge invariance at the 
boundaries 
comes from, as this term depends on $Q$. We can rephrase the 
subtleties 
of 
this system as follows: There is nothing special about the action---or 
the 
constraint, as it is simple  enough, $P\approx 0$.  However, 
{\em our insistence on peculiar boundary 
conditions 
makes the addition of a gauge dependent boundary term necesssary 
for 
the 
existence of an extremum. Therefore gauge invariance at the 
boundaries is 
lost.}

 This illustrates the fact that it is not the form of the constraint only 
that 
matters, but also what is fixed at the boundaries: the recipe for no 
trouble 
is 
that {\em the constraint should be linear in the coordinates conjugate to 
whatever 
is 
fixed at the boundaries.}

%
%
\subsection{$RPS$ analysis for the relativistic particle}
\label{sec:RPSRPle}

For the  non-relativistic particle, we have a phase space 
described by the coordinates $t, x, p_x $ and $p_t $, and a constraint 
$
\Phi = p_t^2-  p_x^2-m^2 \approx   0$.
A possible approach is to try to immediately perform a canonical    
transformation such that $\Phi$    becomes a momentum, and to 
proceed 
as 
above. We will look at this approach in section~\ref{sec:constant}.

One approach is to say that the  gauge invariant functions are those 
that 
satisfy 
$\{C_\Phi (q^\alpha, p_\alpha) ,\Phi\} \equiv 0$, i.e.,  
\beq
0 \equiv  (2 p_t {\partial \over \partial t }-2 p_x {\partial \over 
\partial 
x }- m^2) C_\Phi (q^\alpha, p_\alpha)
\eeq
which is solved by 
\beq
C_\Phi = g(t p_x + x p_t,p_x, p_t) e^{t p_x + x p_t - m^2 {x \over 2 p_x}}
\eeq
A more symmetric way to write this is \beq
C_\Phi = g(t p_x + x p_t,p_x, p_t)e^{t p_x + x p_t + {m^2\over p_x p_t} (t p_x 
- x p_t) 
}
\eeq
For the general interacting case it is harder to write the solution, of 
course.

 One really only needs to ask for a weak equality, i.e., that the 
bracket be 
zero on the constraint surface,
\beq \{C_\Phi (q^\alpha, p_\alpha) ,\Phi\} \approx 0\eeq

In this case the constraint surface can be understood to mean either 
the full 
constraint suraface---both branches included---or only one branch.  
For the 
interacting case we have $ \{C_\Phi , \Phi_{EG} \} = \{C_\Phi , (\Pi^0 - B_+ ) g_{00} 
(\Pi^0 - B_- 
)\} = $
\beq
\{C_\Phi , (\Pi^0 - B_+ ) \} g_{00} (\Pi^0 - B_- ) + \{C_\Phi , (\Pi^0 - B_- ) \} g_{00} 
(\Pi^0 - 
B_+ ) = 0
\eeq
where
$$ B_\pm \equiv -\Pi^i \tilde{g}_{0i} \pm \sqrt{ (\Pi^i 
\tilde{g}_{0i})^2 - \Pi^i 
\tilde{g}_{ij} \Pi^j+m^2/g_{00}} $$
We  see that if only one branch is taken the equation is just
$$
\{C_\Phi, (\Pi^0 - B_+ ) \} g_{00} (\Pi^0 - B_- )=0$$
a simpler equation than the ones above! This corresponds to 
``choosing a 
branch''. On the other hand, demanding weak equality on the {\em 
full} 
surface yields \beq
\{C_\Phi, (\Pi^0 - B_+ ) \} \approx \{C_\Phi , (\Pi^0 - B_- ) \} \approx 0
\eeq

Let us now consider fixing the gauge with a  $\tau$-dependent 
gauge, 
$ \chi_\alpha = t - f(\tau) $, where $\alpha \equiv \dot{f}(\tau)$. As 
before, 
we need 
$
\dot{\chi}_\alpha = 0 = {\partial \chi_\alpha  / \partial \tau} +
 \{\chi_\alpha, H_{E'} \}
$ which implies $ v=\alpha /2p_t$, so we have
\beq
H_{E'} = {\alpha \over 2p_t} \Phi  =  {\alpha \over 2p_t} ( p_t^2-  
p_x^2-m^2 
)
\eeq
and the equations of motion in this gauge are
\ba
 \dis \dot{t} = \alpha &\dis \dot{x}  = -{\dot{f} \over p_t} 
p_x\\[.5cm]
\dis\dot{p}_t   = 0 & \dis \dot{p}_x = 0 
\ea 
Using \beq
p_t =  \pm\sqrt{ p_x^2 + m^2}\eeq
we have
\beq
\dot{x}  = \pm\{x,\alpha \sqrt{ p_x^2 + m^2}\} =\pm\dot{f}\{x,\sqrt{ 
p_x^2 + 
m^2}\}
\eeq
so the system is reduced to the coordinates $x, p_x$, with a 
hamiltonian 
\beq
H_\alpha =  \pm\alpha \sqrt{ p_x^2 + m^2}
\eeq

{\em Notice that the hamiltonian comes with two signs: thus propagation
back and forth in time are both  described in this system}

The Dirac bracket computation in this gauge is easily done, since $\{ 
\chi_\alpha, \Phi\} = 2p_t $: let $ \chi_j = \chi_\alpha, \Phi $  as 
before. 
Then  $  \{A,B\}_* =  $
\beq
\{A,B\} - ( \{ A,\chi_\alpha \},\{A, \Phi \}) 
\left( \begin{array}{cc} 0 & -{ \{ 
\chi_\alpha, \Phi\}^{-1}}  \\ { \{ 
\chi_\alpha, \Phi\}^{-1}} & 0 \end{array}   \right)
\left( \begin{array}{c} \{\chi_\alpha,B\} \\ \{\Phi ,B\} \end{array} 
\right)                        
\eeq
and this yields
$$ \{x,p_x\}_* = 1,\: \{t,p_t\}_* = 0,\:
\{x,p_t\}_* = \pm \{ x,  \sqrt{p_x^2 + m^2}\}$$
with all others zero.

Recall that we run into a subtlety with the 
Dirac bracket description of the dynamics. In this gauge, for functions of $x,\, p_x$,  
the 
correct dynamics are described by  equation~\ref{eq:easydyn}
$$
{dF\over d\tau } = {\partial F \over \partial\tau } + \{F, H_\alpha \}
$$                                                         
 which follows from equation (\ref{eq:gdyn}) 
(\cite{Evans}, and see  ex. 4.8 in reference \cite{TeitelBook}). 

In the gauge $\alpha =1$ the connection to the unconstrained case of 
section~\ref{sec:unconstrained} (equation~(\ref{eq:uroot})) with the 
hamiltonian $h$ is clear, but let us 
look at the interpretation 
of the more general gauge-fixings. First notice that  equation 
(\ref{eq:easydyn}) has a very simple interpretation: One can simply 
factor 
out the term
$ \dot{f} $ to read 
$$
{d F \over d t} = {\partial F \over \partial t} + \{F, \pm \sqrt{p_x^2 + 
m^2}\}
$$
Let us show, anyhow,  that this reduced phase space is the same as 
the 
phase space for the unconstrained non-relativistic particle in 
section~\ref{sec:unconstrained}, in 
some unusual $\tau$-dependent coordinates. Recall that the idea 
here is that 
when one picks a gauge, a coordinate system for the reduced phase 
space is 
implicitly chosen---or an active transformation occurs, depending on 
your 
point of view---and that with $\tau$-dependent choices the 
dynamics will 
look different. Let us go back to the unconstrained system of 
section~\ref{sec:unconstrained}: there, $t$ was the time, not $\tau$. 
However, 
as we know from our later point of view $t$ and $\tau$ are one and 
the 
same in the gauge that leads to this
system. So keep in mind in what follows that $t=\tau$. What 
canonical 
transformation corresponds to
$$
H =  \sqrt{p_x^2 + m^2} \longrightarrow H_\alpha = \alpha 
\sqrt{p_x^2 + 
m^2}
\mbox{ ? } 
$$
The full canonical transformation is indeed given by 
\beq
W = x p_\alpha  + \epsilon G = x p_\alpha  + (\alpha -1 ) \sqrt{p_x^2 
+ m^2} 
{(t-t_0 
)}
\eeq
then it follows that
\ba\dis
p_x  = {\partial W \over \partial x} =p_\alpha  &\dis q_\alpha  = 
{\partial 
W\over 
\partial p_\alpha}  = x + (\alpha -1 ){\partial G \over \partial 
p_\alpha}\\ 
\dis
H_\alpha = H + {\partial W\over \partial t} =  \alpha H & \mbox{}
\ea

The choice $\alpha = 0$ solves
\beq
H(q,p\! = \! {\partial W/ \partial q})+{ \partial W\over \partial t}=0
\eeq
Again, this is the so called Principal function equation, the equation that 
defines the 
generating function for the canonical transformation that makes the 
hamiltonian zero (here we have actually  described the method for 
any 
hamiltonian independent of the $q$'s, $H=H(p)$).


How about  the interacting case? With the gauge-fixing choice
$ \chi_\alpha = t - f(\tau) $, where $\alpha \equiv 
\dot{f}(\tau)$---as
before---we have $v= \dot{f} / \{ t, \Phi_{EG} \}$. 
We can next compute the bracket \beq \label{eq:bracket} \{ t, 
\Phi_{EG} \} =
2g^{00} (p_0 - A_0) +  2g^{0i} (p_i - A_i) = 2 g^{\mu 0} \Pi_\mu \eeq
The extended hamiltonian is given then by \beq H_{E'} = { \dot{f} \ov 
2 
\Pi_\mu g^{\mu 0}} \Phi_{EG} = { \dot{f} \ov 2 \Pi_\mu g^{\mu 0}} 
(\Pi_0 - 
B_{+} ) g^{00} (\Pi_0 - B_{-} ) \eeq
where $+$ and $-$ stand for the branches: $$ B_\pm \equiv -\Pi_i 
\tilde{g}^{0i} \pm \sqrt{ (\Pi_i \tilde{g}^{0i})^2 - \Pi_i 
\tilde{g}^{ij} \Pi_j+m^2/g^{00}} $$ 
and $\tilde{a} = a/g^{00}$.

Let us now look at the general equation of motion. The idea is to look 
at the 
branch decomposition of the constraint above.  One has to choose, 
ultimately, 
a branch in which to ``be'', for example when the initial conditions 
are 
chosen. It is then easy to work out the general equation of motion,
\beq
\label{eq:branchdyn}
\dot{A} = \left. {\partial A \ov \partial \tau} + \dot{f} \{ A, \Pi_0 - B_\pm \} 
\right|_{\Pi_0 = B_\pm} 
\eeq
\begin{quote}
{\em 
The conclusion is that at the end one of the branches gets chosen, and 
with 
the above gauge fixing in $t$ the system in $x_i , p_{x_i}$ behaves as 
that 
corresponding to a 
reduced system   with a hamiltonian $h= -A_0 - B_\pm $, just as in 
the non-relativistic case. }
\end{quote}

The Dirac bracket computation can  now be
done;  it yields
$$ \{x_i,p_j\}_* = \delta_{ij},\: \: \{t,p_t\}_* = 0,\:\: 
\{f(x_i,p_i),p_t\}_* \approx \{f(x_i,p_i), A_0 + B_\pm\}$$
with all others zero, where a branch has been chosen.
Thus, the behavior with respect to changes of gauge-fixing---and 
their
effects on the $RPS^*$---is just as in the free case. The interacting
case does not bring in any new conceptual problems, although the 
dynamics
are more complicated. The only confusing aspect may be that  it is 
$\Pi_0$ that comes with two signs. Using the Legendre transform it is easy to
see, however, that this combination is indeed $\dot{t}$---and this is what comes 
with
two signs. Thus, the interpretation of the presence of  two branches as a 
representation
of back and forth motion in time remains valid.

\subsection{$RPS$  analysis for minisuperspace}
\label{sec:RPSmini}
This analysis in this section leads to the same situation as in the 
relativistic 
case, since 
mathematically the relativistic particle is just as minisuperspace. As 
mentioned above the constraint in this system is 
\beq
\Phi_M \equiv P_A P_B g^{AB} - U(Q) \approx 0
 \eeq
for a gauge fixing function of the form $ \chi = Q^0 -f(\tau)$ ---where 
we 
tentatively assign the ``time-keeping'' role to the scale of the 
universe, it is 
useful to rewrite the constraint as
\beq
\Phi_M \equiv (P_0 - {\cal B}_+ ) g^{00}(P_0 - {\cal B}_- ) \approx 0
\eeq
where
\beq
{\cal B}_\pm \equiv 
\pm\sqrt{ (P^I 
\tilde{g}_{0I})^2 - P^i \tilde{g}_{IJ} P^J +\tilde{U} }
\eeq
With the bracket
\beq
\{\chi , \Phi_M\} = 2 g_{A0} P^A \eeq
the Dirac Bracket computation yields
$$ \{Q^I,P_J\}_* = \delta_{IJ},\: \: \{Q^0 ,P_0\}_* = 0,\:\: 
\{f(Q^I,P_J),P_0\}_* \approx \{f(Q^I,P_J), {\cal B}_\pm \}$$
with all others zero, where a branch has been chosen,
and the  dynamics are once more given---for functions on
the reduced coordinates---by 
$$
 \dot{F} = {\partial F \over \partial\tau} +\dot{f}\; \{ F,{\cal 
B}_\pm\} 
$$ 
Notice that once more if one tries to force the hamiltonian philosophy 
on a 
system of this type we find that we need two hamiltonians---with 
the above 
gauge fixing. This unusual situation will be examined further when 
we study the
quantization of these systems. 

Finally, notice that we have the following interpretation for the 
appearance of the 
two branches: \\
{\em Classicaly we have that the effective hamiltonian in the 
reduced phase space 
comes with two signs---one for each branch; this produces a forward 
and backward 
propagation in ``time'', the
conjugate variable to the hamiltonian.}
  
\newpage\section{BRST extended phase space}
\label{sec:BFVc}
Let us now review the BRST treatment of a system with a constraint 
$\Phi$ 
\cite{TeitelBook,BFV}.  We will consider some of the above simple 
systems as 
examples of 
straightforward applications of this formalism.The first objects that 
are 
introduced into the 
extended 
phase space are $\lambda$  and its conjugate momentum $\pi$ , $\{ 
\lambda 
,\pi \} =1$. Since $\lambda $ is arbitrary, its momentum is 
constrained, 
$\pi 
\approx 0$. We thus have two constraints. To each constraint, the 
rules 
say 
we 
must associate a conjugate pair of ghosts, $\eta_0 ,\rho_0 $ and 
$\eta_1 
,\rho_1$  with $\{ \eta_0 ,\rho_0 \} =1$ and $\{ \eta_1 ,\rho_1 \} 
=1$, 
other 
(super)brackets being zero. The total extended phase space is thus 
described 
by $t, p_t , x, p_x , \lambda , \pi , \eta_0 , \rho_0 , \eta_1 , \rho_1 $.  
The 
next 
thing to do is to define the BRST generator, which generates the gauge 
transformations in extended phase space: 
\beq
\Omega= \eta_0  \Phi  + \eta_1  \pi 
\eeq
 We also need a gauge-fixing function, ${\cal O}$. The dynamics are 
then 
generated by the hamiltonian ${\cal H}  = h + \{ {\cal O}, \Omega\} $   
($= 
\{ 
{\cal O}, \Omega\} $, since the original first class hamiltonian $h$ is 
zero.) 
The 
first order action is then
\beq
S = \itif (\dot{q}^\alpha p_\alpha - {\cal H} ) d\tau  =  \itif (\dot{t} 
p_t + 
\dot{x} p_x  + \dot{\lambda}  \pi  + \dot{\eta}_0 \rho_0  + 
\dot{\eta}_1  
\rho_1  - \{ {\cal O}, \Omega\} ) d\tau 
\eeq
We will use the ``non-canonical'' gauge fixing 
${\cal O}_{NC} = \rho_1 f(\lambda ) +\rho_0  \lambda $  and also the
``canonical'' one ${\cal O}_C = \rho_1  \chi    + \rho_0  \lambda $, 
where
$\chi$  is a function of the original phase space  variables only---the 
Dirac
bracket gauge-fixing function: $\chi = \chi (t,x, p_x  , 
p_t ,\tau )$. The terminology will become clearer as we go on.

Consider 
first 
the 
non-canonical gauge for the non-relativistic particle. The hamiltonian 
is $  {\cal 
H}_{NC} = \{ {\cal O}_NC, \Omega\}  = $ 
 \beq
\{ \rho_1 f(\lambda ) +\rho_0  \lambda , \eta_0  \Phi  + \eta_1  \pi 
\}  = 
\rho_1 \eta_1  f'(\lambda )+ \pi f(\lambda )+ \lambda  \Phi  + 
\rho_0 
\eta_1 
\eeq
Let us assume that the proper boundary conditions are imposed so 
that all 
surface terms that arise under the variation of the action vanish. 
Then the 
canonical equations of motion are
\ba \dis 
\dot{x} = \{ x,{\cal H}_NC\}  = {\partial {\cal H}_{NC} \over \partial 
p_x } = 
{\lambda  p_x \over m }\; \;\;    &\dis \dot{p_x }  =0  \\ \dis  
\dot{t}  = \lambda   &\dis  \dot{p_t }=0  \\ \dis 
\dot{\lambda }  = f(\lambda )  & \dis \dot{\pi } = - \rho_1 \eta_1  
f''(\lambda 
)- 
\pi f'(\lambda )- \Phi  \\ \dis 
\dot{\eta_0 } =   \eta_1    &  \dis \dot{\rho_0 } = 0 \\ \dis 
\dot{\eta_1 }  = 0     & \dis 	\dot{\rho_1} = 0 
\ea
This can be summarized by saying that the coordinates $t,x,p_t ,p_x $ 
and 
$\lambda , \pi$  have dynamics independent from the ghosts, and 
that it 
is 
clear that $dx/dt = p_x /m = constant$---just as in the unconstrained 
case, 
$\dot{\Phi }  = \dot{\pi } =0$ for the right initial conditions---and 
that we 
have 
the gauge fixing $d^2 t/d\tau^2 = f(\lambda )$. So despite all the new 
formalism, the dynamical bottom line is always the same.

For the ``canonical'' gauge we have  $ {\cal H}_C = \{ {\cal O}_C, 
\Omega\}  = $
\beq 
\{ \rho_1  \chi   + \rho_0  \lambda , \eta_0  \Phi  + \eta_1  \pi \}  =
\rho_1  \eta_0  \{ \chi , \Phi \}  + \pi  \chi   + \lambda  \Phi  + 
\rho_0
\eta_1  \eeq
and the equations of motion are 
\ba  \dis 
\dot{x} = \{ x,{\cal H}_C\}  = {\partial {\cal H}_C \over \partial p_x } 
= \rho_1 \eta_0 { \partial \{ \chi , \Phi \} \over \partial p_x } + \pi  
{\partial \chi  \over \partial p_x}   + {\lambda  p_x \over m} \;\;\; & 
\dis 
 \dot{p_x }  = -\rho_1 \eta_0  {\partial \{ \chi , \Phi \} \over \partial 
x} - 
\pi  {\partial \chi  \over \partial x}    \\ \dis 
\dot{t}  = \rho_1 \eta_0  {\partial \{ \chi , \Phi \}\over \partial p_t } 
+ \pi  
{\partial \chi \over \partial p_t }  + \lambda & \dis 
\dot{p_t}=  -\rho_1 \eta_0  {\partial \{ \chi , \Phi \} \over \partial t} 
- 
\pi  {\partial \chi  \over \partial t}   \\ \dis 
\dot{\lambda }  = \chi     &  \dis \dot{\pi } = - \Phi  \\ \dis 
\dot{\eta_0 } = \eta_1      &  \dis \dot{\rho_0 } = -\rho_1  \{  \chi , 
\Phi \}  
\\ \dis 
\dot{\eta_1 }  = \eta_0  \{ \chi , \Phi \}    &	 \dis \dot{\rho_1} = - 
\rho_0 
\ea
 For illustration purposes  let us use the gauge $\chi  = t-\alpha\tau 
-t_0$, 
so 
that $\{ \chi , \Phi \}  =1$. Then the equations of motion become
\ba  \dis 
\dot{x}   = \lambda  {p_x \over m}\;\;\; & \dis  \dot{p_x }  = 0  \\ 
\dis 
\dot{t} = \lambda    &   \dis   \dot{p_t } = - \pi  \\  \dis 
\dot{\lambda }  = \chi  & \dis  \dot{\pi } = - \Phi   \\ \dis 
\dot{\eta_0 } =   \eta_1        & \dis  \dot{\rho_0 } = -\rho_1  \\  \dis 
\dot{\eta_1 }  = \eta_0   &  \dis  \dot{\rho_1 } = - \rho_0 
\ea
Again, we can summarize this by writing $dx/dt = p_x /m =$ 
constant as 
before, $d^2t/d\tau^2 = \chi $  is the gauge-fixing , and $ d\Phi 
/d\tau  = 
-\pi   
=0$  if the initial conditions satisfy  $\Phi  = 0 = \pi $ , since $d\pi 
/d\tau  
= 
-\Phi $.  So, again, the dynamics stay  the same.  

Consider next the relativistic case with a  
non-canonical gauge. Again, the hamiltonian is
 \beq
{\cal H}_{NC} = \{ {\cal O}_{NC}, \Omega\}  = 
\{ \rho_1 f(\lambda ) +\rho_0  \lambda , \eta_0  \Phi  + \eta_1  \pi 
\}  = 
\rho_1 \eta_1  f'(\lambda )+ \pi f(\lambda )+ \lambda  \Phi  + 
\rho_0 
\eta_1 
\eeq
Let us assume that the proper boundary conditions are imposed so 
that all 
surface terms that arise under the variation of the action vanish. 
Then the 
canonical equations of motion for the free case are
\ba \dis 
\dot{x} = \{ x,{\cal H}_{NC}\}  = -\lambda 2 p_x 
\; \;\;    &\dis \dot{p_x }  =0  \\ \dis  
\dot{t}  = \lambda 2 p_t  &\dis  \dot{p_t }=0  \\ \dis 
\dot{\lambda }  = f(\lambda )  & \dis \dot{\pi } = - \rho_1 \eta_1  
f''(\lambda 
)- 
\pi f'(\lambda )- \Phi  \\ \dis 
\dot{\eta_0 } =   \eta_1    &  \dis \dot{\rho_0 } = 0 \\ \dis 
\dot{\eta_1 }  = 0     & \dis 	\dot{\rho_1} = 0 
\ea
This can be summarized by saying that the coordinates $t,x,p_t ,p_x $ 
and 
$\lambda , \pi$  have dynamics independent from the ghosts, and 
that it 
is 
clear that $dx/dt = -p_x /p_t =  constant$---as in the unconstrained 
case again, 
$\dot{\Phi }  = \dot{\pi } =0$ for the right initial conditions---and 
that we 
have 
the gauge fixing $d^2 t/d\tau^2 = f(\lambda )$. So despite all the new 
formalism, the dynamical bottom line is always the same. 

For the ``canonical'' gauge we have ${\cal H}_C = \{ {\cal O}_C, 
\Omega\}  =$
\beq
 \{ \rho_1  \chi   + \rho_0  \lambda , \eta_0  \Phi  + \eta_1  \pi \}  =
\rho_1  \eta_0  \{ \chi , \Phi \}  + \pi  \chi   + \lambda  \Phi  + 
\rho_0
\eta_1  \eeq
and the equations of motion are 
\ba  \dis 
\dot{x} = \{ x,{\cal H}_C\}  = \rho_1 \eta_0 { \partial \{ \chi , \Phi \} \over \partial 
p_x } +
\pi   {\partial \chi  \over \partial p_x}   + - 2 \lambda  p_x  \;\;\; & \dis 
 \dot{p_x }  = -\rho_1 \eta_0  {\partial \{ \chi , \Phi \} \over \partial 
x} - 
\pi  {\partial \chi  \over \partial x}    \\ \dis 
\dot{t}  = \rho_1 \eta_0  {\partial \{ \chi , \Phi \}\over \partial p_t } 
+ \pi  
{\partial \chi \over \partial p_t }  + 2\lambda p_t & \dis 
\dot{p_t}=  -\rho_1 \eta_0  {\partial \{ \chi , \Phi \} \over \partial t} 
- 
\pi  {\partial \chi  \over \partial t}   \\ \dis 
\dot{\lambda }  = \chi     &  \dis \dot{\pi } = - \Phi  \\ \dis 
\dot{\eta_0 } = \eta_1      &  \dis \dot{\rho_0 } = -\rho_1  \{  \chi , 
\Phi \}  
\\ \dis 
\dot{\eta_1 }  = \eta_0  \{ \chi , \Phi \}    &	 \dis \dot{\rho_1} = - 
\rho_0 
\ea
 For illustration purposes  let us use the gauge $\chi  = t-\alpha\tau 
-t_0$, 
so 
that $\{ \chi , \Phi \}  =1$. Then the equations of motion become
\ba  \dis 
\dot{x}   = \lambda  {p_x \over m}\;\;\; & \dis  \dot{p_x }  = 0  \\ 
\dis 
\dot{t} = \lambda    &   \dis   \dot{p_t } = - \pi  \\  \dis 
\dot{\lambda }  = \chi  & \dis  \dot{\pi } = - \Phi   \\ \dis 
\dot{\eta_0 } =   \eta_1        & \dis  \dot{\rho_0 } = -\rho_1  \\  \dis 
\dot{\eta_1 }  = \eta_0   &  \dis  \dot{\rho_1 } = - \rho_0 
\ea
Again, we can summarize this by writing $dx/dt = -p_x /p_t  =$ 
constant as 
before, $d^2t/d\tau^2 = \chi $  is the gauge-fixing , and $ d\Phi 
/d\tau  = 
-\pi   
=0$  if the initial conditions satisfy  $\Phi  = 0 = \pi $ , since $d\pi 
/d\tau  
= 
-\Phi $.  So, again, the dynamics stay  the same.

Finally, let us remind the reader that in previous work we showed 
that a 
constraint rescaling in the BRST action is equivalent---via a canonical 
tranformation---to  a rescaling of  the gauge fixing function. We will 
elaborate when we discuss the BFV path integral (see 
section~\ref{sec:BFVpi}).

Notice that in this formalism there are no constraints---as is usually 
said, ``the local 
gauge invariance is made rigid". The gauge-fixing has been 
implemented in the dynamics, {\em
and} the dynamics of                       BRST-invariant functions do not 
depend on the 
gauge-fixing. This is what sets BRST apart from Dirac.  We can use  a 
gauge-fixed action in the
Dirac case as well, but we have to use the Dirac bracket---solve 
the constraints---or alternatively, find functions that commute 
with both the constraints and the gauge-fixing. Here the 
ghosts take care of this, i.e.,  of the reduced simplectic geometry (we will 
see this explicitely when we study the path integrals.)
And we don't even have to reduce. Gauge-fixing does not 
interfere with BRST invariance.

  \newpage\section{Conclusions, summary} 

In this chapter I have begun by reviewing the classical  aspects
of the actions for the non-relativistic
and  relativistic particles as well as minisuperspace, and studied their  invariances
at the lagrangian level. We have seen that the actions are invariant only up to
boundary terms \cite{Teitelpap}.

 Some of the actions, we have seen, carry a  representation of the full
reparametrization group, including its disconnected part  ($Z_2$). The disconnected
part is broken by a background electromagnetic field, however. Continuing with the
review,   Dirac's formalism has been applied to  our parametrized systems, and we have
seen  that they are constrained systems of a special kind: because the actions  are
 invariant only up to boundary terms, the constraints are not linear in the
momenta,  and, moreover,  in the more interesting cases of the relativistic particle
and minisuperspace, the constraint surfaces are disconnected. These are both
well-known facts, and they are the source of serious conceptual and technical problems
in the  quantization process of parametrized systems.

 I have then argued that to be consistent at the classical level  we have to pick a
branch in the cases where  the constraint surface splits---the relativistic case as
well as minisuperspace. The  two branches correspond to two hamiltonians in the
reduced phase space that  generate either forward or backward time displacements.  

I have then made use of the 
reparametrization invariance to  construct---with  the Dirac bracket---the reduced 
phase spaces, and showed that this construction   depends on the gauge-fixing
employed---a  disturbing new feature of these ``gauge'' systems, and a direct
consequence of the actions' lack of invariance at the boundaries. 
However,I have showed that  these reduced phase spaces, $RPS^*_\chi$,  are related
 by time-dependent canonical transformations. 

Finally, I have reviewed the BRST phase phace construction, emphasizing that 
it possesses the advantadge of incorporating both the constraint and the gauge-
fixing in the action---the ghosts take care of the ``reduced phase space''  simplectic 
properties.


\chapter{Canonical Quantization}
\label{sec:quantum}
In this chapter I  will review   the different quantization schemes 
that we 
will apply to our systems. I will pay special 
attention to the approach in which the reduced phase space is 
quantized, 
that is, to the {\em constrain then quantize} method, since it is the 
most 
intuitive and immediate. I will also give a formal introduction to 
the {\em 
quantize then constrain} approaches, but  the more difficult 
questions of the 
precise definition of the Hilbert spaces involved and the fundamental 
questions of the definition and properties of the inner product and 
unitarity 
will be delayed until  the next chapter. The application of all these           
ideas to the 
construction  of path integrals will be considered in the chapter after 
that.

  For the 
relativistic case two approaches will be taken. In the first one, one 
branch of 
the constraint will be chosen. The solution space of the general 
constraint for 
the relativistic particle is given by  equation (\ref{eq:branches}),\beq
\Pi_0 = -\Pi_i \tilde{g}^{0i} \pm \sqrt{ (\Pi_i \tilde{g}^{0i})^2 - \Pi_i 
\tilde{g}^{ij} \Pi_j +m^2/g^{00}} \eeq
where $ \Pi \equiv p - A  $, and $ \tilde{a} \equiv a/ g^{00} $. The 
situation 
for minisuperspace is very similar since the constraint is  essentially the same 
as the 
above one without the electromagnetic field.

In the  second approach we will  look at the possibility of quantizing 
both 
the branches at the same time.
 

\newpage\section{Constrain, then quantize: $RPS$ quantum 
mechanics}
\label{sec:cq}

Reduced phase space quantization is the simplest and most 
transparent 
quantization approach. If one knows what the true degrees of 
freedom in the 
system are it should be simple to quantize them. There are,  in the 
present 
situation some  problems which make the systems we have been 
studying 
interesting: one of them is the issue of  lack of invariance at the 
boundaries.
In the previous chapter I explained that different gauge 
fixings correspond to a different choice of coordinates to describe the 
reduced 
phase space. These correspond to different coordinate choices in the 
unconstrained case, and as Dirac explains canonical transformations 
correspond 
to unitary transformations in quantum mechanics.  The only 
complication 
may  arise from the time dependence of the transformation. 

Other problems in the qunatization process involve the branching of the constraint
surface and  unitarity.

Let us first look at the non-relativistic case. There are no branches 
and the 
$RPS$ hamiltonian is always unitary.
This example will teach us about  the relation that exists between 
{\em 
unitary transformations,  `` pictures'' , canonical 
transformations and different gauge fixings}. 
 
In the $RPS$ approach we  have---after the gauge fixing
         $$\chi _\alpha =t-f(\tau )$$ some coordinates/operators
 ${\bf q_\alpha , p_\alpha} $ and a hamiltonian $\bH_\alpha$, acting 
on a 
Hilbert space of states $|\psi \r  _\alpha $. Time evolution is then 
given 
by the 
Schr\"{o}dinger equation 
\beq
 i{\partial \over \partial \tau}  |\psi \r  _\alpha  = \bH_\alpha  |\psi 
\r  _\alpha 
\eeq
Taking the point of view that different gauge fixings correspond to 
canonical 
transformations of the unconstrained system in 
chapter~\ref{sec:unconstrained} 
generated by
\beq 
W_\alpha  = q_1   p_\alpha  + (\alpha -1) {p_\alpha^2(t-t_0 )\over 
2m} 
\equiv q_1   p_\alpha  + G_\alpha 
\eeq
and remembering---see for example \cite{DiracBook}---that this 
corresponds 
to a unitary transformation of the form
\beq 
 \bf O\longrightarrow  O_\alpha  = U_\alpha O U_\alpha ^{-1} 
\approx  O -i  
[       G_\alpha , O    ]
\eeq
with \beq {\bf U_\alpha  = e^{iG_\alpha} }  \eeq
we see that these correspond to  ( $\alpha =1$ is the reference 
system, e.g., 
$x_1=x$)
\beq {\bf
x\longrightarrow  x_\alpha  = U_\alpha xU_\alpha^{-1} = U_\alpha 
(U_\alpha^{-1}
 x - [U_\alpha^{-1},x])= x +} (\alpha -1)(t-t_0 ){\bp_x \over m}
\eeq just as in the classical Poisson bracket formulation. This 
transformation also affects the eigenstates of operators. 
The new basis states for the operator    ${\bf B}$, for instance, change as
\beq {
|b\r  _1 \longrightarrow|b\r  _1 ={\bf U_\alpha} |b\r  _\alpha  }.
\eeq
For example, the case $\alpha  = 0,$ which we saw corresponds to 
$H=0$ 
classically, leads to $$x_0 =  x - (t-t_0 )p/m , p_0 = p$$ Then the 
system is 
described by the eigenstates of $x_0$, 
\beq
                 \bx_0|x\r  _0 = \bx|x\r   
\eeq
which we can check are 
\beq
|x\r  _0 = e^{-i (t-t_0 ){\bp_x^2\over 2m}} |x\r  _1 = {\bf U}_0 |x\r  _1
\eeq
since  
\beq
\bx_0|x\r  _0 = {\bf U_0 x U_0^{-1} }|x\r  _0 =  {\bf U_0 x}|x\r  _1 = 
x|x\r  _0
\eeq
Notice also that 
\beq 
\l p|x\r  _\alpha  = e^{ -i   p_x x  - i(t-t_0 ){p_x^2 \over 2m}}
\eeq
In the old coordinates ( $\alpha = 1, G=0$), the new basis states are
\beq
\l x|q\r  _0 =\left({m\over 2\pi i(t-t_0 )}\right)^{1/2} e^{ i(x-
q)^2m\over 2(t-t_0 ) 
}
\eeq
These states form a complete set and they are also orthonormal,
\ba \dis 
I = \int dq \;  |q\r  _0 \; _0\l q| \\ \dis
\l x|x'\r   = \int  dq\;  \l x|q\r \!  _0\; _0\! \l q|x'\r   = \delta (x-x')
\ea
and $\; _0\! \l q|q'\r\!  _0 = \delta (q-q')$\\

As for the wave-functions we can take two different points of 
view---continuing
with the case $\alpha  = 0$:
\begin{quote}
i) they are unaffected. $|\psi \r  _0 = |\psi \r  $    , 
 i.e., $$ i{\partial \over \partial 
t}|\psi \r  _0 = \bH |\psi \r  _0.$$
Then
\beq
 {d\over dt}\; {}_0\! \l q|\psi \r   ={d\over dt}\l q|{\bf U_0^{-1}} e^{ -i(t-
t_0 ) 
{p_x^2 \over 
2m} 
}  |\psi_{in}\r   = {d\over dt}\l q|\psi_{in}\r    = 0
\eeq
This is `` all we have done is a change of coordinates''  point of view.
\end{quote}
\begin{quote} 
ii) The states do change, $|\psi \r  _0 = {\bf U_0^{-1}} |\psi \r  $.
Then $i{d\over dt}|\psi \r  _0 = 0$, but \beq
i{d\over dt}( _0\l q |\psi \r  _0 ) = i{d\over dt}\l q |\psi \r   = 
{\bp_x^2\over 2m} \l q |\psi \r  
\eeq
which connects nicely with the $RPS$ description above (write 
$\tau$  for $t$)
\end{quote}
Regardless of the point of view, it is easy to see now that this 
formalism 
corresponds to the 
Heisenberg picture of quantum mechanics. 
The second point of view makes this more obvious perhaps, as the 
states are 
frozen. \begin{quote}
{\em
The punchline is that the quantization of the reduced phase space 
yields the 
quantum mechanics of the unconstrained case, although possibly in 
different 
coordinates---or pictures. } \end{quote}

Let us now study the relativistic case.
For the relativistic case the situation can be  similar to the    
non-relativistic 
case above.  The reduced phase 
space 
structure is complicated because of the branches, it is disconnected; 
in 
consequence one avenue is to quantize one (or both) of the branches 
separately. This would seem the lowest efford extension of the ideas 
in the 
simpler case above. Indeed we can proceed in this manner for the 
free case, 
as well as for some other situations which we will study below.  Let 
us look 
at the free case first. 

As before, in the $RPS$ approach we  have (after the gauge fixing
         $\chi _\alpha =t-f(\tau )$) some coordinates/operators
 ${\bf q_\alpha , p_\alpha} $ and a hamiltonian $\bH_\alpha$, acting 
on a 
Hilbert space of states $|\psi \r  _\alpha $ whith time evolution  
given 
by the 
Schr\"{o}dinger equation 
\beq
 i{\partial \over \partial \tau}  |\psi \r  _\alpha  = \bH_\alpha  |\psi 
\r  _\alpha 
\eeq
Taking the point of view that different gauge fixings correspond to 
canonical 
transformations of the unconstrained system in 
chapter~\ref{sec:unconstrained} 
generated by
$$
W_\alpha  = q_1   p_\alpha  + (\alpha -1) \sqrt{p_x^2 + m^2}\; (t-t_0) 
\equiv q_1   p_\alpha  + G_\alpha 
$$
and remembering---see for example \cite{DiracBook}---that this 
corresponds 
to a unitary transformation of the form
\beq 
{\bf O\longrightarrow  O_\alpha  = U_\alpha O U_\alpha ^{-1} 
\approx  O -i  
[ \bf G_\alpha ,O ]  }
\eeq
with \beq {\bf U_\alpha  = e^{iG_\alpha} }  \eeq
we see that these correspond to  ($\alpha =1$ is the reference 
system, e.g., 
$x_1=x$)
\beq {\bf
x\longrightarrow  x_\alpha  = U_\alpha xU_\alpha^{-1} = U_\alpha 
(U_\alpha^{-1}
 x - [U_\alpha^{-1},x])= x +} (\alpha -1)(t-t_0 ) {\bp_x 
\ov\sqrt{p_x^2+m^2}}
\eeq
and as before the new basis states for the operator ${\bf B}$
\beq {
|b\r  _1 \longrightarrow|b\r  _1 ={\bf U_\alpha} |b\r  _\alpha  }.
\eeq
For example, the case $\alpha  = 0,$ which we saw corresponds to 
$H=0$ 
classically, leads to $x_0 =  x - (t-t_0 )p_x/\sqrt{p_x^2+m^2}, p_0 = 
p$. Then 
the system is 
described by the eigenstates of $x_0$, 
\beq
                 \bx_0|x\r  _0 = \bx|x\r   
\eeq
which we can check are 
\beq
|x\r  _0 = e^{-i (t-t_0 ) \sqrt{\bp_x^2+m^2}} |x\r  _1 = {\bf U}_0 |x\r  
_1
\eeq
Notice also that 
\beq 
\l p|x\r  _\alpha  = e^{ i(  p_x  - (t-t_0 )\sqrt{p_x^2+m^2})}
\eeq
In the old coordinates ( $\alpha = 1, G=0$), the new base states are
\beq
\l x|q\r  _0 = \int dp_x e^{i(p_x (x-q) - (t-t_o)\sqrt{p_x^2+m^2} )}
\eeq
These states form a complete set and they are also orthonormal, just as before, and
as for the wave-functions, we can, as before,  take two different 
points of 
view---continuing
with the case $\alpha  = 0$:
\begin{quote}
i) they are unaffected. $|\psi \r  _0 = |\psi \r  $    , 
 i.e., $$ i{\partial \over \partial 
t}|\psi \r  _0 = \bH |\psi \r  _0.$$
Then
\beq
 {d\over dt}{}_0\l q|\psi \r   ={d\over dt}\l q|{\bf U_0^{-1}} e^{ -i(t-
t_0 ) 
\sqrt{\bp_x^2+m^2} 
}  |\psi_{in}\r   = {d\over dt}\l q|\psi_{in}\r    = 0
\eeq
the `` all we have done is a change of coordinates''  point of view, or
\end{quote}
\begin{quote} 
ii) The states do change, $|\psi \r  _0 = {\bf U_0^{-1}} |\psi \r  $.
Then $i{d\over dt}|\psi \r  _0 = 0$, but \beq
i{d\over dt}( _0\l q |\psi \r  _0 ) = i{d\over dt}\l q |\psi \r   = 
\sqrt{\bp_x^2+m^2} \l q |\psi \r  
\eeq
which connects nicely with the $RPS$ description above (write 
$\tau$  for $t$)
\end{quote}

Now, there are some problems with the use of a square root 
hamiltonian,  like causality. The free case---as well as some others we will 
discuss 
later---can be first-quantized fairly easily. Unitarity is not a 
problem. 

Let us look at the general case. The constraint is essentially \beq
\Pi_0 = -\Pi_i \tilde{g}^{0i} \pm \sqrt{ (\Pi_i \tilde{g}^{0i})^2 - \Pi_i 
\tilde{g}^{ij} \Pi_j +m^2/g^{00}} \eeq
We could try to choose a branch. However, now the problem could be  
unitarity. We 
saw that the hamiltonian in the reduced phase space is 
\beq  h=-A_0-B_\pm = \mp 
\sqrt{(\Pi_i \tilde{g}^{0i})^2 - \Pi_i 
\tilde{g}^{ij} \Pi_j +m^2/g^{00}}\eeq
Now this hamiltonian will become an operator---with some suitable 
ordering.  Under what circumstances there exists an ordering such 
that the 
hamiltonian operator is hermitean? We can ignore the first term, 
$A_0$, as it 
is already hermitean.  
It is in fact sufficient to find an ordering such that \beq
(\Pi_i \tilde{g}^{0i})^2 - \Pi_i 
\tilde{g}^{ij} \Pi_j +m^2/g^{00} \eeq
is hermitean, and hopefully   positive definite, since the square root 
of a 
positive definite 
hermitean operator is hermitean. For example, consider the case of a 
flat 
background. It is easy to 
see then that the above hamiltonian is hermitean, no matter what 
the gauge 
potential happens to be. The problem, however, is that we are not 
assured at all that
the resulting theory will be covariant. As we discussed earlier,
in the unconstrained situation---which is where we are after 
reduction,
it may not be possible to find a covariant and hermitean ordering.
It is possible, however, if there is a frame in which the electric field is
zero.
 
In the case of a curved background there are many questions.
What about our foliation choice? Our choice of time coordinate? Do
these affect the resulting physics? Well, our discussion before 
indicates
that if we change the gauge the resulting theories will be related
by a canonical transformation. But, of course, the starting point
occurs when we choose a branch. It is unclear, in the 
general case,  what will result if we
solve the constraint equation in a different way---i.e., by
choosing the other branch, or by solving for $p_x$ instead, say.
For the flat background case we have only Lorentz transformations
to worry about.

\subsection{Quantization of both branches}
\label{sec:rpsqboth}
Let us look next at the full constraint---at both the branches. 
We can start by constructing one quantum theory with each
branch. We can then take the direct sum of the theories by
keeping the inner product and hamiltonian and the rest of the 
operators
``diagonal''. 
For example, we represent the states by arranging them into two-vectors, 
with one entry for each branch. The inner product prevents that the 
two sectors talk to each other, and time evolution doesn't allow a 
transition either.

This is the trivial construction. It is consistent as
long as there are no covariance issues (like demanding that the 
theories
thus constructed by different observers be the same, which will fail
to be true in complicated metric cases, or if there is an electric field).

But we can also try to allow for interaction.
We would like to keep our one-branch energy eigenstates
in our new theory.  However, we run into an 
immediate problem. If the resulting ``universal'' hamiltonian
is to be hermitean in this new ``universal'' inner product, then
it follows that  eigenstates of the hamiltonian 
corresponding to different eigenvalues will have to be orthogonal.
Thus, for hermicity of the hamiltonian and unitarity of the theory
we need $$(\Psi_{E}, \hat{H} \Psi_{E'})=E'(\Psi_{E},   
\Psi_{E'})=E(\Psi_{E},  \Psi_{E'})=0$$ if $E'\neq E$.
This implies that time evolution cannot take you from one sector to
the other, and the theory is unitary within each sector already unless 
the sectors overlap---unless they have states with the same 
eigenvalues. For the free case there is no such overlap. There is no 
overlap either if the zero component of the electromagnetic potential 
is zero.

As a first example, consider the free case. The two sectors, as 
explained,
are made to decouple to preserve unitarity. Moreover, the theory is 
Lorentz invariant
as long as we define
the inner product right, because a Lorentz transformation cannot 
change the
sign of the energy (or, paths going forward in time in one frame
also go forward in other inertial frames.) All inertial
observers will agree on the construction of this theory.

Consider next the particle in a flat background but
with an electromagnetic field. We can try to do the trivial 
construction, but in general we    run into the problem that 
within each sector the theory is not space-time covariant, as we discussed earlier.
We have to abandon the picture in which the old ``square-root'' hamiltoinian
eigenstates are going to be eigenstates of the new diagonal 2x2 hamiltonian.
We have to invent a new hamiltonian.  This is the route to the two component
formalism (see \cite{Baym}).

Can we attach any special significance to the case where
there exists a frame in which 
\beq
[ \partial_0 - A_0 , \partial_i - A_i] = 0= F_{0i} = E_i\eeq
 i.e.,
 the 
situation in which there is no particle creation in the corresponding field theory?
 Notice that Lorentz invariant statements (and particle
number is a Lorentz invariant concept) in a field theory will show
particle creation corrections only throgh the use of Lorentz
invariant quantities regarding the electromagnetic field.
The quantities \beq  {\vec{ E} \cdot\vec{ B} }, \:\:
E^2 - B^2 \eeq are Lorentz  invariants, and if there
is a frame with zero electric field one of these invariants immediately
vanishes, while the other is forever negative.

  For one thing, solutions to
the square-root Schr\"{o}dinger equation  \beq
i\partial_0 \psi = -A_0 \mp \sqrt { m^2 + (-i\partial_j - A_j)^2}\:\psi 
\eeq are
also solutions to the Klein-Gordon equation \beq [ (-i\partial_0 - 
A_0)^2 
- (-i\partial_j - A_j)^2] \phi = 0 \eeq
As a trivial example, we know that if we have the free case solution, 
and we perform a gauge transformation on the electromagnetic 
potential, \beq A_\mu \longrightarrow A_\mu +\partial_\mu 
\Lambda \eeq  the wavefunction changes in a simple way, \beq \psi 
\longrightarrow e^{i \Lambda} \psi \eeq

Consider the Klein-Gordon inner product, \beq
(\psi_a,\psi_b) = -i \int \psi_a^* \;{1\over 
2}\left( \overrightarrow{D}_0 - \overleftarrow{D}_0 \right)
 \psi_b\: d^3x  = 
\int \psi_a^* \;{1\over 2}\left( {\Pi}_0 + {\Pi}_0^\dagger
\right)
 \psi_b\: d^3x  \eeq
This is a very nice, covariant inner product. Notice that if 
\beq [ \partial_0 - A_0 , \partial_i - A_i] = 0\eeq  then the solutions 
to the 
Klein-Gordon equation match those of our Schr\"{o}dinger equation. 
Moreover, we can use the above inner product, since it decouples the 
two sectors: 
\beq
(\psi_{E},\psi_{E'}) = \int \psi_{E}^* \;{1\over 2}\left( {\Pi}_0 + 
{\Pi}_0^\dagger
\right)
 \psi_{E'}\: d^3x  = \eeq  $$
\int \psi_{E}^* \;{1\over 2}\left( \hat{H} + \hat{H}^\dagger
\right)
 \psi_{E'}\: d^3x  =
\int \psi_{E}^* \;{1\over 2}\left( {E'} + {E} \right)
 \psi_{E'}\: d^3x \sim \left( {E'} + {E} \right) \delta_{E'E}
 $$
So the  Klein-Gordon inner product provides us with the decoupling 
relativistic  inner product we were looking for.

Essentially, the vanishing of the above commutator enables us to 
speak of ``energy eigenstates'', to describe the solution space of the 
Klein-Gordon equation in terms of the eigenfunctions of the 
hamiltonian operator. Schematically, we solve the equation
\beq D_0 \phi = H (D_i) \phi \eeq in terms of the solution to \beq 
H (D_i) \phi = E\phi \eeq i.e., solve then   
\beq D_0 \phi = E\phi \eeq

We can apply this criterion to the more general case where there is  a 
gravitational  background. What is required for the decoupling to 
occur? Here we also have a conserved Klein-Gordon inner product. 
Recall that the Klein-Gordon equation is given by 
\beq
(D_\mu D^\mu -m^2 +\xi R)\; \phi = 0\eeq  which we can ``foliate''  as 
(assume that 
$g^{00} =1 ,\: g^{i0}=0$) 
\beq
 ( D_0 D_0 + g^{ij} D_i D_j -m^2 +\xi R)\; \phi = 0\eeq
The covariant derivative here stands for the fully covariant one---both 
gravitational and electromagnetic.
The theory then decouples if we have \beq \label{eq:decoupler}
[D_0, \:g^{ij} D_i D_j +\xi R]=0\eeq This will yield a decoupled situation, 
no 
particle creation in some sense, unitarity within the one particle sector, etc. The 
Klein-Gordon inner product, \beq
(\psi_a,\psi_b) = -i \int \psi_a^* \;{1\over 
2}\left( \overrightarrow{D}_0 - \overleftarrow{D}_0 \right)
 \psi_b\: d^3\Sigma    \eeq
provides us with a nice inner product in the one-branch approach. 
A solution is provided by a zero  electromagnetic field and by a static 
metric (torsion free\footnote{$[\nabla_a, \nabla_b] = 0 $} situation), 
or the existence of a time-like killing vector field.

The Klein-Gordon equation can then be rewritten in the form
\beq \left( i D_0 - \sqrt{g^{ij} D_i D_j -m^2 +\xi R} \right) \left(  i D_0 
+ \sqrt{g^{ij} D_i D_j -m^2 +\xi R} \right) \phi=0 \eeq

How about minisuperspace? 
Recall that the constraint there was 
\beq
\Phi_M \equiv P_A P_B g^{AB} - U(Q) \approx 0
 \eeq
and that for a gauge fixing function of the form $ \chi = Q^0 -f(\tau)$ 
---where 
we 
tentatively assign the ``time-keeping'' role to the scale of the 
universe, it is 
useful to rewrite the constraint as
\beq
\Phi_M \equiv (P_0 - {\cal B}_+ ) g^{00}(P_0 - {\cal B}_- ) \approx 0
\eeq
where
\beq
{\cal B}_\pm \equiv 
\pm\sqrt{ (P^I 
\tilde{g}_{0I})^2 - P^i \tilde{g}_{IJ} P^J +\tilde{U} }
\eeq
We can proceed much as above. 



\newpage\section{Quantize, then constrain}
\label{sec:qc}

\subsection{Dirac's formalism}
\label{sec:Dirac}
In this formalism  \cite{Dirac} we do not assume the 
existence of a classical
reduced phase space. We do begin with first class constraints and a 
first class
hamiltonian, that is, with constraints that commute with the 
hamiltonian as
well as among themselves. This
 ensures---as we will see---that the physical states
can be defined consistently: once defined ``physical'' at a certain time  they will  stay 
physical
under time evolution. Physical states are defined  by the 
conditions\beq
\hat{\Phi}_\alpha \psi = 0 \eeq Notice that having first class 
constraints and
hamiltonian ensures that the operator algebra is well represented,
$$ [\hat{\Phi}_\alpha,\hat{\Phi}_\beta ] \psi = 0$$ (the commutator 
better be a
linear combination of the constraints!)  and that time evolution 
respect
physicality, $$ \hat{\Phi}_\alpha e^{-i\hat{H} T} \psi = 0 $$

In this approach we start by ignoring the fact that there are 
constraints, and 
{we 
quantize all the degrees of freedom, both physical and gauge. Then 
we select a 
subspace\footnote{Well, not really a subspace, as I will explain.}  of 
the full 
Hilbert space: the kernel of the constraint operator.}

  One interesting---and simple---way to look at this
quantization scheme is to  use the gauge  invariant coordinates of
section~\ref{sec:constant}. In that case the states  in 
the 
coordinate representation are initially of the form\footnote{I remind 
the reader 
that $Q$ stands for the gauge coordinate, $q$ for the physical one, 
etc.}  $\psi 
(Q,q)$, but after 
imposing the ``physicality'' condition
\beq
 {\bf \Phi} \; \psi (Q,q) = {\bf P} \; \psi (Q,q)=0, 
\eeq
we are left with the states $\psi(q)$. Here I have implicitely used the
representation $$
{\bf P}\sim -i{  \partial \over \partial Q}$$ I could have used 
$$
{\bf P}\sim -i{ \partial \over \partial Q } - f(Q)$$
instead, and  get  the states   \beq \psi (q) e^{i\int_{Q_0}^{Q}  f(Q') 
dQ'} \eeq
The first problem we run into is that the states  that satisfy the
constraint are not in the original Hilbert space. Indeed those  are given by  
$\psi \neq
\psi(Q) $, which have infinite norm in the original Hilbert  space, $$
\int_{-\infty}^\infty dQ dq \; \psi(q)^* \, \psi(q) = \infty$$ and  
gauge fixing is needed (although one can also start from a 
compact 
coordinate 
space---by imposing periodicity conditions, say.)   
The inner product in the ``reduced'' Hilbert
space can  then be 
defined by
 \beq
(\phi, \psi) \equiv \int_{-\infty}^\infty dQ dq \: \delta(Q) \: \phi(q)^* 
\psi(q)
\eeq
or by \beq
(\phi, \psi) \equiv {1\over V_Q } \int_{-\infty}^\infty dQ dq  \: 
\phi(q)^* 
\psi(q)
\eeq 
where $V_Q$ is the $Q$-volume, or more generally by 
\cite{TeitelBook}
\beq
(\varphi ,\psi ) = \int dQ dq \: \varphi^*(q)\;  \delta \, (\hat{\chi}) |\{ { 
\hat{\chi} 
,\hat{P}}\} | \; 
\psi(q)  =
\int dq \: \varphi^*(q)\,   \psi(q)  
\eeq
after using the gauge $ \chi  = Q - f(q,p,P).$

This is clearly equivalent to the reduced phase space 
quantization---in the 
gauge 
$\alpha  = 1$. The gauge invariant operators---i.e., those that 
commute 
with 
the constraint---are by definition, those that commute with ${\bf 
P}$, and 
after 
eliminating $Q$ dependence from the states, they are reduced to 
operators 
in 
the physical coordinates only.
So it   is easy to see that this approach is the same as the reduced 
phase 
space method {\em for simple constraints and topologies}. For the  
general 
situation the equivalence of the methods is not clear,  although we 
will try to 
shed some light with the cases at hand.

We will discuss these issues at length in the next chapter.

Observables are defined by operators that commute weakly with the 
constraints, 
i.e., 
first class operators. This ensures that we can work with them within 
the 
physical space. For example, let $\hat{A}$ be a first class 
operator. Then, by defifinition 
\beq  [\hat{\Phi}_\alpha,\hat{A}] = 
\hat{C}^\beta\hat{\Phi}_\beta\eeq
Then we can ask, if $|\psi\r$ is a physical state, 
$$\hat{\Phi}_\alpha |\psi\r= 0$$ is $\hat{A}|\psi\r$ physical? Well, 
\beq 
\hat{\Phi}_\alpha \hat{A}|\psi\r = [\hat{\Phi}_\alpha,\hat{A}]|\psi\r 
= 
\hat{C}^\beta  \hat{\Phi}_\beta |\psi\r =0\eeq
indeed.

Notice also that if $\hat{A}$ is an observable, then $\hat{A} + 
\hat{C}^\alpha 
\hat{\Phi}_\alpha$ is also observable and {\em physically 
indistinguishable} 
from 
$\hat{A}$. These operators are related by a gauge transformation. 
One can 
gauge 
fix by demanding that the observables also commute with   an extra 
set of 
constraints---which are chosen not to commute with the original 
constraints, in 
the sense that the matrix formed by their Poisson bracket is                  
non-
singular. This is 
very similar to the Dirac bracket approach in the classical 
development of the 
previous chapter.

  
\subsection{Fock space quantization}
\label{sec:Fock}

Let us introduce the Fock quantization method. To begin with, we 
will need 
an even number of constraints---a recurring theme.  Let me 
state why immediately: we will, in essence, pair the constraints and assign to each 
  combination opposite sign norm states so that their effect in the theory cancels after
we select the physical space.
Thus, gauge--degrees of freedom effectively disappear from the theory. The key new ingredient
here is the appearance of  states with negative norms.

We assume for 
now  that 
the 
constraints can be canonically transformed to 
momenta\footnote{This is 
always true locally, but, in general, not globally.}, $$P_1\approx 0 
\approx P_2$$ A very important assumption is that we will 
represent these 
as hermitean  operators---as we will see. Let us now define the 
operators
\beq
\hat{a} = \hat{P}_1 + i\hat{P}_2 , \:\: \hat{a}^\dagger = \hat{P}_1 
-i\hat{P}_2\eeq and\beq \hat{b}=-{i\over 2}(\hat{Q}^1 +i\hat{Q}^2), 
\:\: \hat{b}^\dagger = {i\over 2}(\hat{Q}^1 -i\hat{Q}^2)\eeq
As we see, we need an even number of constraints.

The commutation relation that follow from this definition 
are$$[\hat{a},\hat{b}^\dagger] 
= 
[\hat{b},\hat{a}^\dagger]  = 1$$ and the rest zero.

{\em Notice that it is implied by the notation here that both 
$\hat{P}_1$ and 
$\hat{P}_2$ are 
hermitean.} For example, $\hat{a}+\hat{a}^\dagger$ is hermitean, and 
is equal  
to $2P_1$. This fact is crucial for the development of the formalism, 
and is 
a subtle assumption---it selects an indefinite inner product when we 
define 
the vacuum.

The states on this space are defined by the following construction:  

a) it is assummed that there is a ``vacuum'' state,   $|0\rangle$, 
satifying the 
conditions 
$$\hat{a}|0\rangle  = \hat{b}|0\rangle = 0 $$ 

a') This state is also assumed to 
have unit 
norm, $\langle 0 |0\rangle = 1$.

b) The rest of these states are defined by acting on the ``vacuum'' 
above with 

the creation operators.

Before we work out some of the consequences of this prescription, let 
us look 
at what it is doing in terms of the $\hat{P}_i$ operators. What is the 
vacuum? We 
need a state that satisfies $$ (\hat{P}_1 +i\hat{P}_2)|0\rangle = 
-{i\over 2}(\hat{Q}^1 +i\hat{Q}^2)|0\rangle = 0$$
In a standard Hilbert space  
there is no 
such state! Although these operators commute, they 
are not 
hermitean with the usual inner product, so we do not expect to construct 
a basis with their eigenstates. This points the way to
a  cure.

Consider the states $(\hat{a}^\dagger + \hat{b}^\dagger) |0\rangle $, 
$(\hat{a}^\dagger - \hat{b}^\dagger) |0\rangle $, and $ 
\hat{a}^\dagger  
|0\rangle $.
 They have positive, negative, and zero norm respectively.  Our 
Hilbert space 
is not a positive definite inner product space.

Now comes the constraint. The constraints above are equivalent 
classically to 
demanding that$$a\approx 0 \approx a^\dagger$$However, we 
cannot 
demand this condition from the states,    since there is no state in our 
construction that will satisfy $\hat{a}^\dagger |\psi\rangle=0$! By 
the way, 
this is another clue that our definition of the Hilbert 
space---our representation---is not the usual one (there is no such 
vacuum 
in the Hilbert space we learned in kindergarden.... There is a solution to the 
above 
equations that define the vacuum, but the solution has infinite 
norm.) 

We can only demand that the physical states satisfy $$\hat{a} |\psi_F 
\rangle = 0$$
 
Now, what of the other ``half'' of the constraint? The physical states 
do not 
satisfy it! However, {\em notice that expectation value of the constraints in 
between physical states in always zero.} So in this very important
sense we are safe. Moreover, we will now see 
that 
effectively---as claimed at the beginning of this section,
 the Fock space is reduced to a space
isomorphic to the  Dirac      
states---the space of states that satisfy the constraint in Dirac      
quantization:  \begin{quote} {\em 

 A physical state  in the Fock space is   either   the vacuum or  a 
linear 
combination of the vacuum and a physical state that has zero inner 
product 
with 
all 
the physical states.} \end{quote}

Indeed, the other physical states are given by  the null states
\beq
f(\hat{a}^\dagger) |0\r
\eeq
since $[\hat{a}, \hat{a}^\dagger]=0$,   which decouple from the physical states, since 
$\hat{a}^\dagger\equiv (\hat{a} 
)^\dagger$.

Let us describe the observables in this representation. 
It is clear that the observables of the Dirac formalism will not work 
here,
unless they are observable in the strong sense. 
For example, if $\hat{A}$ is again an observable in the Dirac sense,
we know that (let $\hat{\Phi}_\alpha = \hat{P}_1,\: \hat{P}_2$)
$$[\hat{\Phi}_\alpha,\hat{A}] = \hat{O}^1_\alpha 
(\hat{a} + \hat{a}^\dag ) + 
\hat{O}^2_\alpha (\hat{a} - \hat{a}^\dag ) $$
Now, 
if $|\psi\r$ is a physical state in the Fock sense, 
$$\hat{a} |\psi\r= 0$$
 is $\hat{A}|\psi\r$ physical? Well, 
$$
\hat{a}  \hat{A}|\psi\r = [\hat{a},\hat{A}]|\psi\r = [\hat{P}_1+i 
\hat{P}_2,\hat{A}]|\psi\r =
$$ 
$$ 
\hat{K}^1
(\hat{a} + \hat{a}^\dag ) |\psi\r +
\hat{K}^2(\hat{a} - \hat{a}^\dag ) |\psi\r   = 
 (\hat{K}^1- \hat{K}^2) \hat{a}^\dag  |\psi\r $$
So, in general $\hat{A} |\psi\r$ is not physical.

{\em Only classical observables that  are strongly observable are 
acceptable in 
general in 
the  Fock scheme.} 

 This is nor entirely surpising, though. In this approach there is no 
need for 
gauge-fixing, which means that in order to set up a map from the 
Fock space to 
the earlier Dirac one, gauge-fixing will be needed in the second. In 
the Dirac 
approach, one can always make a weak observable strong by adding 
to it a 
linear 
combination of the constraints. Such observables are physically 
equivalent  
gauge 
cousins, and after fixing the gauge only one will remain. Thus at the 
end of the 
day we do have an isomorphism between the two approaches. After 
we are 
done 
with the constraining we end up with the same reduction as in the 
Dirac case. 
However, in the intermediate steps things will look very 
different---as we will 
discuss later---and things can get to be very tricky with theories that 
are not 
completely gauge-invariant, like the ones we will discuss here. We 
will 
continue 
this discussion in the next chapter.

 
\subsection{BRST quantization}
\label{sec:BFVq}

In BRST \cite{TeitelBook} the states we have are originally in the 
extended 
space, so in this respect the philosophy is as in the Dirac approach.
One may start asking if one shouldn't reduce classically even in this 
approach---
and then quantize. As remarked, however, there are no constraints in 
the BRST 
approach---``the invariance has been made rigid''---so it makes 
sense to 
quantize as usual when there are no constraints.

Recall the following from the classical development of a system with 
a 
constraint 
$\Phi$ 
\cite{TeitelBook,BFV}.   The first objects that are 
introduced into the 
extended 
phase space are the multiplier $\lambda$  and its conjugate 
momentum $\pi$,    
i.e., $\{ \lambda 
,\pi \} =1$. Since $\lambda $ is arbitrary, its momentum is 
constrained, 
$\pi 
\approx 0$. We thus have two constraints. To each constraint, the 
rules say we 
must associate a conjugate pair of ghosts, 
$\eta_0 ,\rho_0 $ and $\eta_1 ,\rho_1$  with $\{ \eta_0 ,\rho_0 \} 
=1$ and $\{ 
\eta_1 ,\rho_1 \} =1$---with the other 
(super)brackets being zero. The total extended phase space is thus 
described 
by $t, p_t , x, p_x , \lambda , \pi , \eta_0 , \rho_0 , \eta_1 , \rho_1 $.  
The 
next 
thing to do is to define the BRST generator, which generates the gauge 
transformations in extended phase space: 
\beq
\Omega= \eta_0  \Phi  + \eta_1  \pi 
\eeq
 We also need a gauge-fixing function, ${\cal O}$. The dynamics are 
then 
generated by the hamiltonian ${\cal H}  = h + \{ {\cal O}, \Omega\} $   
($= \{ 
{\cal 
O}, \Omega\} $, since for our systems the original first class 
hamiltonian $h$ is 
zero.) 

The BRST generator has the crucial property that $$\{\Omega, 
\Omega\} = 0$$

Now, the above is translated into the quantum recipe in the usual 
way. 
Observables become operators, and the (super)Poisson bracket 
structure is 
translated into the (super)commutator language in the usual way, $$
\{A,B\} \longrightarrow i\hbar [\hat{A},\hat{B}]
$$  In this case we have both commutators and anticommutators. For 
example, 
we 
will need \beq
[ \hat{\Omega},\hat{\Omega} ] =\hat{\Omega}^2 =  0 
\eeq

 In the 
particle case this means that we start with the states $   |  \Psi  \r $ 
with the 
basis $ | t,x,\lambda ,\eta_0 ,\eta_1  \r  $, say.
In the ``coordinate''  representation we have
\beq
 \l  t,x,\lambda ,\eta_0 ,\eta_1   |  \Psi  \r  \equiv   
\Psi   = \psi  +\psi_0 \eta_0    + \psi_1 \eta_1  + \psi_{01} \eta_0 
\eta_1  
\eeq
where the  $\psi $'s are functions of $x,t,\lambda $.
The inner product in this original extended space is given by
\beq
(\Sigma ,\Psi ) \equiv \int dt dx d\lambda  d\eta_0  d\eta_1 \;   
\Sigma^* 
(z^A)\:\Psi (z^A)
\eeq
Now,  in order to get to the BRST physical space we need to do two 
things, 
and in both the central object  is the BRST generator $\bOmega$ and 
its  
properties\footnote{The first one is a subtle assumption about the 
representation 
of this algebra. See below. }:\\
a) $\hat{\Omega}^\dagger =  \hat{\Omega}$,  \\
b) $\hat{\Omega}^2=0$, \\
which it inherits from the classical description:  $\Omega$ is real, and 
$\{ \Omega, \Omega\}  =0$ \cite{TeitelBook,BFV}.
The BRST physical space is defined by: \begin{quote}
i) the BRST physical condition
\beq
\hat{\Omega}  |  \Psi  \r _{Ph}  \equiv 0
\eeq
where recall that the BRST generator is ${\Omega}= {\eta_0  \Phi } + 
{\eta_1  \pi }$  \\
ii) we need the BRST cohomology, i.e we need to identify
\beq
 |  \Psi  \r _{Ph} \sim   |  \Psi  \r _{Ph} + \hat{\Omega} | \Delta \r 
\eeq
since the factor $\hat{\Omega} | \Delta \r $ is physical 
($\hat{\Omega}^2=0$), but 
has 
zero inner 
product with any physical state\footnote{see last footnote...} 
($\hat{\Omega}^\dagger 
=  
\hat{\Omega}$,  
$\hat{\Omega}  |  \Psi  \r _{Ph}  = 0$).
\end{quote}

Consider, for example,  a gauge theory with constraints $G_a \approx 
0 $ and 
algebra\beq
[G_a, G_b] = C_{ab}^c G_c\eeq with the $C_{ab}^c$ constants. This is 
what reference \cite{TeitelBook} calls  ``Constraints that close 
according to a 
group'', since
the above are the structure constants and the Poisson algebra
reproduces the Lie algebra of a group. The Jacobi
identity of the Poisson bracket implies that for consistency
the constants satify the Jacobi identity---as they do for a 
Lie Group,\beq  C_{ab}^c C_{cd}^e + C_{bd}^c C_{ca}^e +
 C_{da}^c C_{cb}^e=0\eeq
In such a case the BRST generator is defined by \beq \Omega = 
\eta^a G_a -{1\over 2}  \eta^b\eta^c C_{cb}^a {\cal P}_a \eeq
Now, \beq [\Omega, \Omega] = 0  \eeq is equivalent to the algebra 
above.
Indeed\footnote{Following the convention in reference 
\cite{TeitelBook}, 
$[\eta^a,{\cal 
P}_b]=-\delta_ 
b^a$.}\beq  [\Omega, \Omega ] = \eta_a \eta_b ( [G_a, G_b] -C_{ab}^c 
G_c
) + {1\over 4} [ \eta^b \eta^c C_{cb}^a {\cal P} _a,\eta^{b'} \eta^{c'}
C_{c'b'}^{a'}{\cal P}_{a'} ]\eeq Now, \beq {1\over 4} [ \eta^b \eta^c 
C_{cb}^a {\cal 
P} 
_a,\eta^{b'} \eta^{c'}
C_{c'b'}^{a'}{\cal P}_{a'} ] = C_{ab}^d C_{cd}^e {\cal P}_e 
\eta_a\eta_b\eta_c
=0\eeq is the Jacobi identity.

What are the equations for physicality?
In this simple case
there are no ordering ambiguities in the writing of the operator 
\beq \hat{\Omega} = 
\hat{\eta}^a \hat{G}_a -{1\over 2}  \hat{\eta}^b\hat{\eta}^c C_{cb}^a 
\hat{{\cal P}}_a\eeq thanks in part to the total antisymmetry of the 
structure 
constants. Let us   look at the solutions to the equation 
\beq\hat{\Omega} 
|\Psi_{BRST}\rangle = 0\eeq For this purpose, let us use the semi-abstract, semi-coordinate
representation (for the ghosts)   expression for the states  that we 
used before. This is indeed the one that results from projecting \beq 
\langle 
\eta^0, \eta^1,\eta^2 | \left(
\sum   |\psi (z), \lambda)\rangle   \otimes 
| f(\eta^0,\eta^1,\eta^2) \rangle \right) \eeq where \beq \hat{\eta}^a 
|\eta^0,\eta^1,\eta^2\rangle =\eta^a |\eta^0,\eta^1,\eta^2\rangle \eeq 
Now, doing 
a Taylor expansion \beq \langle \eta^0, \eta^1,\eta^2 |
\sum   |\psi (z), \lambda)\rangle   \otimes 
| f(\eta^0,\eta^1,\eta^2) \rangle  = |\psi\rangle + |\psi_a\rangle \eta^a 
+|\psi_{ab}\rangle \eta^a\eta^b + |\psi_{abc}\rangle \eta^a\eta^b\eta^c \eeq
$\equiv |\Psi \r$ and where, w.l.g., we assume that the wavefunctions 
are totally 
antisymmetric in their indices. Then $$  
\hat{\Omega} |\Psi \r = 
\left( \hat{\eta}^a \hat{G}_a -{1\over 2}  \hat{\eta}^b\hat{\eta}^c 
C_{cb}^a 
\hat{{\cal P}}_a\right ) \left(|\psi\rangle + |\psi_a\rangle \eta^a 
+|\psi_{ab}\rangle \eta^a\eta^b + |\psi_{abc}\rangle \eta^a\eta^b\eta^c
\right) =  $$ \beq \eta^a \hat{G}_a|\psi\rangle +
 \eta^a \eta^b \left( \hat{G}_a   |\psi_b\rangle +{i\over 2} C_{ba}^c 
|\psi_c\rangle\right) +
\eta^a \eta^b \eta^c \left( \hat{G}_a   |\psi_{bc}\rangle +i C_{cb}^d  
|\psi_{da}\rangle  \right)\eeq The general solution is then given by 
\beq
\hat{G}_a|\psi\rangle = \hat{G}_a   |\psi_b\rangle -\hat{G}_b   
|\psi_a\rangle  
+i  C_{ba}^c 
|\psi_c\rangle =\sum_{antiSymm}\left( \hat{G}_a   |\psi_{bc}\rangle 
+i 
C_{cb}^d  
|\psi_{da}\rangle\right)=0 \eeq or
more simply by \beq \epsilon^{abc}\hat{G}_a|\psi\rangle 
=\epsilon^{abc}\left(
 \hat{G}_a 
  |\psi_b\rangle -\hat{G}_b   |\psi_a\rangle  
+i  C_{ba}^c 
|\psi_c\rangle\right) =\epsilon ^{abc}\left( \hat{G}_a   
|\psi_{bc}\rangle +i 
C_{cb}^d  
|\psi_{da}\rangle\right)=0 \eeq with $|\psi_{abc}\rangle$ totally 
unrestricted. 

To check the above, multiply the above equation by $\eta_a$, 
$\eta_b$ and 
use
\beq \eta_{i_1} \cdots\eta_{i_N} T^{i_1 \cdots i_N} = \eta_1 \cdots 
\eta_N 
\epsilon^{i_1\cdots i_{N}} T^{i_1 \cdots i_N}\eeq  when $i=1,...,N$.

Evidently, the above also means that $|\psi\rangle$ satisfies the 
constraints 
(i.e., it is a Dirac state). These two sectors represent the extremes. 
The 
``middle'' sectors fall somewhat in between. An example is 
given by the state $$|\eta_0 \!=\! \eta_1 \!=\! G_2\!=\! 0\rangle \sim 
\eta_0\eta_1 \varphi(z,\lambda)_{G_2\!=\! 0} $$ It is clear that this 
state is 
annihilated 
just by looking at the BRST generator---and using the antisymmetry 
of the 
structure constants.

In the abelian case, it is usually stated that the states we end up 
with---i.e., 
{\em 
the BRST cohomology}---are
$$
\Psi   = \psi  +\psi_0 \eta_0    + \psi_1 \eta_1  + \psi_{01} \eta_0 
\eta_1  
$$
with 
\beq
 \hat{\Phi}   \psi   = \hat{\pi}  \psi  = \hat{\Phi}  \psi_0  =  \hat{\pi}  
\psi_0 = 
\hat{\Phi}\psi_1  =  \hat{\pi}  \psi_1 = 
\hat{\Phi}\psi_{01}  =  \hat{\pi}  \psi_{01} = 0, 
\eeq
in other words, with the wave functions independent of $Q, \lambda 
$  (in 
the 
$\lambda $ representation, say), where $Q$ is the hypothetical 
coordinate 
conjugated to the constraint.... This, as we will see in the next 
chapter, is 
incorrect. 

The hamiltonian is given by ${ \hat{\cal H}} = \{ \hat{ \cal O}, 
\hat{\Omega} \}$ , 
for 
some gauge fixing operator $\hat{ \cal O}$, and we could expect that 
it has no 
effect on physical 
states 
due to the two conditions above. We will see that this is also false.

If we use the above ``cohomology'' states, 
the inner product on the physical space and in the
coordinate  representation 
can 
be defined by
\beq
(\Sigma ,\Psi ) \equiv \int dt dx d\pi  d\eta_0  d\eta_1   \eta_0 
\eta_1  
\: \Sigma^*(z^A)  \; \delta (\hat{\chi} ) | \{ \hat{\chi} ,\hat{\Phi} \} |  
\; \Psi(z^A) 
\eeq
Gauge fixing is needed, although here the situation is different 
than 
the one in the Dirac case,  since there the divergence of the 
unregularized 
inner product was genuine, whereas here there isn't a true 
divergence (recall 
that the ghosts carry  ``negative degrees of freedom''), but a $\delta 
(0) 
\times \infty$ situation. We will motivate this definition later. We will see, though,
that choosing the correct states in the cohomology will solve the 
regularization problem.

Although  it is unclear yet how to connect naturally  the 
BFV 
path integral with  the BRST cohomology canonical description, it is easy to 
write
the path integral when one works in the full BRST space, using 
resolutions 
of 
the identity like
\ba  \dis 
I & \dis = \int dt dx d\lambda  d\eta_0  d\eta_1   |  t,x,\lambda 
,\eta_0 
,\eta_1  \r \l 
t,x,\lambda  ,\eta_0 ,\eta_1   |   \\ \dis 
& \dis = \int dp_t  dp_x  d\pi  d\rho_0  d\rho_1    |  p_t ,p_x 
,\pi  ,\rho_0 ,\rho_1   \r \l  p_t ,p_x ,\pi  ,\rho_0 ,\rho_1    |     
\ea
projections like
\beq
\l  t,x,\lambda ,\eta_0 ,\eta_1   |  p_t ,p_x ,\pi   ,\rho_0 ,\rho_1   \r  
= 
e^{i(tp_t 
+ 
xp_x  + \lambda \pi  + \eta_0 \rho_0  + \eta_1  \rho_1 )}  , 
\eeq
and a hamiltonian like ${ \hat{\cal H}} = \{ \hat{ \cal O}, 
\hat{\Omega} \} $.
This is formally obvious: recall that the action is \beq 
      S =  \itif (\dot{t}p_t + \dot{x}p_x  + \dot{\lambda} \pi  + 
\dot{\eta}_0 
\rho_0  + \dot{\eta}_1  \rho_1  - \{ {\cal O}, \Omega\} ) d\tau\eeq    

We will study this issues in better and finer detail in the next 
chapter.

\subsection{BRST-Fock quantization}
\label{sec:BFock}
In this approach the basic BRST formalism is not changed (unlike in 
the 
transition 
from Dirac to Dirac-Fock.) The physical condition is still 
$\hat{\Omega} |\psi\r = 
0$. However, the representation of this Hilbert space is different than 
the one 
we 
used before.

 Again we define the operators 
\beq
\hat{a} = \hat{P}_1 + i\hat{P}_2 , \:\: \hat{a}^\dagger = \hat{P}_1 
-i\hat{P}_2\eeq and\beq \hat{b}=-{i\over 2}(\hat{Q}^1 +i\hat{Q}^2), 
\:\: \hat{b}^\dagger = {i\over 2}(\hat{Q}^1 -i\hat{Q}^2)\eeq where as 
before we 
assume that we have an even number of   
              constraints---which is true with the multipliers.

Now we add the ghosts to this picture:
\beq
\hat{c} = \hat{\eta}_1 + i\hat{\eta}_2 , \:\: \hat{c}^\dagger = 
\hat{\eta}_1 
-i\hat{\eta}_2\eeq and\beq \hat{\bar{c}}= {i\over 2}(\hat{\rho}^1 
+i\hat{\rho}^2), 
\:\: \hat{\bar{c}}^\dagger = {i\over 2}(\hat{\rho}^1 
-i\hat{\rho}^2)\eeq
 The BRST generator now reads \beq
\hat{\Omega} = \hat{c}^\dagger \hat{a}  + \hat{a}^\dagger \hat{c} 
\eeq
The Hilbert space is constructed as usual, starting from a vacuum 
that is 
annihilated by all the destruction operators. Then the rest of the 
states are 
created with the creation operators, just as before, except that we 
have the 
ghost 
part too now. These  degrees of freedom are fermionic (the creation 
operators 
squared are zero, as follows from the brackets.)
Thus, the BRST condition is equivalent (in this Fock representation) to 
asking 
that 
the states be annihilated by the annihilation operators of the ghosts 
and of the 
non-ghost parts.

As discussed in reference~\cite{TeitelBook}, in this representation 
any physical 
state is either the vacuum or $\hat{\Omega}$ of something, i.e., BRS 
exact. This 
is 
similar to what we found in 
Dirac-Fock, i.e., that a physical state there was either the vacuum or 
null. Note, 
however that all the classical observables here are good within the 
physical 
space.

Finally, there are no inner product regularization problems here, no 
hermicity 
problems, and the operator  cohomology duality theorems are also 
true of the 
states---since the only nontrivial state cohomology class is at the 
zero ghost 
number \cite{TeitelBook}.

Moreover, the invariance of the physical states amplitude under 
changes in 
gauge-fixing $\hat{K}$ is direct, \beq
\l \Psi | e^{[\hat{\Omega}, \hat{K} ] }|\Psi'\r = 
\l \Psi |  \Psi'\r \eeq
because there are no hermicity questions.

We will see that this representation leads to the Feynman propagator
in the relativistic case. This is essentially because in the Fock
representation the multiplier is imaginary and half-ranged (see
section~\ref{sec:FFock}).

\section{Conclusions, summary}

This has been mainly a review chapter in which I have introduced the different
quantization schemes that I will apply to the systems under consideration. I have also
described the problems associated with each quantization approach, thus setting the
stage for the developments in the next chapter. The main point to come back to is the
definitions of the state spaces and their inner products, which are not clearly
spelled out by the different prescriptions.

I have, however, completely carried out the reduced phase space quantization, the
``constrain, then quantize'' approach, for the one-branch situations---which I have
argued are the only legitimate situations for the reduced phase space approach. I have
showed that since the construction of the classical reduced phase spaces can only be
completed up to time-dependent canonical transformations, we will get, upon
quantization,   quantum theories    in different pictures. 

I have also introduced the idea that a path integral can be immediately developed in
the BRST quantization approach in the full extended phase space, since it is,  in
essence,  an approach in which the invariance is made ``rigid''---there are no
constraints. I will come back to this path integral in the next chapters---the BFV
path integral.



\chapter{The physical inner product}
\label{sec:product}
In this chapter I study the problem of the development of an inner 
product in the 
physical space.
 This applies to  the ``quantize, then constrain'' approaches  I  have  
introduced in 
the previous chapter (the problems in the
``constrain then quantize'' approach were already discussed,
 where we saw that there is a relationship between different gauge 
choices
and different representations). 

The ambiguity in the definition of the 
inner product in the unconstrained case was also pointed out in the 
first chapter. This ambiguity is present anytime we have to define a 
new inner product.

I explained, in the previous chapters, what the problems are in quantization of
parametrized systems, namely, the definition of the state spaces, the inner products
and the subsequent introduction of path integrals. I will explain how to solve these
problems.

\newpage\section{Introduction}
In this section we will look at the question of the development of the inner product,
 starting with some general  comments on the problems which
have clouded this issue. Until this point we have not been able to
give a satisfactory, well-motivated  definition for the inner product  
in the 
Dirac 
or BRST quantization approaches. We will do so now. We will discuss 
the 
problems that arise because of the arbitrariness that exists in 
defining the 
constraint, and because of the fact that the states that satisfy the 
constraint 
are in general not normalizable. 

Consider the Dirac quantization scheme, with a given constraint 
$\Phi$.
Recall that in the Dirac approach we start with a full Hilbert 
space---that
is, we ignore the fact that there is a constraint and we make the 
transition from classical to quantum as usual. The classical constraint, 
which
is assumed to be a first class constraint, is
made into an operator and  then imposed on the states, that is, 
a subspace of the Hilbert space is selected by demanding that the 
states
in it be zero eigenvectors of the constraint,
\beq  {\bf \Phi} |\psi \rangle = 0\eeq 

 Notice, however, that the definition of the constraint is not unique. 
Two constraints $\Phi\approx 0$, and $\Phi'\approx 0$ are equally 
valid 
starting points for the quantization if they have equivalent solution 
spaces. 
There is no way to choose one over the other in general, so we should 
understand what happens if we change the form of the constraint. Do 
we 
get different quantum theories? Let us start by observing that 
if $\psi$ satisfies ${\bf \Phi} \psi = 0$, it follows that the state
${\bf v}\psi$ does as well, provided that $[ {\bf v, \Phi}] = 0$. In 
fact, the constraints $\Phi\approx 0$ and $v\Phi\approx 0 $ are 
equivalent
 in the quantum sense
as long as they commute as operators  and as long as $\bf v$ has no 
zero 
modes.
 
How do we define an inner product on this subspace? Well, if 
we had an inner product in the original, big, Hilbert space it should
follow that the states selected by the constraint can use it. After
all, the constraint selects states in the original Hilbert space, doesn't
it? Well, it should, but in general it doesn't. The simplest example
is the locally general one, $\Phi = P$. The problem is 
that \\
{\fbox{\em momentum states
 are not normalizable when the coordinate space is infinite}}

To see what kind of problems this will lead us to, consider for 
example the 
quantity\beq 
\langle P \! = \! 0| [{\bf Q,P}] |P\!=\! 0\rangle\eeq 
What is it? We will reach a contradiction whether we assume that 
the 
coordinate space is infinite, finite---there are no $P=0$ states in such 
a 
case---or periodic. This is a problem, because the states $|P\!=\! 
0\rangle$ 
are the 
starting point in the Dirac quantization approach! If we must have 
them, 
then either $\bf P$ cannot be  a hermitean operator (infinite range  case), 
or $\bf 
 Q$ does not exist (periodic boundary conditions). {\em It is 
dangerous to 
assume in general that there exists a definition of inner product in 
the full 
space that we can use with the physical states.} This observation 
applies also 
to the BRST quantization approach. It is because of this that we 
basically have 
to ``redefine''---in general---the inner product in the physical space, 
the only 
exception being the periodic boundary condition case and the Fock 
approach.

Let us look at this problem a bit closer. A Hilbert space is described 
by a set 
of linear operators and a vector space of states, together with a 
good\footnote{i.e.,  finite!} 
definition of 
inner product. Consider then a Hilbert space  with the operators $\bf 
Q,P $ 
such that 

i)  $\bf P=P^\dagger$, $\bf Q= Q^\dagger$ 

ii) $[{\bf Q,P}]=i$

\noindent and let us discuss possible state spaces and inner 
products. 

It is not to hard to see that this Hilbert space cannot contain any 
eigenvectors of either one of the above operators. Indeed, suppose 
for 
example that
$|P_a\rangle$ is an eigenvector of $\bf P$. Then (i is false), \beq 
\langle 
P_a| 
{\bf [Q,P]} |P_a\rangle = i\langle P_a| P_a \rangle\eeq  but also (then 
ii 
is false), 
\beq 
\langle P_a| {\bf [Q,P]} |P_a\rangle = (P_a - P_a) \langle P_a|{\bf Q}| 
P_a 
\rangle=0\eeq 
The point is that if one includes such  states in the Hilbert space the 
operators are not hermitean. In the infinite range case the inner products
diverge.

As a different situation, consider for example  the states representing  
a 
particle in a box.  There are then $\bf Q$ eigenstates---although they 
are not 
normalized to one.
There are no $\bf P$ eigenstates in the coordinate representation. 
This is 
easy to see since in the coordinate representation this is a first order 
differential operator and we have too many boundary conditions. In 
fact, 
$\bf P$ takes states out of the Hilbert space. 

We can also consider the periodic  boundary condition case. Then 
there exist 
momentum eigenstates, and they are normalizable. However, the 
operator 
$\bf Q$ is not well-defined: it takes states outside the Hilbert space.

In the situation with no boundary conditions and an infinite 
coordinate space 
we do have momentum eigenstates. These are not normalizable, 
though, and 
 more to the point, the $\bf P$ operator is not really hermitean with  
respect 
 to 
them\footnote{Unless you are willing to say that $\delta'(0) \cdot 0 = 
\infty$!}.  Yet another  point is that when we quantize the 
constraint {\em we will ask for normalizable momentum eigenstates.} 
It is usually ok to work with momentum states that are not normalizable and
really  not in the Hilbert space, because they are just mathematical tools. In the
quantization of constrained systems they are  not treated as tools---they {\em} are
the building blocks of the  physical Hilbert space!

We 
will have to be careful then that we don't assume that the constraint 
is a hermitean operator in such an inner product space.

\begin{quote} {\em 
The main thing to remember is that if one insists on having 
states of
 well-defined momentum---say---as
 well as a finite inner product,   the corresponding 
momentum operator 
will not 
be 
hermitean. This is not necessarily a problem if we keep it in mind 
and are 
careful with the algebra. In the 
Dirac formalism, for example, if the constraint is $P\approx 0$, the 
states 
satisfying the constraint make the operator non-hermitean. 
Similarly, 
in BRST, 
$\bOmega$ will not be hermitean in the physical sector. However, in 
both 
cases the {\em operators have the hermitian property when used 
between 
``conjugate states'', like $\langle Q_a| {\bf P} | 
P_a\rangle$.}}\end{quote}

Since the gauge coordinates are not going to be to essential in the 
resulting 
theory---once we take care of them---there is a way to deal with 
this, 
although the interpretation of the
method was, until now,  unclear.  I will motivate here the definition 
of the Dirac inner product, as well as point out and clarify some  
important aspects that have created confusion in the literature.

One of these aspects is an   ambiguity that we can 
trace to the definition of the constraint:    
\beq 
(\psi_a,\psi_b) \equiv \int \psi_a^*\;{\bf  |\{ \bchi 
,\Phi\}|}\delta(\bchi)\; 
\psi_b\;\; dV
\eeq 
where the gauge-fixing condition is\footnote{As 
usual, we will have $q$ represent the physical degrees of freedom 
and $Q$ 
the gauge ones.} $\chi \approx 0$, and 
$dV=dqdQ$. If the constraint is multiplied by a function the inner 
product will  indeed be different, and one would reason that the 
same inner product and quantum theory should result from 
constraints that are equivalent.

I will also point out that we can further rewrite this inner product as 
\beq
(\psi_a,\psi_b)\equiv 
\int dV d w dc d\bar{c}\; \psi_a^*\; e^{i( w \bchi + c{\bf 
|\{\bchi,\Phi\}|}
\bar{c})}\;\psi_b \eeq
This leads to the BRST quantization approach. 

 As for the BRST inner product we would expect that  it may reduce  
to the 
Dirac 
inner product
given above. Recall that the BRST states for a system with one 
constraint as above are described by four sectors: one no-ghost
sector, two one ghost sectors and finally one double
ghost sector. Recall also the duality theorems; essentially, we
find that the sectors are all isomorphic in pairs. However, we will 
find that  they are not all isomorphic to the Dirac states.

The inner product can then be formally defined in the following way 
(see \cite{marnelius} and also \cite{TeitelBook}
 p. 325, and ex. 14.23):\beq
  (\Psi_a, \Psi_b) \equiv \int d\rho_1 d\eta_0 d\pi dV \: \Psi_a^*\; 
e^{i\{K,\Omega\} }\; \Psi_b \eeq
How are the states defined? Physical? Zero ghost? Isn't the exponent 
just one 
when evaluated between physical states---states annihilated by 
$\bOmega$? 
Can we motivate this definition? How does it 
compare to the Dirac inner product? Can 
we tie this to the path integrals? What 
about the gauge fixing function $K$,  is it arbitrary? We will address  
all 
these questions.

Yet another approach we will discuss is  Dirac-Fock   
quantization. In this approach the constraints are not imposed on the 
states, and  there is no normalization problem. All the states have a 
well defined norm from the beginning to the end. Only, for some this 
norm will be negative. We will compare this quantization apprach to 
the other ones, and establish their equivalence for well-defined 
gauge systems. For parametrized systems something peculiar 
happens, of course.

At this stage we have that both the Dirac approach and the BRST 
approach suffer from
the same problem. {\em It is not clear at all how to implement the
constraints by starting with a bigger quantum space, and obtain from 
a
quantum reduction a well defined quantum theory. } The basic
reason for this is that one usually works with an infinite-range 
coordinate 
space;
in such a situation the states selected by the constraint are in general 
not 
normalizable, and, as we will see, it is hard to implement the physical 
condition through a projecting operator---but not impossible.

We will look at the finite coordinate space situation, and also
at the unbounded one. Both  cases will be shown to be suitable
for quantization.

As for the path integrals, if one hopes to interpret the BFV path 
integral `a la 
Dirac'---by
looking at the state cohomology, inner product, etc.---it seems 
reasonable to 
expect that one should also understand
the Faddeev path integral in such terms, from the Dirac state space, 
and this 
has not been done  yet:
{\em It is unclear at this point if the Dirac quantization scheme (and 
the 
corresponding quantization scheme in the BRST formalism) can be  
related to 
the 
Faddeev (BFV) path integral, or if this path integral is really related
only to the quantized reduced phase space  (this relation has 
already
been established).}  We will clarify these connections, and show that 
the Faddeev path integral can be constructed within  the Dirac 
approach.

\newpage\section{Dirac quantization and quantum gauge transformations}

The key idea in the Dirac approach to constrained systems is that  
{\em a 
first class constraint introduces arbitrariness in the dynamics}, 
something 
that
is reflected in the arbitary term in the hamiltonian, $H_E = h + 
v\Phi$. Now,
in the quantum world---in the Schr\"{o}dinger picture---this means that 
will have 
an extra 
term in the Schr\"{o}dinger equation; the general solution  will be\beq
|\Psi\rangle  = e^{-i t{\bf H}_E}\;  |\Psi_0\rangle= e^{-i t\bf h}\, e^{-i t\bf 
v\Phi}\; |\Psi_0\rangle  
\eeq
where we have used the fact that the constraint is first class and 
commutes 
with the original hamiltonian $\bf h$. One has to be careful about the 
$\bf v$ 
function/operator as well.

From a different perspective,  a momentum operator $\bf P$ can be 
understood to produce a gauge transformation
in the sense that\beq  e^{-i a\bf P} \, |Q\rangle  = |Q+a\rangle \eeq  
since \beq  
 {\bf Q} \, e^{-i a \bf P}\,  |Q\rangle  =\left( [{\bf Q}, e^{-i a \bf P}] +
 e^{-i a \bf P} Q\right) |Q\rangle  =( a+Q)\, e^{-i a \bf P}\, |Q\rangle  \eeq 
This also implies that in the coordinate representation\beq 
 e^{-i a \bf P} \, \psi(Q) = \psi(Q-a)\eeq  In a sense, what is being said 
here 
is that 
we have \beq | \psi \rangle \sim | \psi \rangle + {\bf P} | Any 
\rangle\eeq since 
the state $ {\bf P} | Any \rangle$ should decouple from any physical 
state:
let us try\footnote{here  $ {\bf P =\Phi}$.} \beq  \langle 
\Phi \!=\! 0|  {\bf \Phi} \, | Any\rangle = \langle \Phi\!=\! 0|  {\bf \Phi}\,  |\Phi 
\!=\! 0 
\rangle \langle \Phi\!=\! 0 | Any\rangle = 0 \cdot \infty \cdot\langle 
\Phi = 0 | 
Any\rangle \eeq  The answer is no in general---we run into 
hermiticity 
problems. {\em  However, if one uses $| Any\rangle=| \chi 
\!=\! 0\rangle$ where $[\chi, \Phi] \neq 0$ then the decoupling 
occurs}.

The quantum gauge transformation  behavior matches the classical 
one. The 
expectation value of the 
position operator behaves as it should classically---whether we let 
the above 
transformation operator act on the states or on the operators. The 
$\bf Q$ 
operator correponds to the classical $Q$---the gauge degree of 
freedom.

Notice, though, that all we needed for the previous
statement was the Heisenberg algebra. 
So this idea provides us with an interpretation for the 
gauge transformation ideas in the quantum version of a theory
with constraints. Let us try to obtain the gauge invariant states and 
an 
inner product for them. 

Gauge invariant states will be defined by \beq
{\bf P} \; |\Psi\rangle  = 0 \eeq
The reason for the use of this definition is that these states will 
clearly be gauge-invariant: \beq 
e^{-i a \bf P}\; |\Psi\rangle  = |\Psi\rangle \eeq 
Whether this condition is too strong or not will be discussed later; for
now let us try to work with it.
Now, there is only one class of states that meet the above criterion:
\beq  |\Psi \rangle  = |P\!=\! 0\rangle \otimes |phys\rangle \eeq 
where the label ``{\em phys}'' refers to the other, {\em physical}
 degrees of freedom.

Where do we need to talk about ``gauge fixing''? Well,
the old inner product definition may run into trouble: \beq 
\langle \Psi_a|\Psi_b\rangle  = \langle P \! = \! 0|P\!=\! 0\rangle 
\langle 
phys_a|phys_b\rangle \eeq 
As mentioned, the states above have infinite norm in the unbounded coordinate 
case: $\langle P \! = \! 0|P\!=\! 
0\rangle 
=\delta(0)$.
{\em We need to fix this normalization  problem,
 as well as provide for a resolution of the
identity in this Hilbert subspace}. 

The resolution of the identity would naively be \beq
{\bf I_{\Phi}}= \sum |P\!=\! 0\rangle \otimes |phys\rangle \langle 
phys | 
\otimes \langle P \! = \! 0| \eeq 
This 
would be a resolution of the identity on the physical subspace; on the 
full 
space it would be a projector into the physical subspace. Indeed, we 
would 
have
$ ({\bf I_{\Phi}} )^2={\bf I_ {\Phi} }  $ if it weren't  for the 
  infinite norm problem. In the next section 
we will
study the situation with periodic boundary conditions. Then we will
look at the infinite coordinate space, and see the connection
between the Dirac quantization approach and the Faddeev path
integral in a direct way.

\subsection{Periodic boundary conditions}

How can we solve the infinite norm problem? It originates with the
 infinite volume in 
coordinate space. {\em  One possible solution is to use a finite volume 
coordinate space}. If we use a finite coordinate space---by imposing 
periodic 
boundary conditions, say---then the norm of the momentum states 
will be 
finite. The momentum operator will now have a discrete spectum, $ 
P_n = n 
{2\pi\over L}$, but otherwise this representation will have the same 
properties  that the infinite volume case has. In the coordinate 
representation, for example, we have the operators $\bf  Q ,P$ with
the commutator $\bf{ [ Q,P]=}i$ represented as $\bP=-
i\partial/\partial Q$
and\footnote{Strictly speaking the coordinate degree of freedom 
should be described by $\exp (\pm i2\pi n {\bf  Q} / L)$.} 
${\bf  Q}=Q$. The $\bP$ e-states are given by $\phi_n= e^{-iP_n 
Q}/\sqrt{L}$,
 where
$P_n = 2\pi n  / L$ as before, and the $\bf  Q$ e-states by $\delta_L 
(Q-
Q_0)$,
where the delta function is also periodic with period L.
Just as in the infinite volume case it is true  that \beq 
e^{-ia\bP}|Q\rangle  = |Q+a\rangle \eeq  and also that \beq  e^{i k {\bf 
 Q}}|P_n\rangle  = |P_n + k\rangle \eeq 
although notice that $k$ is immediately restricted to be of the form 
$k=2\pi 
n/L$, since all the operators in the theory come from classical 
functions 
defined
in the periodic coordinate space---that is, periodic functions! At
any rate, the gauge transformation ideas still apply in this context. 
Notice that the state $P_n = 0 $ is still in the theory.

The resolution of the identity in such a space is given by \beq 
I= \sum |P_n\rangle \langle P_n| = \int_0^L dQ |Q\rangle \langle Q| 
\eeq

{\em How do the earlier definitions of inner product match with this 
philosophy? }We see that here there is no need for gauge fixing! 
However, 
we can artificially introduce a ``gauge-fixing normalization effect'''  
by 
altering the normalization of the gauge invariant states, say by 
multiplying them by a constant factor, or by a function. 
This is ok, the representations are isomorphic, although not unitarily 
so 
(recall that there are different ways to normalize states even in 
the 
unconstrained case).

Notice also that if the constraint has multiple solutions the solution        
states---the physical states---can always be chosen to be orthogonal 
in the 
original bigger space if the 
constraint is hermitean. For example, say $\Phi = (P-a)
(P-b)$. Then the solution states will be orthogonal in the original 
bigger 
space if the original inner
product in the full space is such that the operator $\bf P$ is 
hermitean. This
is the case for the relativistic particle in some
situations, as in the free case. In a
more complicated case this will not apply, of course. In the present 
discussion, orthogonality is preserved under quantum reduction.

How do we describe the dynamics? As explained before, the general 
solution to the Schr\"{o}dinger equation is given by \beq 
|\Psi\rangle  = e^{-i t{\bf H}_E}\,  |\Psi_0\r = e^{-i t\bf h}\, e^{-i t\bf 
vP}\; |\Psi_0\rangle  \eeq 
Consider the transition amplitude
\beq
\langle \Psi_f|\Psi_i\rangle  =\langle \Psi_f|  e^{-i t\bf h}e^{-i t\bf 
vP} 
|\Psi_i\rangle \eeq
 If the initial or final  
states 
satisfy the constraint the amplitude will not 
depend on $\bf v$. If not, just insert the projector ${\bf I_{\Phi}}$.
Notice that this definition of physical/non-physical states is an 
orthonormal 
definition if the constraint is hermitean---or equivalent to a 
hermitean one. The physical and non-physical sectors decouple, and 
this 
decoupling is respected by the---first class---hamiltonian.

Let us now 
consider the following question: 

{\em Can we recover the Faddeev path integral?} 

We can certainly use \beq  {\bf I_\Phi} \equiv 
|P\!=\! 0\rangle \langle P \! = \! 0| = \sum \: \delta_{0P_n} 
|P_n\rangle 
\langle P_n|
\eeq 
and insert this in the  propagation amplitude. 
Consider then the physical amplitude
\beq  \langle Phys| e^{-i t{\bf H}_E}\; |Phys \rangle\equiv 
U_{Phys}(q,q';t)\eeq where
$q$ represents the physical degrees of freedom, so that 
$|Phys\rangle \sim 
|P=0,
q\rangle$. Then \beq 
\langle P=0, q_f| e^{-it{\bf H}_E}\;|P=0,q_i\rangle = \eeq  
\beq 
   \langle  q_f|e^{-it{\bf H}_E} \;|   q_i\rangle 
\eeq 
This doesn't look like a good way to get to the Faddeev path integral 
at all!

Consider instead the amplitude
\beq \langle \psi | e^{-i t{\bf H}_E}  \,\delta( {\bf P})\;  |\psi \rangle\equiv 
U_{Phys} \eeq
We have projected the transition to the physical states. However, this 
is not a gauge-invariant object, because the coefficients of the 
projection, $\langle P \! = \! 0|\psi\r $, are not. This is the boundary 
effect that we will discuss later---such is the situation in the particle 
case. Now, we can use the projector above, but this  will not be a 
very illuminating experience. Notice that, for a constraint of the form 
$ {\bf\Phi =  P}-a =0$,
\beq
\left( |P\!=\! a\rangle \langle P \! = \! a|  Q\!=\!Q_0\r \l Q\!=\! Q_0|             
\right) \cdot
|P\!=\! a\rangle \langle P \! = \! a|  = \eeq
\beq    |P\!=\! a\rangle \langle P \! = \! a|\eeq 
and \beq 
\left( |P\!=\! a\rangle \langle P \! = \! a|  Q\!=\!Q_0\r \l Q\!=\!Q_0|             
\right) ^2 = \eeq  \beq 
\left( |P\!=\! a\rangle \langle P \! = \! a|  Q\!=\!Q_0\r \l Q\!=\!Q_0|  
\right) \equiv \bf K \eeq  Using these two properties, $\bf K    
I_\Phi =
I_\Phi$, and $ {\bf K }^2 = \bf K $,  we are ready to build a path 
integral, using the resolution of the identity 
\beq
\hat{I} = \int dQdq \: | Q,q\r \l Q,q| = \sum _{ P_n} |P_n\r\l P_n|
\eeq The amplitude can be written as 
\beq \langle \psi | e^{-i t{\bf H}_E} \, \delta( {\bf P})\; |\psi \rangle\equiv 
\eeq \beq 
\int DQ\prod \{ \sum _{ P_n} \} \int DqDp \:\delta(P_0-a )\; \left( \p 
\delta(P_i-a )\delta(Q_i-f(\tau_i)\right) e^{i\int [d\tau P\dot{Q} + 
p\dot{q} -h]}
\eeq 

{\em Example: Dirac quantization  of the particle with periodic boundary 
conditions}

The constraint is $\Phi= p_t + p_x^2/2m = 0$. Assume that the $t$ 
coordinate space is finite, with periodic boundary conditions.
Now, this operator  $\bf \Phi$ is conjugate to $t$. We will represent
it by $-i\partial_t-\partial_x^2$.
 The solution space to the constraint
 in this coordinate representation
 is given by the states \beq  \psi(x,t) = e^{i t p_t +i x p_x }
 {1\over \sqrt{T L}}\eeq 
where $p_t = 2\pi n' /T$, and  thus we need \beq 
p_t = -{p_x^2\over 2m} = -{(2\pi n L)^2 \over 2m} = -2{ \pi^2\over 
L^2 m} 
n^2 \equiv 2{\pi\over T} n' \eeq 
We
see that it is hard to have both space and time bounded. Indeed, we
are then looking for the solutions to the above differential equation
in a torus. I guess we could put restrictions on the allow mass,
but still, we would not have all the $p_t$ momenta.

We could go to the constant of the motion coordinate system, and put
periodicity there.   

So, anyhow, let the $x$ coordinate be unbounded. Then we see: the 
solution
space to the constraint is in one-to-one correspondence to the $x$ 
momentum
states, i.e, in one-to-one correspondence to the $x$ coordinate Hilbert 
space.
And these Dirac states are correctly normalized. We can use the 
inner product
\beq (\psi_a,\psi_b) = \int_0^T dt\int dx\; \psi_a^*\; 
\psi_b= \int dx \;\psi_a^*\;\psi_b\eeq 
This is because there are no normalization problems to begin with.

\subsection{Unbounded gauge coordinate space}

In this section we will produce the connection between the Dirac
inner product and the Faddeev path integral. This reasoning
will also take us to the BRST inner product and the rest of the
BRST quantization approach, including the BFV path integral. 

The key to the Fadeev path integral is in the Dirac inner product.
We must obtain a resolution of the identity in the physical space, 
and for that we will use the Dirac inner product. Let us start with
the quasi-projector \beq {\bf Y} \equiv
|P\! = \! 0 \rangle \langle P \! = \! 0|\eeq
This operator will take
any state into the physical space (quite clearly). We
can look at it in the coordinate representation, using
the resolution of identity: 
\beq
{\bf Y} = \int dQdQ' \: |Q\rangle\langle Q| P\!=\! 0\rangle\langle P \! = \! 
0|
Q'\rangle\langle Q'|
= {1 \over 2\pi} \int dQdQ'\: |Q\rangle\langle Q'|
\eeq
 where the normalization $ \langle Q|P\!=\! 0\rangle = 
1/\sqrt{2\pi}$
has been used. Notice that the state $ {\bf Y}|\psi\rangle$
is clearly left unchanged by a gauge transformation.

Now, I have been calling the above operator a projector, and
indeed it would be a projector in the case where the coordinate
$Q$ is defined with boundary conditions, but in the infinite
case $\bf Y$ is not a projector. By defintion a projector $\bf K $
satisfies the property $ \bf K  K =K$, and instead we have ${\bf Y Y
=Y }\delta(0)$. Here is the source of all our headaches.

Consider then the projector
\beq {\bf K_\Phi}\equiv
 |P\!=\! 0\rangle\langle P \! = \! 0|\delta({\bf  Q-a})
= \left(\int dQ\:|Q\rangle \right)\langle Q=a| \eeq
which we can 
also rewrite as \beq {\bf K_\Phi} = \delta({\bf P})\delta({\bf  Q-a})= 
|P\! 
=\! 0 
\rangle 
\langle Q=a |\eeq
It satisfies \beq \bf K_\Phi K_\Phi = K_\Phi\eeq
However, this operator is not hermitean, since $\bf  Q$ and $\bf P$ 
don't
commute. Notice that the basic trick is to realize that \beq \langle 
Q|P\!=\! 
0\rangle \sim  1\eeq  (ignoring normalization factors) and 
\beq  \langle P \! = \! 0|\delta({\bf  Q-a})|P\!=\! 0\rangle  \sim  
1\eeq 
For this reason we can think of $\delta({\bf  Q-a})$ as a regularizing
term---which is not needed in the case of periodic boundary 
conditions.

\begin{quote}
{\em \noindent The operator $\bf K_\Phi$ has the properties that
 
a) It leaves physical states (kets) unchanged

a') $\delta({\bf  Q-a}) \bf K_\Phi   =\delta({\bf  Q-a})$

b) and $\:\: \bf K_\Phi K_\Phi = K_\Phi$}
\end{quote}

Marnelius' idea \cite{marnelius} is along similar lines. The constraint 
imposes us to work with the states $|P\! =\! 0\rangle$. With these we 
construct the 
space. But to define an inner product we use their duals, $ \langle 
Q_a|$. For 
physical quantities, however, we want to define an inner product 
{\em within the 
physical space alone}---the $|P\rangle$'s. Solution: find an operator 
that maps 
the first into the second \beq  |P\rangle \longleftrightarrow \langle Q 
|\eeq This is 
achieved by \beq |P \rangle \longleftrightarrow  \delta({\bf Q}-a)  
|P\rangle\eeq for example, since \beq \langle P\! =\! 0| \delta({\bf 
Q}-a) 
|P\! =\! 
0\rangle =1 \eeq  {\em Notice, though, that there are many more 
choices for the 
operator that will do this!}

This duality map role is also taken by  the BRST exponential operator---as we will
see--- $\exp 
[\bf \Omega, K]$: we 
can understand it as a ``duality map'' operator. In the case of discrete 
spaces it 
reduces to the unity---trivial duality map.

  Consider the {\em physical} states $|\psi_a\rangle$ and 
$|\psi_b\rangle$,
and let us try to build the {\em physical} amplitude ``$\langle \psi_a| 
{\bf O}
|\psi_b\rangle$'',  where the operator $\bf O$ is physical---i.e., it
commutes with the constraint $\bf \Phi = P$, i.e., it is independent of 
$\bf 
 Q$.
As mentioned, this amplitude needs regularization:
\beq  \langle \psi_a|{\bf O} \delta({\bf  Q-a}) |\psi_b\rangle\eeq 
because physical states are $\bf P$ eigenstates. 

Let me summarize:
{\em We define the inner product between physical 
states to be}
\beq
( \psi_a|\psi_b) \equiv 
\langle \psi_a|  \delta({\bf  Q-a}) |\psi_b\rangle
\eeq
Notice that {\em 
physical 
operators will be hermitean in  the regularized inner product if 
they 
were hermitean to begin with in the bigger space.} This is because 
we can 
write such an operator in the form \beq {\bf O} ={\bf O(P,q,p) =  
O'(q,p) + 
O''(P,q,p) P}\eeq  with $ \bf O''$ regular in $\bf P$. The first term is 
hermitean, and the second one is as well---it drops out.

We can now proceed to construct the path integral. The resolutions
of the identity we need to use are the usual ones\footnote{
As usual, the coordinates $Q,P$ refer to the gauge
degrees of freedom, and $q,p$ to the physical ones.}, \beq 
{\bf I} = \int dQ dq\: |Q,q\rangle\langle Q,q| = 
\int dP dp \: |P, p\rangle\langle P,p|\eeq 
which we insert in the amplitude
\beq {\cal A}=
\langle \psi_a | e^{-i \tau ({\bf h}+v{\bf P})} \delta({\bf  Q-a}) 
|\psi_b\rangle 
= 
\langle \psi_a | e^{-i \tau ({\bf h}+v{\bf P})} \delta({\bf  Q-a}) {\bf 
K_\Phi 
K_\Phi ...K_\Phi}|\psi_b\rangle\eeq
to obtain \beq {\cal A} = \int dQdqdPdp \: \delta(Q-Q(\tau))
\delta (P) e^{-i\int d\tau (P\dot{Q}+p\dot{q} - h(q,p))}\eeq

Before  getting into more details, notice also the following
trick: if we have a conjugate pair like $ Q,P$
it follows that \beq
\langle {P} \! =\! 0 | e^{i {\bf Q A}}|P\!=\! 0\rangle = \delta({\bf 
A})\eeq
where the operator $\bf A$ is understood to commute with $\bf 
Q,P$.
We will use this trick to produce the delta functions in the theory.
The phase space keeps getting bigger.  Consider also the 
fact that for two ghosts, $\eta_1,\eta_2$ and their 
conjugate momenta $\rho_1, \rho_2$ we have 
\beq
\langle \rho_1\! =\!\rho_2\! =\!0|e^{i\mbox{\boldmath$\eta_1 {\bf 
O} 
\eta_2$}}
 |\rho_1\! =\!\rho_2\! =\!0\rangle
= det({\bf O})
\eeq
(all these can be seen using the coordinate resolutions of the 
identity, as we will explain in a second), so finally we can write that 
the 
physical quantities
are to be obtained by using the amplitudes
\beq
\langle \psi_a,\pi\! =\! \rho_1\! =\!\rho_2\! =\!0| {\bf A} 
e^{i{\mbox{\boldmath$\lambda \chi+ \eta_1 |\{\chi,\Phi\}|\eta_2$}}}
\;|\psi_b, \pi\! =\!\rho_1\! =\!\rho_2\! =\!0\rangle
 \eeq
which is the equivalent to our earlier expression with the Dirac
inner product in ghost form. To see this we insert
\beq
{\bf I} = \int dqdQ d\lambda d\eta_1 d\eta_2   \:\:
| q,Q,\lambda, \eta_1,\eta_2     \rangle \:     \langle q,Q,\lambda, 
\eta_1,\eta_2  |
\eeq
 and use the projections 
\beq
 \langle  q,Q,\lambda, 
\eta_1,\eta_2|    p,P,\pi, \rho_1 , \rho_2 \rangle = e^{i( 
qp+QP+\lambda\pi+   
\eta_1\rho_1+ \eta_2 \rho_2 )}
\eeq
It is assumed here that the terms 
in the 
exponent commute as operators, otherwise write 
\beq
\langle \psi_a,\pi\! =\! \rho_1\! =\!\rho_2\! =\!0| {\bf A} 
e^{i{\mbox{\boldmath$\lambda \chi $}}}
e^{i{\mbox{\boldmath$  \eta_1 |\{\chi,\Phi\}|\eta_2$}}}
\;|\psi_b, \pi\! =\!\rho_1\! =\!\rho_2\! =\!0\rangle
\eeq instead.

This is essentially the BRST inner product and the
point of departure for the BFV path integral;  we will discuss this 
more 
throughly in the next section. 
 
\noindent {\bf Example\footnote{This is essentially exercise 13.5 in 
Henneaux \& Teitelboim's book 
\cite{TeitelBook}.
} 1: Constraints of the form $G_i=p_i - \partial V/\partial 
q_i 
=0$}  \\
We  consider the constraints \beq G_i=p_i - \partial V/\partial q_i 
=0\eeq
for some subset of the indices $i$, and where   $V$ depends on the 
$q_i$ only.   
{\em a)} The gauge transformation generated by the constraints is given by 
\beq A 
\rightarrow A + \delta_\epsilon A= A + \epsilon^i  [A, p_i - \partial 
V/\partial q_i]\eeq where the 
generator of the canonical transformation is just \beq {\cal G} = 
\epsilon^i
G_i, \:\:\: A\rightarrow A+[A,{\cal G}] \eeq
 This is complicated in general. However, we have \beq
q_j \rightarrow q_j + \delta_\epsilon  q_j= q_j + \epsilon^i [q_j, p_i - 
\partial 
V/\partial q_i] = q_j + \epsilon^i \delta_{ij}\eeq
{\em  b)} The Dirac state condition is given by \beq 
\left( -i{\partial \over \partial q_i } - {\partial V\over \partial q_i}  
\right) 
\psi(q) = 0 \eeq which implies that \beq \psi (q_j + \delta_\epsilon  
q_j ) = 
e^{
i\delta_\epsilon  q_k \:{\bf \hat{p}_k}}
\psi (q_j) = e^{i\delta_\epsilon  q_k \partial V/\partial q_k}
\psi (q_j)\eeq so we have ``classical'' invariance up to a phase.\\  
{\em c)} Let us study this Hilbert space, including the inner product. We 
will 
compare the result with the reduced phase space approach.

The Dirac inner product for these states is given by \beq 
(\varphi_a,\:
\varphi_b) = \int dV \varphi_a^*\: \left(\prod\delta(q_i)\right) \: 
\varphi_b
\eeq say. We have that the $q_i$ degree of freedom if absent...just as 
if we  had started by quantizing a theory with the constraints 
$G_i=0$ as 
above, and $\chi_i = q_i=0$. The solution to the  Dirac 
conditions 
equation is given by\beq\psi(q) \sim e^{i V(q)}\tilde{\psi}(q\! 
\neq\! q_i)\eeq
With this 
additional piece of information it is  clear that the inner product will 
not 
depend on the gauge fixing $\chi$.

What 
about the 
expectation value of {\em observables}? Observables need to be 
hermitean and have expectation values that are not gauge 
dependent.  Thus they  need to commute with both the constraints 
and the gauge-fixing. The solution to \beq [\hat{A}, \hat{\Phi}] 
\approx 0 \eeq is given by \beq
\hat{A} = e{-i \hat{V}} \hat{O}  e{-i \hat{V}} + \hat{\alpha} \hat{\Phi} 
\eeq
The correct interpretation 
is that, 
again, when we change gauge-fixing
we change representations. A unitary transformation is involved: for 
a given 
gauge-fixing we have the states\beq \psi(q)_\chi \sim \left. e^{i 
V(q)}\right|_{\chi\!=\! 0} \tilde{\psi}(q\! 
\neq\! q_i)\eeq  for another\beq \psi(q)_{\chi'} \sim \left. e^{i 
V(q)}\right|_{\chi'\!=\! 0} \tilde{\psi}(q\! 
\neq\! q_i)\eeq The difference is a unitary (canonical) 
transformation.
In the same way, for the purposes of this isomorphism, the operators  
must
transform.  Of course, the question to go back to is: what is my 
reference 
point, or, what is the inner product, or, how do I choose my 
observables?

To summarize,  we have the following rules \\
a) use {\em observables} in the usual sense ($[\bf A,G] = 0$), for they 
map the 
physical space on itself, and they lead to gauge-fixing invariant 
expectation values, and \\  
b) pick the right $ \hat{A} \sim \hat{A} + \lambda \hat{G}$ in the 
equivalence class to ensure hermiticity with respect to the chosen         
gauge-fixing.\\  
\begin{quote}{\em
Gauge-fixing picks an element in the 
equivalence class of observables, and   this is tied to the hermicity 
properties of the 
observables.}
\end{quote}

\noindent{\bf Example 2: the non-relativistic particle and similar cases}\\
As discussed, in  the situation where we do not impose boundary 
conditions  the
solution to the Dirac constraint is not normalizable. The solution
is  given by $\psi = e^{it p_t+i x p_x}$, with the condition
$p_t +p_x^2/2m = 0$, but there are no further conditions.
And the inner product/norm for these states yields immediately
 a $\delta(0)$ from the $dt$ integration. This is the reason why
some ``gauge fixing'' is introduced---to take care
of the ``gauge infinite volume''. The inner product for the
physical states can be defined by\footnote{Here there is no need to 
reorder 
the 
inner product operators, since they will turn out to be hermitean 
immediately.} 
\beq (\psi_a,\psi_b) = \int dx dt
\:\psi_a^* \:\delta({\bf K}(x,t,\tau))\;{\bf  |\{ K,\Phi\}|} \:\psi_b\eeq 
This we can rewrite as\beq 
(\psi_a,\psi_b) = {1\over 2\pi} \int dxdtd\lambda dcd\bar{c} \:
\psi_a^* \;e^{i\lambda {\bf K} + i c|\{{\bf K,\Phi}|\}\bar{c} }\; 
\psi_b\eeq 

As long as we choose $\bf K$ so that $K=0$ can be rewritten
as $t=t(x,\tau)$ we are ok. However, notice that the
 final form of the inner product will depend on
the form of the constraint---not on the gauge-fixing. This ambiguity
is related to the above discussion on the possible different 
representations. 
Indeed, this inner product reduces to the inner product
of the unconstrained case we discussed above---with the same 
``ambiguity''.

  The constraint is 
\beq 
\Phi  = p_t  + {p_x ^2\over 2m }
\eeq 
so, according to Dirac \cite{Dirac}, the states are defined by starting 
with a full Hilbert space,
\beq 
|x,t> ,\; |\psi > ,\; <x,t|\psi  > = \psi (x,t,\tau ),
\eeq 
and then by imposing the condition
\beq
(\bp_t  + {\bp_x^2\over 2m})|\psi > = 0
\eeq
In this Hilbert space the operators are $\bx,\bt, \bp_t , \bp_x  $ with 
the 
usual 
commutation relations $[\bx, \bp_x ] = i = [\bt, \bp_t ]$ with the 
others 
zero.  
The hamiltonian in this  system is zero, so  the Schr\"{o}dinger equation 
of 
motion is just 
\beq 
{\partial \over \partial \tau} |\psi > = 0.\eeq 
 The states are frozen with respect to the $\tau$  parameter, and are 
thus 
gauge invariant. In this case this means that the physical states are 
solutions 
to 
the Schr\"{o}dinger equation in $t$-time: just use the definition of 
\beq \bp_t  = -ih{\partial \over \partial t}\eeq   
in the position representation. These states are in one-to-one 
correspondence 
with the states of the unconstrained non-relativistic particle after we 
impose 
some gauge condition---like $\chi  = \chi _\alpha  = 0.$
Indeed, the inner product is defined to be
\beq
(\varphi ,\psi ) = \int dx dt \: \varphi^* \, \overbrace{\delta ({\chi }) |\{ \chi 
,\Phi \} | }
\;
\psi 
\eeq
and with the above gauge it reduces to the inner product of the 
unconstrained 
case. Notice that the inner product is independent of `` time'', i.e., 
independent 
of gauge fixing. 
Also, from the point of view of this inner product, gauge fixing has to 
be 
chosen 
so that the operator $ |\{ \bchi ,\bPhi  \} | $ has no zero modes, e.g, 
$\chi  
=t$ is ok, but  $\chi  = x$ is not.
This description can be seen to correspond to the Schr\"{o}dinger 
picture.

Let me be more explicit. Consider the constraint 
\beq
\hat{\Phi} = \hat{p}_t + \hat{A}(\hat{x},\hat{p}_x)
\eeq
where $\hat{A}$ is a hermitean operator (in the full space sense).
The states that solve the Dirac physical condition, $\hat{\Phi} \, |\psi^D\r =0$ are given
by\footnote{These can be {\em rewritten} as $ 
|\psi^D\r =\exp( -i \hat{t}\, (\hat{A}-k_t) ) | \varphi(x) \r\! \otimes\! | p_t \!=\!k_t 
\r$ } \beq
|\psi^D\r = e^{-i \hat{t}\, \hat{A}} \; | \varphi(x) \r\otimes | p_t \!=\!0  \r \eeq 
 To see this, write a general state in the form $|\psi\r= \exp(-i
\hat{t}\, \hat{A})\, |\eta\r$ and use  \beq [\hat{p}_t + \hat{A}, e^{  -i \hat{t}\,
 \hat{A}  }]  =  e^{ -i \hat{t}\,  \hat{A}  }  \hat{A} 
\eeq
which means 
\beq
\hat{p}_t + \hat{A}(\hat{x},\hat{p}_x)\; e^{ -i \hat{t}\,  \hat{A}  } = 
e^{ -i \hat{t}\,  \hat{A}}  \hat{p}_t 
\eeq
so we need the zero eigenstate of $\hat{p}_t$---because of the way we chose to write the
solution.  Again, life is easy when $[\hat{p}_t, \hat{A}]=0$.
In the
coordinate representation, with $\psi(t,x) \equiv \l t,x  |   \psi \r$, $\hat{p}_t \sim
-i\partial_t $, etc., the Dirac states  are given by 
\beq
\psi^D (t,x) = e^{-i \hat{A} \, t}\; \varphi(x)
\eeq 
Then the Dirac inner product is given by ($\chi = t -f(\tau)$)
$$
(\psi^D_a, \psi^D_b) \equiv \l\psi^D_a|\, \overbrace{
\delta ({\chi }) |\{ \chi 
,\Phi \} |
}    \,  |\psi^D_b \r = 
$$
$$
\int dtdx\,\left(\psi^{D}_a\right)^*  \, \delta(t-f(\tau)) \, \psi^D_b = 
\int  dx\,\left(\psi^{D}_a\right)^* \,  \psi^D_b=$$ \beq
\int dx\, e^{+i \hat{A} \, t}\; \left(\varphi_a(x)\right)^* \, e^{-i \hat{A} \, t}\;
\varphi_b(x) =\int dx\,  \left(\varphi_a(x)\right)^* \,  \varphi_b(x)
\eeq because of the hermiticity of $\hat{A}$. This is 
the usual inner product of the unconstrained case.

\noindent {\bf  Example 3: the Klein-Gordon inner product---one 
branch only.}

We begin with the phase-space described by the coordinates 
$t,x,p_x,p_t$
and the constraint (which we write relativistically) \beq
\Phi = p_t^2-p_x^2-m^2 \approx 2\sqrt{p_x^2+m^2}
\left( p_t-\sqrt{p_x^2+m^2}\right)
\eeq The last (weak) equality follows when we restrict ourselves to 
the 
positive $p_t$ branch---which we do in this simple example.
That is, we will pick a branch in the decomposition\beq
\delta(\Phi)= \delta\left(2\sqrt{p_x^2 +m^2}\left(p_t-
\sqrt{p_x^2+m^2}\right)
\right) + \delta\left( 2 \sqrt{p_x^2+m^2}\left(p_t+\sqrt{p_x^2 + m^2} 
\right)
\right) \eeq
which is equivalent to working with  the constraint expressed as 
\beq 
\Phi' = \left(p_t - \sqrt{p_x^2 + m^2}\right) \sqrt{p_x^2 + m^2} \eeq 

We now go to the quantum theory, where the variables above 
become
operators in the usual way. For example, in the coordinate 
representation
the variable $p_t$ becomes the operator $-i\partial_t$.

The physical states are then defined beginning with the full Hilbert 
space---the
one corresponding to the above phase-space---and then imposing the 
physical
condition \beq {\bf \Phi} |\psi\rangle =0\eeq just as before.
The states are as in the previous example, since the multiplicative operator does
not have any zero modes. They are thus given by
\beq
|\psi^D\r = e^{-i \hat{t}\, \hat{A}} \; | \varphi(x) \r\otimes | p_t \!=\!0  \r 
\eeq
with $ \hat{A} = \sqrt{\hat{p}_x^2 + m^2}$, or in the coordinate representation as
\beq \psi(t,x) = e^{-i t\sqrt{\hat{p}_x^2 +m^2}}\; \varphi(x) \eeq

With this choice we will find that the inner product is given by
\beq
(\psi_a,\psi_b)_{\Phi'}= \int d^4 x \; \psi_a^*  \,  \sqrt{\hat{p}_x^2 +m^2}   
\; \psi_b\;
 \delta(t-f(\tau))
  \eeq
For the states that satisfy the constraint 
we have \beq  {  \sqrt{\hat{p}_x^2+m^2}}\;\psi = i {\partial \over 
\partial t}\psi\eeq  
so
this form of the inner product is equivalent to the Klein-Gordon 
inner
product, \beq
(\psi_a,\psi_b) = \int d^3x\; \psi_a^* \;{1\over 
2}\left(i\overrightarrow{\partial}_t 
- 
i\overleftarrow{\partial}_t
\right)\,
 \psi_b   \eeq
In general we can say that the effect of rescaling the constraint on the inner
product has been that with   $\Phi' =\Gamma(q,p) \Phi$ we have 
that the new inner product   reads
\beq (\psi_a,\psi_b)_{\Phi'}= (\psi_a,   \hat{\Gamma}\,  \psi_b )_{\Phi}
\eeq


Let us now anticipate the BRST inner product. The above states are 
now
embedded in the extended space: they are given by
\beq  |\Psi\rangle\equiv |\psi^D \rangle \otimes |\lambda \! = \!  
\rho_1\! = \! \rho_2\! = \! 0\rangle \eeq 
and the inner product is given by
\beq 
\langle \Psi_a|\: e^{i  \hat{\pi} \hat{\chi}   }  \;
 e^{i \hat{\eta}_1  \,{\scriptstyle \overbrace{ |\{\chi,\Phi\}| } } \,
\hat{\eta}_2   }\:
|\Psi_b\rangle = \langle \psi_a|\; 
\overbrace{ \delta(\hat{\chi})\; | \{ \chi,\Phi\}| }\,
   \: |\psi_b\rangle \eeq  
which we further rewrite as 
\beq 
 = \int dtdx\: \psi_a^*(t,x)\;
\overbrace{ \delta(\hat{\chi})\, |\{\chi,\Phi\}| } 
\; \psi_b(t,x)
=
\int dx \: \psi_a^*(t,x)\: \sqrt{{\hat{p}_x^2} +m^2}\: \psi_b(t,x)
\eeq  or finally, again as 
 \beq =
 \int dx\: \psi_a^*(t,x)\: {1\over 2} \left( i
\overrightarrow{ \partial}_t -i\overleftarrow{\partial}_t
\right)\: \psi_b(t,x)
\eeq 
which is the BRST inner product (notice that in the last equality
we have used the fact that the states satisfy the constraint). The
gauge $\chi = t-f(\tau)$ leads to this result very directly, but
any gauge that can be rewritten in the form $t=t(p_t,p_x,x)$ will 
give the same answer. The only restriction, as usual, is that
$\{\chi,\Phi\}\neq 0$.

Also notice that the result does depend on the way we write
the constraint---which is by far not unique. But this 
arbitrariness can be eliminated by demanding that some
observables be hermitean---which is what Ashtekar  does 
\cite{ashtekar}.
But how do you pick your observables?

The thing to keep in mind is that our choice in the way
in which we write the constraint will affect the normalization. 
 Notice, however, that even in the unconstrained case one 
is always free to change the states by \beq
|\psi\rangle \longrightarrow  |\psi'\rangle \equiv {\bf \balpha^{-1}}|
\psi\rangle\
 \eeq
as long as the inner product and expectation values are modified 
accordingly,\beq
\langle \psi |{\bf A}|\psi\rangle  \longrightarrow \langle \psi ' |{\bf 
\balpha^\dagger}{\bf 
A}{\bf \balpha}|\psi ' \rangle  \eeq
In fact the solution spaces to the constraint\beq 
{\bf \Phi } |\psi\rangle  = 0 \eeq 
and
\beq  {\bf \Phi v }|\psi\rangle  = 0 \eeq  are isomorphic as long as 
the 
operator 
$\bf v$
 has no 
zero modes---if $|\psi_0\rangle  $ is a solution to the first then ${\bf 
v^{-
1}}|\psi_0\rangle $ solves the second---and viceversa. This 
equivalence/ambiguity 
appears in the definition of the constraint and then in the
above definitions of the inner product---it appears in the 
determinant.

One can fix this ambiguity by demanding that certain operators in 
the theory 
be hermitean \cite{ashtekar}---although this procedure just 
transfers the 
ambiguity from the inner product to the definition of observables.

\begin{quote} {\em
The punchline is that we are free to normalize as we wish. There is 
extensive
freedom on the way we represent the Heisenberg                                   
algebra---something that 
applies to constrained as well as unconstrained systems. The physical 
quantities in theory must remain the same, of course. } \end{quote}

\subsection{Dirac inner product for the two  
branches case: ordering problems and unitarity}

What problems do we  encounter when  the relativistic constraint's 
full 
solution space is used? Let us review the description of the 
inner product and the 
observables and their hermiticity 
properties.

But first we need {\em a good inner product: } 

a) it satisfies $\left( ( \psi_a |\psi_b ) \right)^*
=                                                     
   ( \psi_b | \psi_a ) $  

b) it is  invariant under changes of gauge-fixing 

c) it is conserved under time evolution---the hamiltonian
is hermitean, and  

d) $( \psi | \psi  ) \geq 0$,      $=0$ only if $| \psi  ) =0$.

\noindent The last requirement can be postponed until we decide to 
equate 
norms with probabilities.

Consider now the definition of observables. \\
{\em i)} Observables are chosen by asking that \beq [\hat{A}_D, 
\hat{\Phi} ] 
\approx 0\eeq  but we will have to be careful with the weak 
equality.\\
{\em ii)} Also, $\hat{A}_D \sim \hat{A}_D+\hat{\lambda}\hat{\Phi}$, 
since 
the effect on physical states is the same, and they are both 
observables.\\
{\em iii)} We can use this freedom to ask that \beq [\hat{A}_D, 
\hat{\chi} ] 
\approx 0\eeq  for some $\hat{\chi}$ with $\left. [\hat{\chi}, 
\hat{\Phi} 
]\right|_{\Phi\! =\! 0} \neq 0$. This is ``gauge-fixing''.\\
{\em iv)} Now we need hermiticity of the observables with respect to 
the 
inner product on physical states given by \beq 
(\psi^D_a, \psi^D_b) = \int dV  \:   
\left( \psi^{D}_a\right)^* \; \overbrace{ \prod_a \delta( \chi_a)
\mbox{sdet}\{ \chi_a,   {G}_b\} } \;\psi^D_b\eeq 
(where, again, the big hat just means that the whole will become an operator---and
that we haven't commited to a specific ordering yet).
But first, we need  to demand that the 
operator 
$\overbrace{ \prod_a \delta( \chi_a)
\mbox{sdet}\{ \chi_a, {G}_b\} }$ be 
hermitean and 
positive definite. 
Let us see why. A good inner product needs to satisfy \beq
[ ( \psi_a , \psi_b ) ]^* = ( \psi_b , \psi_a ) \eeq as we 
wrote above. 
This ensures that states have real norms, \beq 
\para \psi \para^2 \equiv ( \psi  , \psi  ) = [ ( \psi  , \psi  ) ]^* \in \IR 
\eeq
 This, for example, tells us that the inner product for the free 
relativistic case 
can be defined to be  the Klein-Gordon inner product,
\beq
(\psi^D_a, \psi^D_b) = \int dV  \: 
\left( \psi^{D}_a\right)^* \; {1\over 2}\left( \delta( t - f(\tau))
\,\hat{p}_t + \hat{p}_t^\dagger\,  \delta( t - 
f(\tau))\right)\;\psi^D_b
\eeq
Why is this the Klein-Gordon inner product?
Consider the  operator
 \beq
\hat{O} = \delta(t-f(\tau)) \left( -i \stackrel{\rightarrow}{d\over dt}\right)
\circ  \eeq
It is easy to see explicitely that its hermitean conjugate is given by
\beq
\hat{O}^\dagger =  \left( -i \stackrel{\rightarrow}{d\over dt}\right) [\delta(t-f(\tau)) \circ ] =
\left( i \stackrel{\leftarrow}{d\over dt}\right) [\delta(t-f(\tau)) \circ ]
\eeq
i.e., $ \hat{p}_t \, \delta(t-f(\tau))$, because the delta function takes care
of any boundary condition problems at $t=\pm \infty$. Then 
\beq
\hat{O}+\hat{O}^\dagger = \delta(t-f(\tau)) \left( -i \stackrel{\rightarrow}{d\over dt}\right) \circ 
+ \left( i \stackrel{\leftarrow}{d\over dt}\right) [\delta(t-f(\tau)) \circ ]
\eeq
After integrating the delta function, the inner product becomes the
Klein-Gordon one, 
\beq 
(\psi^D_a, \psi^D_b) =\left.  \int dx \; 
\left( \psi^{D}_a (t,x ) \right)^* \;
 {1\over 2}\left(  -i \stackrel{\rightarrow}{d\over dt} + i \stackrel{\leftarrow}{d\over
dt}\right)\;\psi^D_b(t,x ) \right|_{t=f(\tau)}
\eeq
which is real, as promised (recall that it is the charge of the Klein-Gordon
field).











Now we need to ensure that the definition does not depend on the 
choice of gauge-fixing. Consider the gauge-fixing $\chi = t-\tau_1$. 
The inner
product with this gauge-fixing reads
\beq (\psi^D_a, \psi^D_b)_1 = \int dV  \: 
\left( \psi^{D}_a\right)^*  \; \left(  \delta(\hat{\chi} ) [\hat{\chi} , 
\hat{\Phi}] +  [\hat{\chi} , 
\hat{\Phi}]^\dagger  \delta(\hat{\chi} ) \right)\; \psi^D_b \eeq
It is not hard to see that it we change the gauge-fixing to $\chi = t-
\tau_2$ the inner product will remain the same as long as the old $
\hat{p}_t ^\dagger$ is the hermitean conjugate of $\hat{p}_t$ in the 
new inner product as well. Indeed, 
 \beq (\psi^D_a, \psi^D_b)_2 = \int dV  \:   
\left( \psi^{D}_a\right)^* \; \left(  \delta(  t-\tau_2 ) [\hat{\chi} , 
\hat{\Phi}] +  [\hat{\chi} , 
\hat{\Phi}]^\dagger  \delta(t-\tau_2 ) \right)\; \psi^D_b = \eeq 
\beq
\int dV  \: 
e^{-i \hat{p}_t^\dagger\: \Delta\tau} \left( \psi^{D}_a\right)^*  \; \left( 
\delta(  t- \tau_1 ) [\hat{\chi} , 
\hat{\Phi}] +  [\hat{\chi} , 
\hat{\Phi}]^\dagger  \delta(t-\tau_1 ) \right)\; e^{ i \hat{p}_t \: 
\Delta\tau}\psi^D_b \eeq
Thus, we just need \beq
\left(  \delta(  \chi ) [\hat{\chi} , 
\hat{\Phi}] +  [\hat{\chi} , 
\hat{\Phi}]^\dagger  \delta(\chi  ) \right) \hat{p}_t  = 
\hat{p}_t^\dagger 
\left(  \delta(  \chi ) [\hat{\chi} , 
\hat{\Phi}] +  [\hat{\chi} , 
\hat{\Phi}]^\dagger  \delta(\chi  ) \right) \eeq
{\em in the physical subspace}.

Observe that this is always true: our inner product definition 
matches the 
Klein-Gordon definition---which we know is conserved.
Let us recall how this Klein-Gordon inner product is produced for 
the 
interacting 
case. We start from a classical action which has the Klein-Gordon 
equation 
as its 
extremizing equation of motion, 
\beq
\int dV \sqrt{g} {\cal L} = \int dV \sqrt{g} \left( \nabla_a \Phi 
\nabla^b 
\Phi^* + 
m^2 \Phi \Phi^*\right) \eeq
and, from the global (or local if $\nabla$ stands for the fully 
covariant 
derivative) $U(1)$ symmetry,
\beq 
\Phi \longrightarrow e^{i \alpha} \Phi \eeq
(and complex conjugate)
 of the action we infer the existence of conserved    Noether current,
\beq
\partial_\mu J^\mu =0 \eeq \beq
J^\mu = \Phi \nabla^\mu \Phi^* - \Phi^* \nabla^\mu \Phi \eeq
  We then use this conserved current to define a conserved inner     
product---by 
integrating the first over some surface.

{\em Now, for the case in which we can factor the Klein-Gordon 
equation and the corresponding solution space, this means that  for 
gauge-invariance the sectors must decouple, because this is is the 
only way  for the operator  $\hat{p}_t^\dagger $   to be the 
hermitean conjugate of $\hat{p}_t $ in the new inner product. Thus, 
unitarity implies decoupling.} Indeed, gauge-invariance of the inner 
product is equivalent to unitarity, (or conservation of the inner 
product), since time evolution is a gauge-transformation for 
parametrized theories. This inner product is not going to yield 
positive norms, though. However, we can stick to one branch if the 
Klein-Gordon 
equation factorizes.

{\em We know that our inner product decouples nicely if the       
Klein-Gordon equation does. This is then a criterium for unitarity in 
the one 
particle 
sector.  We already saw that this will occur if \beq
[D_0, D_i] = 0 \eeq}

Now, as mentioned, for the purposes of a probabilistic interpretation 
we would also like to have an inner product that gives states a 
positive norm. When the inner product decouples the course to take 
is clear: just stick to one branch. If the ordering that we                  
choose---say, because of minimal coupling and/or                   
space-time covariance---and the interactions don't allow decoupling then we are 
in trouble in this respect. We will not have space-time covariance and/or 
minimal 
coupling 
and unitarity in the one particle sector.
We will come back to this
point later.

Let us go back to the observables. They need to be hermitean in this 
new inner 
product.
This  will happen provided they were hermitean with respect to the 
old inner 
product (in the physical subspace is sufficient), and provided
they commute with $\overbrace{ \prod_a \delta( \chi_a)
\mbox{sdet}\{ \chi_a,   {G}_b\} }$. 
Conditions
{\em i)} and {\em iii)} cover this last point, together with the Jacobi 
identity.

We can look at the problem from the point of view of our old 
projector 
$\bf 
K_\Phi $. 

Let us start by playing with the simpler constraint \beq  \Phi= P^2-
a^2  
\approx 
0 \eeq 
where $a$ is positive. Then 
\beq \delta(P^2-a^2)  ={ \delta(P -a )\over a} + {\delta(P+a ) \over a}\eeq  

How do we define the projector  here? Following our previous 
discussion, the 
inner product between physical states is defined by 
\beq
( \psi_a, \psi_b) \equiv 
\langle \psi_a|  {1\over 2} \left( \delta({\bf  Q -Q_0})  {\bf P} +   {\bf 
P}^\dagger\delta({\bf  Q -Q_0})  \right)  |\psi_b\rangle
\eeq
This is the inner product  that we had before (Klein-Gordon).  Notice 
that 
the 
two sectors decouple. Now we need the projector (which leaves the 
physical 
states alone.) A first guess is given by \beq
{\bf K_\Phi} \equiv 
\delta({\bf P^2-a^2}) 
{1\over 2} \left( \delta({\bf  Q -Q_0})  {\bf P} +   {\bf 
P}^\dagger\delta({\bf  Q -
Q_0 }) 
\right) 
\eeq
{\em This operator, however,  is not the identity on  the physical 
states}\footnote{It  does satisfy
$\bf K_\Phi K_\Phi K_\Phi = K_\Phi$.  }, 
\beq
{\bf K_\Phi} |P\! = \! \pm a\r = \pm |P\! = \! \pm a\r 
\eeq
We can form a projetor by simply taking the absolute value of
this operator---see the next chapter:
\beq
\bf K'_\Phi= |K_\Phi| = \delta(\Phi) {1\over 2}
| \delta(Q-Q_0) P + P^\dagger\delta(Q-Q_0)| \eeq

Consider 
now $$ {\bf K_\Phi } = {1\over 2} \delta(\bP^2-a^2) \: |[{\bf  Q -
Q_0 } ,\bP^2-a^2]|\: \delta({\bf  Q -Q_0})  = $$ \beq{1\over 2} 
\left( \delta(\bP -a )+ \delta(\bP+a ) \right)\delta({\bf  Q -Q_0}) 
\eeq  where
now  
we 
have inserted an absolute value. Is this a projector?
 Only if we insert $\Theta(\bf P)$, which is equivalent to picking a branch. 

 The reason I mention these   cases 
is 
because 
they will help clarify later some results in the Faddeev path 
integral. We will come back to the search for a projector in the
next section.

Let us now study a bit more 
the interacting relativistic 
particle.
What happens if the determinant itself doesn't commute with 
the 
gauge-fixing? Well, as we saw this  is as if we were computing the 
expectation value 
of an operator that is not hermitean: {\em our inner product doesn't 
even 
have to be real, it will  be a bad inner product. Moreover, 
the 
hermiticity properties of the rest of the operators are then in 
question}.

 Now, 
for hermicity we need 
commutation of the observables---which already commute with the 
constraint---with the operator \beq \overbrace{ \prod_a \delta( \chi_a)
\mbox{sdet}\{ \chi_a,  {G}_b\} }\eeq In the particle case this 
corresponds to 
asking 
commutation with $\delta(\hat{t}-f(\tau)) \hat{p}_t$---and 
$\hat{\Phi}$ of 
course. Easily 
done---use operators in the $x$  coordinates. For interacting case it may prove
harder to  find 
such ``hermitean'' operators.

As we discussed (and there is already the Klein-Gordon solution to 
this 
problem), however, we can solve part of our problems if we come up  
with a 
hermitean ordering like \beq      i \; \overbrace{ \prod_a \delta( \chi_a)
\mbox{sdet}\{ \chi_a,   {G}_b\} } = 
 \delta(\hat{\chi}_a) sdet[\hat{\chi}_a, \hat{G}_b]/2+   
sdet[\hat{\chi}_a, 
\hat{G}_b]^\dagger \delta(\hat{\chi}_a^\dagger)/2\eeq   For 
the 
free case, this decouples the worlds---it yields minus  one-half 
$\delta(\hat{t} - f(\tau)) i\partial_t+( i\partial_t)^\dagger 
\delta(\hat{t} 
- f(\tau))$.

For the interacting case it yields\beq 2\hat{g}^{\mu 0} (\hat{p}_\mu - 
\hat{A}_\mu) 
\delta(\hat{t} - f(\tau)) + \delta(\hat{t} - f(\tau)) \{2\hat{g}^{\mu 0} 
(\hat{p}_\mu - \hat{A}_\mu)\}^\dagger  \eeq 
which must be Klein-Gordon in curved space, if we started with a covariant 
ordering of the constraint, of course.

Then we can try to find observables that commute with such an 
object. 

Some concepts I discussed can be confusing. One of them is unitarity. 
Unitarity 
means that the hamiltonian is hermitean, that the inner product is 
conserved 
under time evolution (or invariant under gauge transformations.) 
This we 
can 
always produce, and its definition is independent of whether there 
is 
particle 
creation. That has to do with the decoupling of the two sectors, 
because the 
conserved inner product that we have is not positive definite: if we 
want 
unitarity within one sector we will get it if the sectors decouple, 
because 
then we 
will have a conserved inner product  within a sector---an inner 
product with 
a 
definite sign.
\subsection{From Dirac quantization to the Faddeev path integral}

Now that we have developed a projector language and a regularized 
inner 
product we can write an expression for the path integral.

For the simplest case, the regularized physical amplitude is just given 
by 
$$
\l P\! =\! 0, \varphi_a(q) | \delta( {\bf Q}  - Q')  \; | P\! =\! 0, 
\varphi_b(q) \r = $$ \beq
\l P\! =\!0, \varphi_a(q) |  Q\! =\!Q', \varphi_b(q) \r =
\l  \varphi_a(q)   |  \varphi_b(q) \r 
\eeq
where I am using the usual old $q, Q$ etc. notation for the gauge 
and 
physical degrees of freedom. Notice that we could write, just as well,
$$
\l Q\! =\!Q', \varphi_a(q) | \delta({\bf  P} )\;  | Q\! =\!Q', \varphi_b(q)  
\r = $$ \beq
\l P\! =\!0, \varphi_a(q) |  Q\! =\!Q', \varphi_b(q)  \r =
\l  \varphi_a(q)   |  \varphi_b(q) \r 
\eeq
Once we have this it is immediate to write the corresponding path 
integral, 
\beq
\l Q\! =\!Q', \varphi_a(q) | {\bf K_\Phi 
K_\Phi ...K_\Phi}\, 
\delta({\bf P} ) \; | Q\! =\!Q', \varphi(q)_b \r \eeq
The key is to have  $\bf K_\Phi K_\Phi = K_\Phi$ and 
$\bf K_\Phi \delta( {P} )  = \delta( {P} ) $, or 
$\bf \delta( {Q } ) K_\Phi  = \delta( {Q} ) $ etc.  

{\em This operator, $\bf  K_\Phi$, is then the composition law 
operator. } 

It is crucial for full gauge invariance that we choose the gauge-fixing 
states/terms to satisfy\beq
\l \chi\!=\! 0 | \Phi\!=\! 0\r =constant
\eeq
If this condition is not met we loose gauge-invariance. In the path 
integral 
this will appear in the form of an action that is not invariant at the 
boundaries.

For the particle we can write
\beq
U\equiv \l t_f, x_f |     \delta( {\bf \Phi})        \;     |t_i x_i\r \eeq
It is easy to see that this is the usual propagator in the                 
non-relativistic case. We can write a path integral by repeated 
insertion of \beq 
\hat{K}_\Phi = \delta( {\bf \Phi}) \delta( {\bf t}-f(\tau) )\eeq

This procedure yields the Faddeev path integral. For the relativistic 
case we 
 also need $
\l t_f, x_f |    \delta( {\bf \Phi})   |t_i x_i\r $. 
 Remember that we tried  
  \beq
{\bf K_\Phi} \equiv 
\delta({\bf P^2-a^2}) 
{1\over 2} \left( \delta({\bf  Q} -Q_0 )  {\bf P} +   {\bf 
P}^\dagger\delta({\bf  Q}-
Q_0) 
\right) 
\eeq
  This operator, however,
is neither a projector nor 
the identity on  the physical 
states, 
\beq
{\bf K_\Phi} |P\! = \! \pm a\r = \pm |P\! = \! \pm a\r 
\eeq
This can be fixed in a number of ways. 
 We can change the amplitude, or the operator. For example, we can work 
with the 
projector\footnote{To see the second equality, think of the projector as 
a diagonal matrix in the p basis with only two non-zero eigenvalues.} (it is a
projector) \beq
{\bf K_\Phi '} = | {\bf K_\Phi}| =sign({\bf P})\; {\bf K_\Phi} 
\eeq
and with any amplitude of the form
\beq
U\equiv \l t_f, x_f |    {\bf  \left( \alpha\:  \delta( {P}-a )  + \beta  \:
\delta( {P}+a )\right)           } |t_i x_i\r 
\eeq
This will work because this projector  leaves
the operator $ {\bf   \alpha\:  \delta( {P}-a )  + \beta  \:
\delta( {P}+a )         }      $  unchanged. However, to build a
composition law we need to work with the amplitude 
\beq
U\equiv\l t_f,x_f | \; sign({\bf P}) \delta ({\bf P^2 -a^2}) \; |
t_i, x_i \r
 \eeq 
so that this amplitude will be the one appearing
in the composition law, after insertions of the projector. The composition law 
here 
is the usual Klein-Gordon one, generated by the
second part of the projector:
 $$ 
{1\over 2} \left( \delta({\bf  Q} -Q_0 )  {\bf P} +   {\bf 
P}^\dagger\delta({\bf  Q}-
Q_0) \right)
$$

Notice that these projectors contain two important ingredients: 
the physical states (represented by some delta function),
and the  inner product.

We have a choice on how to look at the projector: which part
represents the state and which the composition law. We can instead
take the ``signed'' part into the composition law, and rewrite
 $$
{\bf K'_\Phi= |K_\Phi| = \delta(\Phi) \, {1\over 2}\; 
| \delta(Q}-Q_0) P + P^\dagger\delta({\em Q}-Q_0)| 
$$
which works nicely with the amplitude 
$$
\l x_\mu |\,  \delta({\bf \Phi}) \; |y_\mu \r
$$
The composition law here is the Klein-Gordon one with an absolute value.

This description also holds---quite clearly---if there is ``decoupling'', i.e.,
if the above constant $a$ is made into an operator that nonetheless commutes with
$\bf P$. This describes, then, the relativistic particle with no electric
field, for example. What about the interacting case? It corresponds, in our
simple description here, 
to considering a constraint of the form\beq
{\bf \Phi = P^2- }f({\bf Q})^2
\eeq
Does our projector description still hold? To answer this
question we need to understand the quantities
\beq
\l   \chi_i | \, \left( \delta({\bf  Q} -Q_0 )  {\bf P} +   {\bf 
P}^\dagger\delta({\bf  Q}-
Q_0) \right)\, | \chi_j \r
\eeq
where $|\chi_i\r $ are the two solutions to the constraint equation.
I don't think it will be easy to construct a projector theory if there 
is no decoupling. Thus, we will also miss an understanding of the Faddeev 
path
integral if there is no decoupling.

We can also consider the ``Feynman'' amplitude. It is 
given by \beq
\l t_f, x_f |     {1\over \hat{\Phi} -i\epsilon}   |t_i x_i\r\eeq
In what sense is this an amplitude in the physical subspace? 

We will be able to interpret this amplitude in the next section. Let us 
investigate it a bit more here, though.

Consider again the simple constraint case. Then it is easy to see that 
\beq 
\l Q|     {1\over \hat{P}-a  -i\epsilon}   |Q'\r = \Theta ( Q-Q') 
\l Q|      \delta(\hat{P} -a)    |Q'\r
\eeq 
(times $2\pi i$...) Similarly, 
\beq
\l Q|     {1\over \hat{P}^2 -a ^2  -i\epsilon}   |Q'\r = 
{1\over a} \Theta ( Q-Q')  \l Q|       \delta(\hat{P} -a)   |Q'\r -
{1\over a} \Theta ( Q'-Q)  \l Q|       \delta(\hat{P} +a)    |Q'\r
\eeq
To see this just use \beq {1\over P ^2 -a ^2  -i\epsilon} = {1\over a} 
\left(
{1\over P-a-i\epsilon} - {1\over P+a+i\epsilon} 
\right)\eeq

 Here we can use  the projector
\beq
{1 \over {\bf \Phi} -i\epsilon} {1\over 2}
\left( \bf \delta(Q-\xi) P + P^\dagger \delta(Q-\xi) \right)
\eeq
 This also works, as this operator is a projector and it leaves
the above amplitude unchanged. 

With this operators and amplitudes we can now build
path integrals. These amplitudes we have discussed are the 
various Green function for the Klein-Gordon equation, provided
we use the proper constraint. 

The following have been discussed repeatedly in the literarature 
\cite{Hal1,Ike,Rudolph}:\\
 The {\bf Hadamard} Green function,\beq
\Delta_1(x -y) = {1 \over (2\pi)^3}\int d^4 k \;\delta(k^2-m^2)\;
e^{ik(x-y)} =\eeq
$${1 \over (2\pi)^4}\int_{-\infty}^{\infty} d\lambda\int d^4 k \; 
e^{ik(x-y)+i \lambda (k^2-m^2)}$$
which is a solution to the Klein-Gordon equation. In term of our projectors it 
can 
be written as \beq
\Delta_1(x-y)  \sim \l x_\mu , \pi\! =\!0 | \; e^{i \hat{\lambda}\hat{\Phi}} | 
y_\mu,\pi\!=\!0\r \sim \l x_\mu | \; \delta(\hat{\Phi}) | y_\mu \r \eeq
as discussed before. As we saw it satisfies a composition law generated by 
\beq {1\over 2} \left| \delta({\bf  Q -Q_0})  {\bf P} +   {\bf 
P}^\dagger\delta({\bf  Q}-Q_0) \right|  \eeq
instead of  the Klein-Gordon one,
\beq
A \cdot B \equiv -i  \int d\sigma^\mu \; A(x,z)
\stackrel{\leftrightarrow}{\partial}_\mu  B(z,x) =  \int d\sigma^\mu \; A(x,z)
[\hat{P }_\mu + \hat{P}^\dagger_\mu]  B(z,x)
\eeq   
The {\bf causal} Green function---also as solution of the Klein-Gordon 
equation, 
\beq
i\Delta(x-y) = {1 \over (2\pi)^3}\int d^4 k \;sign(k_0) \delta(k^2-m^2)\;
e^{ik(x-y)} \sim \l x_\mu | \; sign({\bf P}) \delta ({\bf P^2 -a^2}) \; |
y_\mu \r  \eeq
We also saw earlier that this ``signed'' amplitude does satisfy the 
Klein-Gordon composition law.\\
The {\bf Feynman} amplitude, \beq
i\Delta_F (x-y) = {1 \over (2\pi)^4}\int d^4 k \; 
{ e^{-ik(x-y)} \over k^2-m^2+i\epsilon} =
  {-i \over (2\pi)^4}\int_0^\infty  d\lambda\int d^4 k \; 
  e^{-ik(x-y)      -\lambda    (k^2-m^2+i\epsilon)        }    \eeq $$
 \sim \l x_\mu | \; {1\over \hat{\Phi} +i\epsilon} \; |
y_\mu \r  $$
which, again, satisfies the Klein-Gordon compostion law,\\
The {\bf Wightman} functions 
\beq
G^\pm (x,y) =  {1 \over (2\pi)^3}\int d^4 k \;\theta(\pm k_0)\; \delta(k^2-
m^2)\;
e^{ik(x-y)}  \eeq which obey the Klein-Gordon composition laws.\\
Finally we have  the {\bf Newton-Wigner propagators}, which are not 
manifestly 
covariant objects
\beq
G_{NW}^\pm   (x,y) =  {1 \over (2\pi)^3}\int d^4 k \;\theta(\pm k_0)\;k_0  
\delta(k^2-m^2)\;
e^{ik(x-y)}  \eeq 
and which satisfy the non-relativistic compostion law instead of the 
relativistic 
one.

We have deduced in a simple way
  the
 composition laws of these amplitudes in the physical space as well as
their  path integral representations in the extended space:
\beq
A = \int dqdpdQdP \; \delta(\Phi_0) \prod K_{i\Phi} e^{i\int d\tau
(p\dot{q} +P\dot{Q}- h(q,p))}
\eeq
Notice also that we can write 
\beq
\l t_f, x_f |     \delta(\hat{\Phi})           |t_i x_i\r =
\l t_f, x_f,\pi\! =\! 0 |  e^{i\hat{\lambda}\hat{\Phi}} |t_i x_i,\pi\! =\! 
0\r 
\eeq
by adding this new degree of freedom.


Again, it is easy to write the path integrals in all these  cases. {\em 
We 
obtain, for the ``decoupling'' cases the Faddeev path integrals discussed in the next
chapter, as well as the BFV path integral---as we will show.}

How about the composition laws for these amplitudes? If we have 
the 
amplitudes above and the correponding nice projectors we have,
schematically,
\beq 
\l t_f, x_f |    \delta(\hat{\Phi})   |t_i x_i\r  = 
\l t_f, x_f |   {\bf K_\Phi} \delta(\hat{\Phi})   |t_i x_i\r = \eeq \beq 
\int d^4 x 
\l t_f, x_f |   {\bf K_\Phi} |t,x \r \l t,x | \delta(\hat{\Phi})   |t_i x_i\r = 
 \int d^4 x 
\l t_f, x_f |   \delta(\hat{\Phi})  |t,x \r {\bf O} \l t,x | 
\delta(\hat{\Phi})   |t_i 
x_i\r \eeq  for some differential operator
$\bf O$. 
It is not very hard to find these composition laws. For example, for 
the 
``signed'' case the compostion law operator is simply $-i \partial_t$, 
and the 
Feynman amplitude satisfies a composition law of the                                              
                 ``Klein-Gordon'' form. 
 
Let me remark once again, that my construction applies only to the decoupling case. If there 
are interesting interactions my projector formalism may noy be applicable, especially 
in the on-shell situation. It may be possible to extend these ideas to the case of 
the causal
projectors.

\newpage\section{The Fock space inner product.}
The state space is defined in terms of the raising and lowering 
operators, just 
as for the harmonic oscillator case. There is a crucial  difference: the 
commutation relations differ by a sign, and they induce an indefinite 
inner 
product (the definition of inner product is inherent in the algebra, 
since one 
discusses the commutation relations of  the {\em hermitean} 
conjugates of 
operators.)
Recall that the algebra is given by the definitions \beq 
\hat{a} = \hat{P}_1 + i\hat{P}_2 , \:\: \hat{a}^\dagger = \hat{P}_1 
-i\hat{P}_2\eeq and\beq \hat{b}=-{i\over 2}(\hat{Q}^1 +i\hat{Q}^2), 
\:\: \hat{b}^\dagger = {i\over 2}(\hat{Q}^1 -i\hat{Q}^2) \eeq 
As explained in reference \cite{TeitelBook}, one needs an even number of 
constraints. 
This is similar to what we usually do in BRST when we add a 
constrained 
multiplier degree of freedom for each of the constraints. 

The commutation relations that follow from this definition 
are\beq [\hat{a},\hat{b}^\dagger] 
= 
[\hat{b},\hat{a}^\dagger]  = 1\eeq  and the rest zero.

It is implied by the notation here that both $\hat{P}_1$ 
and 
$\hat{P}_2$ are 
hermitean. For example, $\hat{a}+\hat{a}^\dagger$ is hermitean, and 
is equal  
to $2\hat{P}_1$. This fact is crucial for the development of the formalism, 
and is 
a subtle assumption---it selects an indefinite inner product when we 
define 
the  vacuum.

We discussed in the previous chapter that 
the states on this space are defined by,  a) it is assummed that there is a
``vacuum'' state,   $|0\rangle$,  satifying the 
conditions 
$ \hat{a}|0\rangle  = \hat{b}|0\rangle = 0$.  This state is also 
assumed to 
have unit 
norm, $\langle 0 |0\rangle = 1$, and b), 
the rest of these states are defined by acting on the ``vacuum'' 
above with 
the creation operators.

Recall also that this vacuum was a puzzling object: we 
need a state that satisfies \beq  (\hat{P}_1 +i\hat{P}_2)|0\rangle = 
-{i\over 2}(\hat{Q}^1 +i\hat{Q}^2)|0\rangle = 0\eeq  
Can we understand what is going on in more pedestrian terms? 
Following Marnelius' work \cite{marnelius} let us map this 
representation 
into coordinate space.  For that  let us map $\hat{Q}^2\rightarrow 
i\hat{Q}^2 , \:\: \hat{P}_2\rightarrow -i \hat{P}_2$. This is a 
canonical 
transformation. However, you may say that in this representation 
these 
operators are not hermitean---but they will be provided we fix the 
inner 
product. The inner product has to become indefinite! 
Consider the mixed 
representation in which we write the states in the form $\varphi = 
\varphi(Q^1, P_2)$. The vacuum is defined by \beq (\hat{P}_1 + 
\hat{P}_2) 
\varphi_0(Q^1, P_2) =(-i {\partial \over \partial Q^1}  +P_2) 
\varphi_0(Q^1, 
P_2) = 0\eeq  This is solved by \beq \varphi_0(Q^1, P_2) = e^{-i P_2 
Q^ 1}\eeq  
This is 
already the vacuum: it  automatically satisfies\beq (\hat{Q}^1 
-\hat{Q}^2)\varphi_0(Q^ 1, P_2) = (Q^1 -i{\partial \over \partial P_2})
\varphi_0(Q^1,P_2) = 0 \eeq 
What do we use for inner product? Let us go back to the description 
in terms of 
the creation and annihilation operators, and then proceed. In fact,
I will now give a  description of the Fock space quantization and 
inner product in terms of:\\
a) The creation and annihilation operators\\
b) The holomorphic representation---continuum, and \\
c) $q$ and $p$ space\\

As I already discussed, we can represent any state  by the 
holomorphic   function of the creation operators that acting on the 
vacuum produces the desired state. Consider for example the 
eigenstates of the destruction operators (not normalized yet), \beq  
|\psi_{\alpha\beta}\rangle = e^{\alpha \hat{b}^\dagger + \beta 
\hat{a}^\dagger}\; |0\rangle \sim |\hat{a}\! =\! \alpha,\; \hat{b}\! =\! 
\beta\rangle\eeq 
The computation of the inner product of these states is 
straightforward:\beq 
\langle\psi_{\alpha\beta}|\psi_{\alpha'\beta'}\rangle =\langle 0| 
e^{\alpha^* \hat{b} + \beta^* \hat{a} }\, 
e^{\alpha' \hat{b}^\dagger + \beta' \hat{a}^\dagger}\; |0\rangle =\eeq  
\beq 
\langle 0| 
e^{\alpha^* \hat{b}}e^{ \beta' \hat{a}^\dagger} \,  e^{ \beta^* \hat{a} }
e^{\alpha' \hat{b}^\dagger }\;  |0\rangle =\eeq 
\beq  \langle 0| 
\left(  [e^{\alpha^* \hat{b}} , e^{ \beta' \hat{a}^\dagger}] +1\right) 
 \left( [e^{ \beta^* \hat{a} }, 
e^{\alpha' \hat{b}^\dagger }]+1  \right) |0\rangle\eeq 
hence we need to compute \beq \langle 0| 
   [e^{\alpha^* \hat{b}} , e^{ \beta' \hat{a}^\dagger}]  |0\rangle =
\langle 0 | \sum_{n=0}^\infty {\alpha^{*n} \over n!} [\hat{b}^n , 
e^{\beta' \hat{a}^\dagger}] |0\rangle =\eeq  \beq  
\langle 0 | \sum_{n=1}^\infty {(\alpha^*\beta')^n \over n!}  |0\rangle=
e^{\alpha^*\beta' } -1\eeq  since \beq  \langle 0| [\hat{b}^n , 
f(\hat{a}^\dagger )] |0\rangle = \langle 0 | {d^n \over 
d\hat{a}^{\dagger n} } f(\hat{a}^\dagger ) |0\rangle = {d^n \over dx^n 
} f(x) |_{x=0} \eeq  The result is then 
\be
\langle\psi_{\alpha\beta}|\psi_{\alpha'\beta'}\rangle = 
e^{\alpha^*\beta' + \beta^*\alpha' }\ee Notice that the spectrum is 
given by the 
whole
complex plane! This is not surprising, since we are dealing with the 
spectrum of 
{\em
normal} operators.  Other useful facts are \be \langle 0 | [\hat{a}^n , 
\hat{b}^{\dagger 
n}]\, |0\rangle = n! \ee and \ba  \parallel \hat{a}^{\dagger  
n}\hat{b}^{\dagger 
n}\parallel
\equiv  \langle 0 |  \hat{a}^n  \hat{b}^n\hat{b}^{\dagger n}  
\hat{a}^{\dagger 
n}\, |0\rangle = \\ \displaystyle
\langle 0 | [\hat{a}^n , \hat{b}^{\dagger n}][\hat{b}^n , 
\hat{a}^{\dagger n}]|0\rangle = (n!)^2 \ea The more general 
statement is that \be
\l 0 | \hat{a}^{n_1}\hat{b}^{n_2}\hat{a}^{\dag  n_3}\hat{b}^{\dag n_4} 
|0\r = (n_1)!(n_2)! \delta_{n_1 n_4}\delta_{n_2 n_3} \ee
Now, for 
\beq  |\psi\r \sim \sum c_{nm} \hat{a}^{\dag  n}\hat{b}^{\dag  m} = 
\psi(\hat{a}^{\dag  },\hat{b}^{\dag } )\eeq  we have \beq
\para\psi \para= \l 0 |  
\sum c_{n_1 n_2}^* \hat{a}^{ n_1}\hat{b}^{n_2}  c_{n_3 
n_4} \hat{a}^{\dag  n_3}\hat{b}^{\dag  n_4} |0\r = \eeq  
\beq 
\sum c_{m n }^*  c_{n m}  n! m! =  \sum {1\over n! m! } 
\left. [ \partial_{ a^\dag }^n \partial_{b^\dag }^m \psi(\hat{a}^{\dag }
,\hat{b}^{\dag } )]\, [\partial_{a^\dag}^m \partial_{b^\dag}^n 
\psi(\hat{a}^{\dag},\hat{b}^{\dag } )]^* \right|_{a^\dag = b^\dag =0}    
\eeq 

Consider next the eigenvalue  equation
\beq \hat{Q}_2\psi_\alpha= (\hat{b}+\hat{b}^{\dagger})\psi_\alpha = 
{\partial \psi_\alpha\over
 \partial a^\dagger} +
  b^\dagger \psi_\alpha \equiv \alpha \psi_\alpha  \eeq  It is solved 
by 
\beq \psi_\alpha
   = A\;e^{-a^\dagger (b^\dagger -\alpha)}\eeq 
where $A$ is an arbitrary function of $b^\dagger$. Without
loss of generality we choose $A$ such that  the resulting
state is an eigenfunction of $ \hat{P}_1 = ({\hat{a} + \hat{a}^\dag 
)/ 2}$. The resulting eigenvalue equation is easily
solved by \be
\psi_{Q^2 P_1} = N e^{-(a^\dag - P_1)(b^\dag - Q^2)} \eeq
These are the candidates. But we have to check that they are 
normalizable.  It is hard   to compute the inner 
product using
the formula above. It will be interesting to try, though. We can compute \beq ({d \over d
a^\dag})^n  \psi_{Q^2 P_1} =
(-(b^\dag - Q_2))^n  \psi_{Q^2 P_1} \eeq  easily enough.   Next, one 
can 
check\footnote{ Because $ d_x^n [g(x) e^{ax}] = e^{ax} (d_x
+a)^n g(x) $.}
 that
\beq  ({d \over d b^\dag})^m
({d \over d a^\dag})^n \psi_{Q^2 P_1} = \psi_{Q^2 P_1}                       
({d \over d b^\dag} -  
(a^\dag - P_1) )^m (-(b^\dag - Q_2))^n  \eeq  which at zero becomes 
\beq 
\left. ({d \over d b^\dag})^m
({d \over d a^\dag})^n \psi_{Q^2 P_1} \right|_{a^\dag = b^\dag=0} = 
\left. N [({-d \over d b^\dag}  + P_1 )^m (-(b^\dag -
Q_2))^n] \right|_{ b^\dag=0} = \eeq  \beq   N  ({d \over d Q_2}  + P_1  
)^m 
(Q_2) ^n    =  N e^{-P_1 Q_2} ({d \over d Q_2} )^m  [
e^{P_1 Q_2}  Q_2  ^n ] \eeq  since it can be easily checked that \beq   
({d 
\over d x}  + P  )^m [f(x)] = e^{- xP} ({d \over d x}  )^m
[e^{ xP} f(x)] \eeq  (see the last footnote). So, finally, we have \be
\left. ({d \over d b^\dag})^m
({d \over d a^\dag})^n \psi_{Q^2 P_1} \right|_{a^\dag = b^\dag=0} = 
N e^{-P_1 Q_2} ({d \over d Q_2} )^m ({d \over d P_1} )^n [
e^{P_1 Q_2}]\ee The inner product is hence given by \beq 
\para\psi \para= \sum c_{m n }^*  c_{n m}  n! m! =  \sum {1\over   n! m! }
\left. [\partial_{a^\dag}^n \partial_{b^\dag}^m \psi(\hat{a}^{\dag  
},\hat{b}^{\dag } )]\, [\partial_{a^\dag}^m \partial_{b^\dag}^n 
\psi(\hat{a}^{\dag},\hat{b}^{\dag } )]^* \right|_{a^\dag = b^\dag =0} 
= \eeq   
\beq  \sum {1\over   n! m! } \left( N e^{-P_1 Q_2} ({d \over d Q_2} )^n ({d \over d 
P_1} 
)^m
[ e^{P_1 Q_2}] \right)^* N e^{-P_1 Q_2} ({d \over d Q_2} )^m ({d \over 
d
P_1} )^n   [e^{P_1 Q_2}]  \eeq   and more generally,
$ 
(\psi_{Q_2 P_1},\psi_{Q_2 ' P_1'} ) = $ \beq \sum {1\over   n! m! } \left( N e^{-P_1 Q_2} ({d 
\over
d Q_2} )^n ({d \over d P_1} )^m [ e^{P_1 Q_2}] \right)^* 
 N e^{-P_1' 
Q_2'}
({d \over d Q_2'} )^m ({d \over d P_1'} )^n   [e^{P_1' Q_2'}] 
\eeq

 We will now try a different
approach. 
Let us ask the following question: {\em can we come up with an 
inner product in the $a,b$ etc. space  that   matches the above inner 
product? } We also want the algebra to be respected, which for us 
now means that the hermiticity properties of the operators must be 
preserved. The commutator algebra is already respected by the 
representation in which the destruction operators become simple 
derivatives and the creation operators act by multiplication.  {\em  The 
answer is yes. } This inner product is given by \be (
\psi_k (a^*, b^*),  \psi_l (a^*, b^*) )\equiv 
\int { dadbda^* db^*\over (2\pi i)^2}  e^{-aa^* - bb^*} \psi_k^* (a^*, 
b^* ) \psi_l (b^*, a^*) \ee where  here I use the notation in which  the 
operator
$\psi_k (a^\dag, b^\dag)$ becomes in this representation the state 
$\psi_k (a^*, 
b^*)$.  Notice the swap: the state $\psi_l (a^*, b^*)$ becomes $\psi_l (b^*, a^*)$ 
in 
the integral.
To calculate this integration it is convenient to go to the variables
$a= r_a e^{i\theta_a},\: a^*= r_a e^{-i\theta_a}$ and $ b= r_b 
e^{i\theta_b},\: b^*= r_b
e^{-i\theta_b}$. Then it is easy to see that \beq 
 { dadbda^* db^*\over (2\pi i)^2}  = {1\over \pi^2}  r_a dr_a d 
{\theta_a}
r_b dr_b d {\theta_b}\eeq   This
expression has an easy interpretation in terms of the eigenstates  of 
the
destruction  operators \beq  |\psi_{\alpha\beta}\rangle =  e^{\alpha
\hat{b}^\dagger + \beta \hat{a}^\dagger}|0\rangle\equiv  |\hat{a}\! 
=\! \alpha,\;
\hat{b}\! =\! \beta\rangle\eeq  if one uses them in a decomposition 
of 
unity as 
follows,
\be 
\hat{I} = \int{  d\alpha  d\alpha^* d\beta d\beta^* \over (2\pi i)^2}
| \alpha \r_{\hat{a}}\otimes |\beta\r_{\hat{b}} 
  \  _{\hat{a}}\l \beta^* | \otimes \ _{\hat{b}}\l \alpha^* |  \: e^{ -\alpha 
\alpha ^* 
-\beta \beta^*}
\ee For one, one can check that the inner product between these 
states
is the same as was calculated before.  What really 
matters, though,  is that we have a  true representation of  the
above  Hilbert space. We can see this in two ways. It is enough to 
check
that  the vacuum has unit norm and that the algebra and hermiticity 
properties are preserved. Or we can simply check that  \beq  
\l 0 | \hat{a}^{n_1}\hat{b}^{n_2}\hat{a}^{\dag  n_3}\hat{b}^{\dag n_4} 
|0\r = (n_1)!(n_2)! \delta_{n_1 n_4}\delta_{n_2 n_3} = (
  a ^{* n_1} b ^{* n_2},  a ^{*  n_3} b ^{* n_4})\eeq  as 
defined above.  This is true, indeed, as
follows from the following model calculation: $ ( a^{*n}, b^{*m} ) \sim $ 
\beq 
\int {da da^* \over 2\pi i} a^n (a^*)^m e^{-a a^*} = {1\over \pi} 
\int_0^\infty r_a dr_a 
\int_0^{2\pi} d\theta_a  r_a^{n+m} e^{i \theta_a (n-m)} e^{-r_a^2} = 
\delta_{nm} n!\eeq 
(see Faddeev and Slavnov's book, in references \cite{Faddeev}.) 
  As a check, we can compute the
earlier inner product, $ (\psi_{Q_2 P_1},\psi_{Q_2 ' P_1'} ) =$ 
\beq
\sum {1\over   n! m! }
\left( N
e^{-P_1 Q_2} ({d \over d Q_2} )^n ({d \over d P_1} )^m [ e^{P_1 Q_2}]
\right)^* N e^{-P_1' Q_2'} ({d \over d Q_2'} )^m ({d \over d P_1'} )^n  
[e^{P_1' Q_2'}]  = \eeq  \beq  \int {dadbda^*  db^* \over (2\pi 
i)^2} 
e^{-aa^* - bb^*}  
 N^* e^{-(a  - P^*_1)(b  - Q^{* 2})}  N e^{-(b^* - P_1)(a^* - Q^2)} = \eeq   
\beq    
{|N|^2\over
\pi^2}\int 
  r_a dr_a d {\theta_a}
r_b dr_b d {\theta_b} e^{-r_a^2 -r_b^2 -2 r_a r_b cos(\theta_a 
+\theta_b) +
r_a(e^{i\theta_a} y^* + e^{-i\theta_a} x)  +
r_b(e^{i\theta_b} x^* + e^{-i\theta_b} y)  - 2Re[xy] }\eeq  a difficult 
integral. Let us try instead yet another approach:
{\em  we can try to map the above situation into the original 
coordinate system.} Recall that the algebra above was given by the 
definitions \beq 
\hat{a} = \hat{P}_1 + i\hat{P}_2 , \:\: \hat{a}^\dagger = \hat{P}_1 
-i\hat{P}_2\eeq and\beq \hat{b}=-{i\over 2}(\hat{Q}^1 +i\hat{Q}^2), 
\:\: \hat{b}^\dagger = {i\over 2}(\hat{Q}^1 -i\hat{Q}^2).\eeq  Consider 
the mixed 
representation in which we write the states in the form $\varphi = 
\varphi(Q^1, P_2)$.   The vacuum is defined by \beq (\hat{P}_1 + 
i\hat{P}_2) 
\varphi_0(Q^1, P_2) =( i {\partial \over \partial Q^1}  +iP_2) 
\varphi_0(Q^1, 
P_2) = 0\eeq  This is solved by \beq \varphi_0(Q^1, P_2) = e^{-  P_2 
Q^ 1}\eeq  
Can we come up with an inner product here that respects the 
algebra, the hermiticity properties of the operators and that gives 
unit norm to the vacuum?  The answer is yes once more. The inner 
product is given by \be
 (\psi_a, \psi_b ) \equiv \int_ {-i\varepsilon}^{i\infty} dQ^1 \int_{-
\infty-i\epsilon}^{ \infty-i\epsilon}dP_2 [\psi_a(Q_1^*, P_2^*)]^* 
\psi_b(Q_1, P_2)\ee
First notice that the vacuum is normalizable to unity: \ba \dis \para
\psi_0\para= \int_{-i\varepsilon}^{i\infty} dQ^1 \int_{-\infty-
i\epsilon}^{ \infty-i\epsilon}dP_2 e^{-2Q_1P_2} = 
 \int_ { -\varepsilon}^{ \infty} idq \int_{-\infty }^{ \infty }dp e^{-
2iq(p-i\epsilon) } = \\[.5cm] \displaystyle
  \int_ { -\varepsilon}^{ \infty} idq       \delta(2q) e^{-2q\epsilon}     
= i\pi\ea or \ba  \dis = i \int_{-\infty }^{ \infty }dp {1\over 
-2(ip+\epsilon)} \left. e^{-2iqp-2q\epsilon} \right|_{-
\varepsilon}^{\infty} = \\[.5cm] \displaystyle 2\pi 
i \int_{-\infty }^{ \infty }dp {1\over  2(ip+\epsilon)}   e^{ 
2i\varepsilon p+2\varepsilon\epsilon}  =i\pi \ea
It is very important to check that the operators are hermitean. This 
could be troublesome, for example, for the operator \beq \hat{Q}_2 
\sim 
i\partial_{P_2} \eeq  Indeed, for the case of the vacuum we need, for 
hermiticity, that the boundary term \beq  
\int_ {-i\varepsilon}^{i\infty} dQ^1 \int_{-\infty-i\epsilon}^{ \infty-
i\epsilon}dP_2  i\partial_{P_2}e^{-2Q_1P_2}\eeq    vanish, and it 
does, 
\ba \dis = -\int_ { -\varepsilon}^{ \infty} dq \int_{-\infty }^{ \infty 
}dp\partial_p e^{-2iq(p-i\epsilon) }= \\[.5cm] \displaystyle
-\int_{-\infty }^{ \infty }dp \partial_p{1\over -2(ip+\epsilon)} \left. 
e^{-2iqp-2q\epsilon} \right|_{-\varepsilon}^{\infty} =\\[.5cm] \displaystyle
-\int_{-\infty }^{ \infty }dp \partial_p {1\over  2(ip+\epsilon)}   e^{ 
2i\varepsilon p+2\varepsilon\epsilon} =0\ea
It is easily checked that $\hat{P}_1$ and $\hat{Q}_1,\: \hat{P}_2$ are
also hermitean in this inner product, as they should.
So we have yet another representation. 

How unique is this inner 
product
definition? Can we deform the paths? The actual path
definition should reflect the
spectra of the operators. If we can deform the paths, then we should 
be able
to also use different spectra. Is this so?  For one thing, 
in the above definition one  can exchange the paths for $Q^1$ and
$P_2$. This will still work, as the vacuum is symmetric in these
variables.

Let us return to the issue of the spectra of the operators. Can we 
produce a
working resolution of the identity? Yes, it is given by 
\beq \label{eq:FockId}
 \hat{I} =  \int_{-i\varepsilon}^{i\infty} dQ^1 \int_{-\infty-
i\epsilon}^{ \infty-i\epsilon}dP_2 |Q^{1*} P_{2}^* \r\l  Q^1 P_2| \eeq   
since \beq  \l \psi |Q^{1 } P_{2}  \r = 
\left( \l Q^{1 *} P_{2*}  | \psi\r \right)^* =
 (\psi( Q^{1*}, P_{2}^*))^* \: \: \l 
Q^{1 } P_{2}   | 
\psi\r = 
\psi(Q^{1 } P_{2})\eeq

Let us investigate some more this algebra. First, let us exchange 
$Q_1$ and 
$P_1$, taking care that the algebra is preserved. This is just a matter 
of 
covenience. Let us write
\beq 
             \hat{A} = (\hat{Q}^1 - i\hat{P}_2 )/\sqrt{2}, \:\: 
\hat{A}^\dagger = (\hat{Q}^1 + i\hat{P}_2 )/\sqrt{2}\eeq and\beq 
                    \hat{B}=(i\hat{P}_1 -\hat{Q}^2)/\sqrt{2}, 
\:\: \hat{B}^\dagger =(-i\hat{P}_1 -\hat{Q}^2)/\sqrt{2}.\eeq 
The algebra that follows is as before, \beq [\hat{A},\hat{B}^\dagger] 
= 
[\hat{B},\hat{A}^\dagger]^\dagger  = 1\eeq  

To understand the properties of this algebra let us define now 
\beq \hat{{\cal Q}}_- = (\hat{A}-\hat{B})/\sqrt{2}\eeq  and 
\beq \hat{{\cal Q}}_+ = (\hat{A}+\hat{B})/\sqrt{2}\eeq  These have 
the nice 
properties \beq  [ \hat{{\cal Q}}_- , \hat{{\cal Q}}_- ^\dagger] = -1 
\eeq  and 
\beq  [ \hat{{\cal Q}}_+ , \hat{{\cal Q}}_+ ^\dagger] = +1 \eeq  Now, 
the fact 
that one 
oscillator comes defined with a minus sign doesn't carry much 
meaning by 
itself (imagine exchanging the notation for creation and annihilation 
operators in the usual oscillator case). What really matters, is that 
the 
vacuum as defined above corresponds here to requiring \beq 
\hat{{\cal 
Q}}_+ 
|0\rangle  = 
0 \eeq  {\em and } \beq \hat{{\cal Q}}_- |0\rangle  = 0 \eeq  In the 
standard 
case the 
second equation  
corresponds to asking that the creation operator have a zero 
eigenstate. This 
state did not exist in your quantum mechanics class because although 
the 
differential equation has a solution in the coordinate 
representation\footnote{$\varphi (x) \sim exp(x^2/2)$}, {\em it is 
not 
normalizable with the standard inner product}. What this means for 
us is 
that we are going to invent a new inner product---not a positive 
definite 
one, as is easily seen by ``creating''  with the operator $\hat{{\cal 
Q}}_-
^\dagger$.  At any rate, we can further define \beq \hat{z}_+ = 
{\hat{Q} 
^1 - 
\hat{Q}^2 \over \sqrt{2} }\:, \:\:\: \hat{P}_{z_+} = {\hat{P} _1 - 
\hat{P}_2 
\over \sqrt{2} }\eeq  and \beq \hat{z}_- = {\hat{Q} ^1 + \hat{Q}^2 
\over 
\sqrt{2} }\:, 
\:\:\: \hat{P}_{z_-} = {\hat{P} _1 + \hat{P}_2 \over \sqrt{2} }\eeq  
Notice 
that 
the pairs $\hat{z}_+\:,\:\: \hat{P}_{z_+}$ are canonically conjugated---and
 the 
``$+$'' and ``$-$'' sides decouple.  Then we have 
\beq  \hat{{\cal Q}}_+ = {1\over \sqrt{2} } (  \hat{z}_+  + i 
\hat{P}_{z_+}) 
\:\:\:
\hat{{\cal Q}}_- = {1\over \sqrt{2} } (  \hat{z}_-  - i \hat{P}_{z_-}) 
\eeq 
{\em Everything decoupled!  It is now easy to write the coordinate 
expression for 
the vacuum and the inner products:} 

---The ``$+$'' side: $ [ \hat{{\cal Q}}_+ , \hat{{\cal Q}}_+ ^\dagger] = +1 
$ 

We have $\hat{{\cal Q}}_+ = (  \hat{z}_+  + i \hat{P}_{z_+})/\sqrt{2} $, 
$\hat{{\cal Q}}_+^\dagger  = (  \hat{z}_+  - i \hat{P}_{z_+})/\sqrt{2} $-
-the 
standard Hilbert space description.

The vacuum is given by\beq \varphi_{0_+} (z_+) = \langle 
z_+|0_+\rangle =  \pi 
^{-{1\over 4}} e^{- (z_+)^2/2}\eeq 
where the inner product is given by \beq \left( \varphi_+ (z_+),\: 
\varphi_+ 
'(z_+)\right) = \int_{-\infty}^{\infty} dz_+ \:\varphi_+^*(z_+) \:\; 
\varphi_+ 
'(z_+)\eeq 

---The ``$-$'' side: $ [ \hat{{\cal Q}}_- , \hat{{\cal Q}}_- ^\dagger] = -1 
$ 

We have $\hat{{\cal Q}}_- = (  \hat{z}_-  - i \hat{P}_{z_-})/\sqrt{2} $, 
$\hat{{\cal Q}}_-^\dagger  = (  \hat{z}_-  + i \hat{P}_{z_-})/\sqrt{2} $---{\em 
not} the standard Hilbert space description.

The vacuum is given by\beq \varphi_{0_-} (z_-) = \langle z_-|0_-
\rangle =  \pi 
^{-{1\over 4}} e^{+ (z_-)^2/2}\eeq 
and is normalized to one. The inner product is now given by \beq 
\left( 
\varphi_- (z_-),\: \varphi_- '(z_-)\right) = \int_{-i\infty}^{i\infty} 
dz_- 
\:\varphi_-^*(z_-^*) \:\; \varphi_- '(z_-)\eeq 

{\em The full vacuum is given by} \beq  \psi_0 = \pi ^{-{1\over 2}} 
e^{- 
(z_+)^2+ (z_-)^2 \over 2 }  =  \pi ^{-{1\over 2}} e^{2Q^1 Q^2}\eeq {\em  
and the full 
inner product by } \beq  \left( \psi,\: \psi '\right)= 
\int_{-\infty}^{\infty} dz_+ \int_{-i\infty}^{i\infty} dz_- \:\: 
\psi(z_+^*,z_-^*)^* \:\:\psi'(z_+,z_-) =\eeq  \beq 
\int  dQ^1  \int  dQ^2 \:\: 
\psi(z_+^*(Q^1 ,Q^2),z_-^*(Q^1 ,Q^2))^* \:\:\psi'(z_+(Q^1, Q^2),z_-(Q^1 
,Q^2)) 
\eeq 
where $ Q^{1*}= Q^2 $.

The next step is to impose the condition \beq  \hat{a}|\psi\rangle = 
0\eeq 
this is the condition that defines physical states. This yields the
vacuum described above plus null states, as we discuss next. 

The next factor in this formalism is that we have the so-called null 
states. 
Any state of the form $\hat{a}^\dagger |any\rangle$ decouples from 
any 
physical 
state. This situation is remarkably similar to the one that we have in 
the BRST 
formalism. They are both  characterized by a definition of 
cohomology:  the 
states are defined by the kernel of some operator. In the BRST case 
the kernel 
is nilpotent, so the cohomology is proper, \beq  H(\hat{\Omega}) 
\equiv 
{Ker\:\: 
\hat{\Omega} \over Im\:\: \hat{\Omega}}\eeq 
For the Fock space case we don't have a nilpotent operator. The 
``cohomology''  is defined by \beq  H(Fock ) \equiv {Ker\:\: \hat{a} 
\over 
Im\:\: 
\hat{a}^\dagger}\eeq 
The similarities between these two formalisms also include the fact 
that in 
both cases the number of constraints is even.

\subsection{The particle in Fock space}

Consider first   the 
case of a constraint of the form\footnote{ See 
exercise 13.14 in Teitelboim's book \cite{TeitelBook}} $ \bf \Phi=  p_t + A(x, p_x) \approx 0$.
This case covers the non-relativistic particle as well as the relativistic
case when we choose a branch (i.e., the square root hamiltonian situation).
We will examine the full relativistic constraint in a moment.

The physical 
states 
are 
defined by
\beq
({\bf a}+ {\bf A} )  | \Psi \r _{Ph} = 0 ,
\eeq 
where ${\bf a}= {\bf p_t}   + i{\bpi }$ and---for the non-relativistic case we have
\beq  
{\bf A}  =  {{\bf p_x^2} \over 2m} 
\eeq 
and for the relativistic one 
\beq
{\bf A}  =  \sqrt{\bp_x^2+m^2}
\eeq
The following reasoning can be carried out also when there is 
an electromagnetic background with no electric field. In such a case
we would define ${\bf a}= {\bf \Pi_t}   + i{\bpi }$, which
preserves the commutator algebra---we keep the original
$\bf b$ definitions---and $\bf A$ to be the 
square root hamiltonian. Indeed, it is vital that \beq
[ \bf a, A] =0
\eeq
which holds when there is no electric field. We also have $[ \bf b, A] =0$.

This equation for the physical states is appropiate because, defining
\beq
\bf M=    a +    A 
\eeq
we have that 
\beq
\bf \Phi = M + M^\dagger
\eeq
Hence
\beq
\ _{Ph}\!\l \Psi | {\bf \Phi} \; | \Psi \r _{Ph} =0
\eeq
As we will see, another inportant property of our definition (which holds
when there is no electric field) is that $\bf M$ is {\em normal}, i.e., 
 \beq
[ \bf M,  M^\dagger] =0
\eeq
The general  solution to the physicality condition is given by (recall ${[\bf b, A]}=0$)
\beq
 | \Psi \r_{Ph} =  f({\bf a} ^\dagger) \, e^{-{\bf A} {\bf b}^\dagger}  \; | 0\r | f\r 
\eeq
with $f({\bf a} ^\dagger )$ totally arbitrary. However, we can rewrite this solution
in the following more useful  way
\beq
 | \Psi \r_{Ph} =   e^{-{\bf A} {\bf b}^\dagger}  \; | 0\r | f'\r  + \sum_{n>0,\, f} {\bf  
 M^{\dagger n} }\, e^{-{\bf A} {\bf b}^\dagger}  \; | 0\r | f\r 
\eeq
simply because we can write an arbitrary function of $x$ in terms of a power series
around any point, $ f(x) = f(x_0) + \sum_{n>1} c_n\, (x-x_0)^n$---and $[ \bf a^\dagger, A]=0$.
Now, {\em any state of the form ${\bf  
 M^{\dagger n} } \, |phys\r $ is clearly null, but it is  also physical because ${\bf M}$  is
normal (no electric field).}

 Up to null states the physical states are thus  given by
\beq
 | \Psi \r_{Ph} = e^{-{\bf A} {\bf b}^\dagger}  | 0\r | f\r 
\eeq
which for the non-relativistic case, for example, are explicitely $  \exp( { - 
{i\over 2}( \hat{ t } 
-i\hat{\lambda} ) {{\hat{ p}_x^2} \over 2m} })\;  | 0\r | f\r$.
The null states are, as explained, are given by 
\beq
({\bf a}^\dagger  + {\bf A} )^k e^{ -{\bf A}{\bf  b}^\dagger }  | 0\r | g\r ,
\eeq
 where $k$ is an integer, $k \neq 0$.

As for the time evolution operator---and the dynamics---it  is given 
by 
\ba  \dis 
 U = _{Ph} \l x' |  e^{i{\bf p_t}   \Delta \tau } \;  | x\r_{Ph}, \\ \dis 
  | x\r_{Ph} = e^{ -{\bf A} {\bf b}^\dagger }\;   | 0\r | x\r
\ea
so   
\beq
U = _{Ph}\l x' |  e^{-i{\bf A}  \Delta \tau } \; | x\r_{Ph}
\eeq
This, again,  follows from \\
a)  ${\bf p_t}   + {\bf A}  = {1\over 2}( {\bf a}+ {\bf A}  + ( {\bf a}+ 
{\bf A}  
)^\dagger  
)$ \\
b) $_{Ph}\l\Psi ' |  ( {\bf p_t}   + {\bf A}  ) \,  | \Psi\r_{Ph} = 0 = 
_{Ph}\l\Psi 
' |  
{\bf \Phi  }\;   | \Psi \r_{Ph}$ \\ 
c) $_{Ph}\l\Psi ' |  {\bf O_x} \; | \Psi \r_{Ph} = \l f' |  {\bf O_x} \; | 
f\r$.

Also, recall
\beq
e^{i{\bf \lambda \Phi}} \;  | \Psi \r_{Ph} =  | \Psi \r _{Ph} +  | null\r
\eeq
which is almost obvious.

To give another explicit result, consider the
 free relativistic case  with a branch choice---it is not clear yet
how to proceed if one doesn't do this first. 
As before, the physical states 
are  defined by
\beq
({\bf a}+ {\bf A} )  | \Psi\r _{Ph} = 0 ,
\eeq 
where  
\beq  
{\bf A}  =  \sqrt{\bp_x^2+m^2}, \; \; {\bf a}= {\bf p_t}   + i{\bpi } ,
\eeq 
 etc., and up to null states they are given by
\beq
 | \Psi \r_{Ph} = e^{-{\bf A} {\bf b}^\dagger}  \; | 0\r | f\r = e^{ - 
{i\over 2}(\hat{ t} 
-i\hat{\lambda}  ) \sqrt{\hat{p}_x^2+m^2}} \; | 0\r | f\r
\eeq
The null states are
\beq
({\bf a}^\dagger   + {\bf A} )^k e^{ -{\bf A}{\bf  b}^\dagger }  \; | 0\r | g\r ,
\eeq
 where $k$ is an integer, $k \neq 0$. 

As pointed out, we can also consider the case of an electromagnetic
field with no electric components.

These state spaces are  thus isomorphic to the space of functions of the 
coordinate 
$x$---the old physical coordinates.  This includes the resulting inner products.

We can also define the propagation amplitude to be 
\beq
\label{eq:Fock}
\l t_f, x_f,\pi\! =\! 0 |  e^{i\hat{\lambda}\hat{\Phi}} |t_i x_i,\pi\! =\! 
0\r =
\l t_f, x_f |     {1\over \hat{\Phi} + i\epsilon  }         |t_i x_i\r  
\eeq
This amplitude is the causal amplitude---leads to the Feynman 
propagator for the full relativistic case, for example.  

This result follows from the discussion earlier on how to represent 
the Fock space in the coordinate basis.

Let us now consider the full constraint. As usual we will consider first an easier case,
\beq
\hat{\Phi} = \hat{P}_1^2 - \hat{A}^2
\eeq
where  we assume that $\hat{A}$  is a hermitean  operator that commutes with $\hat{a},
\hat{b}$---zero
electric field, as before. The  strategy  will be as before: find a normal operator $\hat{M}$,
$[\hat{M}, \hat{M}^\dagger]=0$, and write the constraint as  a sum of this operator and its
hermitean conjugate, \beq
\hat{\Phi}= \hat{M} + \hat{M}^\dagger
\eeq
 This will ensure two important things: \\
{\em i)} $ \ _{Ph}\!\l \Psi |  \hat{\Phi} \; | \Psi \r _{Ph} =0$\\
{\em ii)} The states $ \hat{M}^{\dagger n} \; | phys\r =0$, $n>0$, are null {\em and} physical.

Let us write the constraint in terms of the new variables,
\beq
  \hat{\Phi} = \hat{P}_1^2 -  \hat{A}^2 = ( { \hat{a} + \hat{a}^\dagger \over 2})^2 - \hat{A}^2 = 
{1\over 4} \left( \hat{a}^2 + \hat{a} ^{\dagger 2}  + 2 \hat{a} \hat{a}^\dagger \right)  - \hat{A}^2
\eeq
  The natural definition is thus
 \beq
\hat{M} = {1\over 4} \left( \hat{a}^2 +   
 \hat{a} \hat{a}^\dagger  \right)  - {1\over 2}  \hat{A}^2
\eeq
which satisfies the above requirements with our assumptions about $\hat{A}$.

To find the physical states we now need to solve the differential equation\beq
[\left({\partial \over \partial b^* }\right)^2 + a^* {\partial \over \partial b^* }-2 \hat{A}^2 ] \;
\psi(a^*, b^*) = 0 \eeq
which is done easily enough
\beq
\psi(a^*, b^*) = g(a^*) \; \exp \left({ -b^* a^*\over 2}  \pm {b^*\over 2} \sqrt{a^{*2} + 8 \hat{A}^2}
\right)
\eeq

Next we need to discuss the  null states.  From the properties of $\hat{M} $ we know that 
the states 
\beq
 \hat{M}^{\dagger n} \;
\exp \left({ -b^* a^*\over 2}  \pm {b^*\over 2} \sqrt{a^{*2} + 8 \hat{A}^2}
\right)
 | 0\r |f\r  
\eeq
 are physical and null. Do these exhaust all the freedom from the function $g$ in the previous
equation? 
If so the physical
space reduces to the usual two branches.

Notice that the operator $
\hat{M}^{\dagger}$ has a zero mode, \beq
\varphi = \exp\left( - b^* ( a^* -2 A) \right) 
\eeq
Let us compute the effect of $\hat{M}^\dagger$ on a physical state---we know that we will
get a physical state!  Now $
\hat{M}^\dagger  |phys\r \sim $ $$
(a^{*2} +a^* {\partial \over \partial b} -8\hat{A}^2) \; 
\exp \left({ -b^* a^*\over 2}  \pm {b^*\over 2} \sqrt{a^{*2} + 8
\hat{A}^2} \right) =  $$ \beq
 ( {a^{*2} \over 2} - 8\hat{A}^2 \pm {a^*\over 2}\sqrt{a^{*2} + 8 \hat{A}^2})
 \; 
\exp \left({ -b^* a^*\over 2}  \pm {b^*\over 2} \sqrt{a^{*2} + 8
\hat{A}^2} \right) 
\eeq
a physical state, as promised. Can we now  find a function such that
\beq
F(\hat{M}^\dagger) \,  |phys\r = \hat{a}^\dagger \;  |phys\r\:?
\eeq
  If so we are set. It is not hard to see that this question is equivalent 
to asking that the function
\beq
f_\pm (a^*) =   {a^{*2} \over 2} - 8\hat{A}^2 \pm {a^*\over 2}\sqrt{a^{*2} + 8
\hat{A}^2}  \eeq
have an inverse. And indeed, it does,
\beq
a^* = \mp {f_\pm + 8\hat{A}^2 \over \sqrt{f_\pm + 2\hat{A} + 8\hat{A}^2} }
\eeq
The function $f(a^*)$ is one-to-one (although not onto).
Thus, the physical space is one in which we have a set of states for each branch,
$|0_+\r\otimes |others\r + |0_{-}\r\otimes |others\r$.


What happens in the fully interacting case? Our decomposition of the constraint,
$\hat{\Phi} = \hat{M} + \hat{M}^\dagger $ is still valid, but with the above
choice,  $\hat{M}$ is not normal. What this means is that the null states in the
theory, $\hat{M}^{\dagger n}\,  |phys\r$ are no longer physical. So we may end up
with many more non-null physical states than we bargained for. This issue is an
important one---what is the correct description of this space in general?
I do not know the answer to this question yet.

\newpage\section{BRST inner product and construction of the path integral} 

Let us now look at the BRST quantization approach---we already 
hinted
above at how the inner product may look in this formalism.  The 
discussion 
will become extremely formal at some points, but we will try to draw 
very 
``unformal'' conclusions at the end. The trouble will always hinge 
around\\
a) The assumption of hermiticity of $\bf \Omega$\\
b) Operator ordering questions 

We will   address all these issues.

The physical space is defined by the condition that
\beq \bOmega | \Psi\rangle = 0\eeq  where recall that \beq 
\Omega = \eta_0 \Phi + \eta_1 \pi\eeq 
Now, there are many ways to write the solution to this equation.
A key property of the BRST generator is that $\bOmega^2=0$, so
any state of the form $\bOmega |\Psi\rangle$ is physical. However,
such a state has also zero inner product with any other physical 
state---since
the BRST generator is by assumption hermitean\footnote{But 
remember that 
it is not hermitean with respect to the physical states.... This is 
indeed 
serious trouble for the cohomology idea. The rule is to always 
compute the commutators first---or first operate and then compute 
the inner 
product.}, 
$\bOmega=\bOmega^\dagger$.
So these states can be factored out---formally. It has been argued 
(see Henneaux's report in \cite{BFV})
that a 
complete set
of physical states is given by 
\beq |\Psi\rangle  = 
|\psi\rangle  + 
|\psi_0\rangle  \eta_0 + |\psi_1\rangle  \eta_1 + |\psi_{01}\rangle 
\eta_0 
\eta_1 \eeq 
where all the kets satisfy $\bPhi, \mbox{\boldmath $\pi$} \: 
|\psi\rangle 
=0$. We will examine this result. Also, notice that this is by no means 
{\em 
the} way to describe the physical 
space.

Consider for example the states
\beq |\Psi_\chi \rangle = |\psi_{\chi =0}, \pi\!=\! \eta_0 \!=\! \rho_1 
\!=\!0\rangle\eeq 
These also satisfy the BRST condition---they are also physical.

These are the states that we will use in the BFV path integral---i.e 
these determine the boundary conditions we will use.
There are many other choices for them, for example
\beq  |\Psi_\Phi \rangle = | \psi_{\Phi= 0} , \lambda \!=\!\rho_0 
\!=\! \eta_1 
\!=\! 0
\rangle \eeq

What was the inner product in the full enlarged space?
The original BRST definition of the coordinate spaces is infinitely 
ranged, and
as such it runs into regularization problems. This inner product, 
when used, 
for example, in the zero 
ghost sector of the physical states described above, is just given 
by\beq (\psi_a,\psi_b) = 
\int dQdq d\eta_0 d\eta_1 d\lambda  \: \psi_a^* \; \psi_b=
\int dQ d\eta_0 d\eta_1 d\lambda dp \int dq \: \psi_a^* \; \psi_b
\eeq 
since physical states do  not depend on  $ Q, \lambda $.
Now, the first part of the integration is where the regularization 
problems appear (we see, however, that these will go away
if the coordinate spaces---including the ghosts'---are finite. In
such a case we can define $\int d\eta \eta = 1$, $\int d\eta = 1/L$, 
so
that the ghost normalizations cancel those of the gauge...or something
like that!) 

 As was previously remarked, the BFV path integral can be
interpreted as coming from a larger {\em quantum} space---with
no constraints. The fundamental reason is that the
 multipliers---which
``impose the constraints''---are dynamical and can be interpreted in 
the 
path integral as legitimate degrees of freedom. Indeed we showed 
how to
obtain the physical amplitude from a full quantum space...the only
place where physicality enters is in the boundary conditions. This is 
the only place where one feels uncomfortable (told me
 Claudio Teitelboim himself!), 
as at first sight the choice of  these boundary conditions seems  {\em 
ad hoc}.
However, as we  will now see,   these boundary conditions are    
easily understood to arise from the required BRST invariance of
the ``end'' states in the amplitude. The BRST boundary 
conditions do 
have an interpretation:
they can be understood in the context of gauge-invariant states. 
States
that implement such boundary conditions are indeed annihilated by 
the 
BRST generator.

Let us look a bit closer into the case where the constraint is
$\Phi = P\approx 0$. The path integral is arguably where BRST in phase space 
started, so let us
begin with that too. As mentioned, we can immediately write this path 
integral if we start from the extended quantum space, which is
spanned by the states\footnote{Again we remind ourselves  that
$\eta_0 \equiv c$, $ \rho_0 \equiv \bar{\cal P}$, $\eta_1 \equiv
-i \cal P$, and $\rho_1 \equiv i \bar{c}$.} 
 $|x,\lambda, c,\bar{c},q\rangle$
where the $q$ degree of freedom denotes the physical, non-gauge
sector. Now the hamiltonian is given by \beq {\cal H}= h+\{ {\cal O}, 
\Omega\}\eeq 
 where $\Omega$ is the BRST generator, \beq 
\Omega = \eta_0 P + \eta_1 \pi\eeq 
formed by combining the two constraints with the ghosts, and 
where $\cal O$ is the ``gauge-fixing'' term.

Now, the propagation amplitude is given by \beq 
\langle {\Psi_a}|e^{-i\tau {\hat{\cal  H}}} | {\Psi_b} \rangle=
\int d\mu(Q)\: \Psi_a^*(Q) \:e^{-i\tau \hat{\cal  H}_Q}\:{ \Psi_b(Q)} 
\eeq 
where the states are physical (boundary conditions).

Now, the BFV path integral provides us with a clue to the 
possible constructions of the extended Hilbert space. The first step
is to look at the boundary conditions that
we used in the path integral: $c,\bar{c}$ and $\pi$
are to vanish at the boundaries. So this defines the physical states.
Now, we can obtain the propagation amplitude by using as the 
propagation
operator \beq 
{\hat{\cal U}} = e^{-i \Delta\tau\hat{\cal H}}\eeq  where the 
hamiltonian 
is the 
extended
(super)hamiltonian: $ {\hat{\cal H}} \equiv \hat{ h} + \{ \hat{O}, \hat{\Omega}\}$. 
In the 
particle cases, for example,  $h=0$. Then we can obtain the correct 
propagation amplitude
from the expression\beq  
U(t_i,x_i,t_f,x_f)\equiv \langle t_f,x_f,c\!=\! \bar{c} \!=\! \pi\!=\!0|\; 
{\hat{\cal U}} \;
|
t_i,x_i,c\!=\!\bar{c}\!=\!\pi\!=\! 0\rangle\eeq   
Notice that our notation anticipates that this amplitude will not
depend on $\tau$, which will be the case. From this expression it
is easy to obtain the BFV path integral by repeated insertion of the
extended space  resolutions of the identity\beq 
{\bf I} = \int dtdxd\pi dcd\bar{c}\: |t,x,\pi, c,\bar{c}\rangle
\langle t,x,\pi,c,\bar{c}| = \eeq $$\int dp_tdp_xd\lambda d{ \bar{\cal 
P}}d{\cal P}
|p_t,p_x,\lambda,{ \bar{\cal P}}, {\cal P}\rangle\langle 
p_t,p_x,\lambda,{\bar{\cal P}},{\cal P}|$$ 
and the projections \beq 
\langle t,x,\pi,c,\bar{c}|p_t,p_x,\lambda, {\bar{\cal P}},{\cal P}\rangle 
=
e^{i\left( tp_t+xp_x+\pi\lambda+c{\bar{\cal P}}+\bar{c}{\cal 
P}\right)}\eeq 
just as in the unconstrained case.

Let us consider the following two types of gauge-fixing terms:\\

a) ${\cal O}_{NC} = \rho_1 f(\lambda) + \rho_0 \lambda$ which 
yields
 \beq \{{\cal O}_{NC} ,\Omega\} = \rho_1 \eta_1 f'(\lambda) + \pi 
f(\lambda)
+ \lambda \Phi + \rho_0\eta_1\eeq 

b) ${\cal O}_{C} = \rho_1 \chi + \rho_0 \lambda$ with \beq 
\{ {\cal O}_{C} ,\Omega\}= \rho_1 \eta_0 \{\chi, \Phi\} +
\pi \chi+ \lambda\Phi+\rho_0\eta_1\eeq 

Notice that the expectation value  of these operators on physical              
states---states
annihilated by the BRST generator---would appear to be  zero, so that 
formally
\beq  \langle {\Psi_a}|e^{ [\hat{{\cal O}} ,\hat{\Omega}]}\;  | {\Psi_b} 
\rangle= 
\langle {\Psi_a}| {\Psi_b} \rangle\eeq  up to hermiticity issues. So 
this is 
where the
invariance of the amplitude under changes in gauge-fixing is 
expected to 
come  from.
However, it is crucial to have hermiticity of the BRST operator---and 
this in 
general we don't have (recall the discussion at the beginning 
regarding the 
Dirac states.)
This also implies that the inner product for the case of zero physical
hamiltonian is all there is to propagation! (recall, however, that  
parametrized systems are more problematic because of the action's 
lack of 
invariance at the  boundaries).  Let us see how this goes. For 
example, take 
the BRST invariant states
$|\Psi_{\Phi}\rangle$ we defined above. Then\footnote{$i\{\:\:,\:\: \} \sim [\:\:,\:\: ]$}  \beq 
\langle \Psi_\Phi | e^{ [\hat{{\cal O}}_C ,\hat{\Omega}]}\;  |\Psi'_\Phi 
\rangle =\eeq $$ \langle \psi_{\Phi=0}, \lambda\!=\! 
\rho_0\!=\!\eta_1\!=\!0|
e^{i 
\left(\hat{\rho}_1\hat{\eta}_0\{\hat{\chi},\hat{\Phi}\}+\hat{\pi}
\hat{\chi}+
\hat{\lambda}\hat{\Phi}+\hat{\rho}_0\hat{\eta}_1 
\right)}
|\psi'_{\Phi=0}, \lambda\!=\! \rho_0 \!=\! \eta_1 \!=\! 0 \rangle$$   
which, 
{\em up to 
ordering},  is
\beq  = \langle \psi_{\Phi=0}| \{\hat{\chi}, \hat{\Phi}\} 
\delta(\hat{\chi}) \; 
|\psi'_{\Phi=0}\rangle\eeq 
i.e., the Dirac inner product. The mentioned ordering questions are 
tricky, 
and we will examine them further below---the Fradkin-Vilkovisky theorem will
justify our guess. 

At any rate, it  
is now  
easy to see now how to proceed to 
form
the BFV path integral: we write \beq
e^{[ {\hat{\cal O}}_C , \hat{\Omega} 
]} = \lim_{N\rightarrow \infty} \left( 1+{1\over N} [  \hat{\cal O} _C 
, \hat{\Omega} 
]\right) ^N \eeq 
and we insert the resolutions of the identity, etc.

We can also consider the states $| \Psi_\chi\rangle$. These 
  will connect  easily with the amplitude  we considered before in the 
Dirac quantization approach.
Notice that up to boundary issues it is just as good to work with the 
constraint
$ Q\approx 0$. The theory at the end will be the same as using
$P\approx 0 $. The BRST formalism does not distinguish between
these two cases.
These states are more to the point, since they
match the path integrals we will evaluate  in the particle case.
It is easy to see, again up to ordering difficulties, that in this case the 
amplitude becomes
\beq \langle \Psi_\chi| e^{ [\hat{{\cal O}}_C ,\hat{\Omega}]}\; 
|\Psi_\chi 
'\rangle
= \langle \psi_{\chi=0}|\delta(\hat{\Phi})\;  |\psi_{\chi '=0} \rangle\eeq 
From this expression it is easy to reach the BFV path integral that
we discussed before.

Finally, we can also consider the non-canonical gauge. For example, 
if we use the  case $f(\lambda)=0$
it is immediate that   $$\langle \Psi_\chi| e^{ [\hat{{\cal O}}_{NC} ,\hat{\Omega}]}\; 
|\Psi_\chi 
'\rangle
= \langle \psi_{\chi=0}|\delta(\hat{\Phi}) \; |\psi_{\chi '=0}\r $$ The 
amplitude $\langle \Psi_\Phi |e^{ [\hat{{\cal O}}_{NC} ,\hat{\Omega}]}\; |\Psi'_\Phi 
\rangle$, on the other hand, appears to be undefined ($0 \cdot 
\infty$).

Let us now discuss less formally the above issues: the state cohomology, the
inner 
product, ordering problems, and finally, the path integral.  


\subsection{Analysis of Cohomology} 

Before we begin, is good to keep in mind that we have three worlds 
to 
compare: 

{\em i)} Dirac; $\hat{\Phi} |\psi_D\rangle = 0 $ 

{\em ii)}  Fock; $\hat{a}|\psi_F\rangle = 0$ 

{\em iii)}  BRST; $\hat{\Omega} |\Psi_{BRST} \rangle = (\hat{\eta}_0 
\hat{\Phi} 
+ 
\hat{\eta}_1
\hat {\pi} )|\Psi_{BRST}\rangle  =0$

\noindent These are all different methods and the spaces under 
consideration are 
also 
different.

Now, solutions to {\em iii)} include\beq |\psi_D\rangle\otimes |\pi\! 
=\! 
0\rangle\otimes |G\rangle,\:\: \:\:
|\psi \rangle\otimes |\pi\! =\! 0\rangle\! \otimes \! |\eta_0\! =\! 0 
\rangle,\:\:\:\:
|\psi \! =\! 0 \rangle \! \otimes \! | f(\pi)\rangle \otimes |\eta_1\! 
=\! 0 
\rangle
\eeq  and \beq |\psi \rangle\otimes |f(\pi)\rangle\! \otimes \! 
|\eta_0\! =\! 
\eta_1\! =\! 
0 \rangle\eeq or
any combination of these.  Now, some of these may not be allowed. 
Or, they 
may differ by a state of the form \beq \hat{\Omega} 
|\Lambda\rangle\eeq 
which---{\em if} $\hat{\Omega}$ is hermitean---is null.  Consider 
the state---in the 
``coordinate''
representation  \beq  
\Psi_{\Omega\! =\! 0 
}  = 
\psi_{\Phi \! =\! \pi \! =\! 0 } +\psi^1_{\Phi \! =\! 0 } \eta_1 
+\psi^0_{\pi\! =\! 0 
}\eta_0 +\psi^{01}\eta_0 \eta_1 \eeq  The standard trick is to argue 
that 
since\footnote{This is trivially true in this simple case where there is 
no 
place for anomalies to appear in the transition 
$\{\Omega,\Omega\}_{PB}=0 
\longrightarrow [\bOmega, \bOmega]=0$}  
$\hat{\Omega}^2 = 0 $ we must identify \beq \Psi \sim \Psi + 
\hat{\Omega} \Lambda\eeq  The reasoning is that the states 
$\hat{\Omega} 
\Lambda$ are physical (true) and that they decouple from other 
physical 
states.
This, however, is not true in general. This is because in the physical 
space
we don't have hermiticity of $\hat{\Omega}$ (with the exception of 
the 
bounded situation, which we ignore for now). That is, the above 
argument is 
based on the assumption that \beq   
\langle \Lambda| \hat{\Omega}^\dagger |\Psi_{\Omega =0}\rangle 
=0\eeq {\em 
This 
puts restrictions on what $|\Lambda\rangle$ can be}.

The trouble spot can be traced to the fact that in general it isn't true 
that \beq 
\langle P\! = \! 0| \bP | Any \rangle =0\eeq   when for example $| 
Any 
\rangle= 
|P\! = \! 0 \rangle$---though this  true in the bounded case.

For this reason it is incorrect 
to assume, for example,  that one can always express a state in the 
form 
\beq \Psi_{\Omega\! 
=\! 0 }  = 
\psi_{\Phi  \! =\! 0\! =\! \pi } +\psi^1_{\Phi  \! =\! 0\! =\! \pi } \eta_1 
+\psi^0_{\Phi  \! =\! 0\! =\! \pi }\eta_0 +\psi^{01}_{\Phi  \! =\! 0\! =\! 
\pi 
}\eta_0 
\eta_1 \eeq by a proper choice of $|\Lambda\rangle$. {\em Such 
asumption 
is not 
consistent with  the existence of an inner product and/or hermicity 
of the 
BRST generator in an unbounded space}. Such a state, 
$|\Lambda\rangle$, does 
not exist that yields 
such a representation and such that $\hat{\Omega}|\Lambda\rangle$ 
is truly 
null. This can be seen in the following way:  we need to find the 
general form 
of 
$|\Lambda\rangle $ such that \beq \langle \Psi_{\Omega =0}| 
\hat{\Omega} 
|\Lambda\rangle =0\eeq  Next we need to find the possible forms of 
the 
physical states that we can reduce from the now validated 
equivalence
\beq |\Psi_{\Omega =0}\rangle  \sim |\Psi_{\Omega =0}\rangle  + 
\hat{\Omega} |\Lambda\rangle \eeq  Writing \beq |  \Psi_{\Omega 
=0}\rangle 
\sim  
\psi_{\Phi \! =\! \pi \! =\! 0 } +\psi_{\Phi \! =\! 0 }^1  \eta_1  
+\psi_{\pi\! =\! 0 
}^0  \eta_0  +\psi^{0 1 }\eta_0  \eta_1\eeq and \beq 
|\Lambda\rangle 
\sim\Lambda  +\Lambda^0  \eta_0  +\Lambda^1 \eta_1  
+\Lambda^{0 1 
}\eta_0  
\eta_1   \eeq 
we demand that  the above inner product vanish: $0=\langle 
\Psi_{\Omega 
=0}| \hat{\Omega} 
|\Lambda\rangle $
\beq 
= \left(  \psi_{\Phi \! =\! \pi \! =\! 0 } +\psi_{\Phi \! =\! 0 }^1  \eta_1  
+\psi_{\pi\! =\! 0 
}^0  \eta_0  +\psi^{0 1 }\eta_0  \eta_1   \: ,\:\:  \hat{\Omega} \:  (
\Lambda  +\Lambda^0  \eta_0  +\Lambda^1 \eta_1  +\Lambda^{0 1 
}\eta_0  
\eta_1  ) \right) \eeq 
\beq = \left(  \psi_{\Phi \! =\! \pi \! =\! 0 }  +\psi_{\Phi \! =\! 0 }^1  
\eta_1  
+\psi_{\pi\! =\! 0 
}^0  \eta_0  +\psi^{0 1 }\eta_0  \eta_1    \: ,\:\:   
\hat{\Phi} \Lambda  \eta_0  + \hat{\pi} \Lambda \eta_1  +(\hat{\pi} 
\Lambda^0  - \hat{\Phi} \Lambda^1 )   \eta_1 \eta_0  \right)  \eeq  
\beq  =
-\left(  \psi_{\Phi \! =\! \pi \! =\! 0 } \: ,\:\:  
    \hat{\pi} \Lambda^0  - \hat{\Phi} 
\Lambda^1  \right) +\left(   \psi_{\pi\! =\! 0 
}^0 \: ,\:\:     \hat{\pi} \Lambda    \right) -
\left(\psi_{\Phi \! =\! 0 }^1    \: ,\:\:     \hat{\Phi} \Lambda  \right) 
=0\eeq  
which will happen if the $\Lambda$ state is dual to the other               
one---for then 
the hermiticity properties of $\bOmega$ will be saved.

Thus, it is possible in general to write any state in the cohomology in 
the 
form  \beq  \Psi_{\Omega\! =\! 
0 }  = 
\psi_{\Phi  \! =\! 0\! =\! \pi } +\psi^1_{\Phi  \! =\! 0\! =\! \lambda } 
\eta_1 
+\psi^0_{\chi\! =\! 0 \! =\!  \pi}\eta_0 +\psi^{01}_{\chi\! =\! 0 \! =\! 
\lambda}\eta_0 
\eta_1 \eeq   
{\em Notice that duality of the sectors is lost if the constraint has more than one branch.}
We could have also obtained   Fock space 
representations, if we had started with a different extended Hilbert 
space at the beginning. There would then be different solutions to 
the BRST equation---see section 3.2.4.

Notice that the original BRST inner product in extended space can be 
used for these states---there 
are no regularization problems!  What has happened? Well, without 
the 
cluttering of the mutipliers we have written the cohomology as 
\beq |\varphi 
\rangle = |P\! =\! 
0\rangle + |Q\! =\! a\rangle \eta\eeq  The BRST inner product forces 
the 
inner 
product to match the space and its dual---because of the role that 
the ghost 
plays in the inner product: \beq   \langle \varphi |\varphi \rangle = 
\int 
d\eta 
\left(  \langle P\! =\! 
0| + \langle Q\! =\! a| \eta     \right) \left(    |P\! =\! 
0\rangle + |Q\! =\! a\rangle \eta  \right)\eeq  \beq = \int d\eta 
\:\eta \:   
      2Re  ( 
\langle P\! =\! 
0|Q\! =\! a \rangle  )= 2Re \langle P\! = \! 0 |Q \! = \! a\rangle \eeq 

We can consider other representations---use one  ghost coordinate 
and one 
ghost momentum as it happens in the path integral.

\subsection{Inner product in the zero ghost sector}
\label{sec:zeroghost}

 So we have an inner product, what else do we want? {\em We want 
an 
inner product defined on the zero ghost sector, which yields positive 
definite 
norms.} For that we are going to 
need a map from one sector its dual sector, so that the above nice 
properties 
can be used. This is where the operator $\exp ( [ \bf K, \Omega])$ 
enters. We 
already discussed how this inner product may be defined. The only 
problem 
was related to operator ordering. When the inner product 
computation was 
performed earlier, I ignored the fact that the operators in the 
exponent don't 
commute. However, as we will see,  the result is still correct.

 Consider the states that we discussed before
in the cohomology discussion,$$
\Psi_{\Omega\! =\! 
0 }  = 
\psi_{\Phi  \! =\! 0\! =\! \pi } +\psi^1_{\Phi  \! =\! 0\! =\! \lambda } 
\eta_1 
+\psi^0_{\chi\! =\! 0 \! =\!  \pi}\eta_0 +\psi^{01}_{\chi\! =\! 0 \! =\! 
\lambda}\eta_0 
\eta_1 $$
We will work with the states in the ``middle sectors''. Notice that 
inner product in the BRST enlarged space is well defined for these
states:
\beq
\int d\eta_0 d\eta_1\:\; \l  \psi^0_{\chi\! =\! 0 \! =\!  \pi}    |           
\psi^1_{\Phi  \! =\! 0\! =\! \lambda } \r \eta_0 \eta_1 
\eeq
as we discussed earlier.  These states have the same ghost number. 
Could we just define an inner product within one of these sectors? For 
definiteness let us 
work with the states $ 
\psi^0_{\chi\! =\! 0 \! =\!  \pi}\eta_0 
 $ and see what we can do. 
These states satisfy\beq \hat{\pi},\;\hat{\chi},\; \hat{\eta}_0,\; 
\hat{\rho}_1 \; \: |\psi^0_{\chi\! =\! 0 \! =\!  \pi\! =\! \eta_0  } \r
=0\eeq Consider now the operator\beq
e^{ [\hat{K},\hat{\Omega} ]} = e^{i\hat{\lambda}\hat{\Phi} + 
\hat{\rho}_0 \hat{\eta_1} } \eeq
for $ \hat{K} = \hat{\rho}_0 \hat{\lambda} $.  

It is easy to see that the new inner product is well defined for these 
states and matches the Dirac one:
$$ (\psi,\psi') \equiv \l    \psi^0_{\chi\! =\! 0 \! =\!  \pi \! =\!   
\eta_0 }| 
\; e^{ [\hat{K},\hat{\Omega} ]}\;
         |  \psi'^0_{\chi\! =\! 0 \! =\!  \pi \! =\!  \eta_0   }  \r 
$$
$$
\int d\eta_0 d\eta_1\:\; \l  \psi^0_{\chi\! =\! 0 \! =\!  \pi}    | 
\; e^{ [\hat{K},\hat{\Omega} ]}\;
 | \psi'^0 _{\chi\! =\! 0 \! =\!  \pi} \r 
\eta_0 \eta_1 =
$$ \beq \label{eq:BFVDIRAC}
\int d\eta_0 d\eta_1\:\; \l  \psi^0_{\chi\! =\! 0 \! =\!  \pi}    | 
\; e^{i\hat{\lambda}\hat{\Phi} + \hat{\rho}_0 \hat{\eta_1} }
 | \psi'^0 _{\chi\! =\! 0 \! =\!  \pi} \r 
\eta_0 \eta_1 =\l \psi^0_{\chi\! =\! 0} | \delta(\hat{\Phi}) | \psi'^0 
_{\chi\! =\! 0} \r
\eeq

This case is in fact going to reappear in the next chapter---these 
states are the ones we will use in the path integrals.

Consider, as another example, the states \beq 
|  \psi_a (t,x) ,  \; \Phi \! =\! 0\! =\! \lambda \! =\! \eta_1 \! =\!  
\rho_0 \r \sim  
\psi^1_{\Phi  \! =\! 0\! =\! \lambda } 
\eta_1 
 \eeq
 which also satisfy the BRST condition, and the gauge-fixing function 
$
\hat{K} = \hat{\rho}_1 \hat{\chi} $, where $\chi = t-f(\tau)$, and the 
constraint is the one for the relativistic particle. It is easy to calculate 
\beq
[\hat{K},\hat{\Omega}] =   
\hat{\rho}_1 \hat{\eta}_0 [\hat{\chi}, \hat{\Phi}] + i \hat{\pi}\hat{ 
\chi }
\eeq
Let us calculate\beq
\l  \psi_a (t,x) ,  \; \Phi \! =\! 0\! =\! \lambda \! =\! \eta_1 \! =\!  
\rho_0 |  \;  e^{ \hat{\rho}_1 \hat{\eta}_0 [\hat{\chi}, \hat{\Phi}] + i 
\hat{\pi}\hat{ \chi } } |\psi_b (t, x) ,  \; \Phi \! =\! 0\! =\! \lambda \! 
=\! \eta_1 \! =\!  \rho_0 \r =\eeq $$ 
\int d^4 x \:  \psi_a^*  (t, x)  \left[ \int d \pi d\rho_1 d\eta_0 
\; e^{  \rho _1  \eta _0 [\hat{\chi}, \hat{\Phi}] + i  \pi \hat{ \chi } 
}\right] \psi_b (t, x) $$
Now, $$ [\hat{\chi}, \hat{\Phi}] = [\hat{t}, \hat{p}^2]= i2 \hat{p}_t $$ 
To compute this integral we make use of the 
  Campbell-Baker-Hausdorff 
theorem\footnote{MILLER, Symmetry Groups +....QA171M52}. That is 
$$
ln(e^A e^B) = A+B+{1\over 2} [A,B] + {1\over 12} [A,[A,B]] - {1\over 
12} 
[B,[B,A]] + ...=$$ \beq  B + \int_0^1 dt\:\: g[ e^{t Ad\;A} \:e^{t Ad\;B}](A) \eeq 
where $ Ad\; A \equiv [A,\:\:]$ is an operator ($n^2 \times n^2$ 
matrix that 
acts on $n^2$-``vectors''), and \beq g(z) = {\ln \: z \over z-1} = 
\sum_{j=0}^
\infty 
{(1-z)^j \over j+1} = 1+ {1 \over 2} (1-z) + {1 \over 3 } (1-z)^3 +...\eeq 
Fortunately  the series ends quickly here, and we have
\beq
e^{  \rho _1  \eta _0 [\hat{\chi}, \hat{\Phi}] + i  \pi \hat{ \chi } }
= e^{  \rho _1  \eta _0 [\hat{\chi}, \hat{\Phi}]  - i \rho_1 \eta_0 \pi   
}\; 
e^{   i  \pi \hat{ \chi } } =\eeq
 $$  \left( 1+   
\rho _1  \eta _0 [\hat{\chi}, \hat{\Phi}]  - i \rho_1 \eta_0 \pi   
  \right)
 e^{   i  \pi \hat{ \chi } } $$ 
The ghost integrations are now easily done, and yield\beq
\int d \pi d\rho_1 d\eta_0 \; 
e^{  \rho _1  \eta _0 [\hat{\chi}, \hat{\Phi}] + i  \pi \hat{ \chi } } =
i \hat{p}_t \delta(\hat{\chi}) + i  \delta(\hat{\chi}) \hat{p}_t
\eeq This is the Klein-Gordon inner 
product, since
 \beq 
\l \psi_a | \delta(\hat{t} - f(\tau)) \hat{p}_t |\psi_b\r = 
\int dx dt\, \l \psi_a | x,t \r \l x,t | \delta(\hat{t}\! -\! f(\tau))
 \hat{p}_t |\psi_b\r\eeq
 which is just 
\beq \int dx dt\, \delta( t\! -\! 
f(\tau)) \, \l \psi_a | x,t\r \,\left( -i{\partial \over
\partial t}\right) \,  \l x, t|\psi_b\r 
\eeq   
Notice that what this inner product is doing is 
giving us the hermitized expression of the
determinant  operator---as discussed
before.  It does not yield an absolute value or anything else. From this point 
of view the Feyman amplitude is the more natural
 object in the theory.

Is this true for the general interacting case? Recall   equation 
\ref{eq:bracket}
$$ \{ t, \Phi_{EG} \} =
2g^{00} (p_0 - A_0) +  2g^{0i} (p_i - A_i) = 2 g^{\mu 0} \Pi_\mu $$
so \beq
[\hat{t} ,\;  [\hat{t}, \; \hat{\Phi} ]] = 2 \hat{g}^{00}
\eeq
In general, things will not be so easy, because, for whatever ordering we 
choose, 
the series will not 
terminate that quickly.

We need\beq
[ \hat{g}^{00} ,\; \hat{g^{\mu 0} \Pi_\mu}   ] =0  \eeq
for whatever ordering is used.

 However, if there is no gravitational background---only an
electromagnetic one---then the result is again immediate:
\beq
\int d \pi d\rho_1 d\eta_0 \; 
e^{  \rho _1  \eta _0 [\hat{\chi}, \hat{\Phi}] + i  \pi \hat{ \chi } } =
i \hat{\Pi}_0 \delta(\hat{\chi}) + i  \delta(\hat{\chi}) \hat{\Pi}_0
\eeq
The reason is that in such a case we have 
\beq
[\hat{t},\; \hat{\Phi}_E ] = 2i \hat{\Pi}_0
\eeq
and 
\beq
[\hat{t} ,\; [\hat{t}, \hat{\Phi}_E ] ] = -2
\eeq
which, again, commutes, and the series terminates.

What about using other gauge-fixing functions? We will consider 
this issue in the next section.

\subsection{The Fradkin-Vilkovisky theorem}
In agreement with the Fradkin-Vilkovisky theorem for the path 
integral   we would expect that the defined inner product  is 
invariant under changes of
 gauge-fixing. In essence we need to show that 
\beq
{\delta\over \delta \hat{K} } \l \Psi | \; e^{ [\hat{\Omega}, \hat{K} ] 
}|\Psi'\r = 0
\eeq provided that the states satisfy $\hat{\Omega} |\Psi\r =0$,
or simply that to first order the amplitude does not depend on  
$\varepsilon$ ,
\beq
 \l \Psi | \; e^{ [\hat{\Omega}, \hat{K} +\varepsilon \hat{J}] }|\Psi'\r = 
\l \Psi | \; e^{ [\hat{\Omega}, \hat{K}]} |\Psi'\r + O(\varepsilon^2) \eeq
This will be a local statement---``large'' gauge transformations are not 
covered.
\begin{prop}
\label{prop2}
For certain operators $\hat{K},\; \hat{J} $ and for BRST invariant states we 
have: 

i) $(\Psi_a, \Psi_b) \equiv \l \Psi_a|\;  e^{[ \hat{\Omega}, \hat{K}] } | \Psi_b\r
$ exists, and 

ii) $\hat{\Omega}$ is hermitean in the new inner product, and 

iii) ${d \over d \epsilon } \l \Psi_a| \; e^{[ \hat{\Omega}, \hat{K} +
\epsilon\hat{J} ] } | \Psi_b\r = 0 $  \end{prop} 
 
This theorem was originally proved within the path integral context
(see references in \cite{BFV}).

As we saw before, as long as we choose the right gauge-fixing for the states 
we 
want to work with the above amplitude exists---the regularization works. So 
let 
us assume that we are taking the operator $\hat{K}$ as we defined it before.  
We 
can, in the same fashion,  check that \beq (\Psi_a, \hat{\Omega} \Psi_b) 
=( \Psi_a, \hat{\Omega}^\dagger \Psi_b) =0\eeq and moreover, \beq
(\Psi_a, \; [\hat{M}, \hat{\Omega}] \Psi_b) =([\hat{M}, \hat{\Omega}]\Psi_a, 
\;  
\Psi_b) =0
\eeq
as long as $\hat{M}$ is regular. This follows, in part, from the fact that  \beq
[ \hat{\Omega}  ,\; [  \hat{\Omega},\, \hat{A}] ] =0 \eeq

Finally, we also have that
 \beq
[ [\hat{\Omega},\, \hat{A}]   ,\; [  \hat{\Omega},\, \hat{B}] ] =
[ \hat{\Omega}  ,\; \hat{C} ]  
\eeq
where $\hat{C} = [ \hat{\Omega} ,\: [ \hat{A} ,\,   [\hat{\Omega}, \hat{B}]           
] 
]  
$

Using the Campbell-Baker-Hausdorff 
theorem mentioned above, it now follows that 
\beq
e^{[\hat{\Omega},\; \hat{A} ] + [\hat{\Omega},\; \hat{B} ]} =
e^{[\hat{\Omega},\; \hat{A} ]} \: e^{[\hat{\Omega},\; \hat{B}' ]} 
\eeq
because all the extra terms in the CBH theorem can be written in the BRS-
exact 
form, $[\hat{\Omega}, stuff]$.  This
simple result, applied to our situation implies that \beq
 e^{[ \hat{\Omega}, \; \hat{K} +
\epsilon\hat{J} ] } =  e^{[ \hat{\Omega}, \; \hat{K} ]} \: e^{[ \hat{\Omega},\; 
\epsilon\hat{J} ] } \approx  e^{[ \hat{\Omega}, \; \hat{K} ]} \:
\left( 1+ [ \hat{\Omega},\; 
\epsilon\hat{J} ]                \right) \eeq
which finishes the argument (in fact, it is true that $\exp ( [\hat{\Omega}, \; 
\hat{K} 
]) = 1 + [\hat{\Omega},\; \hat{L}]$ see exercise  14.13 in \cite{TeitelBook}).

\subsection{The composition law and the BFV path integral}

This inner product definition was of course 
inferred 
from the BFV path integral, from the term \beq e^{i \int d\tau\: \{ K, 
\Omega\}}\eeq 
  In the path integral derivation we use a trick: 
we 
expand the exponent to first order, compute the expectation value, 
and        
re-exponentiate. That is, we can say that to compute the expectation 
value (for simplicity I use $\tau$-independent gauge-fixing) 
of
 \beq 
\langle \Psi_\Phi |  \; e^{[  {\hat{\cal O}}_C , \hat{\Omega} 
]       \Delta\tau }|\Psi'_\Phi 
\rangle = 
\eeq 
$$ 
\langle \psi_{\Phi=0}, \lambda\!=\! \rho_0\!=\! 
\eta_1\!=\! 0|\;
e^{i\Delta\tau 
\left(-i\hat{\rho}_1\hat{\eta}_0 [\hat{\chi},\hat{\Phi}]+\hat{\pi}
\hat{\chi}+
\hat{\lambda}\hat{\Phi}+\hat{\rho}_0\hat{\eta}_1 
\right)}
|\psi'_{\Phi=0}, \lambda \!=\!  \rho_0 \!=\!  \eta_1 \!=\!  0 
\rangle=
$$  
\begin{eqnarray*}
\lefteqn{\langle \psi_{\Phi=0}, \lambda\!=\! \rho_0\!=\! \eta_1\!=\! 0| } \\ 
&&
\lim_{N\rightarrow\infty} \left( 
1+{i\Delta\tau \over N} 
\left(-i\hat{\rho}_1\hat{\eta}_0 [\hat{\chi},\hat{\Phi}]+\hat{\pi}
\hat{\chi}+
\hat{\lambda}\hat{\Phi}+\hat{\rho}_0\hat{\eta}_1 
\right) 
\right)^N
|\psi'_{\Phi=0}, \lambda \!=\!  \rho_0 \!=\!  \eta_1 \!=\!  0 \rangle 
\end{eqnarray*}
Next, we insert the resolutions of the identity. Thus we obtain the 
path integral---in this case the one with ``momentum'' 
BRS-invariant boundary conditions. By the same technique it is easy to 
check that the 
naive result for the above amplitudes is indeed correct.

However, we would also like to write a composition law exclusively 
in the physical subspace. For that we will have to develop an analog 
of the projector formalism in the Dirac quantization case.  For the 
states we consider ($\chi=0$ states) this operator is provided by (let here $
  [\hat{K},\hat{\Omega} ]   =  i\hat{\lambda}\hat{\Phi} + 
\hat{\rho}_0 \hat{\eta_1}$   
for $ \hat{K} = \hat{\rho}_0 \hat{\lambda} $, and think $\chi_a = t - 
f(\tau_a) $, $Q\sim t$ being the gauge degree of freedom)
\beq
\hat{\cal{ K}}_{\chi_J} = \delta(\hat{\eta}_0) \delta(\hat{\rho}_1) 
\delta(\hat{\pi})
\delta(\hat{\chi}_J ) (-i)[\hat{\chi}_J,\hat{ \Phi} ] 
e^{i [\hat{K} , \hat{\Omega} ] }
\eeq 
which satisfies 
\beq
\l    \psi^0_{\chi_f \! =\! 0 \! =\!  \pi \! =\!   \eta_0 \!=\!\rho_1}|\; 
e^{i[\hat{K},\hat{\Omega} ]} 
         |  \psi^0_{\chi_i \! =\! 0 \! =\!  \pi \! =\!   \eta_0 \!=\!\rho_1}  \r  =
\eeq
$$
\l    \psi^0_{\chi_f\! =\! 0 \! =\!  \pi \! =\!   \eta_0 \!=\!\rho_1}| \;
e^{i[\hat{K},\hat{\Omega} ]}\hat{\cal{ K}}_{\chi_J} 
         |  \psi^0_{\chi_i\! =\! 0 \! =\!  \pi \! =\!   \eta_0 \!=\!\rho_1}  \r  =
$$
$$
\l    \psi^0_{\chi_j\! =\! 0 \! =\!  \pi \! =\!   \eta_0 \!=\!\rho_1}| \;
e^{i[\hat{K},\hat{\Omega} ]}
\delta(\hat{\eta}_0) \delta(\hat{\rho}_1) \delta(\hat{\pi}) 
\delta(\hat{\chi}_J) 
(-i)[\hat{\chi}_J,\hat{ \Phi} ] e^{i [\hat{K} , \hat{\Omega} ] }
         |  \psi^0_{\chi_i \! =\! 0 \! =\!  \pi \! =\!   \eta_0 \!=\!\rho_1}   \r  =
$$
$$
\int \! dQ dq\, \l    \psi^0_{\chi_f \! =\! 0 \! =\!  \pi \! =\!   \eta_0 \!=\!\rho_1}| 
\,
e^{i[\hat{K},\hat{\Omega} ]}|  \psi^0_{  0 \! =\!  \pi \! =\!   \eta_0 \!=\!\rho_1}  
\r \,
\l    \psi^0_{  0 \! =\!  \pi \! =\!   \eta_0 \!=\!\rho_1} |\, \delta(\hat{\chi})  (-
i)[\hat{\chi}_J,\hat{ \Phi} ]
   \, e^{i[\hat{K},\hat{\Omega} ]} |  \psi^0_{\chi_i \! =\! 0 \! =\!  \pi \! =\!   
\eta_0 \!=\!\rho_1}  \r 
$$
which after insertion of the full resolution of the identity yields the 
composition law,
\beq
\int dQ dq\;      \l    \psi^0_{\chi_f \! =\! 0} | \;
\delta(\hat{\Phi}) |  \psi^0, Q  \r 
\l   \psi^0,Q | \; \delta(\hat{\chi}_J)(-i)[\hat{\chi}_J,\hat{ \Phi} ]\;   
\delta(\hat{\Phi}) |\psi^0_{\chi_i\! =\! 0} 
\r 
\eeq
 We can obtain, for example, the Klein-Gordon 
composition law, since $$ \l \psi_a | \delta(\hat{t} - f(\tau)) \hat{p}_t 
|\psi_b\r = 
\int dx dt\; \l \psi_a | x,t \r \l x,t | \delta(\hat{t} - f(\tau)) \hat{p}_t 
|\psi_b\r =$$ 
\beq \int dx dt\delta( t - f(\tau)) \; \l
\psi_a | x,t\r  \,\left( -i {\partial \over \partial t}\right) \,  \l x, t|\psi_b\r \eeq just as we did 
previously in the Dirac quantization  approach. We
have to be careful about ordering for the jacobian---as usual---and for the 
relativistic case, for example, instead of the above
the operator we should use---remember that we will get the on-shell or 
Feynman amplitudes---is 
 \beq
\hat{\cal{ K}}_\chi = \delta(\hat{\eta}_0) \delta(\hat{\rho}_1) 
\delta(\hat{\pi})
\left( \delta(\hat{\chi}) (-i)[\hat{\chi}_J,\hat{ \Phi} ] +(-i)[\hat{\chi}_J,\hat{ 
\Phi} ] \delta(\hat{\chi}) \right)
e^{i [\hat{K} , \hat{\Omega} ] }
\eeq 
for the half range case, or
\beq
\hat{\cal{ K}}_\chi = \delta(\hat{\eta}_0) 
\delta(\hat{\rho}_1)\delta(\hat{\pi}) 
\delta(\hat{\chi})   \;|  (-i)[\hat{\chi}_J,\hat{ \Phi} ] | \;
e^{i [\hat{K} , \hat{\Omega} ] }
\eeq 
for the full range one.
Notice, though,  that there is no way to get the absolute value for the 
jacobian  in BFV.  From this
point of view the Feynman propagator is the more natural amplitude in the 
theory, as we already mentioned in section~\ref{sec:zeroghost}.

\subsection{ Other questions}

 \noindent {\em How about the BRST observables, are they hermitean 
with
respect to this inner product?}  

Recall that an observable 
is defined by demanding that it commute with the BRST generator.
Now, if an observable $\hat{A}$ is to be hermitean it needs to be
hermitean with respect to the full inner product and commute
with the term $\{{\hat{\cal O}}, \hat{\Omega }\}$. Now \beq 
\{ \hat{A}, \{{\hat{\cal O}} , \hat{\Omega} \} \}= - \{\hat{\Omega}, \{ 
\hat{A}, {\hat{\cal O}} \} \}\eeq 
by virtue of the Jacobi identity and the fact that $\hat{A}$ is an 
observable.
Hence, for physical states we have, provided that we have hermicity 
of the BRST generator \beq 
\langle \Psi | \{\hat{ A}, \{ {\hat{\cal O}},\hat{\Omega}\}\} 
|\Psi'\rangle =0\eeq 
and we have hermiticity of the observables with respect to
the regularized inner product\footnote{Again, hermiticity of $\bf 
\Omega$ 
has been assumed, i.e., we are using the above regularized inner product in 
general.}.

\noindent{\em Does the BRST formalism fix the arbitrariness in the 
quantum 
theory
that arises from writing the constraint in different ways?}

The inner product suffers from the same ambiguity we described 
before,
since, indeed,  it reduces to the Dirac inner product. How do the BRST 
observables
behave under a rescaling of the constraint? Notice that no matter
what we do we need to preserve the hermiticity of $\bOmega$.
There is a cohomology definition for operators---just as there is one
for the states.

\noindent{\em How about BRST-Fock?}

 There never was an inner product problem there.

\newpage\section{Conclusions, summary} 
We started with the Dirac condition on the states, and we saw that 
unless the 
original gauge coordinate was bounded (compact gauge group) the 
Dirac 
states are not normalizable in the original bigger space.
I emphasized that this is 
 the ``inner product problem'': we       
cannot---in general---extract from the original big Hilbert space an 
inner 
product for the Dirac subspace.

As a cure we could use compact gauge coordinates---we could force 
something like periodic boundary conditions.

If we want to use a full coordinate space,
 we could use the ``dual space trick'' 
of 
Marnelius. 
I introduced instead a projector, and showed that this projection
procedure  leads to the Faddeev  path 
integral, and is equivalent to the reduced phase space method for 
constraints 
that can be made into a momentum. It is also cumbersome because 
we don't 
really have what Marnelius calls an ``inner product space''.

This discussion was tied to the proper definition of the Dirac inner product,
which I argued must be gauge-invariant and    also yield states with real norms.
This lead to the derivation of the Klein-Gordon inner product for the free relativistic
case.

 The projection procedure becomes unclear when the constraint  is
quadratic and unfactorizable. This is not surprising, 
because the existence of the
projector implies a composition law, and   in general there is no such composition
 law for the
on-shell propagators.

We can also use a Fock representation. A weaker condition is 
imposed than 
the full constraint, so the physical states are regular in the original 
inner 
product, which can be used in the physical subspace. However, Fock 
space is 
compact---a discrete space. This corresponds to ``half the 
constraints'' being 
imposed, as described below. The gauge degrees of freedom disappear
partly because of the appearance of null states. I showed that this is true of both the
simple case in which the constraint can be made into a momentum and the case where we have
a quadratic constraint that factorizes. If this condition is not met it is not
clear at all that  there are enough physical null  states  
in the theory to reduce the physical subspace
 to the form $|vac_1\r \otimes |others\r + |vac_2\r
\otimes  |others\r$.

 In the BRST formalism  an even weaker condition is imposed on the 
states.  I argued that in the cohomology discussion one must be careful about the allowed
states, which must respect the hermicity properties of the BRST generator. The extended
space inner product is then well defined for the  states in this cohomology.
 These states are in fact the natural
boundary conditions in the construction of the path integral (moreover, with them one can
then  derive the Klein-Gordon inner product, as described below).
 Another consequence of this
restriction is that the usual duality statements of the different ghost sectors do not
apply for the case of quadratic constraints. 

  However, 
and still 
with 
the intention of having an inner product space and because one 
wants to 
work with zero ghost states, a duality operator is used, the famous 
$\exp ([  
\hat{K}, \, \hat{\Omega}])$ (the motivation for this factor comes from the BFV path
integral, where it appears). This inner product, when used with the proper set of states in
the above cohomology yields the Klein-Gordon inner product in the relativistic case, for
example. I showed that this definition of inner product leads to the BFV path   
integral---modulo 
operator 
orderings---by simple insertions of the resolution of the identity in full space. 
BRST state space is not necessarily compact,  unlike the 
Fock case---although one can also use the BRST-Fock representation.

I also developed a projector formalism in BRST akin to the one  I constructed for the Dirac
case, through which one can derive a composition law. Again, the Klein-Gordon case was given
as an example.

Moreover, consistency of the quantum reduction leads to two 
different  
possible representations of the multipliers: fully-ranged real multipliers---this 
leads to 
states that satisfy the constraints---and half-ranged imaginary 
representations which lead to states satisfying  the Fock
condition---``half the constraints''.  

I also showed how to construct the BFV path integral strictly in the physical space,
through
the development of an analogous projector formalism to the one I developed for the 
Dirac/Faddeev case. This complements my earlier description of the construction of the
path integral in the fully extended BRST space.

\chapter{Path integrals}
\label{sec:PIs}

In this section I will discuss the different path integral approaches
to quantization of constrained systems.

We will discuss first the path integrals in phase space---the ones 
more 
directly 
connected with the Hilbert space description---and then the path 
integrals 
in 
configuration space, which should be  derivable from the first by 
integration 
of 
the momenta.

I will not worry too much, for the most part, about the problems
associated specifically with theories in curved backgrounds and how
to skeletonize the path integrals (in phase-space or configuration 
space) in such cases, although I will review some important results 
in section~\ref{sec:curvedPI}.

\newpage\section{Path integrals in phase space}
\label{sec:PIPS}

We will look first at Faddeev's path integral 
\cite{TeitelBook,Faddeev}; 
then we 
will study the more complex BFV path integral \cite{TeitelBook,BFV}. 
These
correspond to the reduced phase space and the BRST quantization 
schemes, respectively. 

The non-relativistic case and single branched cases
 will be discussed first, and
then the more  complex 
interacting cases will be examined. We will pay particular attention 
to the 
problems arising from having a constraint that cannot be made into a 
momentum by a well-defined canonical transformation, and to 
possible path 
integral recipes in such a situation.

\subsection{The Faddeev path integral: one branch}
\label{sec:Faddeev}

I will now introduce the Faddeev path integral \cite{TeitelBook,Faddeev}
 from the point of view
of 
reduced 
phase space quantization.

 Consider starting from a well defined reduced phase 
space. Let 
this space be described by the coordinates  $q,p$   and some 
hamiltonian 
$h(q,p)$. The unphysical part of the phase space will be described by 
the 
variables $Q,P$ and the constraint $P \approx 0$. The 
extended hamiltonian is then $H_{E} = h(q,p)+\lambda P$. In this 
simple situation the system effectively decouples into the 
physical and unphysical parts and we can drop the gauge part.
With boundary conditions on the physical coordinates 
the path integral in this reduced phase space is then just like the 
unconstrained case in section~\ref{sec:unconstrained}, and we can 
write
\beq
\Gamma(q_i , q_f ; \tau_i , \tau_f)
= \int Dq Dp \; e^{i \itif d\tau ( p\dot{q} - h(q,p) )} 
= \int Dq Dp\;  e^{i \itif pdq - h(q,p) d\tau }
\eeq
where the Liouville measure is 
\beq 
DqDp = {d p_0 \over 2\pi} \p {d q_i d p_i \over 2\pi}
\eeq
Notice that if $h(q,p)=0$ we immediately have that
$\Gamma(q_i , q_f ; \tau_i , \tau_f) = \Gamma(q_i , q_f)$ only 
(indeed $\Gamma = \delta (q_f - q_i )$ !!), since the integrand
becomes independent of the parameter. {\em This illustrates the 
relationship 
between reparametrization invariance---or parametrization 
indifference---and 
a zero hamiltonian}, which
Dirac described  so well \cite{Dirac}.

Now we can add the unphysical degrees of freedom $ Q, P$  by the 
use of the 
identity
\beq
1= \int DQDP \; \delta (Q-Q(q,p,\tau ))\delta (P) e^{i\itif d\tau  P 
\dot{Q}} 
\eeq
where the measure is defined\footnote{$DQ DP\; \delta (Q-Q(q,p,\tau 
)) 
\delta 
(P) \equiv \p  dQ_i dP_i \delta (Q_i-Q(q,p,\tau )) \delta (P_i)$ also 
works }
 by
\beq
DQ DP \; \delta (Q-Q(q,p,\tau )) \delta (P) \equiv dP_0 \delta (P_0) 
\p  dQ_i 
dP_i 
\delta (Q_i-Q(q,p,\tau )) \delta (P_i)
\eeq
This ``1'' can then be inserted in the path integral above:
\beq
\Gamma (q_i,q_f;\tau_i , \tau_f )  = \int DqDpDQDP \delta (Q-
Q(q,p,\tau )) 
\delta (P)  e^{i\itif d\tau  ( P\dot{Q} + p\dot{q} - h(q,p) )} 
\eeq
\beq
= \int DqDpDQDPD\lambda  \delta (Q-Q(q,p,\tau ))  e^{i\itif d\tau  ( 
P\dot{Q} 
+ p\dot{q} - h(q,p) + \lambda  P )}
\eeq
with 
\beq
D\lambda  = {d\lambda_0 \over 2 \pi}\p  {d\lambda_i \over 2\pi }
\eeq
This is the simplest situation. One can also write\footnote{ note the 
absolute 
value!}
\bea \dis 
\delta (Q-Q(q,p,\tau )) \delta (P)& \dis =\delta (f(Q,q,p,\tau 
))\;|{\partial 
f\over 
\partial Q}| \;\delta (P) \\ \dis 
 & \dis = \delta (f(Q,q,p,\tau ))\;|\{ f,P\} | \;\delta (P), 
\eea
so the path integral can be written as
\ba \dis 
\Gamma (q_i,qf,\Delta \tau ) 
  & \dis =\int DqDpDQDP \delta (f(Q,q,p,\tau )) \delta (P)|\{ f,P\} |  
e^{i\itif
d\tau (  P\dot{Q} + p\dot{q} - h(q,p))} \\[.5cm] \dis 
 & \dis  = \int DqDpDQDP \delta (\chi ) \delta (\Phi ) |\{ \chi ,\Phi \} |  
e^{i\itif  d\tau  (P\dot{Q} + p\dot{q} - h(q,p))}\\[.5cm] \dis 
   & \dis = \int DqDpDQDP D\lambda D\pi Dc D \bar{c}\:
e^{i\itif d\tau ( P\dot{Q} + p\dot{q} - h(q,p)+\lambda\Phi + 
\lambda\chi + c 
|\{ \chi ,\Phi \} | \bar{c} )}
\ea
with the obvious identifications. To be perfectly clear, let us write 
out the 
full 
measures,
\ba \dis 
D\mu & \dis \equiv   DqDpDQDP \delta (\chi ) \delta (\Phi ) |\{ \chi 
,\Phi
\}| \\[.5cm] 
  \dis & \dis \equiv \left(  {dp_0 dP_0\over 2\pi}  \p  {dQ_i dP_i 
dq_i 
dp_i\over        
          2\pi }   \right)  
    \left( \delta_0 (P) \p \delta_i (Q-Q(q,p,\tau )) \delta_i (P)
 \right)\\[.5cm] \dis 
& \dis =  \left(  {dp_0 dP_0\over 2\pi}  \p  {dQ_i dP_i dq_i dp_i\over 
2\pi
} \right) 
    \left( \delta_0 (\Phi) \p \delta_i(\chi) \delta_i (\Phi ) |\{ \chi 
,\Phi \} |_i 
\right)
\ea
and also
\beq
D\lambda D\pi Dc D \bar{c} \equiv {d\lambda_0 \over 2 \pi} \p 
{d\pi_i  
\over 
2 \pi}dc_i d \bar{c}_i  
 \eeq

Let us do a {\em canonical} change of variables in this integral, 
$(q,p,Q,P) 
\rightarrow (z^\alpha ,w_\alpha )$: a change of variables that is also 
a 
canonical transformation. This implies $dqdpdQdP = dz^\alpha 
dw_\alpha $, 
since the jacobian for the change of variables is the Poisson bracket 
of the 
new 
variables---which is one by definition of canonical transformation. 
After this 
change of variables the path integral becomes
\beq
\Gamma (q_i,qf;\tau_i ,\tau_f ) =
\int Dz^\alpha Dw_\alpha  \delta (\chi ) \delta (\Phi ) |\{ \chi ,\Phi \} 
| 
e^{i\itif 
d\tau ( \dot{z}^\alpha w_\alpha  - H(z,w) + {dW\over d\tau})}
\eeq
where $Dz^\alpha Dw_\alpha \equiv DqDpDQDP$,  and to be clear
\beq
Dz^\alpha Dw_\alpha  = {dP_0 dp_0\over 2\pi} \p  
{dz^\alpha_i dw_{\alpha i}\over 2\pi }   
\eeq
$W$ is the generator of the canonical transformation, i.e.,
 \beq
P\dot{Q} + p\dot{q} -h(q,p) =  \dot{z}^\alpha w_\alpha 
-H(z,w)+{dW\over 
d\tau} 
\eeq
It is important to note that the form of the identity one uses is not 
unique by 
any means, nor is unique what we define as the gauge momentum or 
coordinate.  {\em The main point is that the path integral, to make 
sense in a 
quantum mechanical framework, must reduce to the original physical 
quantity 
$\Gamma $,  which lives in a well-defined physical phase space.}

  Let us apply these ideas mindlessly to the coordinate system of 
section~\ref{sec:constant}. Recall that the hamiltonian is zero and the 
constraint 
$P \approx 0$,  with 
\ba \dis 
Q= t-t_0  &  \dis  P= p_t + {p_x^2\over 2m} & \mbox{\em gauge degrees 
of 
freedom,}\\  \dis 
q= p_x (t-t_0) - mx & \dis  p= - {p_x \over m}  &  \mbox{\em physical 
degrees of 
freedom }
\ea
We then obtain the result  
\ba \dis 
\Gamma (q_i,q_f) 
& \dis = \int DQDPDqDp \delta (Q-Q(q,p,\tau )) \delta (P)  
e^{i\itif d\tau  ( P\dot{Q} + p\dot{q} ) } \\[.5cm] \dis 
  & \dis = \int DqDp e^{i\itif  pdq} \\[.5cm]
& \dis  = \delta (q_f - q_i)  
\ea
There is nothing wrong with this result, it's just that the action we 
are using 
is 
not the original one. As  we saw in section~\ref{sec:constant} the 
action 
$\itif d\tau  ( P\dot{Q} + p\dot{q} ) $ {\em differs from the original 
one by a 
surface term that depends on the gauge coordinates}, the $\itif dW$  
above. 
Indeed $\Gamma '$ differs from $\Gamma$  precisely by this gauge 
dependent 
boundary term in the action, where
\beq
\Gamma ' = \int DtDxDp_t Dp_x  \delta (\chi ) \delta (\Phi ) |\{ \chi 
,\Phi \} |  
e^{i\itif d\tau  ( p_x  \dot{x} + P_t \dot{t} )}
\eeq
with
\beq
 DtDxDp_t Dp_x \equiv dp_{t 0} dp_{x 0} \p { dt_i dp_{t i} dx_i dp_{x 
i} \over 
(2\pi )^2}
\eeq
and 
\beq
 \delta (\chi ) \delta (\Phi )|\{ \chi ,\Phi \} | \equiv
 \delta_0(\Phi ) \p  \delta_i(\chi ) \delta_i(\Phi )|\{ \chi ,\Phi \} |_i
\eeq
since the 
measure `` boundary effect''
$$
dP_0 dp_0 = dp_{t 0} dp_{x 0} { \partial (P_0,p_0)\over \partial (p_{t 
0}, 
dp_{x 
0})} = -{1\over m} dp_{t 0} dp_{x 0} 
$$
is absent except  for the constant factor $(-1/m)$---which we will 
ignore---
and 
in fact, using the $\tau$-dependent gauge $t=t(\tau )$, $\{ \chi ,\Phi 
\}  = 1$,  
we obtain the blatantly `` gauge dependent''   result
\beq
\Gamma '= \left( {m\over 2\pi i \Delta t} \right)^{1\over 2} 
e^{ i {\Delta x^2\over 2 \Delta t} }
\eeq
where gauge dependence comes in  through the factor 
$\Delta t= t(\tau_f)-t(\tau_i)$---the boundary conditions.  With the 
measure 
I 
defined above the dependence comes in from the boundary 
conditions, not 
the 
gauge fixing condition, since there is no gauge fixing at the 
boundaries. 
Please 
note that this is just a matter of semantics. We could have indeed 
defined 
the 
measure  to be \cite{Teitelpap}
\ba \dis 
D\mu^* & \dis \equiv DqDpDQDP \delta (\chi ) \delta (\Phi ) |\{ \chi 
,\Phi \} | 
 \\[.5cm] \dis 
 & \dis \equiv \left(  {dp_0 dP_0\over 2\pi}  \p  {dQ_i dP_i dq_i 
dp_i\over        
          2\pi }   \right)  \cdot \\[.5cm] \dis 
 & \dis \:\: 
    \left( \delta_0 (P) \p \delta_i (Q-Q(q,p,\tau )) \delta_i (P) \right)
dQ_0 \delta_0(Q-Q(q,p,\tau ))\\[.5cm] \dis 
& \dis =  \left(  {dp_0 dP_0\over 2\pi}  \p  {dQ_i dP_i dq_i dp_i\over 
2\pi }          
     \right) 
    \left( \delta_0 (\Phi)  \delta_0 (\Phi ) |\{ \chi ,\Phi \} |_0 \p 
\delta_i(\chi) 
\delta_i (\Phi ) |\{ \chi ,\Phi \} |_i \right)
\ea
which in the present case would add the integration $dt_0 \; \delta 
(t_0 
-t(\tau_0))$,  a term that overrides the boundary conditions (or a 
term that 
implements them!) This is the point of view of 
reference~\cite{Teitelpap}, 
which implicitly uses this measure $D\mu^*$. The result would be 
the same 
as long as the gauge matches the intended boundary conditions. {\em 
The 
advantadge of the measure $D\mu$ I previously defined is that it 
makes the 
result gauge-fixing invariant, unlike $D\mu^*$.} But this is all a 
matter of 
semantics. If you know what the boundary conditions are for the 
path 
integral, and wish to use the measure $D\mu^*$, just choose the 
gauge-fixing 
function to match the boundary conditions, that's all.

 Regardless of the measure one uses, $\Gamma '$ cannot be reduced 
to the 
gauge invariant reduced phase space path integral, as their actions 
differ by 
the variant surface term. However it is easy to see that $\Gamma '$ 
can be 
reduced to something else. Let us use the measure defined just 
above, 
$D\mu^*$, and let us perform the `` gauge''  integrations; then
\ba \dis
\Gamma ' &\dis = \int Dx Dp_x    e^{i\itif d\tau  p_x  \dot{x}  -  t 
\left. {p_x^2 
\over 
2m}\right|_{\tau_i}^{\tau_f} }
\\[.5cm] &\dis = \int Dx Dp_x    e^{ i\itif d\tau  (p_x  \dot{x} - 
{p_x^2\over 
2m} 
{dt\over d\tau } ) - (t-t(\tau )) \left. {p_x^2 
\over 
2m} \right|_{\tau_i}^{\tau_f} }
\ea
With gauge and boundary choices satisfying $\left.  t-t(\tau )
\right|_{\tau_i}^{\tau_f} = 
0$, 
we have
\ba \dis
\Gamma ' &\dis = \int Dx Dp_x    e^{ i\itif d\tau  (p_x  \dot{x} - 
{p_x^2\over 
2m} {dt\over d\tau } ) } \\[.5cm] 
&\dis = \int Dx Dp_x    e^{ i\int  p_x  dx - {p_x^2\over 2m} dt } 
\ea
so the system reduces to the unconstrained case. It can also formally 
look 
like 
the reduced phase space of section~\ref{sec:rpsc}---in the gauge 
$t=f(\tau)$:
\ba \dis
H_\alpha  &\dis = \alpha  {p_x^2\over 2m} = H_1 + {\partial 
W_\alpha \over 
\partial t},\\[.5cm]   \Gamma ' &\dis = \int Dx Dp_x  e^{i\itif d\tau  
(p_x  
\dot{x} - 
\alpha { p_x^2 \over 2m}) }
\ea
We thus see here the same phenomenon as before: changing gauge 
choices 
corresponds to canonical transformations in $RPS^*$, with the 
generator of 
the 
canonical transformation playing its basic role as a surface term in 
the 
action. 
This is no longer surprising: {\em all the gauge dependence of the 
action 
comes 
from a surface term, and we know that actions that differ by a 
surface term 
can be understood to differ by a canonical transformation}. 

Continuing with the measure $D\mu^*$, if on the other hand we don't 
have 
$\left. t-
t(\tau ) \right|_{\tau_i}^{\tau_f} = 0$, then we can just eliminate that 
surface 
term---as is done in reference \cite{Teitelpap}---substracting it off the full 
action:
\ba \dis
\Gamma ''  & \dis = \int DtDxDp_t Dp_x  \delta (\chi )\delta (\Phi )|\{ 
\chi 
,\Phi 
\} | e^{i\itif d\tau (p_x  \dot{x} + p_t \dot{t} ) + \left. (t-t(\tau )) {p_x 
2 \over 
2m} \right|_{\tau_i}^{\tau_f} }\\[.5cm]
& \dis = \int  Dx Dp_x    e^{i\itif d\tau  (p_x  \dot{x} - {p_x^2 \over 
2m}{dt\over d\tau } ) } \\[.5cm]
&\dis  = \int  Dx Dp_x    e^{i\itif p_x  dx - {p_x^2\over 2m }dt}
\ea
This is the point of view taken in reference~\cite{Teitelpap}. The 
reason for 
adding such a term should be clear now: to get rid of the gauge 
dependence 
of 
the action. Let us take a closer look at this action:
 \beq
A''  = \int d\tau  (p_x \dot{ x} + p_t\dot{t} + \lambda  \Phi ) + (t-
t(\tau )) \left.
{p_x^2 \over 2m} \right|_{\tau_i}^{\tau_f}
\eeq
 Recall that in the gauge invariant variables of 
section~\ref{sec:constant} we 
saw that \ba \dis 
& \dis \itif ( P\dot{Q} + p\dot{q} - w(\tau )P) d\tau  \\ \dis
= & \dis \itif ( p_x \dot{ x} + p_t\dot{t} - w(\tau )\Phi ) d\tau  - (t-
t_0 ) \left. {p_x 
^2\over 2m} \right|_{\tau_i}^{\tau_f} = A'- \left.  
B\right|_{\tau_i}^{\tau_f}
\ea
 where $B = (t-t_0 ) p_x^2/2m = QP^2 m/2$ is the generator of the 
canonical 
transformation.

Now, we can therefore {\em split the original action into a gauge 
invariant 
and 
a gauge dependent part}:\ba \dis 
\itif (p_x \dot{ x} + p_t\dot{t} - w(\tau )\Phi ) d\tau  = \\ \dis
\itif ( p_x \dot{ x} + p_t\dot{t} - w(\tau )\Phi ) d\tau  - (t-t_0 ) \left. 
{p_x^2\over 
2m} \right|_{\tau_i}^{\tau_f} + (t-t_0 )\left.  {p_x^2\over 2m}
\right|_{\tau_i}^{\tau_f} \ea
The last term is the one that will change when we change gauge 
fixings. We 
can 
substract it off, and add it already implementing the boundary 
conditions: 
we 
finally have the gauge-fixing-at-the-boundaries-independent action
\beq
\itif ( p_x \dot{ x} + p_t\dot{t} - w(\tau )\Phi ) d\tau  - (t-t_0 ) \left. 
{p_x^2\over 
2m} \right|_{\tau_i}^{\tau_f}
\eeq
This is the action of reference~\cite{Teitelpap}, and it is to be 
employed if 
one 
wants to gauge-fix at the boundaries, i.e., use the measure $D\mu^*$:
\beq
DtDxDp_t Dp_x   = dp_{t0} dp_{x0} \p { dt_i dp_{t i} dx_i dp_{x i} 
\over 
(2\pi)^2}
\eeq
with
\beq
 \delta (\chi ) \delta (\Phi ) = \delta_0(\chi ) \delta_0(\Phi ) \p  
\delta_i(\chi ) 
\delta_i(\Phi ).   \eeq
If one  uses the measure $D\mu*$, however, there is no need to 
modify the 
action, as the formalism already is  gauge-fixing invariant. The merit 
of 
proceeding as in reference~\cite{Teitelpap}  and using $D\mu^*$, is 
that it 
appears that one is closer to the reduced phase space approach---
after all, 
the 
unphysical degrees of freedom are taken care of by delta functions at 
all 
``$\tau$imes''. 

  It can be criticized that the surface term that one adds to take care 
of the 
gauge fixing dependence is not unique, as one could add a surface 
term of 
the 
form $\left.\gamma (\tau ) p_x^2/2m \right|_{\tau_i}^{\tau_f}$---i.e., 
subtract the 
term
$(t(\tau )+\left. \gamma (\tau ) -t) p_x^2/2m 
\right|_{\tau_i}^{\tau_f}$ instead---but one 
could also add such a term in the other approach. The action is 
determined 
classically only up to some surface terms,
 but what we are doing here 
is 
more than that, as it affects the dynamics, and the path integral as 
well.

  But are we closer to the reduced phase space philosophy? 
The bottom line is that there is no absolute reduced phase space to 
frame a 
quantization scheme if we insist on using gauge-dependent boundary 
conditions. But now we recognize the gauge-fixing dependence as 
being equivalent to  
  ``time''  dependent canonical transformations in the reference reduced 
phase 
space of our choice---and the existence of  different pictures
in quantum mechanics. If we are able to define quantum 
mechanics for the true observables of the system, i.e., constants of 
the 
motion, 
we are in business. But how does one do that?  For the 
cases at hand it seems that such a 
world 
 is pretty boring!

At any rate, we have seen that the one-branch Faddeev approach yields a path
integral that  is equivalent to the one in the reduced phase space
approach. We can obtain, for example, the propagator for the 
Schr\"{o}dinger equation for constraints of the form
$\Phi \approx p_t -H$, with $\hat{H}$ the hamiltonian operator
in some ordered form. Ordering is equivalent to a choice of 
$H(q,p)= \l q| \hat{H}|p\r$, as usual. 
Again, by Schr\"{O}dinger equation I mean any equation of the 
form $i\partial_t \psi = \hat{H} \psi$. This includes the non-relativistic
particle and the relativistic square-root forms, after picking a branch of
the quadratic constraint.

\subsection{The Faddeev path integral: multiple branches }
\label{sec:Faddeevbr}
In the previous section we discussed the Faddeev path integral in the 
simplest case. In fact it appears that only in such a case the path 
integral is 
unambiguoulsy defined. When there is more than one branch in the 
$RPS$ 
the simple recipe
\ba \dis 
\Gamma (q_i,qf,\Delta \tau ) 
  & \dis =\int DqDpDQDP \delta (f(Q,q,p,\tau )) \delta (P)|\{ f,P\} |  
e^{i\itif
d\tau (  P\dot{Q} + p\dot{q} - h(q,p))} \\[.5cm] \dis 
 & \dis  = \int DqDpDQDP \delta (\chi ) \delta (\Phi ) |\{ \chi ,\Phi \} |  
e^{i\itif  d\tau  (P\dot{Q} + p\dot{q} - h(q,p))}\\[.5cm] \dis 
   & \dis = \int DqDpDQDP D\lambda D\pi Dc D \bar{c}\:
e^{i\itif d\tau ( P\dot{Q} + p\dot{q} - h(q,p)+\lambda\Phi + 
\lambda\chi + c 
|\{ \chi ,\Phi \} | \bar{c} )}
\ea
cannot be mindlessly applied. 

There are several pitfalls in the mindless approach. We can compute 
the 
above path integral with or without an absolute value on the 
determinant.
Or we can not use a ``hermitean'' skeletonization for it.... We will
illustrate below.

Having two branches, in the simplest 
interpretation,  means having two distinct reduced phase 
spaces. A more general way to write the path integral follows from 
the study 
of quantum mechanics with complex topologies:
\ba \dis 
\Gamma_{top} (q_i,qf,\Delta \tau ) 
 & \dis  = A_+\int DqDpDQDP \delta (\chi_+ ) \delta (\Phi_+ ) |\{ 
\chi_+ 
,\Phi_+ \} |  
e^{i\itif  d\tau  (P\dot{Q} + p\dot{q} - h(q,p))}  \\[.5cm] &\dis \:\:
 + A_-\int DqDpDQDP \delta  (\chi_- ) \delta (\Phi_- ) |\{ \chi_- ,\Phi_- 
\} |  
e^{i\itif  d\tau  (P\dot{Q} + p\dot{q} - h(q,p))} 
   \ea
Included in this prescription is the possibility of using one branch 
only. The 
philosophy of this 
path integral is that the two worlds described by the branches don't 
``talk''
to each other. In fact, it is not hard to see that consistency
of the formalism will demand decoupling for the above 
prescription---the
composition law.

 What we 
have is essentially two quantum systems that don't interact. Another 
way of 
looking at this is to say  that the number of degrees of freedom 
doubles.

This is the simplest interpretation---two disconnected worlds.
Interesting things will happen when we try to connect these two 
worlds. Whether they are or are not connected depends on the 
inner product we bring into the theory, as we discussed 
earlier. As we will see next, this is
in turn reflected in the properties of the determinant
in the path integral. 

Unitarity will depend at the end on the inner product we
 choose---within
each branch and across branches---and on whether we connect the 
two
worlds consistently. We expect that this will put serious constraints 
on
the inner product between branches. The symmetries of the theory
may make things easier.

Let us look at the determinant. For the full form of the constraint the 
determinant 
is not a 
simple numerical operator. It becomes a full-fledged operator, and 
the usual 
issues of ordering come up. For example, for the free case this 
determinant is 
\beq
 |\{\chi, \Phi\}|   = 2  |E|
\eeq
where I have included the absolute value. With the skeletonization in 
the 
path integral---which corresponds to the operator ordering in the 
canonical 
formalism---this can mean different things; for example $E=E_i$, or 
$(E_i + 
E_{i-1})/2$ with or without absolute values. The guiding theme here 
has to be 
hermiticity, but 
with respect to what inner product? The problem is that in the 
physical 
space the operator $E$ is not hermitean, as we saw.

Let us see how things can go wrong. Let us take the mindless route
of just applying blindly the Faddeev prescription. Here are two
examples of strange results (which will cease to strange in a 
moment).

If we just use $E_i$ for the determinant, the path integral yields 
an oscillating result---depends on the number of bones in the 
skeletonization.

If we use $|E_i|$ the result is $$A=\int {dEdp_x\over 2\pi} 
\delta(E^2-
p_x^2 -m^2) e^{ip_x \;\Delta x}$$

As for a nice result, if we still use the mindless approach but use
the ``hermitean skeletonization''
with the absolute value,  det$=|E_i +E_{i+1}|/2$ we get
\cite{Ike}
$$A=
\int { dEdp_x \over 2\pi} \delta(E^2-p_x^2-m^2) e^{i E\;\Delta t 
-ip_x\Delta x}
= \Gamma_{top}$$ just as defined above. The reason is that this 
choice
of determinant decouples the branches---it corresponds to the
 Klein-Gordon
inner product, as we will discuss later.
Without the absolute value the branches still decouple, but
the ``negative'' branch part changes sign with each iteration in the
skeletonization.

We already provided  an interpretation for the existence of 
the two
branches.
The basic idea is that the two branches represent the two possible 
time
orientations of the paths: positive energy means positive orientation, 
negative energy means negative orientation---going back in time. In 
fact,
it is not hard to realize that the only way in which we can 
represent a particle going back in time in this theory
 is by associating to it
a negative hamiltonian.

Now, the above decomposition of the path integral, although a priori
consistent, lacks imagination. As mentioned, it corresponds to taking
the {\em direct sum} of two Hilbert spaces, 
with an inner product that decouples them. The most general way to 
write the
path integral is, schematically $$
\Gamma_{\uparrow\downarrow}
= a_\nearrow\int (\nearrow) +a_\searrow \int (\searrow) +
a_{\nearrow\searrow} \int(\nearrow\searrow) 
+a_{\searrow\nearrow}
\int(\searrow\nearrow) + ...
$$

Consistency (the composition law) will fix these coefficients.
We need to show that a  positive definite  inner product is 
in general not compatible with this path integral.
Let the composition law be given by$$
\Gamma(x_f, t_f;x_i,t_i) = \int dxdt\:\Gamma(x_f, t_f;x_a,t_a)
\;\hat{O}\;\Gamma(x_a,t_a;x_i,t_i)$$ Now, let us
expand---writing this 
in a very economic fashion$$= \int_a
\left(\int_{fa}(\nearrow)+\int_{fa}(\searrow)+\int_{fa}(\nearrow
\searrow) +
 \cdots  \right)
\hat{O}
\left(\int_{ai}(\nearrow)+\int_{ai}(\searrow)+\int_{ai}(\nearrow
\searrow) +
 \cdots  \right)
$$ $$=\int_a \left(
\int_{fa}(\nearrow)\hat{O}\int_{ai}(\searrow)+
\int_{fa}(\nearrow\searrow)\hat{O}\int_{ai}(\nearrow) +
\int_{fa}(\nearrow)\hat{O}\int_{ai}(\searrow\nearrow) + \right.$$
$$ \;\;\;\;\;\;\;\;\;\;\;\;\;\;\;\;\;\;\;\;\;\; \left.
\cdots + (\nearrow \mbox{exchange}\searrow)
 \right)$$
As was already discussed in the previous chapter, it isn't clear that
one can make sense of the Faddeev path integral
when there is no decoupling, as an operator constructive
interpretation is lacking.In
general,we have a consistent situation, as follows.

 In we restrict the theory to   one  branch, 
   forcing  decoupling, sometimes the theory will also be consistent with 
other 
requirements, like space-time covariance, and sometimes it won't---as we 
have discussed. 
The results in reference 
\cite{Ferraro} (see also reference \cite{Parker79}
 for the configuration space version of this path integral) are another 
example of this 
conflict between unitarity and {\em space-time covariance} when one tries to 
stay in  one 
branch (the gravitational background case). 
Staying in one branch means that we have a unitary
theory  (if we order 
the hamiltonian to be hermitean) and   an inner product with  yields 
positive 
definite norms for the states.  But we may have to 
give up some symmetries to do this. 
This is really an issue regarding  the ordering requirements that we 
start with.  
Decoupling   means
unitarity within the one branch sector---i.e., the one particle
sector.   

For example, we know that the Faddeev path integral
for the one-branch square-root relativistic particle  hamiltonian
solves the corresponding Schr\"{o}dinger equation, and hence,
when there is no electric field and the
space-time is flat (decoupling) it also solves the
Klein-Gordon equation (this is a general feature of the
decoupling case)---a covariant equation.

If we try to work with both  branches, in general we do not have  
decoupling. Then the
interpretation of the Faddeev path integral is lost. As far as I know, this path
integral is an object that properly belongs to reduced phase space 
quantization, or to
Dirac quantization when it is equivalent to reduced pahse space quantization. It could
be that one can show that there is a proper place for
this path integral in a more general setting, though.
From our understanding of the free case we can say that the
hermitized form of the determinant and its absolute value will be necessary.


We studied all these cases in the operator formalism already---we 
know how to 
derive these path integrals from reduced phase space quantization 
and from Dirac quantization as well.

\subsection{The Faddeev-Fock path integral} 
\label{sec:FFock}
We have already remarked that one can also obtain an amplitude
by using the Fock representation.  Recall 
equation~\ref{eq:Fock}:
\beq
 {\cal A}  = 
\l t_f, x_f,\pi\! =\! 0 |  e^{i\hat{\lambda}\hat{\Phi}} |t_i x_i,\pi\! =\! 
0\r =
\l t_f, x_f |     {1\over \hat{\Phi} + i\epsilon  }         |t_i x_i\r  
\eeq
This is easy to see if one uses the resolution of the
 identity---equation \ref{eq:FockId},
$$
 \hat{I} =  \int_{-i\varepsilon}^{i\infty} dQ^1 \int_{-\infty-
i\epsilon}^{ \infty-i\epsilon}dP_2 |Q^{1*} P_{2}^* \r\l  Q^1 P_2| 
$$
Here $Q_1 = \lambda$, and $P_2 = p_t$ are the proper 
identifications.

Let us look at the easier case first  $P_2=P \approx 0$.
Consider ($P_1 = \pi $, $Q^1 = \lambda $, $Q^2 = Q$)
$$
\l Q _f, \pi \! =\! 0 | \; e^{i \hat{\lambda  } \hat{P } } | Q_i ,\pi \! =\! 0 
\r  =  $$
$$
\int_{-i\varepsilon}^{i\infty} d\lambda \int_{-\infty-
i\epsilon}^{ \infty-i\epsilon}dP  \; 
\l Q_f, \pi \! =\! 0 | \; e^{i  \lambda   P  }|   \lambda, P\r \l   
\lambda^*, P^* |   Q_i ,\pi \! =\! 0 \r  = $$
$$
 \int_{-\infty-i\epsilon}^{ \infty-i\epsilon}dP  \; 
e^{ i Q_f P } {1 \over P +i\epsilon} e^{-i Q_i P }
$$

As mentioned, this amplitude is the causal amplitude---leads to the 
Feynman 
propagator for the full relativistic case, for example, and it will  lead 
us to the 
causal
amplitude, the Feynman amplitude.  An important question arises.
 Can we interpret this amplitude within the Fock quantization 
approach? We did 
use the Fock space representation, in which the multiplier appears 
with half the 
range and is imaginary. But where is the Fock quantization 
condition? Can we 
interpret it, for instance, as
$$
\l Q _f, \pi \! =\! 0 | \; \delta(\hat{a})  | Q_i ,\pi \! =\! 0 \r
$$ or 
$$
\l Q _f, \pi \! =\! 0 | vac\r\l vac | Q_i ,\pi \! =\! 0 \r 
$$ perhaps?

At the moment it seems that the most natural interpretation for this 
amplitude is: 
it is the amplitude that corresponds to the BRST amplitude (see 
equation 
\ref{eq:BFVDIRAC}) when the Fock representation is used:
$$ (\psi,\psi') \equiv \l    \psi^0_{\chi\! =\! 0 \! =\!  \pi \! =\!   
\eta_0 }| 
\; e^{ [\hat{K},\hat{\Omega} ]}\;
         |  \psi'^0_{\chi\! =\! 0 \! =\!  \pi \! =\!  \eta_0   }  \r 
$$
$$
\int d\eta_0 d\eta_1\:\; \l  \psi^0_{\chi\! =\! 0 \! =\!  \pi}    | 
\; e^{ [\hat{K},\hat{\Omega} ]}\;
 | \psi'^0 _{\chi\! =\! 0 \! =\!  \pi} \r 
\eta_0 \eta_1 =
$$ $$
\int d\eta_0 d\eta_1\:\; \l  \psi^0_{\chi\! =\! 0 \! =\!  \pi}    | 
\; e^{i\hat{\lambda}\hat{\Phi} + \hat{\rho}_0 \hat{\eta_1} }
 | \psi'^0 _{\chi\! =\! 0 \! =\!  \pi} \r 
\eta_0 \eta_1 =\l \psi^0_{\chi\! =\! 0} | {1\over \hat{\Phi}+i\epsilon} 
| \psi'^0 
_{\chi\! =\! 0} \r
$$
i.e.,{\em  it is the amplitude that appears in the BRST-Fock approach (recall
section~\ref{sec:BFock})}. It can also be said that 
this object follows from an action that carries a {\em trivial } representation
of the disconnected $Z_2$ part of the reparametrization group. Paths going
back are weighted the same as their forward going counterparts. The sign of
the lapse carries this information---it is the only one that can in these path integrals.

This amplitude should thus   be called the ``BRST-Fock amplitude" (and path integral).

 This path integral is otherwise very similar to the earlier ones.
The only difference is in the contours of integration---which
are those associated to the Fock space inner product.

\subsection{The BFV path integral: one branch }
\label{sec:BFVpi}
As explained at the end of section~\ref{sec:BFVq}, the path integral in the BRST 
extended
phase  space \cite{TeitelBook,BFV,Gomis} is given by
\beq
\Gamma ' = \int Dt Dp_t  Dx Dp_x  D\lambda  D\pi  D\eta_0  D\rho_0  
D\eta_1  
D\rho_1   e^{iS}
\eeq
where $S$ is given by 
\beq
S =  \itif (\dot{t}p_t + \dot{x}p_x  + \dot{\lambda }\pi  + \dot{\eta}_0 
\rho_0  + 
\dot{\eta}_1  \rho_1  - \{ {\cal O}, \Omega\} ) d\tau 
\eeq
where ${\cal O}$ is the gauge fixing function. 
The boundary conditions are that at $\tau  = \tau_i, \tau_f$
 \beq
\pi  = \rho_1  = \eta_0   = 0, 
\eeq
and as usual we fix $x$ and $t$ as well\footnote{In the literature the 
variables 
$C= \eta_0  ,i\bar{C} = \rho_1  ,  \bar{P} =\rho_0  , -iP = \eta_1 $  
  are often used. Notice that this is a canonical transformation.}.

The Fradkin-Vilkovisky theorem assures us that the path integral is 
independent of the choice of this function ${\cal O}$, although there 
are 
some 
caveats (see reference \cite{Govaerts} for example). Let us look at 
two different 
types of gauge 
fixing.

  {\em Non-canonical gauge fixing:} ${\cal O}_{NC} = \rho_1 f(\lambda 
) 
+\rho_0  
\lambda $ 

With this choice we have 
$$
\{ {\cal O}_{NC}, \Omega\}  = 
\{ \rho_1 f(\lambda ) +\rho_0  \lambda , \eta_0  \Phi  + \eta_1  \pi 
\}  = 
\rho_1 \eta_1  f'(\lambda )+ \pi f(\lambda )+ \lambda  \Phi  + 
\rho_0 
\eta_1 
$$
and the action $S$ is given by
\beq
S =  \itif d\tau (\dot{t}p_t + \dot{x}p_x  + \dot{\lambda }\pi  + 
\dot{\eta}_0 
\rho_0  + \dot{\eta}_1  \rho_1  - \rho_1 \eta_1  f'(\lambda)- \pi 
f(\lambda 
)- \lambda  \Phi  - \rho_0 \eta_1  ) 
\eeq
For simplicity let us look 
at the non-relativistic free case using the gauge $f(\lambda ) =  0 $ (but see 
ref.~\cite{Gomis}).

 { \em Result:} 
\beq
\Gamma '=( {m\over (2\pi i \Delta t}) )^{1\over 2} e^{ i (\Delta x)^2 
\over 2 
\Delta t }
\eeq
when the range of the lapse is the full one \cite{Teitelpap}---just as 
for 
the 
unconstrained particle of section~\ref{sec:unconstrained}. If the 
range is 
$(0,\infty )$ we get the Green's function for the Schr\"{o}dinger equation 
\cite{Gomis}. The last  $\lambda $  integration is indeed
%
\beq
    {\Delta \tau \over 2\pi } \int dp_{x0} d\lambda_0 \delta (\Delta t 
- 2m 
\Delta \tau  \lambda_0) e^{ i(p_{x0} \Delta x- \lambda_0 p_{x0}^2 
\Delta 
\tau )}
\eeq
Hence, with  the $(0,\infty)$ range for $\lambda$   we get the usual 
propagator 
times a Heaviside function, $\Theta(\Delta t)$, 
\beq
\Gamma'_G =  {\Theta(\Delta t)\over 4\pi m}\int dp_{x0} e^{ i(p_{x0} 
\Delta 
x- 
\Delta t {p_{x0}^2 \over 2m}) }= \Theta(\Delta t)  \Gamma '
\eeq
It isn't hard to see that this is a Green's function for the Schr\"{o}dinger 
operator---see section~\ref{sec:unconstrained}.
It is interesting to note that we can do all but the $x$, $p_x $ 
integrations 
to 
yield the unconstrained path integral once more,
\beq
\Gamma ' = \int Dx Dp_x    e^{ i\itif d\tau  (p_x  \dot{x} - {p_x^2 
\Delta 
t\over 
\Delta \tau  2m} )}
\eeq
or
\beq
\Gamma'_G = { \Theta (\Delta t)\over 2m} \int Dx Dp_x    e^{ i\itif 
d\tau  
(p_x  
\dot{x} - {p_x^2 \Delta t\over \Delta \tau  2m} )}
\eeq
if we have chosen the half range for the ``lapse''  $\lambda $.

In fact, this type of result is fairly general. Consider the 
general constraint case in the simple gauge above. 
$\Gamma_{f=0} ' =$ 
$$
  \int Dt Dp_t  Dx Dp_x  D\lambda  D\pi  D\eta_0  
D\rho_0  
D\eta_1  
D\rho_1  \; e^{i
\itif d\tau (\dot{t}p_t + \dot{x}p_x  + \dot{\lambda }\pi  + 
\dot{\eta}_0 
\rho_0  + \dot{\eta}_1  \rho_1  - \lambda  \Phi  - \rho_0 \eta_1  ) 
} = 
$$
$$ 
\int d\lambda_0 \;\Delta\tau 
\int Dt Dp_t  Dx Dp_x   \;   e^{i
\itif d\tau (\dot{t}p_t + \dot{x}p_x )  
- \lambda_0  \itif d\tau (\Phi  ) } $$
which, as I will discuss in the next section, is just
$$  \l x_\mu  |\;  \delta( \hat{\Phi}) | 
y_\mu  \r= 
\l x_\mu, \pi\!=\! 0 |\;  e^{i \hat{\lambda}  \hat{\Phi}} | 
y_\mu, \pi\!=\! 0 \r$$
  which are the Dirac (Fock) amplitudes---full range case (half-range) of
the previous chapter. Thus, for the one-branch case we have a solution to 
the (general, modulo ordering) Schr\"{o}dinger equation.\\
{\em Canonical gauge fixing:} ${\cal O}_C = \rho_1  \chi ' (t,x,p_x 
,p_t 
,\tau )  
+ 
\rho_0  \lambda $ \\
Here we have
\ba
{\cal H}_C &= \{  {\cal O}_C, \Omega\}   \\ 
&=\{  \rho_1  \chi '  + \rho_0  \lambda , \eta_0  \Phi  + \eta_1  \pi  
\}  = 
\rho_1 
\eta_0  \{ \chi ', \Phi \}  + \pi  \chi  ' + \lambda  \Phi  + \rho_0 
\eta_1 
\ea
{ \em Result:} \\
Let $\chi' \equiv \chi / \epsilon $. Taking the limit $\epsilon 
\rightarrow 0 $ the path integral reduces to the Faddeev path 
integral of 
section~\ref{sec:Faddeev}, as we expect by now, with our new 
understanding of 
the BRST operator formalism. If the limit of integration is taken at the 
end,
the  result is still the same.

Let us see how this works. With this choice of gauge fixing the action 
$S_C$ is
\beq
S_C = \int d\tau(\dot{t}p_t + \dot{x}p_x  + \dot{\lambda} \pi  + 
\dot{\eta}_0 
\rho_0  + \dot{\eta}_1  \rho_1  - \rho_1 \eta_0  \{ {\chi \over 
\epsilon}, 
\Phi 
\} - \pi  {\chi \over \epsilon} - \lambda  \Phi  - \rho_0 \eta_1 ) 
\eeq
Now consider the change of variables
\ba
\pi  &\longrightarrow & \pi  \epsilon \nonumber \\
\rho_1 &\longrightarrow &\rho_1  \epsilon 
\ea
Notice that the measure is unaffected by this change of variables, 
since 
the 
boundary conditions on the path integral mean that there are as 
many 
$\pi$   
as $\rho_1$   integrations: the jacobian for this change of variables is 
thus 
one.

Now let us take the limit $\epsilon \rightarrow 0 $:
\beq
S \longrightarrow\int d\tau (\dot{t}p_t + \dot{x}p_x  + \dot{\eta}_0 
\rho_0   
- 
\rho_1 \eta_0  \{ \chi ,\Phi \}  - \pi  \chi  - \lambda  \Phi  - \rho_0 
\eta_1 )
\eeq
After the ghost integrations we formally obtain the Faddeev path 
integral---for 
the full range of the lapse case---the only tricky points being that 
there is 
no 
chance that the jacobian $ \{ \chi , \Phi \} $ will get an absolute 
value, and 
that 
it isn't clear, of course, that one can interchange these limits in 
general---although things work here. At any rate, 
this 
gauge 
fixing is not really canonical as it isn't operating at the boundaries: 
we are 
really using the measure $D\mu $ as the boundary conditions imply 
that 
there 
is no gauge fixing at the boundaries. 

We can also  think about more general gauge fixings and the 
resulting 
path 
integrals in configuration space after the momenta are integrated. 
This 
will 
be 
discussed shortly.

\subsection{The BFV path integral: multiple branches}
\label{sec:BFVpirel}

The path integral is again given by 
\beq
\Gamma ' = \itif Dt Dp_t  Dx Dp_x  D\lambda  D\pi  D\eta_0  D\rho_0  
D\eta_1  
D\rho_1   e^{iS}
\eeq
where $S$ is given by 
\beq
S =  \itif (\dot{t}p_t + \dot{x}p_x  + \dot{\lambda }\pi  + \dot{\eta}_0 
\rho_0  + 
\dot{\eta}_1  \rho_1  - \{ {\cal O}, \Omega\} ) d\tau 
\eeq
where ${\cal O}$ is the gauge fixing function, just as in the previous 
section. In 
fact, since we did not specify right away the form of the constraint, a 
good 
number of the equations in the previous section hold here, but for 
convenience I 
will write them again.
The boundary conditions are that at $\tau  = \tau_i, \tau_f$
 \beq
\pi  = \rho_1  = \eta_0   = 0, 
\eeq
and as usual we fix $x$ and $t$ as well.\\
 {\em Non-canonical gauge fixing:} ${\cal O}_{NC} = \rho_1 f(\lambda 
) 
+\rho_0  
\lambda $ \\
With this choice we have 
$$
\{ {\cal O}-{NC}, \Omega\}  = 
\{ \rho_1 f(\lambda ) +\rho_0  \lambda , \eta_0  \Phi  + \eta_1  \pi 
\}  = 
\rho_1 \eta_1  f'(\lambda )+ \pi f(\lambda )+ \lambda  \Phi  + 
\rho_0 
\eta_1 
$$
and the action $S$ is given by
\beq
S =  \itif d\tau (\dot{t}p_t + \dot{x}p_x  + \dot{\lambda }\pi  + 
\dot{\eta}_0 
\rho_0  + \dot{\eta}_1  \rho_1  - \rho_1 \eta_1  f'(\lambda)- \pi 
f(\lambda 
)- \lambda  \Phi  - \rho_0 \eta_1  ) 
\eeq
For simplicity let us use the gauge $f(\lambda ) =  0 $ (but see 
ref.~\cite{Gomis}).

 { \em Result:} when the full range of the lapse is used,
 we get the   Hadamard  Green function,\beq
\Delta_1(x -y) = {1 \over (2\pi)^3}\int d^4 k \;\delta(k^2-m^2)\;
e^{ik(x-y)} =\eeq
$${1 \over (2\pi)^4}\int_{-\infty}^{\infty} d\lambda\int d^4 k \; 
e^{ik(x-y)+i \lambda (k^2-m^2)}$$
which is a solution to the Klein-Gordon equation.  If we use instead 
the half-range for the lapse we get
the   Feynman amplitude, $$
i\Delta_F (x-y) = {1 \over (2\pi)^4}\int d^4 k \; 
{ e^{-ik(x-y)} \over k^2-m^2+i\epsilon} =
  {-i \over (2\pi)^4}\int_O^\infty  d\lambda\int d^4 k \; 
  e^{-ik(x-y)      -\lambda    (k^2-m^2+i\epsilon)        }   $$\beq 
 \sim \l x_\mu | \; {1\over \hat{\Phi} +i\epsilon} \; |
y_\mu \r  \eeq
which  satisfies the Klein-Gordon compostion law. 
This corresponds to  an action that carries a  {\em trivial} representation of the
$Z_2$ part of the reparametrization group, as opposed to the previous one, which is faithful.
The half range case can be pictured by a lapse with an absolute value---providing
a trivial representation (the lapse changes sign, but the absolute value destroys this effect.)

It is not  possible to get all the other Green functions we discussed, 
because it is 
not possible 
to write some of those amplitudes in the form
\beq
\l x_\mu , \pi\!=\! 0 | \; e^{i \hat{\lambda}\hat{G}} |y_\mu , \pi\!=\! 
0\r 
\eeq for
some version of the constraint. We can get the one branch cases, of 
course.\\
{\em Canonical gauge fixing:} ${\cal O}_C = \rho_1  \chi ' (t,x,p_x ,p_t 
,\tau )  
+ 
\rho_0  \lambda $ \\
Here we have
\ba
{\cal H}_C &= \{  {\cal O}_C, \Omega\}   \\ 
&=\{  \rho_1  \chi '  + \rho_0  \lambda , \eta_0  \Phi  + \eta_1  \pi  
\}  = 
\rho_1 
\eta_0  \{ \chi ', \Phi \}  + \pi  \chi  ' + \lambda  \Phi  + \rho_0 
\eta_1 
\ea
{ \em Result:} \\
Let $\chi' \equiv \chi / \epsilon $. As before, taking the limit 
$\epsilon 
\rightarrow 0 $ the path integral formally reduces to the Faddeev 
path integral of 
section~\ref{sec:Faddeev}. However, notice that there isn't  an 
absolute value on 
the determinant. Recall that the Faddeev path integral did not 
converge in such a 
situation. Thus, once we have taken this limit, this path integral 
converges only 
with a half-range of the lagrange multiplier, and then it yields the 
Feynman 
propagator. 

What this is telling us is that the Fradkin-Vilkovisky theorem 
conditions do not 
allow for this ``degenerate'' gauge-fixing, if the full range of the 
lapse 
representation is desired. See reference \cite{Govaerts} for extensive 
discussions 
on this topic.  

Finally, let me point out that using the non-canonical gauge-fixing, 
and using the 
special case $f(\lambda)=0$, the action in the path integral is 
\beq
S =  \itif d\tau (\dot{t}p_t + \dot{x}p_x  + \dot{\lambda }\pi  + 
\dot{\eta}_0 
\rho_0  + \dot{\eta}_1  \rho_1  - \lambda  \Phi  - \rho_0 \eta_1  ) 
\eeq
and the path integral can be considerably simplified (\cite{Hal2}):
$\Gamma_{f=0} ' =$ 
$$
  \int Dt Dp_t  Dx Dp_x  D\lambda  D\pi  D\eta_0  
D\rho_0  
D\eta_1  
D\rho_1  \; e^{i
\itif d\tau (\dot{t}p_t + \dot{x}p_x  + \dot{\lambda }\pi  + 
\dot{\eta}_0 
\rho_0  + \dot{\eta}_1  \rho_1  - \lambda  \Phi  - \rho_0 \eta_1  ) 
} = 
$$
$$ 
\int d\lambda_0 \;\Delta\tau 
\int Dt Dp_t  Dx Dp_x   \;   e^{i
\itif d\tau (\dot{t}p_t + \dot{x}p_x )  
- \lambda_0  \itif d\tau \, \Phi    }= $$
\beq
 \Delta\tau 
\int Dt Dp_t  Dx Dp_x   \;  e^{i
\itif d\tau (\dot{t}p_t + \dot{x}p_x )  }\;
\delta( \itif d\tau  \, \Phi   )\eeq

What is this expression?
The answer is simply---for the general constraint case
$$
\Gamma_{f=0} ' =  \l x_\mu  |\;  \delta( \hat{\Phi}) | 
y_\mu  \r= 
\l x_\mu, \pi\!=\! 0 |\;  e^{i \hat{\lambda}  \hat{\Phi}} | 
y_\mu, \pi\!=\! 0 \r=
$$
$$
\Delta\tau \int_{-\infty}^{\infty} d\lambda \; \lim_{N\rightarrow 
\infty} 
\l x_\mu  |\;  \left( 1+ i{ \Delta\tau \over N} \hat{\Phi} \right)^N    | 
y_\mu \r = $$
$$
\Delta\tau \int_{-\infty}^{\infty} d\lambda \int
dp_0^4 \left( \prod_{i=1}^N  dx^4_i dp^4_i \right)
\; \lim_{N\rightarrow \infty} 
\l x_\mu  |\;  \left( 1+ i{ \epsilon_1} \hat{\Phi} \right)  
| p_{\mu 0} \r \l p_{\mu 0} |\cdot
$$
$$\:\:\:\:\:\:\:\: | x_{\mu 1} \r 
\l x_{\mu 1} | ...
\left( 1+ i{ \epsilon_N} \hat{\Phi} \right) | y_\mu \r =
$$
\beq \label{eq:teitel}
\int d\lambda  \;\Delta\tau 
\int Dt Dp_t  Dx Dp_x \;    e^{i
\itif d\tau (\dot{t}p_t + \dot{x}p_x )  
- \lambda   \itif d\tau \,\Phi    } \eeq
Indeed, we recover our old Dirac amplitude, which of course satisfies 
the 
constraints for the full range of the lapse (see \cite{Hal2,Hal3}), since
\beq
\l x_\mu  | \; \hat{\Phi}\;  \delta(\hat{\Phi}) | y_\mu \r =0 \eeq

For the half-ranged case, the BRST-Fock representation, we
get the amplitude
\beq
\Gamma_F = \l x_\mu | {1\over \hat{\Phi} + i \epsilon} | y_\mu\r
\eeq

\subsection{Constraint rescalings and canonical transformations in 
BRST extended 
phase space}
\label{sec:rescalings}
This section clarifies the issue of constraint rescalings, interpreted 
within the BRST 
formalism. One would
expect  by now that 
rescaling the constraint 
\beq
	      \Phi  \longrightarrow \Omega^2 \Phi  \equiv   \Phi '
\eeq
may  affect the path integral, as we saw a boundary effect in the 
Faddeev path integral, even though  the constraint surface is
unaffected. We already interpreted this effect as 
being equivalent to performing a transformation in the state space.
Let us see what happens in BRST.

 Indeed, consider again
the simple situation in section~\ref{sec:Faddeev}. Since  the 
gauge degrees of freedom are added inserting an identity like 
$$
1= \int DQDP \; \delta (Q-Q(q,p,\tau ))\delta (P)e^{i\itif d\tau  P 
\dot{Q}}
$$
where the measure is defined by
$$
DQ DP \delta (Q-Q(q,p,\tau )) \delta (P) = dP_0 \delta (P_0) \p  dQ_i 
dP_i 
\delta 
(Q_i-Q(q,p,\tau )) \delta (P_i)
$$          
in the physical reduced phase space path integral, using a rescaled 
constraint 
would have not affected things: we would have used instead
$$
1= \int DQDP \delta (Q-Q(q,p,\tau ))\delta ( \Omega^2 P) \Omega^2 e^{ 
i\itif 
d\tau  P \dot{Q}} 
$$
with the measure
$$
DQ DP \delta (Q-Q(q,p,\tau )) \delta (P) \Omega^2 = $$ $$
dP_0 \delta ( \Omega^2 P_0) \Omega^2 \p  dQ_i dP_i \delta (Q_i-
Q(q,p,\tau )) 
\delta ( \Omega^2 P_i) \Omega^2
$$
In fact, the safest thing to do seems to be to use the measure
\ba \dis 
D\mu^* & \dis \equiv DqDpDQDP \delta (\chi ) \delta (\Phi ) |\{ \chi 
,\Phi \} | 
\\[.5cm] \dis 
 &  \dis \equiv \left( {dp_0 dP_0\over 2\pi  }\p  dQ_i dP_i {dq_i 
dp_i\over 
2\pi  
}\right) \cdot \\[.5cm] & \dis \:\:\:\: \:
\left( \delta (P_0)  \p 
\delta (Q_i-Q(q,p,\tau )) \delta (P_i) \right)  
    dQ_0\;\delta (Q_0-Q(q,p,\tau ))
\ea
as it is impervious to constraint or gauge fixing rescalings. However, 
as 
remarked in section~\ref{sec:BFVpi}, the measure that appears in the 
BFV 
formalism---with the boundary conditions we used there!---is not 
$D\mu^*$ 
but $D\mu$. So the path integral there is seen to depend on 
$\Omega$. In 
fact, 
it isn't hard to see that under the rescaling $\Phi  \rightarrow 
\Omega^2 
\Phi   
\equiv  \Phi' $
the path integral changes by the rule 
\beq
\Gamma'_\Phi  \longrightarrow \Omega^{-2} \Gamma'_\Phi= 
\Gamma'_{\Omega^2 \Phi } 
\eeq
There is, as we mentioned,  an interpretation for this effect within 
BFV, however. 
If in the 
path 
integral one performs the canonical transformation (notice that the 
Poisson 
bracket is unaffected)
$$
\lambda  \longrightarrow  \lambda' \equiv \Omega^{-2} \lambda ,
$$ 
and 
$$ 
\pi      \longrightarrow \pi '    \equiv \Omega^2 \pi 
$$
the action---as we will now show---doesn't change, except  that the 
constraint and 
the 
gauge 
fixing function are rescaled. Here $\Omega$ has to be a constant, 
otherwise other  
brackets will be affected.

 Consider the more general gauge-fixing
 \beq
 {\cal O}_{NC} = \rho_1 \chi(\lambda, x^\mu ) 
+\rho_0  
\lambda \eeq 
With this choice we have 
$$
\{ {\cal O}-{NC}, \Omega\}  = 
\{ \rho_1 \chi(\lambda, x^\mu ) +\rho_0  \lambda , \eta_0  \Phi  + 
\eta_1  \pi \}  
=   $$  \beq
\rho_1 \eta_1  \{\pi, \chi(\lambda, x^\mu )\} + 
\rho_1 \eta_0 \{ \Phi ,\chi(\lambda, x^\mu ) \} + \pi \chi(\lambda, 
x^\mu ) + 
\lambda  \Phi  + \rho_0 
\eta_1 
\eeq
and the action  is given by
\begin{eqnarray}
S &= & \itif d\tau \left[ \dot{t}p_t + \dot{x}p_x  + \dot{\lambda }\pi  
+ 
\dot{\eta}_0 
\rho_0  + \dot{\eta}_1  \rho_1  - \rho_1 \eta_1  \{\pi, \chi(\lambda, 
x^\mu )\} -  
\right. \nonumber \\  & &
  \:\:\:\:  \left. \rho_1 \eta_0 \{ \Phi ,\chi(\lambda, x^\mu ) \} -\pi 
\chi(\lambda, x^\mu ) - \lambda  \Phi  - \rho_0 \eta_1  \right]    
\end{eqnarray}
  Now, let us perform the above transformation. The action becomes
 \begin{eqnarray}  
S' &= & 
 \itif d\tau \left[ \dot{t}p_t + \dot{x}p_x  + \dot{\lambda }\pi  + 
\dot{\eta}_0 
\rho_0  + \dot{\eta}_1  \rho_1  -
 \rho_1 \eta_1  \{\pi, \chi(\Omega^{-2}\lambda, x^\mu )\} \Omega^2 
- \right.  
\nonumber \\  & &   \left.
 \:\:\:\rho_1 \eta_0 \{ \Phi ,\chi(\Omega^{-2}\lambda, x^\mu ) \}   - 
\Omega^{2}  
\pi 
\chi(\Omega^{-2}\lambda, x^\mu ) - \Omega^{-2}\lambda  \Phi  - 
\rho_0 \eta_1  
\right]   
\end{eqnarray}

Now let us define a new constraint $   \Phi' = \Omega^{-2} \Phi $ and 
a new 
gauge-fixing function $
\chi' =\Omega^2 \chi(\Omega^{-2}\lambda, x^\mu ) $. We see that the 
have our old 
action with
these new constraint and gauge-fixing:
\begin{eqnarray}  
 S' & =&   
 \itif d\tau \left[ \dot{t}p_t + \dot{x}p_x  + \dot{\lambda }\pi  + 
\dot{\eta}_0 
\rho_0  + \dot{\eta}_1  \rho_1  -  \rho_1 \eta_1  \{\pi,
\chi'\} 
 - 
\right.  \nonumber \\  & &  \:\:\:\:  \left.
\rho_1 \eta_0 \{ \Phi' ,\chi ' \}   -    \pi 
\chi' -  \lambda  \Phi'  - \rho_0 \eta_1  \right] 
\end{eqnarray} 
Thus, the result of this active canonical transformation on the action 
is to rescale 
the 
constraint and the  gauge-fixing function.
 Now consider the path integral, and recall that we have boundary 
conditions on 
$\pi$.
Let us now perform the  above canonical transformation on the 
action.
We can now\\
a) obtain an equivalent path integral with a rescaled constraint and 
gauge-fixing, 
or\\
b) absorb the transformation on the measure, up to a boundary term, 
because 
there is one more
$\lambda$ integration than $\pi$.\\
Since the Fradkin-Vilkovisky theorem 
\cite{TeitelBook,BFV} says that the action is invariant under gauge-
fixing 
changes (and this is obvious in our path integrals), {\em all that the 
canonical 
transformation does in {\em a)}  is to rescale the constraint}. Notice 
that this is 
essentially 
due to the boundary conditions on $\pi$. If we didn't have such 
boundary 
condition the measure would be $D\mu^*$, and there wouldn't be 
such an 
effect.

{\em  Thus, a canonical transformation in the action is equivalent, via 
the 
Fradkin-Vilkovisky (FV) theorem, 
to a rescaling of the constraint and of the amplitude.}

This transformation is not unitary in the physical space, but as I explained earlier (Chapter 1),
this effect can be understood as a simple change of normalization of the basis kets (e.g., the
Newton-Wigner states versus the covariant ones for the relativistic case), or is a manifestation
of the fact that one must choose the inner product and observables in the theory by using some
external input---like physics! 

To summarize: 
we started with an action that is not gauge invariant at the 
boundaries, so 
if 
we 
insist on gauge-fixing invariance (FV theorem) we cannot gauge-fix 
at 
the 
boundaries, and as a result we have this rescaling of the constraint 
effect.
  
If we don't insist on gauge-fixing invariance, but demand 
``covariance'' 
only---understanding the gauge-fixing dependence as an effect 
of
canonical  transformations in the $RPS^*$---then the measure one 
uses is
$D\mu^*$,  and  there won't be such a rescaling effect (however, the 
ambiguities
in the quantized reduced phase space inner product are always going 
to be there.)

\subsection{Skeletonizations and curved space-time}
\label{sec:curvedPI}

In this subsection I want to quickly review some of the work that 
has already been
done on the issue of how to skeletonize path integrals in the case 
where there
is a curved background \cite{Hal4,Kuch83,Ferraro,Parker79,Parker81}.

 Consider first the case of a path integral  in configuration space.
Without the guidance of a canonical (operator) scheme one needs 
some other principles to skeletonize the action. In this case this 
requirement is space-time covariance. 

Two ingredients
enter the path integral: the skeletonized action and the skeletonized 
measure.  For the first it is natural to use the action of the classical 
path that connects the endpoints of the skeletonization.  This is the
Hamilton-Jacobi function, and it is a scalar in both its arguments, the 
endpoints---provided we started with an invariant action, of course.

Of the measure one requires space-time covariance. Unfortunately, as we will 
now see, this does not fix
the measure uniquely.

 Parker \cite{Parker79} provides
us with a beautiful connection between this path integral and the 
wave equation, which I will summarize here.

 Consider the path integral
\beq
\l x,s|x',0\r = \int d[x(s')]\, [\Delta^p] \;\exp\left( 
{i\over \hbar} \int_0^s d\tau  \; \{ 
{1\over2} m g_{\alpha\beta} \dot{x}^\alpha \dot{x}^\beta -{\hbar^2 
\over 2m} [\xi +{1\over 3}(p-1)] R(x)  
\} \right)
\eeq
Parker shows that this amplitude---which I will define more 
precisely in a moment---satisfies the equation 
\beq
i\hbar {\partial \over \partial s } \, \l x,s|x',0\r = [ -{\hbar^2 \over 2 
m} g^{\alpha\beta}(x) \nabla_\alpha \nabla_\beta + {\hbar^2 \over 2 
m}\xi R(x) ]\; \l x,s|x',0\r 
\eeq
and moreover 
\beq
\lim_{s\rightarrow 0} \; \l x,s|x',0\r = [g(x)]^{-{1/ 2}}  \delta(x-x')
\eeq

The measure in the path integral is given by 
\beq
\left(  {m \over 2\pi i \epsilon} \right)^{ n/2} \prod_{j=1}^N  d^n x_j  
\sqrt{g(x_j) } \, [\Delta(x_{j+1},x_j) ]^p \left(  {m \over 2\pi i \epsilon} 
\right)^{ n/2}
\eeq
and where
\beq
 \Delta(x_{j+1},x_j) = [g(x_{j+1})]^{-1/2} det[-
{\partial^2\sigma(x_{j+1},x_j)\over \partial x_{j+1} \partial x_j}] \;
[g(x_{j })]^{ -1/2}
\eeq
is easily seen to be scalar, since $\sigma(x_{j+1},x_j) $ is defined to 
be the proper length---a scalar.

Parker uses a geodesic to compute the skeletonized action. 

The main point is that there is an ambiguity in the measure that is
associated with the ambiguity in the transition from classical to 
quantum---ordering ambiguities. 
In this case this is reflected by the appearance (or modification)
of the curvature term in the wave-equation. 

Similarly, there is an ambiguity in the phase-space constructions. 
The 
measure in such a case is  uniquely defined, but there is is again a 
one
parameter family of skeletonizations  of the action \cite{Hal4,Kuch83}.

This ambiguity is troublesome, and there have been attempts to fix 
it.
For example, Halliwell \cite{Hal2,Hal4}, in the context of 
minisuperspace, has proposed that the conformal value of the 
curvature parameter is the right one because it ensures
invariance under constraint rescalings. Indeed, consider rescaling the 
constraint---classically---and then quantizing it. How do we order it? 
We can  interpret the new constraint as having a new metric and a 
new potential and 
order  covariantly with respect to this new metric, as Halliwell 
assumes we will do.  Then, he shows, the conformal value for the 
curvature term insures that this paradigm is invariant under such 
rescalings.
This argument assumes that one will take the  factor in the quadratic 
term of the constraint and  call it the supermetric.

The choice of the conformal value  in the Klein-Gordon equation 
is interesting because it implies that if the metric is rescaled and the 
equation is rewritten  covariantly  with respect to the new metric, 
the
new solution is related in a simple way to the old one---by a 
rescaling. There is nothing in the theory so far that demands the
use of conformally covariant amplitudes, though.  Perhaps conformal transformations are
ultimately tied to the concept of time and the probabilistic interpretation.

  
\newpage\section{Path integrals in configuration space (PICSs)}
\label{sec:PICS}
In this section we wil first review some of the path integral 
approaches in 
configuration space. Then  we will compare them with the ones that we 
will obtain 
from integrating out the momenta in the phase space path integrals.
\subsection{Review of path integral formalisms in configuration 
space}

\subsubsection{Fadeev-Popov} 
Consider the path integral in configuration space
\beq
A= \int Dz \; e^{i S[z (\tau)]}
\eeq
where we assume that both the action and the measure are invariant 
under 
gauge-transformations. What does this statement mean for the 
measure? It 
means that for an arbitrary functional
$F$ we have\footnote{This will hold if it holds in the skeletonized 
sense, of 
course.}
\beq
\int Dz\; F[ z^{\cal G} ] = \int Dz\; F[ z ] \eeq or
\beq
D(z^{{\cal G}^{-1}}) = Dz\eeq
where $z^{\cal G}$ is the result of operating on the coordinates with 
the group. 
This is also called the Haar masure. 

Now consider the identity \beq
1=\int D\omega \; \delta(F[z^\omega (\tau)]) \Delta_{FP} \eeq
where  \beq
\Delta_{FP} = |         {  \delta F[z^\omega (\tau)]  \over \delta \omega 
}    | 
\eeq
Notice that for this to be an identity we have to be careful that 
$D\omega \sim 
\delta\omega$. 
Both are invariant measures expressed in terms of whatever 
parameter we use to 
describe the group. This equation means that
$\delta(F[z^\omega (\tau)]) \Delta_{FP} $ is invariant, because 
$D\omega$ and $1$ 
are. It also means that $\left.
\Delta_{FP} \right|_{F=0} = \left. \Delta_{FP}\right|_{\omega_0}   $ is 
also 
invariant, since we have  \beq 1=\left.
\Delta_{FP}\right|_{\omega_0} \int D\omega \; \delta(F[z^\omega 
(\tau)])   \eeq
because any change in $\omega_0$ can be compensated by the 
measure.

We now insert this identity in the above path integral and change 
the orders of 
integration,
\beq
A= \int D\omega \int Dz \; \delta(F[z^\omega (\tau)]) \Delta_{FP}\;  
e^{i S[z (\tau)]}
\eeq
It is not too hard to see now that \beq {\cal A} =
\int Dz \; \delta(F[z^\omega (\tau)]) \Delta_{FP}\;  e^{i S[z (\tau)]}
\eeq
does not depend on $\omega$. This is because $Dz$, 
$\delta(F[z^\omega (\tau)]) 
\Delta_{FP}$, and $S[z (\tau)]$ are all invariant. One can further 
rewrite this  as 
\beq
{\cal A} =
\int DzD\pi DcD\bar{c}  \; \delta(F[z^\omega (\tau)]) \;  
e^{i S[z (\tau)] + 
i \int d\tau \; (\pi F[z^\omega (\tau)]  +c \,\Delta_{FP} \, \bar{c} )}
\eeq
where---have to be careful about $d\tau$ factors
\beq 
D\pi = \prod d\pi_i d\tau_i,\: \;DcD\bar{c} = \prod dc_i 
d\bar{c}_id\tau_i 
\eeq
This is the Faddeev-Popov path integral. 

For the situation at hand, the only tricky point is to find the correct 
group 
measure, to compute the determinant correctly. 
Let the reparametrizations be generated by $\omega(\tau)$:
\beq
z(\tau) \longrightarrow z(\tau + \omega(\tau)  )\approx 
 z(\tau)  + \omega\; { d \over d\tau } z(\tau) 
\eeq
Under reparametrization changes, the quantity ${ d \over d\tau } z$ 
acts as a 
vector (of the form ``$v_\mu $''---a covariant vector).  
Since $z$ is a scalar, it follows that $\omega$  
behaves as a vector 
of the other kind (``$v^\mu$''---contravariant---with the index fixed here, $\mu=1$ only).  Now we need to know what the 
proper measure 
for these objects is.
Consider\footnote{See references \cite{Fujikawa} for more on this 
approach to 
defining path integrals.} the quantity \beq
\int \left( \prod_a du_a \right) e^{i \sum_a u_a u_a } =const
\eeq 
which is a number---no matter how $u_a$ tranform. 
Let $u_a = v^\mu_a \lambda _a^{3/2} d\tau_a^{1/2}$. Then we have
\beq
\int \left( \prod_a v^\mu_a \lambda _a^{3/2} d\tau_a^{1/2} \right) 
e^{i \sum_a 
v^\mu_a \lambda _a^2 v^\mu_a\; \lambda _a d\tau_a} =const
\eeq 
Now, the integrand is a scalar ($\int \lambda d\tau \; v^2 \lambda 
^2$),  the 
whole integral is also a scalar, so the measure must be as well. Thus, 
a proper 
measure for $\omega$ is
\beq D\omega =
\prod_a d(\omega_a \lambda _a^{3/2} d\tau_a^{1/2} )
\eeq
We also will need 
\beq
D\lambda = \prod d(\lambda_i^{-1/2} d\tau_i ^{1/2} \lambda_i)
\eeq
which we can obtain from \beq
 const= \int 
\prod d(\lambda_i^{-1/2} d\tau_i ^{1/2} \lambda_i) e^{i\int 
(\lambda_i^{-1/2} 
d\tau_i ^{1/2} \lambda_i)^2} =
\int 
\prod d(\lambda_i^{-1/2} d\tau_i ^{1/2} \lambda_i) e^{i\int \lambda 
\; d\tau} 
\eeq
For a scalar $z$, we can form the invariant
$\int d\tau \; \lambda z^2 $, from which we infer the measure $d(z  
\lambda^{1/2} \, d\tau ^{1/2} )$.

 For the free relativistic particle, for example, we can now form the 
path integral 
\beq
{\cal A} = \int Dx^\mu D\lambda D\pi DcD\bar{c} \; e^{i\itif d\tau \{
( {1\over 2} { x\cdot x \over \lambda} + \lambda m^2 ) + \pi 
(\dot{\lambda} 
-\chi) + c \,\Delta_{FP} \, \bar{c}  \} }
\eeq
where the gauge-fixing is $  \dot{\lambda} -\chi(\lambda, x^\mu)  
=0$, and  
\beq
\Delta_{FP} = {\delta_{\omega}  \over \delta (\omega_a \lambda 
_a^{3/2} 
d\tau_a^{1/2} ) } (\dot{\lambda} -\chi) \eeq
now, to calculate this let us review the following transformation laws: 
under a 
change of parametrization generated by $\omega(\tau)$ we have
\beq x(\tau) \longrightarrow x(\tau+ \omega) \approx  x(\tau) 
+\dot{x} \omega 
\eeq 
$$
\lambda(\tau) \longrightarrow \lambda(\tau+ \omega) (1+ 
\dot{\omega}) 
\approx  \lambda(\tau) +{d\over d\tau} (\lambda \omega) 
$$
$$
\dot{\lambda} \longrightarrow \lambda +{d^2\over d\tau^2} 
(\lambda \omega) 
$$
Thus \beq
\delta_\omega (\dot{\lambda} -\chi) 
= \delta \dot{\lambda} - {\partial \chi \over \partial \lambda} 
\delta \lambda - 
{\partial \chi \over \partial x^\mu} \delta x^\mu 
=\eeq
$$
{d^2\over d\tau^2} (\lambda \omega) 
    - 
{\partial \chi \over \partial \lambda}   {d\over d\tau} (\lambda 
\omega)  -
{\partial \chi \over \partial x^\mu}  \dot{x}^\mu \omega $$
Hence
\beq
\Delta_{FP} = {\delta_{\omega}  \over \delta (\omega_a \lambda 
_a^{3/2} 
d\tau_a^{1/2} ) } (\dot{\lambda} -\chi)  = \eeq
$$
{d^2\over d\tau^2}  
 {\circ \over \lambda ^{1/2} d\tau^{1/2} }  - 
{\partial \chi \over \partial \lambda}  {\circ \over \lambda ^{1/2} 
d\tau^{1/2} } -
{\partial \chi \over \partial x^\mu}  { \circ \over \lambda ^{3/2} 
d\tau^{1/2} }
$$
 Finally  ($\epsilon_i = d\tau_i$) 
\beq
{\cal A} = \int \epsilon_0 d\lambda_0 \prod_i^{N-1} \left(
(\lambda_i \epsilon_i )^{d/2} d\lambda_i \epsilon_i d^d x_i d\pi_i 
dc_i d\bar{c}_i
\right) 
 e^{i\itif d\tau \, \{
( {1\over 2} { x\cdot x \over \lambda} + \lambda m^2 ) + \pi 
(\dot{\lambda} 
-\chi) + c \,\Delta_{FP} \, \bar{c}  \} }
\eeq
This is essentially the BFV path integral, again up to a normalization 
constant.

 {\em Notice that we can extract factors of $\lambda_i \epsilon_i$, as we 
defined the gauge transformations and 
  the skeletonizations, $\epsilon_i$,  so that these factors are constant.}
 Hence, up to this gauge-invariant constant the path integral reduces to 
\beq
{\cal A} = \int \epsilon_0 d\lambda_0 \prod_i^{N-1} \left(
  d\lambda_i \epsilon_i d^d x_i d\pi_i 
dc_i d\bar{c}_i
\right) 
 e^{i\itif d\tau \, \{
( {1\over 2} { x\cdot x \over \lambda} + \lambda m^2 ) + \pi 
(\dot{\lambda} 
-\chi) + c \,\Delta_{FP} \, \bar{c}  \} }
\eeq
This is the path integral that will reappear below.

\subsubsection{The geometric path integral}
\label{sec:geo}
I will now review  the geometric path integral construction 
\cite{Geo,Fujikawa} and show that it 
is 
equivalent to the Faddeev-Popov one, up to a numerical, gauge 
invariant 
measure factor.

We will discuss the relativistic free particle. The amplitude
we want to compute is 
\beq
A(x^\mu_i, x^\mu_f) = \int [d\lambda][dx^\mu]\, \exp(-S[\lambda, x])
\eeq 
where the action is that of the free relativistic particle witha lagrange
multiplier,
\beq
S = {1\over 2} \itif d\tau \, [\lambda^{-1} \dot{x}^2 + \lambda m^2]
\eeq
in imaginary time (here I follow closely the discussion in Cohen et al, 
in reference \cite{Geo}).

The path integral is based on a 
(super)geometric definition of the measure. The basic
idea in this construction is illustrated by the following example, in which
the measure for the coordinate functions for the particle is defined.
 
The 
measure 
$Dx$, 
is  defined by first constructing an invariant inner product for 
$\delta x(\tau ) = \delta x_\tau$, which is a vector in the tangent 
space to
the  (super)manifold of all functions $x(\tau )$:
\beq
<\delta z(\tau ), \delta y(\tau )>_P \equiv \itif d\tau\,   \delta z(\tau ) 
\, \delta 
y(\tau )\,  \lambda (\tau )  
\eeq
$P$ denotes a point in the (super)manifold, $P = (x(\tau ),t(\tau ), 
\lambda 
(\tau ))$. It isn't hard to see that this inner product is gauge 
invariant\footnote{Under a transformation $\tau  \rightarrow f(\tau 
)$, 
we 
have that 
$\delta z(\tau )\rightarrow \delta z(f(\tau ))$,   the same 
transformation 
law 
as 
for $z(\tau )$, and $\lambda $ transforms as 
$\lambda (\tau ) \rightarrow \lambda (f(\tau )) df/d\tau $.}. Then  
the 
path
integral measure in the  tangent 
space is defined by
\beq
\label{eq:gmeasure}
1 \equiv \int D(\delta x(\tau ))\;  e ^{i <\delta x(\tau ), \delta x(\tau 
)>_P}
\eeq
which forces the measure to be a gauge-invariant object.
Notice that with a skeletonization $\tau_i$ this just means
\beq \label{eq:geomeo}
 D(\delta x(\tau )) = \p  d(\delta x(\tau ))_i \left({d\tau_i \lambda 
(\tau_i)\over 
\pi }\right)^{d/2}
\eeq
where $d$ is the number of space-time dimensions.
From a  mathematical point 
 of view (again, see \cite{Geo})  one is defining an inner product (metric)
in the tangent space to a (super)manifold. Using this inner product one
 then defines a volume form in the tangent space (a measure). This
in turn is also used for the definition of a volume form in the manifold
itself. If the measure in the tangent space is of the form
$f(q) da^1\wedge  \cdots\wedge  da^n$, where $q$ denotes a point in the
 manifold, then 
the form $ f(q) dq^1  \wedge \cdots\wedge dq^n$ is a well-behaved
 volume form for 
the manifold. 

Similarly, the invariant measure for the lagrange multiplier
is defined by the following inner product:
\beq
\l \delta\lambda_1(\tau), \delta\lambda_2(\tau)\r_P =
\itif d\tau \, {\delta\lambda_1 (\tau)\, \delta\lambda_2 (\tau) \over
 \lambda(\tau)}
\eeq
and then by demanding
\beq \label{eq:mlapse}
1= \int [d(\delta\lambda)]\; \exp(-{1\over 2}
 \parallel \delta \lambda \parallel^2)
\eeq

It is not too hard to see that $\lambda$ can be alternatively described
 by a pure gauge part and a pure non-gauge part, $\lambda\sim (c, f(\tau))$, 
where $c= \itif d\tau\, \lambda$ is an invariant. Indeed, any 
$\lambda$ can be written in the form $\lambda= c \dot{f}$, where
$f(\tau)$ satisfies $f(\tau_i)= \tau_i$ and $f(\tau_f)=\tau_f$ (or viceversa).

An important  point in this approach \cite{Geo} is that one can show that 
the 
measure for $\lambda$ can then be split into a gauge dependent and a gauge 
invariant 
part---with a gauge invariant jacobian.

Indeed, one finds that
$$ 
\parallel \delta \lambda \parallel^2= {\delta c^2 \over c} - \itif
d\tau\sqrt{g}\,  g \, \xi \Delta \xi  =
$$ \beq \label{eq:ortho}
{\delta c^2 \over c} - \itif
d\tau\sqrt{g}\,  g \, \xi \left(  c^{-2} \lambda^{-1} {d^2\over d\tau^2} \lambda \right) \xi 
\eeq
where $g=\lambda^2$ is the metric on the line, 
\beq
 \Delta = g^{-1} {d\over d\tau} {1\over \sqrt{g}} {d\over d\tau} \sqrt{g}
\eeq
is the laplacian associated with this metric, and the corresponding covariant
derivative---which on vectors is $(\xi^i)_{;j} = {d \xi\over d\tau} + 
{1\over 2} g^{-1} {dg\over d\tau}$,  for example, 
and the quantity $\xi
= \delta f\circ f^{-1}$ is the vector field associated with the diffeomorphism
$\tau \rightarrow f(\tau)$.

The  norm (or, with the obvious generalization,inner product) for $\xi$ is now defined to be
\beq
\label{eq:xi} \parallel \xi\parallel ^2 =\itif d\tau\, \sqrt{g} \; \xi \, g \, \xi
\eeq 
We can now rewrite 
equation \ref{eq:mlapse} by using equation \ref{eq:ortho} and by writing 
\beq
[d(\delta\lambda)] = J(\lambda)\, d(\delta c)\,  [d(\delta\xi)]
\eeq
i.e., $1=\int [d(\delta\lambda)]\; \exp(-{1\over 2}
 \parallel \delta \lambda \parallel^2 )= $
$$
\int J(\lambda)\, d(\delta c)\,  [d(\delta\xi)]\; \exp\left(-{1\over 2}
{\delta c^2 \over c} + {1\over 2} \itif
d\tau\sqrt{g}\,  g \xi \Delta \xi \right) =
$$ 
$$J(\lambda)
\int d(\delta c)\,  [d(\delta\xi)]\; \exp\left(-{1\over 2}
{\delta c^2 \over c} + {1\over 2} \itif
d\tau\sqrt{g}\,  g \xi \Delta \xi \right) =$$ \beq
 J(\lambda) c^{1/2} \mbox{det}^{-1/2} \left( -c^{-2} \Delta \tau^2{d^2\over
d\tau^2} \right)
\eeq
 as well as equation \ref{eq:xi},
the norm definition for the diffeomorphisms.  

We now have to compute the determinant, $ \mbox{det'} \left( -c^{-2} \Delta \tau^2{d^2\over
d\tau^2}\right)$---where the prime stands for ``zero mode excluded''.  The boundary conditions for
the functions into which the operator acts are that they vanish at the boundaries. The
eigenfunctions of this operator   are $\sin (n\pi \tau/\Delta \tau)$ , with eigenvalues  
   $ n^2 \pi^2\over c^2$. To compute the determinant
we next use the trick $ \log \det A = \mbox{tr} \log A$---i.e., the logarithm of the product of the
eigenvalues is the sum of the logarithms of the eigenvalues,
\beq
\log \mbox{det'}= \sum_{n>0} \log({n^2 \pi^2\over c^2})=-\left.{d\over ds}\right|_{s=0} 
\sum_{n>0} \left({n^2 \pi^2\over c^2}\right)^{-s}\eeq
since $-\left.{d\over ds}\right|_{s=0} \alpha^{-s} = \log \alpha$, and now this is just
\beq
-2 \zeta(0) \log c + \mbox{const}= \log(c) + \mbox{const} \eeq
where
\beq
\zeta(s)= \sum _{n>0} \left({1\over n}\right)^{ s}
\eeq is Riemann's $\zeta$-function.

Thus, we find that, up to a normalization constant the determinant equals \beq
  \mbox{det'} \left( -c^{-2} \Delta \tau^2{d^2\over
d\tau^2}\right) = c
\eeq

 So we have the measure
\beq
J= c^{-1/2} \mbox{det}^{1/2} \left( -c^{-2} \Delta \tau^2{d^2\over
d\tau^2}\right) = 1
\eeq
(up to a normalization constant), 
which is gauge invariant as
promised as it depends only on $c$---trivially in this case. 

 Consider now the original amplitude that we wanted to compute. 
The action itself is already 
invariant, 
so 
the 
gauge volume neatly separates and can be thrown out.
Indeed, 
$$
A(x^\mu_i, x^\mu_f) = \int [d\lambda][dx^\mu]\, \exp(-S[\lambda, x]) =
$$ \beq
\int  J \, d(\delta c)\,  [d(\delta\xi)]\; [dx^\mu]\, \exp(-S[c , x])
\eeq
with ($\lambda = c/\Delta \tau$) \beq 
 S[c , x] = {1\over 2} \itif d\tau \, [\left({c\over \Delta\tau}\right)^{-1} \dot{x}^2 + 
{c\over \Delta\tau} m^2]
 \eeq 
Factorizing the gauge group out we end up with 
\beq
A\sim \int    d(\delta c)\,    [dx^\mu]\, \exp(-S[c , x])
\eeq
The $[dx^\mu]$ integrations can now be performed to yield again a power of our famous determinant
and a gauge-invariant exponent (what else could they yield?),
\beq 
\int dc\,  \int         [dx^\mu]\, \exp(-S[c , x]) =\int dc\,  \exp\left(
-{(\Delta x)^2 \over 2 c} - {m^2 c\over 2 }\right)
\left( -c^{-2} {d^2\over d\tau^2}\right)^{-d/2} 
\eeq
The final ``lapse'' integrations yield the Feynman propagator or the on-shell amplitude---and no
other Green functions---depending on whether we use half the range for $c$ or the full one. 

 The ambiguities 
in this 
method lie in the definition of:  a) the inner product, and b) in  
equation~(\ref{eq:gmeasure}), 
as instead of the ``$1$'' we could actually use any gauge invariant 
quantity. At the end we will regularize, at any rate, so as long as the constant is
invariant it doesn't matter. But we will see that the BRST phase space path integral needs
no regularization.

For 
more on  this approach see the references in \cite{Geo}.

\subsection{PICS from the Faddeev path integral}
\label{sec:FaddeevC}
For simplicity, let us consider the non-relativistic case. The main lesson from this
example is that the phase space integral takes care of  regularization automatically,
and this yields a non-trivial measure in coordinate space. 
If we integrate the momenta in the Faddeev path integral---with the 
measure 
$D\mu$, say---we obtain
\beq
\Gamma' = \int DtDx ({2\pi  dt\over m})^{1\over 2} \p  \{
\delta_i (t-f(\tau))  ({2\pi  dt\over m})^{-{1\over 2}}\} e^{i\itif d\tau  
( m{\dot{x}^2 \over  2\dot{t} })}
\eeq 
where $dt \equiv  t_i-t_{i-1}$ is  the
skeletonization, etc. We do obtain here a ``gauge-fixed'' configuration space path integral.

\subsection{PICS from the BFV path integral}
\label{sec:Faddeev-Popov}
Here I will relate the BFV path integral to the described configuration space path integrals---by
integrating the momenta out.
 Consider, as an introduction, the non-relativistic case.
Performing  the momenta integrations in the BFV path integral we get 
$
\Gamma' =$ 
$$ \int DtDxD\pi DcD\bar{c}   (8\pi im dt)^{-{1\over 2}} \p  
(8\pi 
im 
dt)^{-{1\over 2}}\cdot $$ 
\beq \:\:\:\:\:\:\:\:\: 
 \exp\left({ i\itif d\tau  (m{\dot{x}^2\over 2\dot{t}} + \pi 
({\ddot{t}\over 2m} - f({\dot{t}\over 2m})) + ic(-{d^2\over d\tau^2} + 
f'({\dot{t}\over 2m}){d\over d\tau} )\bar{c})} \right)
\eeq
where the more general non-canonical gauge fixing
$$
 {\cal O}_{NC} = i\bar{C} f(\lambda ) + \bar{P} \lambda  
$$
has been used ( recall that  $C= \eta_0 , i\bar{C} = \rho_1  , \bar{P}
=\rho_0  , -iP  = \eta_1 $ ).
This is essentially the Faddeev-Popov path integral above, up to
a gauge-invariant normalization constant. 

Let me now describe two related points. I discussed ealier that the BFV path integral
reduces---after a smart gauge choice---to the form (equation \ref{eq:teitel})
$$
\Gamma_{f=0} ' =  
\int d\lambda  \;\Delta\tau 
\int Dt Dp_t  Dx Dp_x \;    e^{i
\itif d\tau (\dot{t}p_t + \dot{x}p_x )  
-  \itif d\tau \,\lambda \,\Phi    }$$
Incidentally, this is the form advocated by Claudio Teitelboim many years ago \cite{Claudio}, and
traces its origins to earlier work by Feynman and Nambu \cite{proper}.
 
This expression needs no regularization. What is the induced measure in configuration space?
It is easy to see that after the (gaussian) momentum integrations we are left with
\beq
\Gamma_{f=0} ' =  
\int dc \;\Delta\tau 
\int {\cal D}t   {\cal D} x   \;    e^{{i\over 2} 
\itif d\tau ({\dot{x}^\mu\dot{x}_\mu \over c} + c m^2 )  }
\eeq
where the measure 
 is \beq
{\cal D}^d x   = \p  d^d   x(\tau )_i \left( 2 \pi i d\tau_i \lambda 
(\tau_i)\right)^{-d/2}
\eeq
i.e., ``the Feynman and Hibbs measure'', \cite{Feynman}, 
which we can compare with geometric measure, equation \ref{eq:geomeo}
$$ 
 D(  x(\tau )) = \p  d( x(\tau ))_i \left({d\tau_i \lambda 
(\tau_i)\over 
\pi }\right)^{d/2}
$$
As I claimed, these are the same up to a gauge-invariant constant.

Let me now derive the Faddeev-Popov path integral for the relativistic case. Let us work with the
non-canonical  gauge-fixing 
  ${\cal O}_{NC} = \rho_1 \chi(t,x,\lambda 
) 
+\rho_0  
\lambda $ 

With this choice we have $\{ {\cal O}_{NC}, \Omega\}  = $
$$
\{ \rho_1 f(\lambda ) +\rho_0  \lambda , \eta_0  \Phi  + \eta_1  \pi 
\}  = 
\rho_1 \eta_1   \partial_\lambda \chi +  2 \rho_1 \eta_0 p_t \partial_t \chi
-2 \rho_1 \eta_0 p_x \partial_x \chi
 \pi \chi+ \lambda  \Phi  + 
\rho_0 
\eta_1 
$$
and the action $S$ is given by
\beq
  \itif d\tau (\dot{x}^\mu p_\mu +  \dot{\lambda }\pi  + 
\dot{\eta}_0 
\rho_0  + \dot{\eta}_1  \rho_1  - \rho_1 \eta_1  \partial_\lambda \chi - 
 2 \rho_1 \eta_0 p_t \partial_t \chi+ 2 \rho_1 \eta_0 p_x \partial_x \chi
-\pi 
\chi - \lambda  \Phi  - \rho_0 \eta_1  ) 
\eeq
It is not hard to perform the gaussian momentum integrations. The result is that
one gets the Faddeev-Popov path integral: the action becomes
\beq
\itif d\tau \,\left( {\dot{x} ^\mu\dot{x}_\mu \over 2 \lambda} +\lambda m^2 /2 + \pi(\dot{\lambda}
-\chi) + 
i  \eta_0 \left( -{d^2\over d\tau^2} + \partial_\lambda \chi {d\over d\tau}+
{\dot{t}\over \lambda} \partial_t \chi+{\dot{x}\over \lambda} \partial_x \chi
 \right)\rho_1
\right)
\eeq
and the measure is ($\epsilon_i = d\tau_i$ are both notations I have employed for the
skeletonizations)  \beq
  \p  d^d x\,
 \epsilon_i d\lambda\,d\pi\,d\eta_0\,d\rho_1  \left({\lambda \epsilon_i\over \pi  }\right)^{-d/2}
\eeq
Again, measures in the different approaches agree up to gauge-invariant normalization.




\newpage\section{The class of paths that contribute}
\label{sec:paths}
Let us first study the situation for the case of the non-relativistic 
parametrized particle.
To begin with,  we are describing the paths by writing them  in the 
form 
$x(\tau ), t(\tau )$, 
so  any paths are allowed that can be written this way. Unlike the 
unconstrained 
case we see that paths going back in time are included in the path 
integral. 
However, the action in 
the 
path integral has the final say on the matter. Indeed, after 
momentum 
integrations we end up with the parametrized action,
$$
S = \itif d\tau  L = \itif d\tau  \: m {\dot{x}^2\over 2\dot{t}}
$$
Now, to see the path going back and forth in time---turning around-
-we 
need 
$$
{dt\over dx }= 0 = {dt\over d\tau }{d\tau \over dx}; $$
with regular parametrizations this implies $dt/d\tau  = 0 $ ($dx/ 
d\tau  
\neq  
0$). Clearly the action blows up for such paths. The two worlds 
$dt/d\tau  
> 
0$ 
and $dt/d\tau  < 0$ are separated and cannot ``talk'' to each other. 
We 
need 
to 
choose one, and the boundary conditions force one of them to be 
realized 
(one 
can picture operator insertions like $\Theta (\Delta t) = \Theta 
(t(\tau_f) - 
t(\tau_i))$ to forever settle the choice). In this case it is easy to see 
in the 
BFV 
path integral that $\lambda$  is essentially $dt/d\tau$, as follows 
from 
the 
$p_t$  integration; this is why the $\lambda$  integration is related 
to the 
appearance of the Heaviside theta function. For this reason, paths 
going 
back 
and forth in time are not allowed---i.e., will not contribute in the 
path 
integral. 
The path integral really divides into two: paths going back and paths 
going 
forward in time. There will be no particle creation in this formalism 
unless 
the 
action is modified so that it doesn't blow up when $dt/d\tau  = 0$ 
(notice 
that 
we are taking the point of view that only differentiable paths 
contribute).

Notice that the original unconstrained action already did not allow 
such 
paths. 
In the original ($DxDt$) unconstrained path integral such paths could 
not 
even 
be described; however, going to the parametrized version of the path 
integral 
brings in principle all the possible paths of the form $x(\tau ), t(\tau 
)$. If 
the 
action doesn't eliminate them---as it does in this case by 
uncontrolled 
oscillatory cancellation---they will contribute.

Notice that this discussion also applies to the electromagnetic
interaction case. It is also the situation for more general 
hamiltonians---as in the square-root cases.

Next consider the relativistic {\em unconstrained} particle (see 
\cite{rel.unc.ple,Ruffini92}, for example), which we saw is 
equivalent---up to normalization---to the reduced phase space 
quantization of one branch at the time. Here we are integrating over 
paths that go forward in time---even in the parametrized case, as 
discussed above.
That it is possible to construct such a single particle theory and still 
be consistent
with special relativity indeed follows from considering what type of 
trajectories one is summing over in the path integral and how these 
change from observer to observer \cite{Hartle88}. A Lorentz 
invariant classification of trajectories in configuration space is 
provided by the light-cone structure. If we say a trajectory stayed 
within the light-cone and then integrate over all such trajectories in 
the path integral, then we are constructing a path integral 
independently of the choice of Lorentz frame---as long as we are 
using a Lorentz-invariant action. If, on the other hand, we allow 
trajectories outside of the light cone, but we ask that they move 
forward in time, our class of paths will be different than if we had 
started in a different frame and chosen the paths with the same 
scheme, as  a space-like section of a path will be moving back in 
time
in some other frames. Of course, in the free case the end result will 
be the same despite this choice of different sets of paths in different 
frames, because one can set up a one-to-one correspondence 
between the sets of 
paths that wonder out of the light-cone in two frames and that
have the same action. This, however, is a peculiarity of the simplest 
case.

Taking only paths that go forward in time corresponds to taking one 
branch only. Only when the Klein-Gordon equation decouples---we 
saw---does this yield a covariant result. Thus, in general it is not 
possible to set up such a one-to-one correspondence between the 
sets of paths that different observers use, {\em because the action 
gets in the way}. For the electromagnetic case with non-trivial 
electric field, for example, the action is  sensitive to  changes in time 
direction. 
The free case, on the other hand,  is characterized by its $Z_2$ 
invariance: a path and its time-reversed one have the same action 
(recall section~\ref{sec:rgeneral}). 

Let us look carefully at this logic for the free case. We are going to 
consider two observers, $O_A$ and $O_B$. They will both start with 
the same path integral---the free relativistic one with one branch, 
say, the Faddeev path integral. They will both consider paths that go 
forward in time (in their respective frames) only, not necessarily 
causal. Thus, they will agree on all their paths except for some of the 
{\em acausal} ones. Not all of them---this depends on the relative 
velocities of the observers. 

Consider one of this problematic acausal paths---in frame $O_A$, 
say---and let it be infinitesimal. Thus, this path is going forward in
 time at a constant velocity grater than $c$, and as far as observer 
$O_B$ is concerned it is a bad path: it is going back in time. However, 
if we change the orientation of this path in the space-time diagram, 
it is going forward in time {\em  at a speed faster than light}. So this 
path, after this transformation, is one of the {\em acausal  paths of 
$O_B$.} If the action is invariant under $Z_2$, the observers will 
agree on the computation of the path integral, and the amplitude will 
be Lorentz invariant.

What about the interacting case? Let us consider  the electromagnetic 
interaction---in a flat background. There is no
longer a saving symmetry. We saw that the breaking term is simply
\beq
e\,\delta x^\mu A_\mu
\eeq
Our observers cannot match each others' paths anymore. Recall, 
however, that an acausal path that goes forward in time for one 
observer 
becomes a path going back in time for another. It is not too hard to 
see 
that a sufficient condition for this case is the vanishing of 
\beq
e \oint dx^\mu A_\mu
\eeq 
for  space-time loops that lay locally  on planes with $dt\wedge dx^i$ area 
elements, because 
in such a situation we can set up the one-to-one correspondence of 
paths: 
again match a path in one coordinate frame to its Lorentz boosted 
and
``inverted'' one in the other frame. Part of the action is still
invariant under $Z_2$. So when the observers compare their
actions for their respective paths they will find that the difference
is indeed $e\oint dx^\mu A_\mu$.
 
 This condition is equivalent to  the  condition of
vanishing curvature we found before, which  leads to  the decoupling 
of the
Klein-Gordon equation, \beq
F_{0i} = [ D_0,\, D_i ]
\eeq 
(use Stokes' theorem to see this).

The bad guys in this picture are the acausal paths. If we eliminate 
them we will always obtain covariant results. What does this mean? 
Can we 
insert  a Heaviside function at each $\tau$ slice restricting
 $(\Delta x^\mu)^2 > 0$?

 As pointed out in reference \cite{rel.unc.ple}, paths that do not 
respect causality contribute in the ``square-root'' action path 
integral, which is not surprising: all that happens when a path goes 
faster than light is that the action becomes imaginary. Thus, such 
paths contribute exponentially rather than in an oscillatory way. 

In the case where both branches enter in the path integral we are 
dealing with paths that go back and forth in time.  Our observers will 
always agree on their path integrals as long as the actions are 
Lorentz invariant. Causality can be enforced by insertion of a 
Heaviside function, or by the use of representations in which the 
lapse is half-ranged.

  How about having a curved background? Ferraro \cite{Ferraro} shows,
for example, that when there is a Killing time-like vector field in the theory,
it is possible to take the square-root of  the Klein-Gordon equation: the
resulting hamiltonian is hermitean. {\em  Remember that when taking the 
square-root it is crucial to get a hermitean, cavariant hamiltonian.} The
hermicity part is not the problem for the 
electromagnetic case, but it is here.
As for space-time covariance, it is not clear what one should demand of 
comparisons by different observers, as, in general, there are no isometries of
the metric analogous to the ones given by Poincar\'{e} group.

  
\newpage\section{Conclusions, summary}

There are two main aspects of the path integrals we discussed that 
are worth noting. One is that one has to be aware of the fact that one 
may not be using actions that are fully invariant---there may be 
some residual gauge  dependence at the boundaries. When this 
happens the connection with a quantized phase space is lost, strictly 
speaking, although for the cases at hand we provided   a connection 
to  the $RPS^*$.

The other tricky aspect is in the determinant in the Faddeev approach. This determinant is  
directly tied to the composition law and inner product in the physical 
space. Do we need an absolute value? Although, from the 
point of view of the quantized reduced phase space it needs to be 
there, in practice it is relevant only if the determinant changes sign. 
This means that it vanishes, and for the cases at hand, this is also 
related to the fact that the reduced phase space is split---in essence 
that there really wasn't a quantized reduced phase space to begin 
with. If there existed a quantized phase space then the absolute value 
would not be necessary. 

This determinant then determines the coupling between the 
branches---for the split $RPS$ case. There are many choices for it, 
starting from the ambiguities resulting from how to write the 
constraint to ordering ones. To resolve these ambiguities one needs  
extra input.

We have succeeded in putting all these path integrals in phase space 
in  some Hilbert space construction. Those we have then related to 
the path integrals in configuration space. They can all be put in three
categories: reduced phase space, Dirac,  or Dirac-Fock quantization.
The bottom line of our analysis is that {\em if we are careful we can
make sense of all the path integrals.}  Then, at the end,  we will get
the (Dirac) amplitude
$$
A_D = \l x^\mu| \, \delta(\hat{\Phi})  \; | y^\mu \r
$$
which satisfies the constraints (i.e., the Schr\"{o}dinger equation or the
Klein-Gordon equation), 
or the (Fock) amplitude
$$
A_F =\l x^\mu| \,{1\over  \hat{\Phi}+i\epsilon}   \; | y^\mu \r
$$
which gives us a Green's function. {\em The only important exception is the Faddeev
path integral in the presence of interactions that do not allow for 
``decoupling''.} It is not clear how to make sense of it, and this is tied up to the fact that 
the on-shell ampltudes do not satisfy in general a composition law. {\em The Faddeev path integral
was designed and is as an object that properly belongs in the quantized reduced phase space
approach, and there is no clear generalization of it other than the BVF path integral when there
isn't a well-defined quantized reduced phase space.}

I have also reviewed the path integral constructions in configuration space and showed
that they are equivalent to the phase space ones up to normalization---up to gauge-invariant
normalization constants. An important point is that the BFV amplitude requires no regularization,
unlike the configuration  space path integrals---Faddeev-Popov and the geometric one---the
aformentioned normalization constant gets chosen just ``right''.

Finally, I connected the discussion on the factorization of the 
Klein-Gordon equation with the discussion on the paths that contribute
in the path integrals. I showed that the group $Z_2$, the disconnected
part of the diffeomorphism group plays an important role in determining
when it is possible to restrict the theories to one branch ans mantain
space-time covariance. 


\chapter{Second quantization and quantum gravity}
\label{sec:second}

What is ``second quantization''? This terminology comes from
the early  days of  quantum
mechanics and the first attempts to
 quantize relativistic wave equations---such as the Klein-Gordon 
equation.
In this context ``second quantization'' stands for the idea of 
find a lagrangian ${\cal L}$ that has as the 
equation of motion the Klein-Gordon equation. The Klein-Gordon
wave-function then becomes a field $\phi$. This
process of second quantization solves two  problems associated
with the use of the Klein-Gordon equation as a wave-function
equation \cite{Baym}.

 Firstly, the Klein-Gordon equation is second order
in time. This is already a serious fault, as the initial value
problem now requires the specification of both the wave-function 
and
its time derivative. 
The fact that the  Schr\"{o}dinger equation 
is first order in time  is central
to  quantum mechanics and its interpretation. For the free
case, however, we can separate the equation into two parts 
that are first order in time and that
don't ``talk'' to each other, the positive and negative
energy sectors, and that have associated with
them unitary quantum theories.

Another problem is that in general one cannot construct
a {\em conserved} and {\em positive} probability density with
the wave-function. A conserved inner product exists, but one 
that doesn't yield a positive probability.

To sum up: there are two branches, but one cannot in general
restrict the theory to one of them without losing
unitarity in the positive sector and/or space-time covariance.

A pretty way to understand what is going on is to first
rewrite the equation as a first order matrix equation, 
with two components. The two components are essentially
the wave-function and its first derivative, the two independent
initial values (like a coordinate and a momentum),
 or some combination of the two.
This equation then looks like a Schr\"{o}dinger equation but with 
a hamiltonian that is not hermitean in general when we restrict to  
one branch (see Baym's book---reference \cite{Baym}---for a 
delightful discussion on these issues, as well as \cite{Feshbach58}). 


\newpage\section{Minisuperspace and (2+1)-dimensional quantum gravity}
What can we say about the issue of unitarity within the 
``one-universe'' sector?  As explained, minisuperspace
models are mathematically very similar to the relativistic
particle in a curved background. Indeed, one can write an action 
for full gravity which is very reminiscent of the particle 
\cite{Hartle86}
\beq
S[g_{ab},N^a] = \int d\tau \int_{\Sigma} d^3 x \sqrt{ gR(U_{ab} U^{ab} 
- U^2)}
\eeq where $U_{ab} \equiv \dot{g}_{ab} - N_{a;b} - N_{b;a}$, and
where $g_{ab}$ is the spatial metric on the spacelike hypersurface 
$\Sigma$
labeled by the time $\tau$, $g= det(g_{ab})$, and $R$ is the spatial 
scalar
curvature on $\Sigma$. Unitarity within the one branch sector
will depend on the supermetric: does it have ``time-like'' 
Killing vector (super) fields---can we factorize the Wheeler-DeWitt
equation? In general the answer is no for the particle and for
minisuperspace \cite{Hajicek86}. For example, consider the 
Robertson-Walker model 
described by the metric
\beq
ds^2 = {- N^2(\tau) \over q(\tau) } dt^2 + q(\tau) d\Omega_3^2
\eeq
where $d\Omega_3^2$ is the metric on the unit three-sphere 
\cite{Hal4}.
The hamiltonian constraint that corresponds to the appropiate
Einstein-Hilbert action with a cosmological term then is 
\cite{Louko87}
\beq
{\cal H} = {1\over 2} ( -4p^2 +\Lambda q -1).
\eeq
Whether we can factorize this equation or not depends on  the 
ordering needed for space-time covariance \cite{Hal2}. In fact, in this case there 
is no ambiguity \cite{Hal4}, and the Wheeler-DeWitt equation is 
\beq {1\over 2}\left( 4{d^2 \over dq^2} + \Lambda q -1 \right) \; 
\Psi =0\eeq Decoupling occurs only for the zero cosmological constant 
case.

The authors in reference \cite{Wald93} have proposed a solution to 
this problem---in the minisuperspace context---based on another 
definition of time (unlike the one considered here, 
``deparametrization'', which as we have seen brakes space-time covariance in 
the general case.)

 The idea is to start from the Wheeler-DeWitt
 equation---with a covariant ordering, of course---and select a subset 
of its solution space that is positive definite with respect to the 
Klein-Gordon inner product. This subset is chosen as the set of 
solutions that are positive frequency with respect to conformal 
transformations---which are a symmetry in the distant conformal 
past. In essence, we are picking a subset of solutions that are 
positive  definite in the ``past'' in a covariant 
way\footnote{Incidentally, we see here another point of view for 
wanting ``time-like'' Killing vector field. If there is one it means that 
we have a ``time-like''  operator that commutes with the constraint. 
This means that the Klein-Gordon equation decouples.}.

A problem of this approach---and of any square-root one---is that 
of
causality with respect to the metric. This is a problem for the 
particle, but not for the universe, as Wald explains.

Next consider (2+1)-gravity. This
is a remarkable system, because 
it is described by a finite number of degrees of freedom. The
constraints eliminate almost all the dynamics.

As discussed by Carlip \cite{Carlip93,Carlip941},
the different quantization schemes available seem to yield
substantially different theories. For example \cite{Moncrief,Carlip90},
it is possible, through a proper choice of ``extrinsic time'', to
explicitely
reduce the theory classically and  the quantize the remaining, finite,
degrees of freedom. The Chern-Simons approach to the theory is
also of the ``constraint then quantize'' kind, and Carlip \cite{Carlip90}
has shown their equivalence, at least for simple topologies.

On the other hand we have the Wheeler-DeWitt approach \cite{Carlip941},
which is 
much less understood. However, and for whatever that is worth, 
the results of this work lead me to 
believe that the correct physics will be described by a theory in which this 
equation will play a
central role. Linear constraints (constraints that can be made globally into a momentum) 
can probably
always be treated in the reduced phase space context. However, the appearance of a quadratic
constraint signals a serious departure from the reduced phase space quantization philosophy---as we
have seen. When the Klein-Gordon equation is second quantized we can say that a constraint becomes a
dynamical equation---one that belongs in a field theory. There is no more talk of constraints or
invariances, either. Perhaps this is the way to go for quantum gravity as well.

In terms of the ideas of the previous chapters, these two approaches correspond to 
the quantized reduced phase space approach and to the Dirac approach. The fact that the 
constraints become much simpler in the case of a specific foliation (a choice of gauge) can be used 
to  solve classically  part of the constraints---the momentum constraints. The rest of the
constraints can then be solved classically, or  through the Dirac/Wheeler-DeWitt approach.
 Also, it isn't clear
either what happens---in this context---to the case of non-zero cosmological constant. It seems, at
least for some minisuperspace models, that the zero $\Lambda$ case is degenerate---akin to $P^2
\approx 0$.

At any rate, it seems that the conformal factor is the analog in quantum gravity  of the time
coordinate in the particle case. 

 (2+1)-dimensional quantum gravity is definitely a promising area for future research.



\chapter{Conclusion}
 The goal of this research was to examine whether it is  
possible 
to quantize   simple parametrized systems.  
The answer is yes, one can quantize these systems
consistently.  However, the development
of a probabilistic interpretation is not always possible, and even when it is 
possible  there are some features  in these 
theories   that make the quantization   process 
difficult. Let me review what was achieved in the present work.

First I pointed out that even in the unconstrained             
non-relativistic case, the quantization process contain some 
ambiguities: there are many choices for the inner product and the 
observables. I also studied the issue of    space-time covariance of the ``Schr\"{o}dinger
square-root'' equation, and showed that covariance can be mantained 
if there  is a frame in which the electric field is zero---a result that
ties nicely with the associated field theory of a Klein-Gordon field in
an electromagnetic  background.

In the next chapter I studied the various systems classically, 
specially   the behavior of the     Dirac  reduced phase spaces  under 
changes of gauge-fixing.  
One of the particular    problems of these constrained systems  is that 
the actions are not 
gauge 
invariant at the boundaries---this due to the boundary conditions 
one 
wants 
to 
use or to the non-linearity of the constraint, your choice---and 
therefore 
the definition of a reduced phase space is trickier than usual.         
I showed that changes of the              
 gauge-fixing can be understood in terms of 
``time''-dependent canonical transformations in the reduced phase 
space. 
The reduced phase space  for the single branch cases is isomorphic to 
the unconstrained space 
of 
chapter~\ref{sec:unconstrained}, although depending on the gauge 
fixing it 
may come in some ``time''-dependent coordinates. The reduced 
phase space for the two branches case is in fact doubled.

The next step was canonical quantization.
The fact that a gauge fixing can be regarded as a canonical 
transformation 
allowed us as well to define the quantized reduced phase space, 
 where the different gauge 
fixings play the role of unitary transformations---or ``pictures''.
Quantization of both branches of the reduced phase space makes only 
sense if the two decouple---if the theory is to preserve unitarity.
This, we concluded, is a nice method, but one that has a very small
range of applicability.

I then discussed  Dirac quantization. The main problem in this 
approach is that the states selected by the ``physicality condition'' do 
not really belong in the original unconstrained inner product space. 
One of the solutions is to introduce a regularized projector.  This is 
not unique, but I reasoned that  as in the unconstrained case, it 
doesn't have to be unique. The inner product in constrained theories 
is always defined up to such ambiguities. One can choose to fix it by 
demanding that certain operators one deems important be 
hermitean, for example, but the ambiguities are there. To fix them one needs exterior input,
like a classical limit   or other experimental facts.

For the one branch systems I found no
differences between the Dirac  quantization approach and the reduced phase space quantization.
  The
constrain/reduce  order is not important: Dirac's method 
yields  the same quantum mechanics as the reduced phase space 
quantization.
Both suffer from the same inner product and ordering ambiguities.

In the two branch theories, the above equivalence was also found to 
hold when decoupling occured in the Dirac quantization. If 
decoupling does not occur, Dirac quantization does not produce a 
unitary theory in the one particle sector, while  reduced phase space 
quantization breaks the required invariances by forcing decoupling.

I also showed  that requiring that the Dirac inner
product be well behaved (real norms) leads to
the Klein-Gordon inner product in the relativistic case.

We studied some of the circumstances under which decoupling 
occurs in the Dirac covariant formulation. A sufficient condition
was (equation \ref{eq:decoupler}) $$ 
[D_0, \:g^{ij} D_i D_j +\xi R]$$ 
where the derivative operators are the fully covariant ones. As an 
example, with no gravitational background,  the electric field has to 
be null 
 $$ [ D_0 , \; D_i ] =0 $$ 
which is also the condition for no particle creation in the 
corresponding second quantized theory.

The Fock space approach is equivalent to the above except in the 
theories we study in this paper---those that are not fully invariant.
I discussed the coordinate representation and found that it leads to 
the causal amplitudes. The states do not satisfy the constraints. This 
representation leads, in the coordinate representation,
 to a lagrange multiplier with 
half the range. 
            This approach is essentially  gauge-fixed from the beginning 
and it needs no further regularization. 

I also showed that for the relativistic free case it yields two branches.

Then the  BRST approach was  discussed.  It is equivalent to either the 
Dirac (or $RPS$ for the simplest situations) approaches, or to Fock, depending on the 
representation we choose. It is also a more elegant treatment of the 
inner product problem. However, it doesn't seem to contain any new 
physics. 
I described the state cohomology, paying careful attention to 
regularization issues. I developed a well defined inner product and a 
physical projector formalism just as in the Dirac case, from which the 
path integral representation was derived. For example, 
I derived the Klein-Gordon inner product within this formlism,
and the associated composition law. I was  able to
prove the  Fradkin-Vilkovisky theorem at the operator level.

When the Fock representation of the states was employed, I showed
that the BRST-Fock amplitude leads naturally to the Feynman propagator---in
the relativistic case, and more generally to amplitudes of the
form
$$
\l {1\over \hat{\Phi} + i\epsilon} \r 
$$
where $\hat{\Phi}$ is the constraint.

Path integrals in phase space were  derived from all the above 
approaches, and from those I showed how to derive the path 
integrals in configuration space: the ambiguities that exist in the 
Hilbert space description do not disappear (though sometimes they 
will hide under assumptions like ultralocality of the supermetric).

For the one branch case  all the path integrals reduce to the 
unconstrained case---up to 
coordinate systems in  the Faddeev $D\mu^*$ case.
For the two branches case, the determinant that appears in the Faddeev  path 
integral determines the coupling between the branches, and it is 
essentially the inner product in disguise.

I concluded that the Faddeev path integral is an object that belongs only in the quantized reduced
phase space approach---it ceases to make sense when the constraint does not decouple as an operator.
For the free case it yields the on-shell propagator if the absolute value of a carefully ordered
determinant is used, $|E_i+E_{i+1}|$.  The ordering leads to decoupling of the branches, while the
absolute value is just what one expects from the reduced phase space approach---essentially, it is
a rule for the Dirac  delta function. Thus, this path integral is an artifact: one is really doing
two reduced phase space quantizations at the same time. If the constraint is more complicated
decoupling can still be imposed by a choice of ordering. But the equation that the amplitude will
satisfy is not one that implements other symmetries that may be relevant (like space-time
covariance).

The BRST path integral is instead the most general one: it yields reasonable results in the fully
interacting cases---the amplitudes are a solution  of the constraint or its causal Green function.
Neither of the phase space path integrals need   regularization---unlike the configuration space path
integrals. Other than differences in gauge-invariant normalizations, though, the Faddeev-Popov and
geometric path integrals are equivalent to the BFV path integral---and explicitely equivalent to the
path integral that results from integrating out the momenta.

The  range of the lapse $\lambda$ is related to 
the class of paths allowed. For example, causal paths only will 
contribute in the   half-ranged case.

We also saw the role that    the nontrivial element of $Z_2$ in the reparametrization
group, 
the disconnected, path-reversing part,  plays
in determining whether it is possible to select a branch and mantain
space-time covariance in the flat case. In the electromagnetic case
the condition that this imposes on the actions is as the condition
for decoupling of the covariant Klein-Gordon equation.

$Z_2$ also plays another role. If the actions carry a faithful representation
of the full reparametrization group we get on-shell physics (Dirac representation). 
If the
representation is trivial we get causal amplitudes (Fock representation).

Finally, I discussed the issues of unitarity and the probabilistic 
interpretation. Unitarity can always be achieved, as long as we have 
a hermitean hamiltonian. The inner product will in general not be 
one that leads to a positive definite inner product space for the case of two 
branches with interactions, so the probability interpretation  will be 
lost.
If decoupling is forced some other features of the theory  may be 
lost, like space-time covariance in the particle case. Minisuperspace models  
are most like  the relativistic case in a gravitational 
background  where there is particle creation. What this means in the 
particle case is that the Klein-Gordon equation  fails to decouple in 
such a background. If decoupling were to be forced, space-time covariance 
would be broken. The analogy holds here for minisuperspace,   
and  in general it may  not be   
possible 
to avoid the jump into ``third quantization'', i.e., the developement 
and quantization of superactions that yield the constraint as an 
equation of motion.  There are, however,  promising alternatives 
to full decoupling \cite{Wald93}.

 \newpage
 \noindent {\bf \large  Acknowledgments}\\
I want to thank my advisors (I am fortunate enough to have
two of them),  Ling-Lie Chau and Emil Mottola, for their patience, help and support. I also 
  thank my teachers Jackie Barab and Rod Reid, as well as 
 Steven Carlip and Joe Kiskis for help at odd, desperate hours.
I dedicate this work to all of them and especially to my son Cristian 
and the rest of my family.
\newpage
%
\newpage
\renewcommand{\baselinestretch}{1} \tiny

\end{sloppypar}
\end{document}